%% file: main.tex
\documentclass[twocolumn,pra,aps,superscriptaddress,longbibliography,nofootinbib,notitlepage,10pt]{revtex4-1}

\usepackage{graphicx}
\usepackage{amsmath, bbm}
\usepackage{amsfonts}
\usepackage{amsthm}
\usepackage{amssymb}
\usepackage{amsthm}
\usepackage{mathtools}
\usepackage{physics}
\usepackage{booktabs}
\usepackage{quantikz}
\usepackage{xcolor}
\usepackage{float}
\usepackage{textcomp}
\usepackage[colorlinks, citecolor=gray]{hyperref}
\usepackage{cleveref}
\usepackage{listings}
\usepackage{ragged2e}

\newcommand{\bs}[1]{\boldsymbol{#1}}

\newcommand{\fSim}{\text{fSim}}
\newcommand{\CZ}{\text{CZ}}
\newcommand{\OTOC}{\text{OTOC}}

\tikzset{uniform_gate/.style={
  minimum width=10em,
  text width=10em,
  align=center
}}

\definecolor{backcolour}{rgb}{0.95, 0.95, 0.92} 
\definecolor{codegreen}{rgb}{0, 0.5, 0}      
\definecolor{codeblue}{rgb}{0.0, 0.0, 0.6}   
\definecolor{codegray}{rgb}{0.5, 0.5, 0.5}      

\lstdefinestyle{mystyle}{
    backgroundcolor=\color{backcolour},
    commentstyle=\color{codegreen}\itshape,
    keywordstyle=\color{codeblue}\bfseries,
    numberstyle=\tiny\color{codegray},
    stringstyle=\color{purple},
    basicstyle=\ttfamily\footnotesize,
    breakatwhitespace=false,
    breaklines=true,                
    captionpos=b,
    keepspaces=true,
    numbers=left,
    numbersep=5pt,
    showspaces=false,
    showstringspaces=false,
    showtabs=false,
    tabsize=2,
    comment=[l]{//},                
    morekeywords={Function, End, For, Do, Return, Input, Output, float, int, in, with}, 
}

\newcommand\sampleA{$^{13}{\rm C}$-Tol\,}
\newcommand\sampleB{$^{13}{\rm C}$-DMPB\,}
\newcommand\sampleC{${\rm d}_{6}$-DMPB\,}
\newcommand\proton{^{1}\text{H}}
\newcommand\carbon{^{13}\text{C}}
\newcommand\deuterium{^{2}\text{H}}

\newcommand{\nn}{\nonumber \\}
\newcommand{\cO}{\mathcal{O}}

\newcommand{\QSimulate}{\affiliation{%
Quantum Simulation Technologies Inc., Boston, MA, USA}}
\newcommand{\GoogleQAI}{\affiliation{%
Google Quantum AI, Santa Barbara, CA, USA}}
\newcommand{\GoogleResearch}{\affiliation{%
Google Research, Mountain View, CA, USA}}
\newcommand{\GoogleDeepmind}{\affiliation{%
Google Deepmind, London, United Kingdom}}
\newcounter{ExtendedFigure}

\begin{document}

\title{Quantum computation of molecular geometry via many-body nuclear spin echoes}


\author{\small C.~Zhang${}^{\#}$}
\affiliation{Department of Chemistry, University of California, Berkeley, Berkeley, CA, USA}

\author{\small R.~G.~Corti\~nas${}^{\#}$}
\GoogleQAI

\author{\small A.~H.~Karamlou${}^{\#}$}
\GoogleQAI

\author{\small N.~Noll}
\GoogleQAI

\author{\small J.~Provazza}
\QSimulate

\author{\small J.~Bausch}
\GoogleDeepmind

\author{\small S.~Shirobokov}
\GoogleDeepmind

\author{\small A.~White}
\GoogleQAI
\QSimulate

\author{\small M.~Claassen}
\GoogleQAI
\affiliation{Department of Physics and Astronomy, University of Pennsylvania, PA, USA}

\author{\small S.~H.~Kang}
\affiliation{Department of Chemistry, University of California, Berkeley, Berkeley, CA, USA}

% ...

\begin{abstract}
    \newpage
    Quantum-information-inspired experiments in nuclear magnetic resonance spectroscopy may yield a pathway towards determining molecular structure and properties that are otherwise challenging to learn.
    We measure out-of-time-ordered correlators (OTOCs)~\cite{aleiner16microscopic, roberts15localized, mi21information, google25constructive} on two organic molecules suspended in a nematic liquid crystal, and investigate the utility of this data in performing structural learning tasks.
    We use OTOC measurements to augment molecular dynamics models, and to correct for known approximations in the underlying force fields.
    We demonstrate the utility of OTOCs in these models by estimating the mean ortho-meta H-H distance of toluene and the mean dihedral angle of 3',5'-dimethylbiphenyl, achieving similar accuracy and precision to independent spectroscopic measurements of both quantities.
    To ameliorate the apparent exponential classical cost of interpreting the above OTOC data, we simulate the molecular OTOCs on a Willow superconducting quantum processor, using AlphaEvolve-optimized~\cite{novikov2025alphaevolvecodingagentscientific} quantum circuits and arbitrary-angle fermionic simulation gates.
    We implement novel zero-noise extrapolation techniques based on the Pauli pathing model of operator dynamics~\cite{google25constructive}, to repeat the learning experiments with root-mean-square error $0.05$ over all circuits used.
    Our work highlights a computational protocol to interpret many-body echoes from nuclear magnetic systems using low resource quantum computation.
\end{abstract}

\maketitle

\makeatletter
\let\orig@addcontentsline\addcontentsline

\newcommand{\stoptocentries}{\renewcommand{\addcontentsline}[3]{}}
\stoptocentries 
\makeatother

\def\thefootnote{\#}\footnotetext{These authors contributed equally to this work.}
\def\thefootnote{\arabic{footnote}} 

Accurate molecular structure determination is key to probing structure-function relationships in many areas of chemistry and biology.
Nuclear Magnetic Resonance (NMR) spectroscopy provides access to structurally rich information embedded within couplings between pairs of spins~\cite{levitt2008spin}.
Dipolar couplings in solid-like systems provide the most direct access to geometric information.
However, the complex spin dynamics generated by these terms increases the challenge of observing~\cite{cho2005multispin} and interpreting~\cite{butler2009dynamics, dumez2012first} an information-containing signal.
Current techniques to overcome these challenges~\cite{baum_multiplequantum_1985,gullion1989rotational,bennett1998homonuclear,COLOMBO1988189,GROMMEK2006404} focus on reducing the effective system size to one or a few spins, where signals remain large and interpretable.
This has been pivotal in structural elucidation efforts, ranging from model systems in solid state~\cite{Raleigh1989,Creuzet1991} and liquid crystal~\cite{Warren1981,Polson1995} to complex, biologically relevant targets such as amyloid fibrils\cite{petkova2002structural,wasmer2008amyloid} and SARS-CoV-2~\cite{medeiros-silva_atomic_2023}.
A methodological gap remains however, as these techniques limit the maximum measurable distance between pairs of spins (e.g. C-C distances are limited to around $\sim$6\AA.~\cite{shcherbakov_angstroms_2022}), and longer-range distance constraints remain more challenging to estimate.

\begin{figure}
    \centering
    \includegraphics[width=\linewidth]{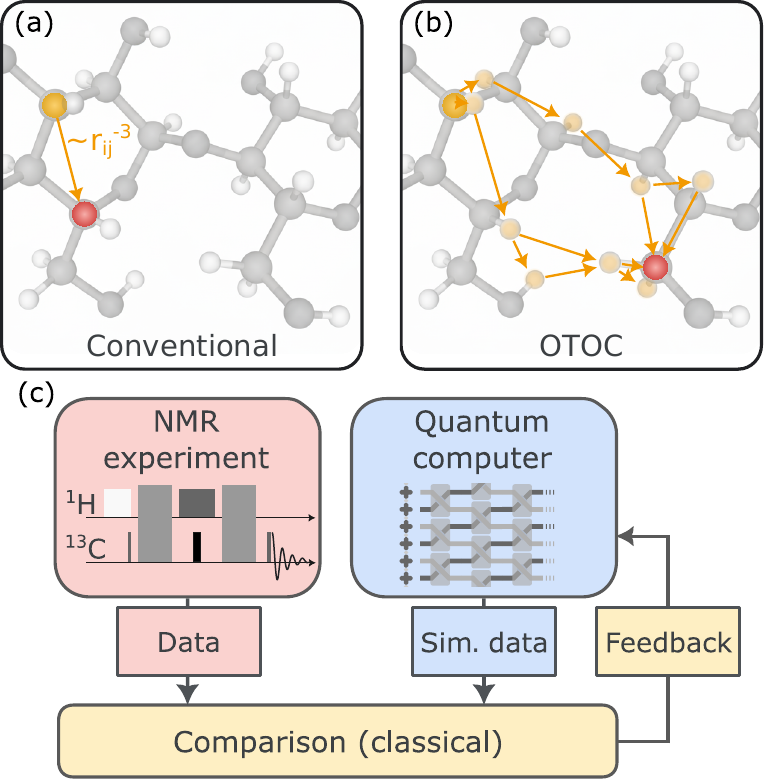}
    \caption{\textbf{Making a longer molecular ruler with an out-of-time-ordered correlator (OTOC).}
    a,b) A comparison between conventional spin transport measurements that infer distance restraints from single couplings, and OTOC measurements, which probe the growth of large quantum coherences through the H spin network.
    By utilizing all the couplings in the spin network, the OTOC is not limited in distance by the $1/r^3$ scaling that limits the distances measurable by conventional techniques.
    c) Our proposal to use a quantum computer to assist in processing OTOC (or other challenging-to-classically-simulate) data from a large spin cluster.
    Following nuclear magnetic resonance (NMR) data collection, the quantum computer provides an artificial system that is iteratively tuned ---via classical feedback--- until it matches experiment.}
    \label{fig:fig1_overview}
\end{figure}

Efficient simulation of many-body spin dynamics, such as in NMR~\cite{sels2020quantum, obrien2022quantum, algaba2022co, schuster2023learning, seetharam2023digital, fratus2025can, marthaler2025good}, has been suggested as an application for quantum computers.
For quantum applications, one targets observables that do not concentrate~\cite{angrisani2024classically,kechedzhi2024effective,schuster2024polynomial} and that are sensitive to microscopic system details~\cite{google25constructive}.
In this work, we propose that this sensitivity is relevant to the NMR practitioner, as quantum-information-inspired experiments could provide access to long range structural information.
Recent efforts have demonstrated quantum advantage in the estimation of out-of-time-ordered correlators (OTOCs)~\cite{aleiner16microscopic,roberts15localized,li2017measuring} on superconducting quantum hardware~\cite{google25constructive}.
The OTOC experiment is based on a many-body echo, in which polarization initially localized on a target spin migrates through the spin network, before a Hamiltonian-engineered time-reversal refocuses to the initial state.
This refocusing is sensitive to perturbations on distant butterfly spins, which allows one to measure the extent of polarization propagation through the spin network [Fig.~\ref{fig:fig1_overview}(a-b)].
We suggest this may help to fill the aforementioned NMR methodological gap.

Herein, we demonstrate a pathway towards determining otherwise challenging-to-compute molecular structure and properties via a hybrid technique of NMR spectroscopy and digital quantum simulation [Fig.~\ref{fig:fig1_overview}(c)].
We measure OTOCs generated by Hamiltonian-engineered pulse sequences on $^{13}$C-labeled organic molecules, [4-${}^{13}$C]-toluene and {[}1-${}^{13}$C{]}-3{'},5{'}-dimethylbiphenyl (DMBP), which we suspend in liquid crystal solvents to suppress intermolecular couplings while partially retaining intramolecular through-space dipolar terms.
We compare experimentally obtained OTOCs from toluene to classical simulations using reference data, and illustrate the sensitivity of OTOCs to molecular structure by simulating stretching the molecule between the ortho and meta carbon atoms, yielding an interpretable NMR signal.
To alleviate the exponential-scaling simulation cost of this approach, we repeat the above analysis on superconducting quantum hardware.
We compile a Trotterized approximation of the all-to-all dipolar Hamiltonian evolution to a swap network of arbitrary-angle fermionic simulation (fSim) gates, and develop novel error mitigation techniques based on the Pauli-path picture of operator dynamics to recreate the learning experiment with high accuracy.
Next, we demonstrate a DMBP structure learning protocol by combining NMR OTOC data with a realistic model obtained through classical molecular dynamics to estimate the distribution of the dihedral angle between the two phenyl rings.
We use the AlphaEvolve coding agent to optimize product formula generation, and implement the resulting circuits on the quantum device.
We compare the resulting estimates of the dihedral angle distribution against multiple quantum coherence (MQC) spectroscopy data from an independent DMBP sample with deuterated methyl groups, experimentally validating the OTOC learning protocol through a secondary unscalable spectroscopic approach.
The combination of careful hardware error-mitigation, algorithmic compilation strategies, and physical chemistry model construction demonstrates the potential of scalable quantum-information-inspired experiments to augment traditional NMR data used for structure determination.

\section*{Results}

\begin{figure*}
    \centering
    \includegraphics[width=\linewidth]{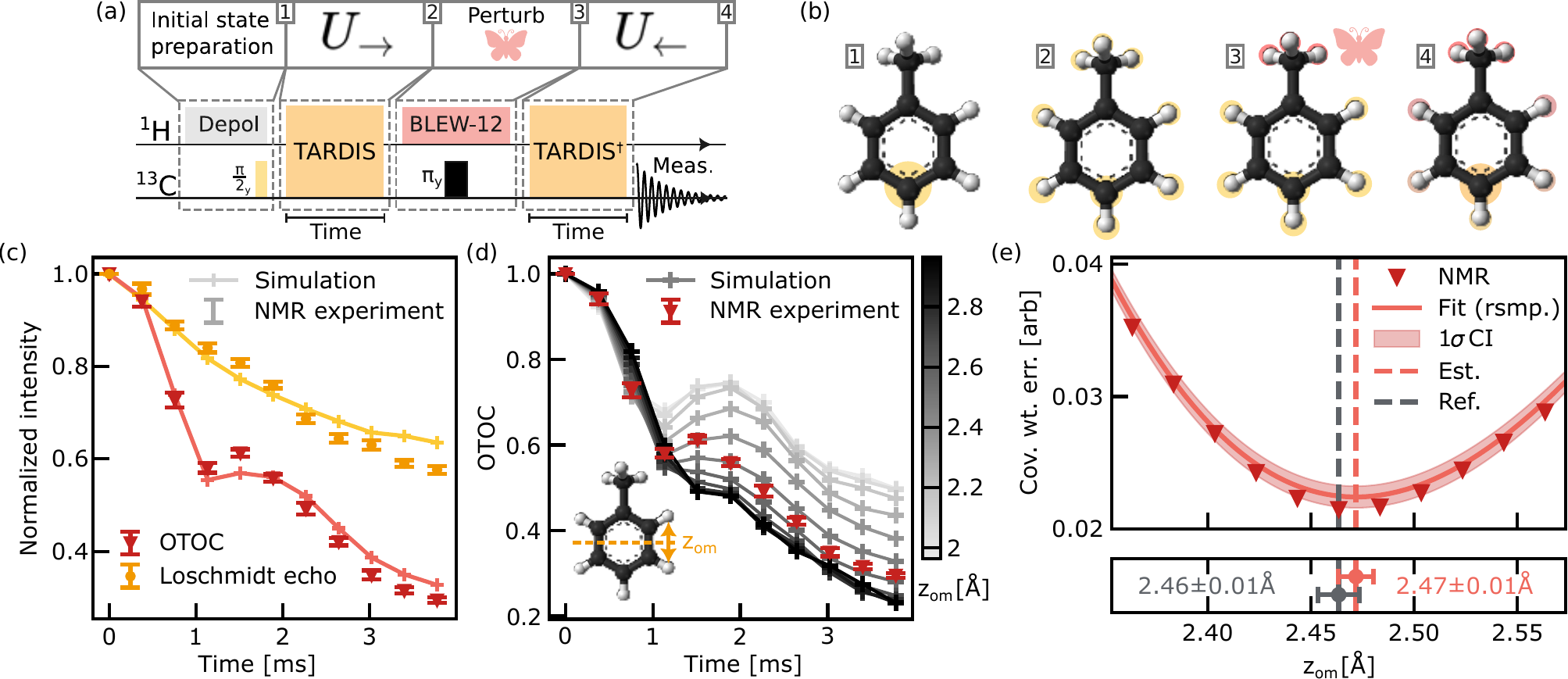}
    \caption{\textbf{Benchmarking the structural sensitivity of out-of-time-ordered correlators (OTOCs) in $[4-^{13}$C$]$-toluene.}
    a) Description of the OTOC protocol (top) as implemented in a nuclear magnetic resonance (NMR) spectrometer (bottom).
    b) Cartoon showing the spread of the spin cluster through the molecule following the sequence in a).
    c) OTOC (red) and Loschmidt echo data (yellow) from the NMR experiment.
    Points show NMR experiment data with 1$\sigma$ confidence intervals (CI), compared to numerical simulations (lines).
    d) Sensitivity of the OTOC experiment to an artificial stretch of benzene between the ortho and meta carbon atoms (inset), with experimental data from b) overlaid (red).
    e) Simulated learning of the ortho-meta H-H distance $z_{\mathrm{om}}$ from OTOC data.
    (top) Red triangles show the covariance weighted mean error between simulation and NMR experiment, line shows a cubic fit with $1\sigma$ CI shaded.
    (bottom) Estimate of $z_{\mathrm{om}}$ (red) compared to reference data from the literature (gray), with bootstrapped $2\sigma$ CI.
    }
    \label{fig:fig2_sensitivity}
\end{figure*}

We illustrate in Fig.~\ref{fig:fig2_sensitivity}(a-b) the encoding of long-range correlations in an NMR experiment using a sample of \mbox{$[4$-${}^{13}$C$]$-toluene} in N-(4-Ethoxybenzylidene)-4-butylaniline (EBBA) liquid crystal as a benchmark system with externally determinable molecular structure.
We perform experiments using the newly-developed Time-Accurate Reversal of Dipolar InteractionS (TARDIS) pulse scheme, that engineers an effective double-quantum Hamiltonian to propagate information between the $x$-polarized $^{13}$C ``measurement'' spin and the methyl proton ``butterfly'' spins.
Due to symmetries in the double-quantum Hamiltonian, the TARDIS sequence is approximately 
invertible, allowing us to refocus the propagated information back to the $^{13}$C spin, which is reflected in the amplitude of the free induction decay (FID).
In the absence of any perturbation between forward and backward evolution this sequence performs a Loschmidt Echo (LE) on the system [Fig.~\ref{fig:fig2_sensitivity}(c), yellow]. 
In principle this LE sequence should yield a constant signal, corresponding to perfect refocusing.
The observed decay is partially due to higher order Hamiltonian terms in the static effective expansion that are imperfectly refocused, and partially due to inhomogeneities in the applied radio frequency (RF) field across the NMR sample (see Supplementary Information for detailed modeling).

To determine the information spread through the toluene molecule during forward evolution, we apply a ``butterfly'' unitary operation on the methyl group before refocusing.
This yields an OTOC [Fig.~\ref{fig:fig2_sensitivity}(c), red], which in the absence of error takes the functional form
\begin{equation}\label{eq:OTOC}
    C(t) = \mathrm{Trace}[X_{\mathrm{{}^{13}C}}(t)\,B \,X_{\mathrm{{}^{13}C}}(t)\, B^{\dagger}],
\end{equation}
where $X_{\mathrm{{}^{13}C}}(t) = U(t)X_{\mathrm{{}^{13}C}}U^{\dagger}(t)$ is the forward-evolved measurement operator.
At short times the $\mathrm{{}^{13}C}$ spin remains undisturbed by the distant butterfly; the OTOC matches the LE.
However, as the spin spreads across the system the butterfly effect grows, and the OTOC decays faster than the LE.
This decay is characteristic of all OTOCs: the information from the measurement spin propagates with a diffusive front, and causes decay upon reaching the butterfly.

The onset and rate of the OTOC decay depends on the dipolar couplings, enabling determination of a model molecular structure corresponding to the thermal average of all molecular configurations.
Due to the simplicity and symmetry of toluene, this average structure can be well approximated by a rigid phenyl ring and a freely-rotating methyl group~\cite{field2003multiple}. 
The liquid crystal environment can be accounted for by an orientational order parameter.
We model OTOCs generated by the exact TARDIS pulse sequence, and include independently-measured RF inhomogeneity.
Then, for illustrative purposes, we can simulate an artificial stretching of the molecule between the ortho- and meta- positions [Fig.~\ref{fig:fig2_sensitivity}(d), inset], measured by the ortho-meta H-H distance $z_{\mathrm{om}}$.
This affects the OTOC decay strongly [Fig.~\ref{fig:fig2_sensitivity}(d)]: we observe that a stretch or contraction of $0.5$\AA~shifts the OTOC by up to $20\%$.
The direction of this shift is counterintuitive: decreasing $z_{\mathrm{om}}$ (which increases coupling strengths) slows the OTOC at later times.
We explain this observation in the Supplementary Information.

To turn the OTOC sensitivity into a tool for learning molecular structure, we construct a cost function to minimize over our target parameter $z_{\mathrm{om}}$.
In Fig.~\ref{fig:fig2_sensitivity}(e) we plot the covariance-weighted error to the data from Fig.~\ref{fig:fig2_sensitivity}(c) and 4 other datasets generated by engineering an onsite field (see Supplementary Information).
Performing a cubic fit and bootstrapping error bars, we obtain an estimate $z_{\mathrm{om,\,c}}=2.47\pm 0.01$~\AA.
This agrees with the reference value $z_{\mathrm{om,\,r}}=2.46\pm 0.01$~\AA~\cite{field2003multiple} to within error bars, and shows comparable precision to this reference data.
We caution however that this fits only a single parameter of the entire toluene molecule, whilst Ref.~\cite{field2003multiple} fits the entire molecular structure.

\begin{figure*}
    \centering
    \includegraphics[width=\linewidth]{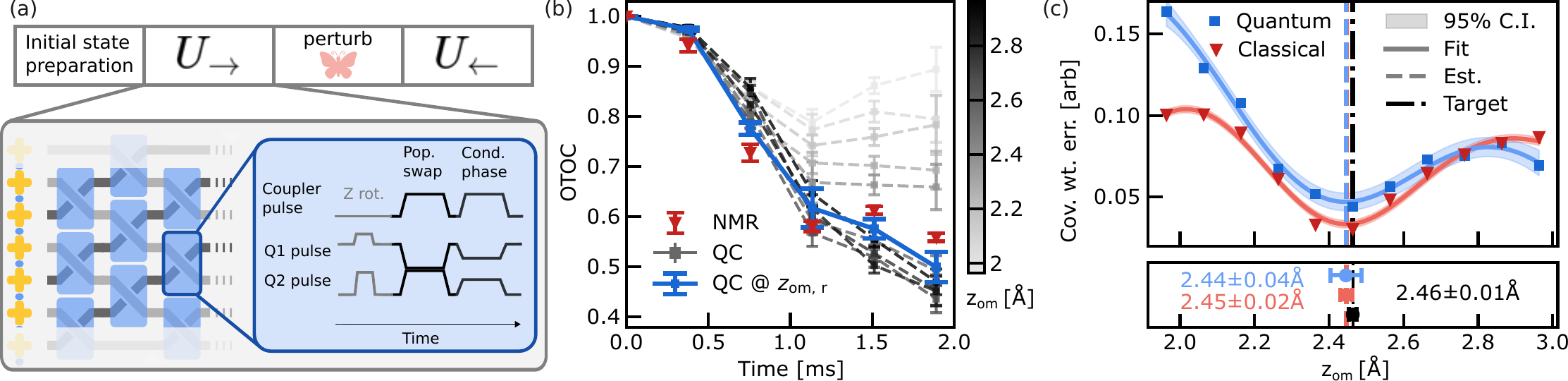}
    \caption{\textbf{Alleviating the exponential cost of out-of-time-ordered correlator (OTOC) simulation with a quantum computer.}
    a) Approximating nuclear spin evolution on superconducting quantum hardware.
    A Trotterized digital quantum simulation of the double quantum Hamiltonian is executed by compiling individual couplings with swap gates that permute spin indices.
    The compiled gate is executed using a pulse train divided into a Z rotation, partial population swap, and conditional phase gate.
    b) Error-mitigated simulations of the first five points of the OTOC curve on quantum hardware, swept over the same range of ortho-meta C-C bond lengths $z_{\mathrm{om}}$ as in Fig.~\ref{fig:fig2_sensitivity}(c).
    The target simulation at $z_{\mathrm{om,\,r}}=2.46$~\AA~is emphasized (solid blue line), error bars are $1\sigma$ confidence intervals (CI).
    c) (upper) Comparison of the learning experiment using quantum and classical data.
    Fits (lines) are from a bootstrapped Gaussian process regression, with $1\sigma$ CI shaded.
    (lower) Comparison of two estimates of $z_{\mathrm{om}}$ to reference data, with bootstrapped $2\sigma$ CI.}  
    \label{fig:fig3_qc}
\end{figure*}

The classical hardness of OTOC simulation prevents scaling our structure learning protocol to larger systems of chemical and biological interest using classical post-processing alone.
We ameliorate this issue by efficiently simulating the strongly correlated dipolar coupling Hamiltonian on superconducting quantum hardware, implementing the scheme envisaged in Fig.~\ref{fig:fig1_overview}(c).
Compared to established applications in quantum simulation of electronic structure, OTOC simulation is a relatively low-cost application for quantum computers due to short scrambling timescales, low measurement complexity, and large error tolerances.
However, the all-to-all-coupled dipolar spin Hamiltonian is challenging to simulate compared to local spin models, due to the large number of terms.
To solve this problem, we implement a first-order Trotterization~\cite{childs2021theory} of the lowest-order Magnus expansion of the full TARDIS sequence, working in the interaction picture~\cite{low2018hamiltonian} on the $^{13}{\rm C}$-para-H coupling, and neglecting all other C-H couplings.
We compile the H-H interactions into a swap network~\cite{kivlichan2018quantum}, and implement the resulting fSim gate in an individually calibrated two-pulse scheme [Fig.~\ref{fig:fig3_qc}(a)].
The resulting circuits use up to 1080 2-qubit pulses to simulate the first six time steps of the toluene OTOC curve.
We further develop a physically motivated zero-noise extrapolation scheme~\cite{temme2017error, li2017efficient} based on the effect of noise in the Pauli-path OTOC picture~\cite{google25constructive, schuster2023learning}.

In Fig.~\ref{fig:fig3_qc}, we repeat the sensitivity experiment across the range of $z_{\mathrm{om}}$ shown in Fig.~\ref{fig:fig2_sensitivity}(c), but replacing the classical computation with quantum simulation performed on a Willow device.
We observe that the quantum device is able to qualitatively replicate the OTOC decay across the $z_{\mathrm{om}}$ range considered.
Comparing our hardware to the classical simulation of Fig.~\ref{fig:fig2_sensitivity}(c), we find a root-mean-square error between the quantum and classical datasets of $\epsilon_{\mathrm{tot}}=0.058$.
This is made up of algorithmic error (i.e. Trotter error) in the quantum circuit ($\epsilon_{\mathrm{alg}}=0.035$) and residual experimental bias after mitigation ($\epsilon_{\mathrm{exp}}=0.050$).
In the Supplementary Information, we provide detailed budgeting of the algorithmic, NMR, and quantum computer error sources.
Motivated by the accuracy in replicating the OTOC decay on quantum hardware, we reproduce the covariance-weighted error curve [Fig.~\ref{fig:fig3_qc}(c), blue], and re-estimate the ortho-meta C-C bond length.
We observe that the dip remains, and obtain an estimate $z_{\mathrm{om,\,q}}=2.44\pm 0.04$~\AA, which suffers a small loss of precision, but continues to agree with the reference data within experimental error.
To separate the effect of the reduced dataset from the algorithmic and experimental error, we repeat this plot using the accurate classical simulation from Fig.~\ref{fig:fig2_sensitivity}, but using only the first $5$ points of the OTOC [Fig.~\ref{fig:fig3_qc}(c), red], achieving an estimate $z_{\mathrm{om,\,c-}}=2.45\pm0.02$~\AA.
The loss of precision in the quantum calculation thus comes partially from the smaller dataset used, and partially from a mix of algorithmic and experimental error.

\begin{figure*}
    \centering
    \includegraphics[width=\linewidth]{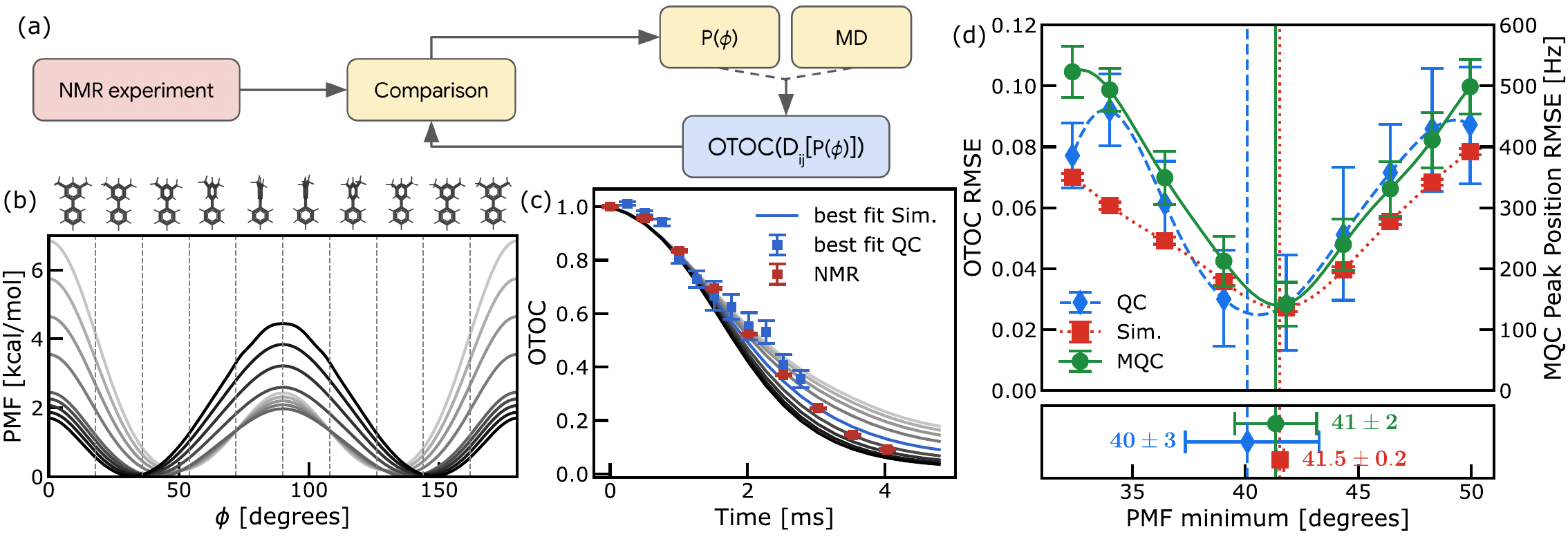}
    \caption{\textbf{Refining a molecular mechanics model of the torsion free energy of {[}1-${}^{13}$C{]}-3{'},5{'}-dimethylbiphenyl (DMBP), validated by multiple quantum coherence (MQC) experiments.} 
    a) Workflow diagram identifying whether a task is performed in the nuclear magnetic resonance spectrometer (red), on the quantum computer (blue), or classically (yellow).
    b) Plots of the 9 potential of mean-force (PMF) candidate functions of the dihedral angle of DMBP.
    c) Out of time ordered correlators (OTOC) of DMBP, comparing experimental NMR data (red points), OTOC simulations run on the Willow chip using the most likely PMF (blue points), and classical simulations with added RF inhomogeneity (lines).
    d) Root-mean-square error (RMSE) of the OTOC or MQC peak position for the 9 candidate free energy surfaces shown in (b); lines are a Gaussian process fit.
    Error bars in all plots are $1\sigma$ CI except for the estimates of the PMF minimum which are shown with $2\sigma$ CI from bootstrapping.
    }
    \label{fig:fig4_DMBP}
\end{figure*}

To demonstrate an OTOC based learning protocol on degrees of freedom common to chemical systems, we perform a learning experiment focusing on the biphenyl dihedral angle $\phi$ of DMBP dissolved in 4-Cyano-4'-pentylbiphenyl (5CB).
The energy barrier to this rotation is sufficient to prevent completely free spinning, but the barrier is insufficient to make the structure rigid at experimental temperatures and pressure, so we consider an ensemble of $\phi$ angles.
This distribution is not precisely captured by molecular dynamics with an approximate molecular mechanics force-field (MD/MM), nor with high-level vacuum electronic structure simulations.
However, of the intramolecular degrees of freedom of the DMBP molecule, the phenyl and methyl bond distances are too rigid to be significantly affected by the interaction with the liquid crystal environment, whilst the methyl torsions are too unconstrained to be affected by this environment.
We can leverage this by capturing the rigid rotations using MD/MM, and re-parameterizing our model such that only the dihedral angle contributions remain.

Our task is thus to learn the free energy surface, or potential of mean-force $\mathrm{PMF}(\phi)$ [Fig.~\ref{fig:fig4_DMBP}(b)], along the dihedral angle $\phi$.
This yields the dipolar couplings via Boltzmann's distribution: $D_{ij}=\int d\phi \exp[-\beta\;\mathrm{PMF}(\phi)]D_{ij}(\phi) / \int d\phi\exp[-\beta\;\mathrm{PMF}(\phi)]$, with $\beta$ the inverse temperature.
We capture the $\phi$-dependent couplings $D_{ij}(\phi)$ by binning and averaging MD trajectories as separated by this collective variable.
Then, we take a simple sinusoidal model with a few free parameters, and interpolate between estimates from MD/MM, an artificial shifted double well potential, and a vacuum phase DFT torsion scan.
This yielded $9$ candidate PMFs (Fig.~\ref{fig:fig4_DMBP}(b)), of which the MD/MM estimate corresponds to a PMF with minimum at $\phi=32.4^{\circ}$, the DFT estimate corresponds to a PMF with minimum at $\phi=41.8^{\circ}$, and the artificial double well with a corresponding PMF minimum at $\phi=50^{o}$.

The learning task can be accomplished by simulating the OTOCs for these candidate PMFs on a Willow quantum chip and selecting the candidate PMF that minimizes the difference between the simulated and experimentally measured OTOC curves.
To generate quantum circuits that match the numerical verification of TARDIS, we use exponentially-costing exact simulation and the AlphaEvolve coding agent to evolve a first-order Trotter formula generator and produce a novel product formula algorithm.
This achieves a low mean error ($0.0084$) to the exact OTOC simulation across the entire PMF range, using no more than 792 CZ gates.

We show the learning experiment on the quantum device in Fig.~\ref{fig:fig4_DMBP}(d), using the first $2.5$~ms of NMR data for the quantum simulation (blue, QC), and the entire $4$~ms decay curve for a comparative classical simulation (red, sim.) that includes accurate modeling of the RF inhomogeneity.
The quantum computer achieves an RMSE of $0.05$ to exact simulation of the input circuits across the entire range of data, matching the performance of the toluene example.
The minima of the two cost functions agree within error bars, $\phi_{\mathrm{QC}}=40^{\circ}\pm 3^{\circ}$, whilst $\phi_{\mathrm{sim}}=41.5^{\circ}\pm 0.2^{\circ}$.
In Fig.~\ref{fig:fig4_DMBP}(c) we plot the OTOC for the QC simulation of the nearest PMF, with NMR experimental data and classical simulations overlaid.

To confirm the OTOC prediction of the dihedral angle model, we simulate MQC spectra from a partially deuterated DMBP sample [(1-$^{13}$C)-3',5'-bis(methyl-d$_3$)biphenyl] using the 9 PMFs, and compare against experimental data. 
We plot [Fig.~\ref{fig:fig4_DMBP}(d), green] the RMSE in the $6$- and $7$-quantum coherence spectra peak positions between experiment and the set of PMF candidates.
The minima of the MQC, QC, and simulation RMSEs agree to within error bars, the error bars are similar in size, and the learned PMF improves the MQC RMSE by a factor $4$ compared to a bare MD calculation.
These results validate both the precision and the accuracy of our OTOC learning technique.

\section*{Outlook}

This work introduces a new quantum-information-inspired framework for determining long-range distance constraints in NMR, with potentially broad applications. 
Long-range distance constraints have proven essential in many contexts in NMR~\cite{shcherbakov_angstroms_2022}, serving as “anchor points” in structural biology~\cite{petkova2002structural,wasmer2008amyloid,mcdermott2009structure,han2010solid}.
In these works, they have been used to discriminate polymorphs or furnish critical restraints for intramolecular and inter-subunit geometries inaccessible to local probes.
Beyond biomolecular systems, they have been critical in validating structural models and connectivity in materials such as metal–organic frameworks~\cite{kong2013mapping}.

OTOCs probe long-range information via the propagation of local polarization across an extended spin network and its sensitivity to weak local perturbations under time-reversal.
This approach offers advantages over traditional time-ordered correlation (TOC) experiments, i.e. protocols that do not involve time-reversal.
Specifically, OTOCs may provide improved sensitivity to spatial correlations and distances in systems where local control is limited~\cite{schuster2023learning}, such as the ${}^{13}$C-proton networks studied within this work.
While the size of the many-body correlations and the accessible distance measurements will be system-specific~\cite{Karabanov2011}, these factors do not change the core applicability of the technique.
Estimates (see Supplementary Information) suggest accessible distances of 20-60~\AA~for OTOC-based measurements, approaching the length scale of F\"orster Resonance Energy Transfer (FRET)~\cite{Stryer1967}. 
This is beyond the reach of state of the art NMR techniques such as proton-driven spin diffusion (PDSD)~\cite{COLOMBO1988189}, rotational echo double resonance (REDOR)~\cite{gullion1989rotational}, and radio-frequency-driven recoupling (RFDR)~\cite{bennett1998homonuclear}, all of which are TOC protocols. 

Additionally, while the experiments presented here focus on single OTOCs, the framework naturally fits within the scope of multi-dimensional NMR.
Multiple OTOCs, excited at distinct molecular sites or under effective Hamiltonians with varying symmetry-selection rules, could enable multi–anchor-point spectroscopy, supporting extensive extraction of structural parameters.
Moreover, OTOCs need not be restricted to pulse-sequence excitation: molecular dynamics themselves may drive butterfly operations, providing access to distinct dynamical timescales and rendering the OTOC a direct probe of molecular motion~\cite{krushelnitsky2013solid}.
The ability of quantum computers to simulate all of these potential experiments opens up future possibilities far beyond the studies undertaken in this work.

Despite the large body of work on the asymptotic analysis of product formulas~\cite{childs2021theory}, less work has gone into the constant factor analysis at beyond-classical system sizes.
The performance of product formula techniques~\cite{low2018hamiltonian,childs2019faster,campbell2019random,zhao2023making,mansuroglu2023problem} can be highly system- and quantity-specific, and metrics such as unitary infidelity are only loose predictors of state- and operator-specific quantities such as OTOCs (see Supplementary Information).
Estimates of Trotter errors suggest that naive methods would require $10^5-10^6$ gates to execute OTOCs on 50-spin systems (see Supplementary Information) --- a large, but not astronomical gap to current hardware requirements given careful problem selection.
This presents a challenge to the quantum algorithms community, as we do not expect this gap to be overcome by physical hardware alone.

Our application of the AlphaEvolve algorithm~\cite{novikov2025alphaevolvecodingagentscientific} presents an interesting new direction for such algorithm optimization research.
Despite successfully overcoming the Trotterization challenge for DMBP, our AlphaEvolve optimization loop relies on evaluating candidate circuits against a complete, classically pre-computed dataset of OTOCs, which is not immediately scalable to beyond-classical system sizes.
However, in this work we leveraged the inherent strengths of the LLM-based agent to write code that constructs Trotterized circuits, rather than generating circuits directly~\cite{jern2025agentqfinetuninglargelanguage}.
We argue that this has two key advantages.
The first advantage is that the resulting code can generate circuits for times and landscape parameters outside its training data set, allowing some degree of generalization and protection against overfitting.
However, while interpolation between the training data points works in limited cases, the generator commonly fails to extrapolate; improving on this is a key target for future research.
The second advantage is that the resulting function and the prompts used can be understood and analysed by humans, which we attempt in the Supplementary Information.
We find that the primary focus of the code (relative to a first order Trotter formula) is a mixture of light cone pruning, term and qubit ordering, adaptive time steps and a distance-based term rescaling, some of which have been suggested before in the literature~\cite{childs2019faster,zhao2023making}.
We suggest that such analysis can allow the ideas identified by AlphaEvolve to be scaled to system sizes inaccessible to exact simulation.

Molecular dynamics estimates are compromised by several sources of error, including finite size effects, force field inaccuracy, and insufficient sampling.
While each of these issues may be addressed in principle, simultaneously addressing all of them is impractical.
Instead, addressing these shortcomings through Hamiltonian learning represents a promising avenue for accurately assessing structural properties, enabled by scalable reproduction of many-body spin dynamics by quantum computers.
Despite the limitations of the current scheme, in part due to experimental uncertainty, the learned mean value of the DMBP dihedral angle improves over predictions from condensed phase molecular mechanics.
These results highlight that experimentally obtained quantum many-body echoes can parameterize highly tuned structural models by providing information that can correct the deficiencies of molecular dynamics simulations.

\section*{Materials and methods}

\subsection*{Materials}

Nematic liquid crystals 4{'}-Ethoxybenzylidene-4-butylaniline ($\geq$99.0\%) [EBBA] and 4-Cyano-4{'}-pentylbiphenyl ($\geq$98.0\%) [5CB] were purchased from Tokyo Chemical Industry America and used as received.
Isotopically enriched {[}4-${}^{13}$C{]}-toluene ($\geq$99.0\%) was purchased from Sigma Aldrich and used without further purification.
Isotopically labeled samples of {[}1-${}^{13}$C{]}-3{'},5{'}-dimethylbiphenyl and 3{'},5{'}-bis(methyl-${\rm d}_{3}$)biphenyl were synthesized according to literature procedures using starting materials purchased from Sigma-Aldrich; full synthetic details are provided in the Supplementary Information.
The purity of synthesized [1-${}^{13}$C]-3{'},5{'}-dimethylbiphenyl and [1-${}^{13}$C]-3{'},5{'}-bis(methyl-${\rm d}_{3}$)biphenyl was confirmed by ${}^{1}$H and ${}^{13}$C NMR spectroscopy, showing no detectable impurities.

\subsection*{NMR sample preparation}

[4-${}^{13}$C]-toluene was prepared by mixing 4.2 wt\% of the compound with EBBA, heating to 358 K to reach the isotropic phase, and sonicating to homogenize. [1-${}^{13}$C]-3{'},5{'}-dimethylbiphenyl and 3{'},5{'}-bis(methyl-${\rm d}_{3}$)biphenyl were prepared by mixing 2\% v/v of each compound with 5CB, followed by heating to 313 K and sonication. Prior to each NMR experiment, all samples were homogenized by sonication above the nematic–isotropic transition temperature, injected into the NMR magnet, thermally cycled several times through the transition point, and then cooled slowly to the target experimental temperature.

\subsection*{NMR experiments}

NMR experiments were performed on a Bruker UltraShield 11.75 T (500 MHz for ${}^{1}$H) magnet equipped with a Bruker Avance I console.
A 5 mm double-resonance BBO smart probe was tuned to facilitate ${}^{1}$H and ${}^{13}$C experiments.
The sample temperature was maintained at \mbox{295 K} for {[}4-${}^{13}$C{]}-toluene$@$EBBA and \mbox{289 K} for [1-${}^{13}$C]-3{'},5{'}-dimethylbiphenyl$@$5CB, [1-${}^{13}$C]-3{'},5{'}-bis(methyl-${\rm d}_{3}$)biphenyl$@$5CB using the internal temperature control unit.
Pulse powers of the ${}^{1}$H and ${}^{13}$C channels were synchronized to achieve a nutation amplitude of $\sim 23.8$ kHz.
All OTOC experiments were preceded by a preparation filter generating pure ${}^{13}$C magnetization via a combination of gradient pulses and Loschmidt echo elements, followed by the application of a TARDIS element for variable durations.
The TARDIS element was designed to generate effective dipolar double-quantum evolution in the presence of homo- and heteronuclear dipolar couplings $(H_{\rm eff}\propto \sum_{i<j}d^{II}_{ij}I^{+}_{i}I^{+}_{j}+\sum_{i,j}d^{IS}_{ij}I^{z}_{i}S^{z}_{j}+{\rm h.c.})$, taking into account finite pulse width and pulse amplitude effects (see Supplementary Information).
A selective perturbation, in the form of a local rotation, was applied to the methyl protons in both target systems.
This was achieved through the application of a BLEW12~\cite{BURUM1981173} element applied to the proton channel, resonant with the methyl spins, followed by a second BLEW12 proton element sandwiched by two carbon $\pi$-pulses.
Time-reversal was achieved through a second $(\pi/2)$-phase-shifted TARDIS element sandwiched by two $\pi$ pulses on the carbon channel. 
Application of the $(\pi/2)$-phase-shift reversed the sign of the double-quantum part of the effective generator ($R^{I}_{z}(\pi/2)I^{+}_{i}I^{+}_{j}R^{I}_{z}(-\pi/2)=-I^{+}_{i}I^{+}_{j}$)~\cite{baum_multiplequantum_1985}, whereas the two $\pi$ pulses on the carbon channel inverted the sign of the heteronuclear part similar to a spin echo.
Finally, readout was performed on the ${}^{13}$C channel.
MQC spectroscopy followed the general scheme outlined in~\cite{field2003multiple}.
Full details of the NMR experiments and all pulse sequence elements are given in the Supplementary Information.

\subsection*{Processing of NMR and SC experimental data for learning experiments}

Each NMR dataset shown is an average over $10$ independent measurements, allowing us to calculate the standard error $\sigma_t=[\sum_{j}(\bar{C(t)}-C_j(t))^2]^{1/2}/N$.
Measurements of datapoints at different times were assumed to be independent.
However, as the measured peak amplitudes are normalized by the OTOC at $t=0$, the uncertainty in this amplitude was propagated to the other OTOC estimates, which led to a covariance matrix $\Sigma_{\mathrm{NMR}}$ with non-zero off-diagonal terms.
$\Sigma_{\mathrm{NMR}}$ was then used to obtain the covariance weighted error plotted in Fig.~\ref{fig:fig2_sensitivity}(e): $\mathrm{Cov.\;wt.\;err}=[(\bs{C}_{\mathrm{NMR}}-\bs{C}_{\mathrm{sim}})^T\Sigma_{\mathrm{NMR}}(\bs{C}_{\mathrm{NMR}}-\bs{C}_{\mathrm{sim}})]^{1/2}/\mathrm{Trace}[\Sigma]^{1/2}$.
Here, we write $\bs{C}$ as shorthand for arrays containing the OTOC data indicated in the subscript.
Error bars in Fig.~\ref{fig:fig2_sensitivity}(e) were obtained by bootstrapping over the initial $10$ points for each NMR dataset.
To combine this data with the output of the quantum computer for Fig.~\ref{fig:fig3_qc} and Fig.~\ref{fig:fig4_DMBP}, we took the covariance matrix $\Sigma_{\mathrm{QC}}$ (which we assume to be diagonal), and combined this with the NMR covariance matrix to yield $\Sigma_{\mathrm{tot}}=[\Sigma_{\mathrm{NMR}}^{-1}(\Sigma_{\mathrm{NMR}}^{-1}+\Sigma_{\mathrm{QC}})^{-1}\Sigma_{\mathrm{QC}}^{-1}]^{-1}$.
This was substituted for $\Sigma_{\mathrm{NMR}}$ in the above definition of the covariance weighted error to give the cost function in Fig.~\ref{fig:fig3_qc}(c).
Error bars in this figure were calculated by bootstrapping over the NMR datapoints, and re-sampling the quantum computer data from a normal distribution.
For Fig.~\ref{fig:fig4_DMBP}(d), we used the root-mean-square error $\mathrm{RMSE}=[\frac{1}{N}\|\bs{C}_{\mathrm{sim}}-\bs{C}_{\mathrm{NMR}}\|]^{1/2}$ as a cost function instead of the covariance weighted error.

\subsection*{Molecular dynamics simulations details}

Starting structures were generated with Packmol version 21.0.0~\cite{martinez2009packmol} at a concentration of 2.05\% v/v 1,3-dimethyl-5-(phenyl-13C)benzene and a starting density of 1.01 g/ml.
The liquid crystal molecules were approximately aligned along the x-axis of the simulation box with dimensions [65\AA, 50\AA, 50\AA] by providing angle constraints in the Packmol input file.
The initial structures were then relaxed by minimizing the system at constant volume and equilibrated with 500 ps of isothermal-isobaric (NPT) dynamics.
All simulations used a 2 fs time step with hydrogen bond constraints and a hydrogen mass of 4 amu.
NPT calculations used a Monte Carlo barostat with updates every 25 steps.
All ensemble averages were evaluated with respect to 8 independent trajectories with production times of 1 $\mu$s each.
All molecular dynamics simulations were performed using OpenMM~\cite{eastman2023openmm}.

Non-bonded parameters from GAFF version 2.11~\cite{wang2004development} were deployed for 5CB and DMBP in combination with bonded parameters from Grappa version 1.3.1~\cite{seute2025grappa}.
Although Grappa was designed for biomolecules, the training set includes a very large dataset of small organic molecules and we have observed empirically that the Grappa force field imparts additional flexibility to the aliphatic chains in the liquid crystal relative to GAFF.
This is consistent with modifications made to GAFF for liquid crystal systems~\cite{zhang2011atomistic}.

The order parameters of 5CB are known to be sensitive to the temperature \cite{horn1978refractive, sherrell1979susceptibilities}, and are difficult to model with molecular dynamics~\cite{cacelli2002stability}.
This necessarily impacts both the mean-value of the dihedral angle of DMBP, and the $\phi$-dependent coupling matrices $D_{ij}(\phi)$.
To minimize this effect, we shift the temperature of the experiment to $289$~K (as reported in the above section), where the order parameters most accurately match simulation, and use this throughout all DMBP experiments.

\subsection*{Classical simulation of toluene}


Toluene in EBBA was modeled as a static benzene ring with a freely rotating methyl group in a fixed plane, in a similar manner to Ref.~\cite{field2003multiple}. 
The motion in the liquid crystal background was treated with the Saupe tensor formalism, assuming all molecular motion occurs on timescales $\ll 1$~ms (see Supplementary Information).
Data for the molecular co-ordinates and all J-couplings was taken from Ref.~\cite{field2003multiple}, except for the J-coupling between the $^{13}$C and the nearest proton (which is significantly larger than all other J-couplings).
The Saupe tensor, chemical shifts, and the remaining J coupling were obtained by fitting to MQC data (see Supplementary Information).
Data was separately validated by comparing to the molecular dynamics simulations of the toluene-EBBA system above.

\subsection*{Trotterized Hamiltonian simulation and machine learning}

\subsubsection*{AlphaEvolve setup}

AlphaEvolve~\cite{novikov2025alphaevolvecodingagentscientific} is an evolutionary coding agent that leverages large language models (LLMs) to discover novel and efficient solutions for a variety of scientific problems.
In this work, AlphaEvolve is used to generate a set of Python functions that produce quantum circuits approximating Hamiltonian evolution across an entire parameter landscape of time and Hamiltonian inputs, as opposed to optimizing individual quantum circuits or circuit parameters directly.
The process was seeded with a human-written program (a first-order Trotter formula) that served as the initial solution to a given task. 
Through a process of mutation, evaluation, and selection, AlphaEvolve ultimately generated a population of programs that produce simulation circuits with significantly lower approximation error against the reference  (improving from a $10.4\%$ mean error to $0.82\%$), while remaining below the gate budget enforced by the device (see Supplementary Information for further explanation).

The MAP-Elites-like setup of AlphaEvolve~\cite{mouret2015illuminatingsearchspacesmapping} was used as a base for this work, optimized for a single RMS error metric targeting the average approximation error across the entire set of landscape parameter and time values evaluated.
By stochastically injecting domain-specific suggestions into the prompt, the optimization algorithm was steered towards granular changes relevant to the task.
Here, task-specific modifications were introduced, such as
a mixture of hand-crafted and LLM-generated task instructions specific to quantum simulation and circuit optimization problems. These led to increased performance of the algorithm, both in terms of achieved metric value and convergence speed, i.e.~number of required LLM samples.
The LLMs are pre-trained on large corpora of scientific literature and thus already possess knowledge of the most commonly used quantum algorithms, quantum simulation, and other relevant domain knowledge.

For each parent program, in addition to the program’s overall RMS error, an associated error matrix was included in a tabular format giving the OTOC error for each point in the parameter landscape and time. This feedback allows the LLM to perform in-context reinforcement learning: by focusing on regions with high error, the model can attempt to generate a more accurate circuit for these regions without sacrificing overall performance.

The algorithm's parallelized nature was leveraged to redistribute resources. To cover a larger portion of the search space, multiple, initially independent AlphaEvolve runs were started. As the optimization progresses, the independent runs were regularly synced by re-seeding with the current best solution.
This was either performed via full re-seeding or by eliminating only the worst-performing run. Empirically, we found that this strategy resulted in a much faster improvement of the fitness function with respect to the number of LLM samples used.

\subsubsection*{Circuit optimization problem setup}

For processing by AlphaEvolve, a problem needs to be split into four well-defined pieces, as described here:

\textbf{Problem Specification}: A Python program implementing a first-order Trotter formula with a swap network for 15 qubits on a chain in a dense brick-wall fashion served as starting point for the evolutionary process. For this initial program,
the input parameters of the function were the total time $t$ and number of Trotter steps to be implemented, as well as the homogeneous/heterogeneous (j) couplings and an on-site terms.
Additionally, we provided the butterfly and measurement indices, the landscape parameter $p$, and a maximum number of allowed CZ gates, despite these being unused in the function initially. 
These additional parameters were included in the function signature to inform AlphaEvolve about the wider context of the program, and to give additional information could be useful for optimization.

\textbf{Prompt and sampling}: In each iteration, AlphaEvolve selected a program from its database to act as a parent.
This parent program was then used to generate a prompt for an LLM, which was tasked with creating a child program that improves upon the parent.
A fixed generic prompt was included, describing quantum simulation, Trotterization, NMR, and OTOCs on a very high level.
Per LLM call, a domain-specific or generic prompt was further added, as well as focus prompts that primarily target code simplification, refactoring and removal of dead code, or to focus on a particular task (lightcone, THRIFT \cite{Bosse2025-ha}, commutator scaling).
In all cases, the signature of the function to be generated remained fixed. 

\textbf{Evaluation}: The obtained function was used to generate quantum circuits for all choices of parameters and time steps.
These circuits were then individually evaluated by averaging the OTOC over 250 randomly-sampled initial states. To ensure that this random evaluation yields only a small approximation error, this approximate OTOC was compared with the exact OTOC computation that uses the full set of $2^{15}$ states. For the latter a new technique using sparse matrix computations on GPU was implemented to produce unitary matrices for large circuits instead of much slower regular einsum-based method.
The seed for the state sampling varied for each time and landscape parameter choice, but remained fixed throughout the optimization run to avoid small random fluctuations influencing the evaluation of mutations while speeding up the pipeline.
The root mean square error between the evaluated OTOCs and the reference data was used as an optimization metric.

\textbf{Database Integration}: Child programs that were both valid and exhibited a sufficiently high fitness score were added to the database of programs.
Programs were discarded, for example, if the circuit they produced exceeded a predefined number of gates.
Counting CZ gates as implemented in the Cirq framework with Cartan's KAK decomposition \cite{tucci2005introductioncartanskakdecomposition} was the dominant bottleneck, with this calculation alone being the dominant part of the step time.
To optimize this, a highly parallel KAK decomposition was implemented, effectively removing the contribution of the compilation step to the total time and compute budget.
Consequently, the evolved program was provided with an option to use CZ counting as a tool (which it did by implementing a two-pass circuit generation, see Supplementary Information).

We provide an additional analysis on the efficiency of different instruction prompts, high-level intuition of what kinds of modifications AlphaEvolve suggests and generated codes and circuits in the Supplementary Information.

\subsection*{Details of superconducting qubit devices and gate calibration}


Quantum simulation was performed on a Willow quantum computer similar to that described in \cite{google25constructive} (see also \cite{google2025QEC_bellow_threshold}). Processors in this lineage are based on a 2D grid of flux tunable superconducting transmon qubits connected by flux tunable couplers between nearest neighbors. In this work, the mean frequency of operation was $\sim6.2$~GHz with anharmonicity $\sim$210~MHz.

As the two molecules addressed in this work require nine and fifteen qubits only, we selected a sub-grid of the 105 qubit Willow with performance better than average in the relevant benchmark metrics. 
Qubits were arranged in a line, with the measurement qubit at one end.
The average lifetime of the single photon excitation in the chosen line of qubits was $T_{1}=114$~\textmu s. The average coherence time of the chosen qubits as measured by a Hahn-Echo sequence was $T_{2E}=130$~\textmu s. The performance of single qubit microwave gates was measured by Clifford randomized benchmarking (RB) error and the associated qubit impurity, which were found to be on average 0.00020 and 0.00015, respectively. The average cycle Pauli cross-entropy benchmarking (XEB) error, including error contributions from two single qubit gates and a single two-qubit gate, was for CZ 0.0015, and for $\sqrt{i\textrm{SWAP}}$ 0.0014. The average readout error of the single measurement qubit was 0.00985. Since the data taking for the quantum simulation of the OTOC-NMR learning sequence took several days, these metrics drifted slightly before completion. While the processor requires recalibration during this time, these numbers are typical and representative of our experiments.

The all-to-all coupling was achieved by a swap network, a technique adapted from \cite{kivlichan2018quantum}.
Here, one compiles swap gates through a depth $N$ ``brick wall'' pattern of two-qubit interactions, such that the swap gates permute floating spin indices through the qubit array. 
This is optimal in terms of both the number of gates and the depth required to execute one interaction between each pair of spins, as is needed for a single Trotter step of the all-to-all coupled dipolar Hamiltonian.
In our quantum simulation of Toluene, we further approximated that the ${}^{13}$C spin only interacted with its nearest proton.
This single coupling was treated in the interaction frame and the ${}^{13}$C spin was removed from the swap network, reducing the number of gates needed by $17\%$.
This approximation was justified as all other ${}^{1}$H-${}^{13}$C couplings were two orders of magnitude smaller for the entire set of toluene geometries considered, and we observed that the interaction frame treatment slightly reduced the Trotter error.

The quantum simulation for Toluene was based on a natural decomposition of the target effective Hamiltonian into Fermionic simulation (fSim) gates \cite{foxen2020demonstrating}.
These gates are parametrized by two two-qubit interaction angles: the swap angle and the conditional phase angle.
Across the entire landscape, the Trotterization of the Toluene evolution required calibrating a set of 80 unique fSim gates. 
Previous work was able to operate with a single pulse fSim \cite{Arute2020SeparatedDynamics,Mi2021TimeCrystal, Neill2021QuantumRing,mi21information, Mi2022NoiseResilient, Morvan2022BoundStates}.
However, achieving arbitrary combinations of swap and conditional phase angles with a single pulse often requires either a long gate duration, which degrades performance by decoherence, or a large interaction coupling strength between qubits, which increases leakage processes outside of the computational manifold.
In this work, due to the range of coupling strengths in Toluene and the coarse-grained Trotterization, the swap angles that needed to be calibrated spanned the entire domain from 0 to $\pi/2$.
To solve this issue, we adapted a two-pulse fSim approach from earlier work in a Sycamore architecture \cite{foxen2020demonstrating}.
Here, a first base band interaction pulse sets the swap angle and induces a spurious $<100$~mrad conditional phase, while a second pulse enacts a conditional phase and a spurious $\sim$ 30~mrad swap angle.
The two-pulse approach is calibration intensive, but provides higher performance and flexibility than a single pulse approach.

The two pulses for each fSim gate were calibrated together as a single gate with an iterative procedure adapting one of two high precision periodic ``Floquet" calibrations previously developed, depending on the target swap angle.
For small target swap angles ($<30$~mrad) we adapted the technique presented in \cite{Mi2021TimeCrystal}.
For larger swap angles and for the conditional phase we adapted the technique presented in \cite{Arute2020SeparatedDynamics}.
To achieve the desired precision in the full range we fine-tuned the number of gates used in the periodic calibration and fitting procedure depending on the value of the expected angle.
The procedure accounts for the aforementioned spurious terms as part of the calibrated final gate, attaining the target interaction angles with a max tolerance of 20~mrad, and typical error $<5$~mrad.
As in previous work \cite{google25constructive, mi21information}, single qubit phases of the fSim gates are measured by XEB fitting and removed by phase matching single qubit gates to yield the total gate. 
The median XEB error for the tuned fSim gates was measured at 0.0026 and max XEB error 0.0045.
Compatible pairs of qubits were calibrated simultaneously, and therefore benchmarks include cross-talk effects.
More details on these experiments can be found in the Supplementary Information.

For the quantum simulation of DMBP, a 15 qubit linear chain with similar performance was used.
In contrast to the Toluene experiments, the AlphaEvolve coding agent generated circuits using a CZ decomposition, which did not require us to use the fSim construction detailed above.

\subsection*{Error mitigation pipeline for noisy quantum circuits}

\begin{figure}[tb]
    \centering
    \includegraphics[width=\linewidth]{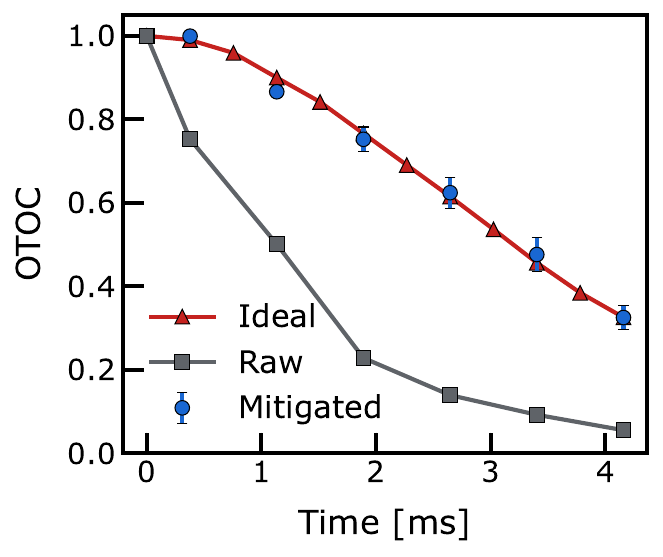}
    \par\vspace{1ex} 
    \begin{minipage}{\linewidth}
        \noindent\justifying 
        \textbf{Extended Data Fig.~1.} \textbf{Benchmark of Pauli Path ZNE error mitigation on the 15-spin molecule at odd time indices.}
     Out-of-time-order correlator (OTOC) measurements for DMBP at $\textit{odd}$ NMR time indices, taken during a different experimental run from that shown in the main text, on a different date, to provide a secondary benchmark.
     The smallest raw signal (gray squares) measured was $0.055 \pm 0.003$ (data includes real-time readout mitigation, double-sided light cone filtering, twirling, and dynamical decoupling). 
     Despite this small signal, the Pauli Path ZNE-mitigated data (blue circles) was successfully extracted, serving as a stringent test for the mitigation technique. 
    \end{minipage}
    \label{fig:extended_data_1}
\end{figure}

The structured character of the NMR OTOC circuits prohibits the use of mitigation strategies used in earlier works that are tailored to random OTOC circuit ensembles \cite{google25constructive}.
The circuits used are also very deep: the deepest 9-spin circuits span $326$ circuit moments and used $540$ two-qubit composite $\fSim$ gates, while the deepest circuits for the 15-spin molecule span $484$ moments and used $792$ $\CZ$ gates.
(These numbers count only relevant gates within the double light cones of the measurement and butterfly operators; we typically remove gates that will have no impact on the final OTOC measurement from the circuit.)
The resulting damping of the noisy OTOC signal was observed in the worst case to be as small as 0.15. 

To reliably and accurately extract data from these noisy quantum circuits, we employ a composite, reconfigurable error mitigation framework designed to operate across different gate sets, using distinct configurations for the $9$-spin ($\fSim$-based) and $15$-spin ($\CZ$-based) molecular $\OTOC$ experiments.
We developed a four-stage mitigation pipeline consisting of algorithmic and physical-level stages: (1) {Circuit generation} at the algorithmic level; (2) {Algorithmic-level error mitigation}, which involved double-sided light cone filtering \cite{google25constructive}, Pauli Path ZNE (described below) \cite{Temme2016ZNEPEC, Li2017ZNE, Cai2022EMReview, Minev2025ZNE}, algorithmic-level DD, and sub-Clifford algorithmic-block twirling \cite{Bennett1995Twirling, Knill2004Twirling}; (3) Native gate decomposition and optimization; (4) Physical-layer mitigation, which involved hardware-adapted DD sequences for idle periods with Clifford twirling \cite{Bennett1995Twirling, Knill2004Twirling,Viola1998,Viola1999}, and real-time randomized readout error mitigation \cite{BergMinev2022TREX}. 
This describes the full pipeline for the 9 spin fSim circuits. The 15 spin CZ-based circuit used the same pipeline, but without the need for algorithmic-level DD, and using full (rather than sub) Clifford group twirling for the two-qubit gates and idle periods. 

As part of this work, we developed a novel Pauli-path zero-noise extrapolation (ZNE) scheme based on the evolution of the measurement operator in the Pauli path picture~\cite{mi21information,angrisani2024classically,rudolph2025pauli,google25constructive}.
This is based on the identification that the noisy expectation value $C(\lambda)$ (with $\lambda$ the noise scaling factor) is the Laplace transform of the underlying Pauli path total Hamming weight distribution $c(H)$~\cite{google25constructive}, where $H$ is the noise accumulation weight.
Though this distribution $c(H)$ can be highly oscillatory, the Laplace transform smooths out high-frequency components, allowing simple, low frequency approximations to $c(H)$ to robustly extrapolate $C(\lambda)$.
Taking a normal approximation of $c(H)$, we derive an analytical functional form for the observed decay which serves as our main model for extrapolation:
\begin{equation}
C(\lambda) = C(0)\frac{\Big(1 + \mathrm{erf}(\tfrac{\bar{H}_c-\lambda\sigma_c^2}{\sigma_c})\Big)}{\Big(1 + \mathrm{erf}(\bar{H}_c/\sigma_c)\Big)} e^{-\lambda\bar{H}_c}e^{\lambda^2\sigma_c^2/2}\;.
\end{equation}
The functional form is parameterized by three key quantities: $C(0)$ is the noise-free extrapolated value, while $\bar{H}_c$ and $\sigma_c$ represent the mean and standard deviation respectively of the Pauli path Hamming weight distribution $c(H)$. These two parameters effectively characterize the noise-induced decay profile. We define the dimensionless parameter $\beta$ as the ratio of the spread to the average noise accumulation $\beta = \frac{\sigma_c}{\bar{H}_c}$, where $\beta>0$ and typically $\beta<0.5$.
For deep circuits where $\beta$ is non-negligible, the quadratic term $e^{\lambda^2\sigma_c^2/2}$ and the error function ($\mathrm{erf}$) correction become essential, as they account for the non-exponential, concave decay profile caused by the complex quantum dynamics, and the physical constraint of non-negative noise weight ($H \ge 0$). This model is conjectured to work even in the regime where Pauli path numeric approximations fail (as is the case in these experiments).

The Pauli Path ZNE framework can be very sensitive to statistical noise when working with deep circuits with low fidelities. To achieve robust extrapolation across the entire time series, we implemented a specialized Bayesian model-fitting update workflow. This time-correlated strategy utilizes data from shallower, less-noisy time steps—where the circuit fidelity is stronger—to establish a Bayesian prior. This prior then iteratively seeds and updates the model fit for deeper, noisier circuits, stabilizing the parameter estimation for the extrapolation. 
We employ a global-parameter optimization algorithm, which combines a local minimization routine with random jumps across the parameter space to overcome local optimization minima and improve robustness to noise, and a rare outlier rejection protocol.
We employ bootstrap resampling over the mitigation circuit instances to evaluate both the mitigated values and their respective uncertainties. 
This combined with the Pauli path theory allows the system to overcome the challenges associated with signal damping and accurately estimate the zero-noise limit even in the deepest regions of the circuit.

\section*{Acknowledgments}
The authors acknowledge useful discussions with \mbox{L. Emsley}, \mbox{L.D. Nielsen}, \mbox{T. Polenova}, \mbox{R. Schurko}, \mbox{R. Tycko}, \mbox{G. Pantaleoni}, and \mbox{N. Wiebe}. We thank Pines Magnetic Resonance Center’s Core NMR Facility at UC Berkeley for spectroscopic resources used in this study. \mbox{D.W. Berry} worked on this project under a sponsored research agreement with
Google Quantum AI.

\section*{Author contributions}
C. Zhang prepared molecular samples and performed the NMR experiments. R.G. Corti\~{n}as and A.H. Karamlou performed the superconducting quantum simulation experiments. C. Zhang, C. Bengs, C. Ramanathan, A. Ajoy, and T.E. O'Brien conceived of the NMR OTOC experiments. C. Zhang and C. Bengs designed the pulse sequence elements and selected the molecular systems. S.H. Kang and R. Sarpong synthesized isoptically-enriched samples of DMBP. J. Provazza, N.C. Rubin, and A. White performed molecular dynamics simulations of the 9Q and 15Q experiments and conceived of the learning experiment. N.C. Rubin, C. Zhang, C. Bengs, J. Provazza, A. White, and T.E. O'Brien performed noisy density matrix simulations of the toluene and DMBP experiments. R.G. Corti\~{n}as, X. Mi, and T.E. O'Brien designed the circuits for the 9-qubit superconducting experiment. J. Bausch, S. Shirobokov, A. Senior, N. Toma\v{s}ev, B. Kozlovskii, F.J.H. Heras, and P. Kohli designed, tested, and ran the AlphaEvolve optimization experiments. R.G. Corti\~{n}as designed and implemented the pulse calibration sequence for the arbitrary-angle fSim gate, with assistance from X. Mi and J.A. Gross. Z.K. Minev, N. Noll, R.G. Cortin\~{a}s, A.H. Karamlou and T.E. O'Brien implemented the coding infrastructure for the superconducting experiments. N. Noll, D. Kafri and Z.K. Minev performed accurate modeling of the superconducting qubit experiments. Z.K. Minev and N. Noll designed and implemented the error mitigation sequences with assistance from T.E. O'Brien and T. Schuster. The Google Quantum AI team designed, built, calibrated and maintained the superconducting qubit chip, along with the dilution refrigeration, the electronics stack, and the superconducting experiment codebase. All authors contributed to the writing and revisions of the manuscript.

\pagebreak
\widetext
\begin{center}
\textbf{\large Supplementary Information: Quantum computation of molecular geometry via many-body nuclear spin echoes}
\end{center}

\makeatletter
\newcommand{\resumetocentries}{\let\addcontentsline\orig@addcontentsline}
\resumetocentries 
\makeatother

\setcounter{equation}{0}
\setcounter{figure}{0}
\setcounter{table}{0}
\makeatletter
\renewcommand{\theequation}{S\arabic{equation}}
\renewcommand{\thefigure}{S\arabic{figure}}

\input{SI}
\end{document}

%% file: SI.tex
\addtocontents{toc}{\protect\setcounter{tocdepth}{3}}

\tableofcontents

\section{Outline}

\input{outline}

\section{Chemical Synthesis Details}\label{sec:synthesis}

\input{synthesis}
\clearpage
\section{NMR details}\label{sec:nmr}

\input{nmr}

\section{Molecular Dynamics model}\label{sec:md}

\input{md}

\section{Trotter scaling estimates}\label{sec:trotter}

\input{trotter}

\section{AlphaEvolve circuit simulation}\label{sec:alphaevolve}

\input{alphaevolve}

\section{Classical resource scaling analysis on a simplified model}\label{sec:classical}

\input{classical}

\section{The Superconducting Quantum Simulator}\label{sec:sc}

\input{sc_expt}

\section{Accurate noise modeling of superconducting experiments}\label{sec:noise}

\input{modeling}

\input{main.bbl}

%% file: outline.tex
In this Supplementary Information, we cover additional details of the NMR and superconducting experiments, classical simulation, and AlphaEvolve optimization that went into this work.
We begin in Sec.~\ref{sec:synthesis} with full details of the synthetic chemistry used in this in this work (excepting details of chemicals bought from external vendors).
Then, in Sec.~\ref{sec:nmr}, we detail the NMR theory required for this experiment, as well as the design of the Time-Accurate Reversal of Dipolar InteractionS (TARDIS) pulse sequences, details of the primary error mechanisms observed (explaining the noisy classical simulation that was used in the main text), and explicit description of all experiments performed in this work.
This includes details of the construction of the reference toluene Hamiltonian used in this work from additional NMR experiments (other than OTOCs) and literature reference data.
We also provide NMR data from experiments not shown in the main text (including in Fig.~\ref{fig:all_toluene_otocs} the data that went into Fig.2(e) of the main text, and in Fig.~\ref{fig:saupe_tensor_learning} a demonstration of learning two components of the toluene directional order parameter), and some physical insight into the observed features in the toluene OTOC decay.
In Sec.~\ref{sec:md}, we provide details of the molecular dynamics simulations used for the DMBP simulation, how this was used to develop the dihedral-angle-dependent coupling matrices used in the main text.
We also expand on various schemes used to correct the DMBP PMF with higher accuracy simulation, and results of MD simulations of Toluene.
In Sec.~\ref{sec:trotter}, we perform an analysis of lower-order Trotter methods for OTOCs in general, and specifically for the DMBP simulations performed in this work.
We find that the error in the OTOC is significantly suppressed in first-order Trotter schemes, and that the DMBP OTOC curves can be replicated with only a few Trotter steps.
Then, in Sec.~\ref{sec:alphaevolve}, we provide extended analysis of the optimization and output of the AlphaEvolve coding agent applied to the DMBP results, in an attempt to make the output human-understandable.
In Sec.~\ref{sec:classical}, we test the performance of two classical computation methods at simulating OTOCs in a simplified 1D model to exemplify the scaling behaviour of both methods.
Then, in Sec.~\ref{sec:sc}, we provide details of the superconducting chip used, the compilation of the time evolution to a SWAP network, and the calibration of the arbitrary angle FSim gate.
And finally, in Sec.~\ref{sec:noise}, we provide accurate noise modeling and budgeting of the superconducting qubit device, and a comparison between noisy simulations of the circuits run and the obtained experimental data.

%% file: synthesis.tex
\subsection{General considerations}

\subsubsection{Reagents and solvents}

Unless stated otherwise, commercial reagents were purchased from Sigma Aldrich, Acros Organics, Ambeed, Fisher Scientific, Matrix Scientific, Chem-Impex, Combi-blocks, Tokyo Chemical Industries, Oakwood Chemical, Strem Chemicals, Spectrum Chemical and/or Alfa Aesar, and used without additional purification. Solvents were purchased from Fisher Scientific, Acros Organics, Alfa Aesar, and Sigma Aldrich. Tetrahydrofuran (THF), diethyl ether (Et${}_{2}$O), benzene (PhH), toluene (PhMe), methanol (MeOH), and triethylamine (Et${}_{3}$N) were sparged with argon and dried by passing through alumina columns using argon in a Glass Contour solvent purification system. Dichloromethane (DCM) was freshly distilled over calcium hydride under a nitrogen (N${}_{2}$) atmosphere prior to each use.

\subsubsection{Experimental conditions}

Unless otherwise noted, reactions were carried out in flame- or oven-dried glassware under a positive pressure of N${}_{2}$ in anhydrous solvents using standard Schlenk techniques. Reaction temperatures above room temperature (22–23 ${}^{\circ}$C) were controlled by an IKA\textregistered~temperature modulator and monitored using liquid-in-glass thermometers. Reaction progress was monitored by thin-layer chromatography (TLC) on Macherey-Nagel TLC plates (60 \AA, F254 indicator). TLC plates were visualized by exposure to ultraviolet light (254 nm), and/or stained by submersion in aqueous potassium permanganate solution (KMnO${}_{4}$), p-anisaldehyde or ceric ammonium molybdate (CAM) stains and heating with a heat gun. Organic solutions were concentrated under reduced pressure on an Heidolph temperature-controlled rotary evaporator equipped with a dry ice/isopropanol condenser. Flash column chromatography was performed with either glass columns using Silicycle SiliaFlash\textregistered~P60 silica gel (40–63 $\mu$m particle size) using ACS grade solvents. All yields refer to chromatographically and spectroscopically (${}^{1}$H and ${}^{13}$C NMR) pure material.

\subsubsection{Analytical Methods}

NMR spectra were recorded using deuterated solvents, obtained from Cambridge Isotope Laboratories, Inc. ${}^{1}$H NMR and ${}^{13}$C NMR data were recorded on Bruker AVQ-400, AVB-400, NEO-500, AV-600 and AV-700 spectrometers using deuterated chloroform (CDCl$^{}_{3}$) as solvent at ambient temperature. Chemical shifts ($\delta$) are reported in ppm relative to the residual solvent signal (CDCl${}_{3}$: $\delta_{\rm H}$ = 7.26 ppm, $\delta_{\rm C}$ = 77.16 ppm). Data for ${}^{1}$H and ${}^{13}$C spectroscopy are reported as follows: chemical shift ($\delta$ ppm), multiplicity (s = singlet, d = doublet, t = triplet, q = quartet, m = multiplet, br = broad, app = apparent), coupling constant (Hz), integration. All raw free-induction-decays (FIDs) were processed, and spectra analyzed, using the program MestReNOVA 11.0 from Mestrelab Research S. L. Melting points were determined using a MEL-TEMPTM apparatus and are uncorrected. High-resolution mass spectra (HRMS) were obtained from the Mass Spectral Facility at the University of California, Berkeley, on a Finnigan/Thermo LTQ-FT instrument (ESI). Data acquisition and processing were performed using the XcaliburTM software.

\subsection{Synthesis of \texorpdfstring{{[}1-${}^{13}$C{]}-3{'},5{'}}{[1-13C]-3',5'}--dimethylbiphenyl}

\begin{figure}
    \includegraphics[trim={2cm 0 2cm 0},clip,width=0.8\columnwidth]{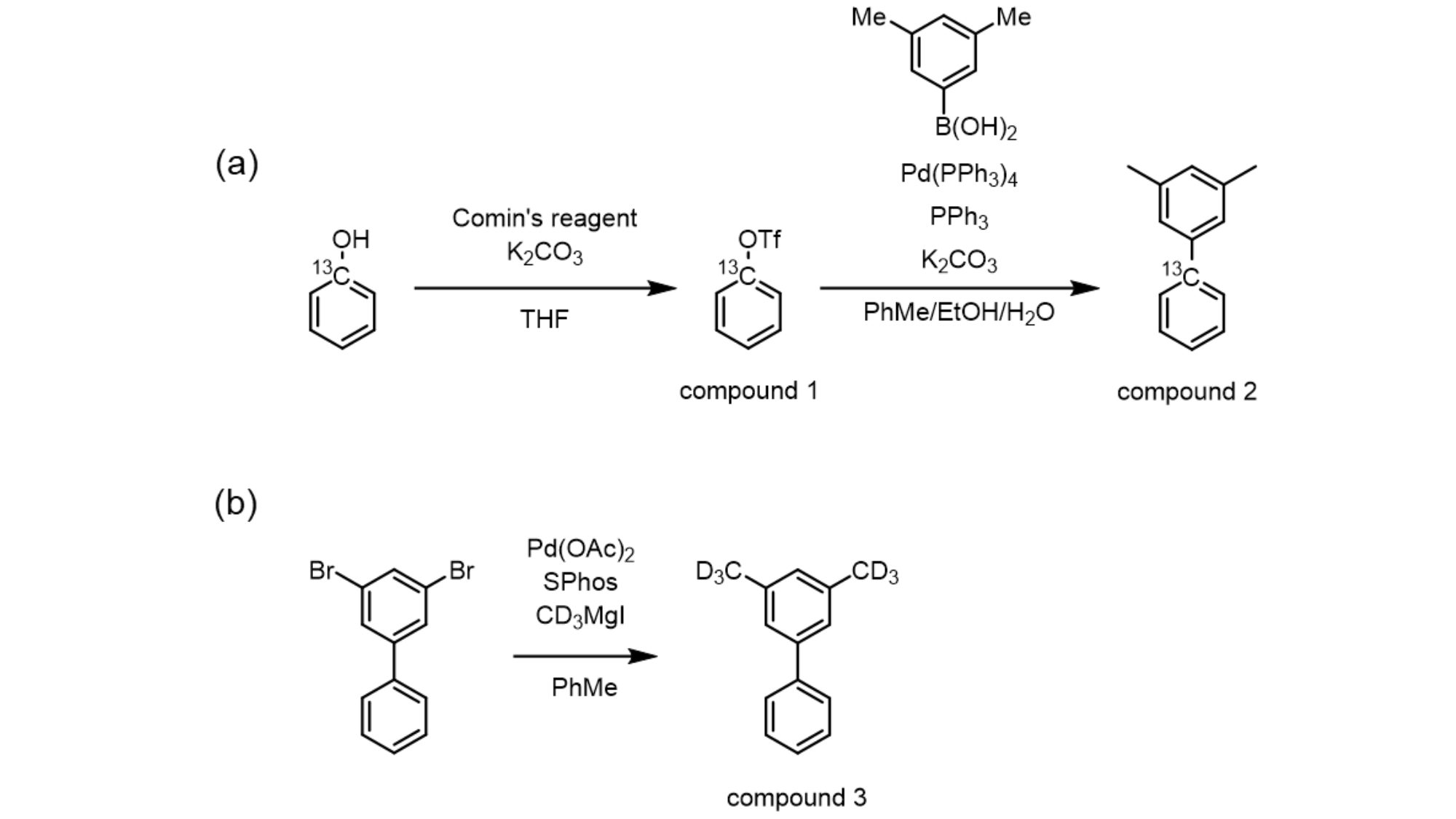}
    \caption{\label{fig:reaction1} (a) Synthesis of {[}1-${}^{13}$C{]}-3{'},5{'}-dimethylbiphenyl (compound 2) via Suzuki coupling. 
    (b) Synthesis of 3{'},5{'}-bis(methyl-${\rm d}_{3}$)biphenyl (compound 3) via Kumada cross coupling.}
\end{figure}

In a 50 mL round-bottom flask, phenol-1-${}^{13}$C (250.00 mg, 1 equiv, 2.63 mmol), Comin’s reagent (1.24 g, 1.2 equiv, 5.26 mmol) were dissolved in anhydrous THF (26 mL). The reaction was stirred under N${}_{2}$ atmosphere overnight. Upon full conversion of the reaction, the crude mixture was concentrated in vacuo and redissolved in dichloromethane (100 mL). The organic phase was washed with H${}_{2}$O (100 mL) and the resulting aqueous phase was extracted with dichloromethane (3 $\times$ 100 mL). The combined organic layers were dried with sodium sulfate and concentrated in vacuo. The title compound 1 was obtained as yellow oil (534 mg, 89\%) after purification with silica column chromatography (4:1 hexane:EtOAc).

In a 100 mL round-bottom flask, phenyl-${}^{13}$C trifluoromethanesulfonate compound 1 (543 mg, 1 equiv, 2.35 mmol) was dissolved in a solvent mixture of toluene (23 mL), ethanol (4 mL), and H${}_{2}$O (4 mL) with tetrakis(triphenylphosphine)-palladium(0) (135.83 mg, 0.05 equiv, 117.54 $\mu$mol), (3,5-dimethylphenyl)boronic acid (352.57 mg, 1 equiv, 2.35 mmol), triphenylphosphine (92.49 mg, 0.15 equiv, 352.61 $\mu$mol), and K${}_{2}$CO${}_{3}$ (2.11 g, 6.5 equiv, 15.28 mmol). A reflux condenser was attached, and the reaction was stirred at reflux overnight. Upon completion of the reaction, H${}_{2}$O was added and the aqueous phase was extracted with dichloromethane (3 $\times$ 100 mL). The combined organic layers were dried with sodium sulfate and concentrated in vacuo. The title compound 2 was obtained as colorless oil (305 mg, 71\%) after purification with silica column chromatography (100\% hexane). 
\newline
\noindent ${\rm R}_{f}$ (10\% EtOAc/hexanes, CAM) 0.90; ${}^{1}$H NMR (600 MHz, CDCl${}_{3}$) $\delta$ 7.59 (dq, J = 6.6, 1.0 Hz, 2H), 7.43 (q, J = 7.6 Hz, 2H), 7.35 (dd, J = 7.7, 1.4 Hz, 1H), 7.24 – 7.22 (m, 2H), 7.01 (t, J = 1.5 Hz, 1H), 2.40 (s, 6H); ${}^{13}$C NMR (151 MHz, CDCl${}_{3}$) $\delta$ 141.63, 138.38, 129.03, 128.77, 127.51, 127.21, 127.12, 125.26, 21.55; HRMS (EI) Exact mass calcd for ${}^{13}{\rm C}{\rm C}_{13}{\rm H}_{14}$ {[}M{]} [M]: 183.1129. Found 183.1128. 

\subsection{Synthesis of 3{'},5{'}-bis(methyl-\texorpdfstring{${\rm d}_{3}$}{d3})biphenyl}

In a 50 mL round-bottom flask, Pd(OAc)${}_{2}$ (21.59 mg, 0.03 equiv, 96.15 $\mu$mol), 2-dicyclohexylphosphino-2,6-dimethoxy-1,1-biphenyl (52.63 mg, 0.04 equiv, 128.21 $\mu$mol), and 3,5-dibromo-1,1’-biphenyl (1.00 g, 1 equiv, 3.21 mmol) were dissolved in anhydrous toluene (25 mL). To this solution, (methyl-d${}_{3}$)magnesium iodide (1.19 g, 7.05 mL, 1.0 M, 2.2 equiv, 7.05 mmol) was added drop-wise. After stirring the reaction at room temperature for 30 min, the reaction was heated to 80 ${}^{\circ}$C and stirred overnight. The reaction was quenched with a saturated aqueous solution of NH${}_{4}$Cl. The aqueous phase was extracted with ethyl acetate (3 $\times$ 100 mL). The combined organic layers were filtered over a plug of silica and concentrated in vacuo. The title compound 3 was obtained as colorless oil (510 mg, 85\%) after purification with silica column chromatography (9:1 hexane:EtOAc).
\newline
\noindent ${\rm R}_{f}$ (10\% EtOAc/hexanes, CAM) 0.90; ${}^{1}$H NMR (600 MHz, CDCl${}_{3}$) $\delta$ 7.66 – 7.61 (m, 2H), 7.49 – 7.44 (m, 2H), 7.41 – 7.35 (m, 1H), 7.27 (dd, J = 7.0, 2.5 Hz, 2H), 7.05 (dt, J = 6.5, 1.7 Hz, 1H); ${}^{13}$C NMR (151 MHz, CDCl${}_{3}$) $\delta$ 141.64, 141.42, 138.27, 129.04, 128.77 (d, J = 2.0 Hz), 127.34 (d, J = 1.7 Hz), 127.21 (d, J = 2.2 Hz), 125.27, 20.72 (dt, J = 39.2, 19.7 Hz); HRMS (EI) exact mass calcd for ${\rm C}_{14}{\rm H}_{8}{\rm D}_{6}$ {[}M{]}: 188.1472, found 188.1472.

\clearpage

\begin{figure}
    \includegraphics[trim={8cm 3cm 8cm 3cm},clip,width=0.8\columnwidth]{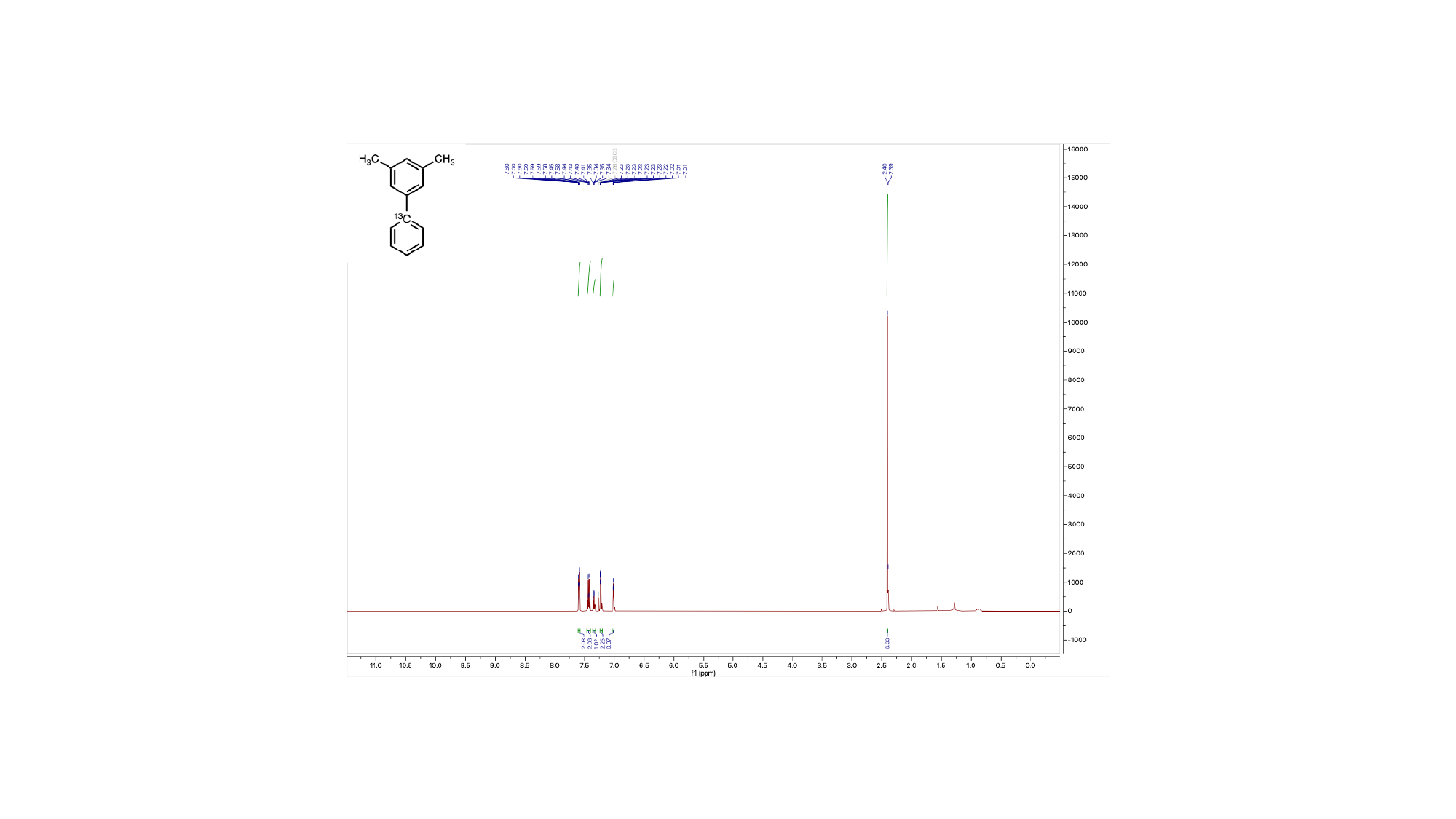}
    \caption{${}^{1}$H NMR (600 MHz, CDCl${}_{3}$) for {[}1-${}^{13}$C{]}-3{'},5{'}-dimethylbiphenyl.}
\end{figure}

\begin{figure}
    \includegraphics[trim={8cm 3cm 8cm 3cm},clip,width=0.8\columnwidth]{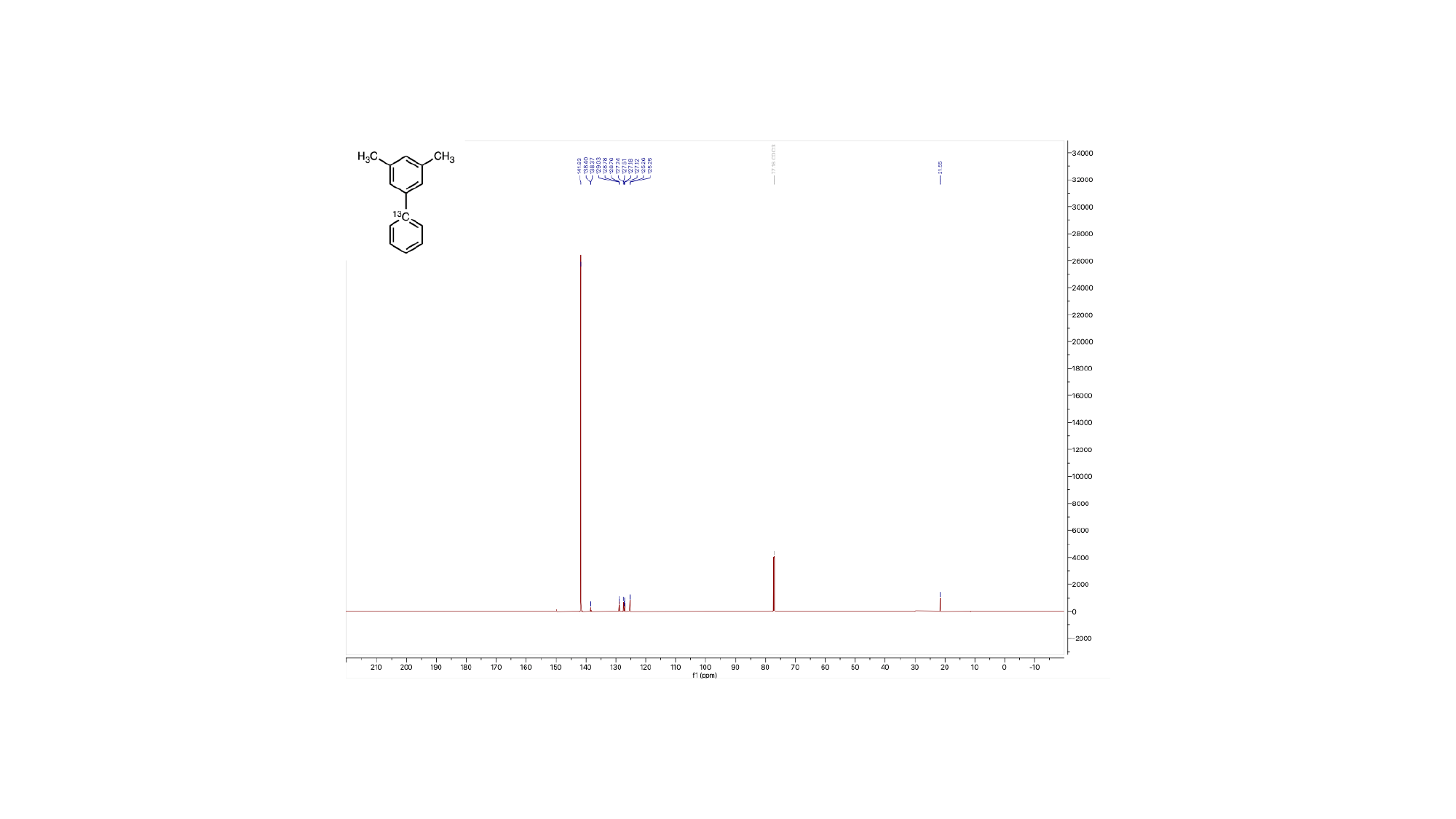}
    \caption{${}^{13}$C NMR (151 MHz, CDCl${}_{3}$) for {[}1-${}^{13}$C{]}-3{'},5{'}-dimethylbiphenyl.}
\end{figure}

\begin{figure}
    \includegraphics[trim={8cm 3cm 8cm 3cm},clip,width=0.8\columnwidth]{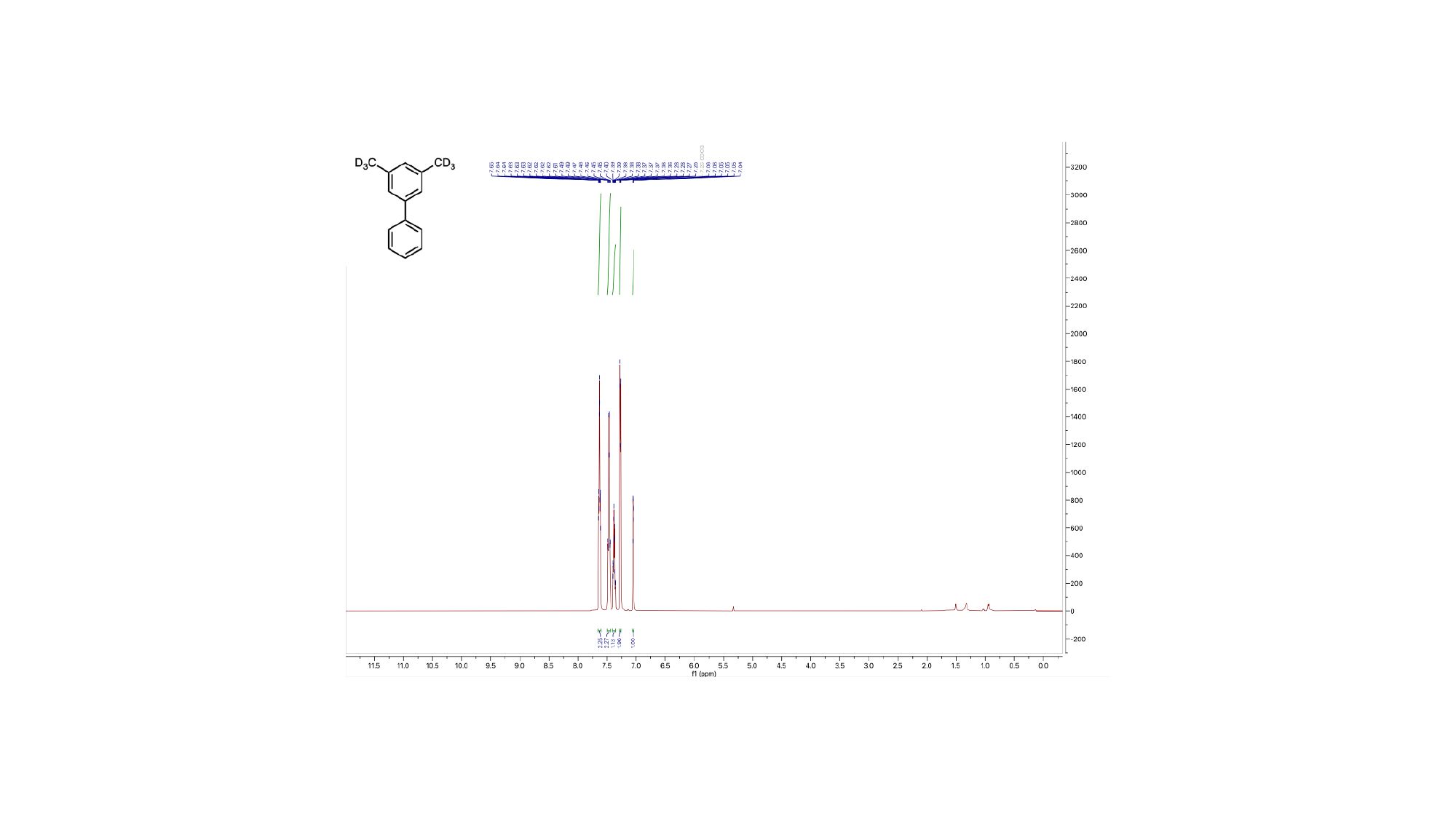}
    \caption{${}^{1}$H NMR (600 MHz, CDCl${}_{3}$) for 3{'},5{'}-bis(methyl-${\rm d}_{3}$)biphenyl.}
\end{figure}

\begin{figure}
    \includegraphics[trim={8cm 3cm 8cm 3cm},clip,width=0.8\columnwidth]{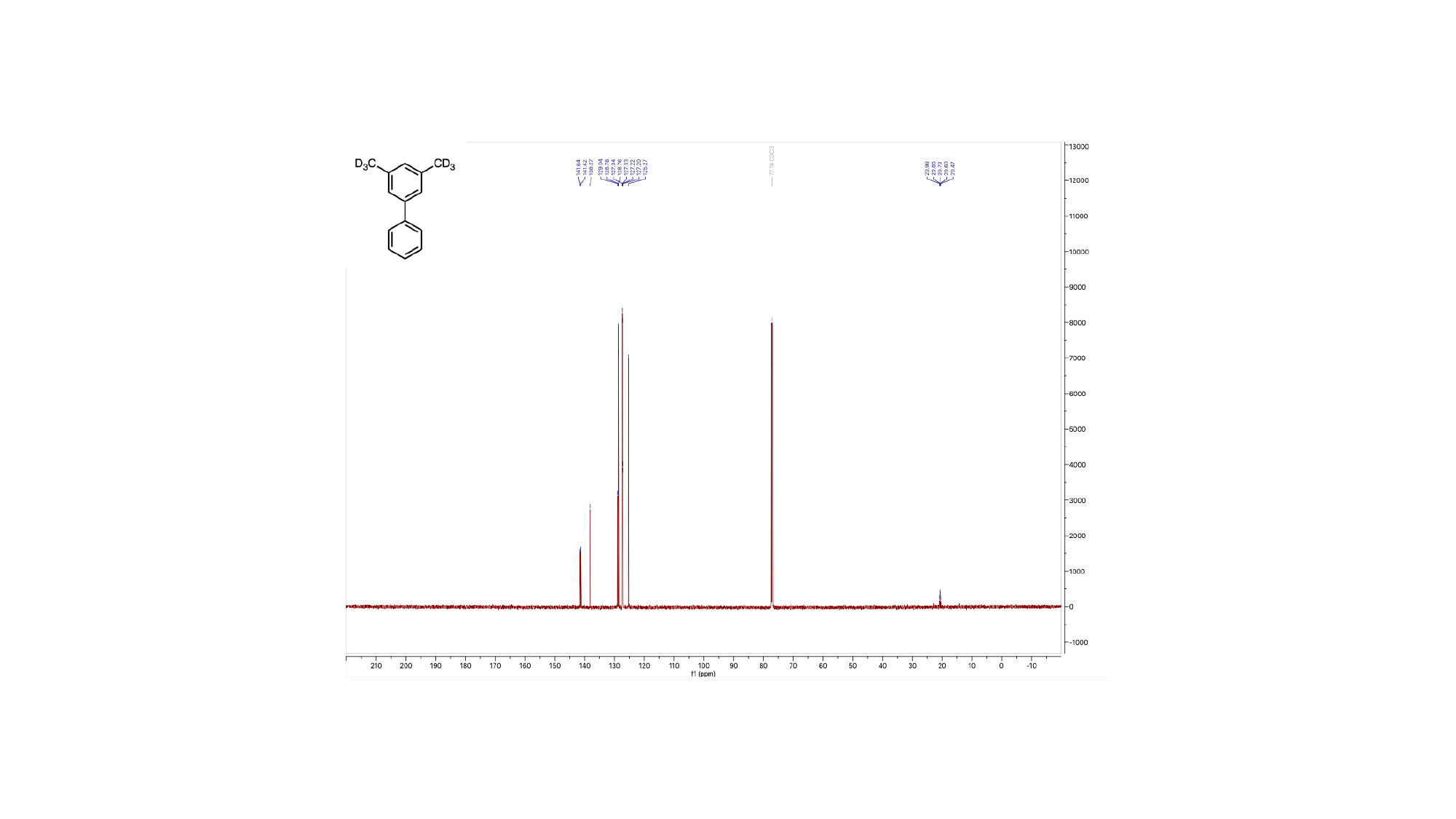}
    \caption{${}^{13}$C NMR (151 MHz, CDCl${}_{3}$) for 3{'},5{'}-bis(methyl-${\rm d}_{3}$)biphenyl.}
\end{figure}

%% file: nmr.tex
In this section, we detail the NMR experiments performed in this work and present additional data not shown in the main text.
We begin with a quick primer noting the differences between common notation in NMR and quantum information.
Next, we review the NMR Hamiltonian, describing how it can be derived from molecular structure at finite temperature, where molecular motion cannot be neglected, and explain the Saupe tensor formalism that was used for modeling toluene in this work.
We then expand on the OTOC experiment shown in Fig.~2(a) of the main text, providing details of the TARDIS-1 and TARDIS-2 pulse sequences used for forward and backward evolution, the filtering sequence used to prepare a high-fidelity initial state and mitigate RF inhomogeneity, and the BLEW12 pulse used as the butterfly excitation. We also calculate the effective Hamiltonians of all models to lowest order in average Hamiltonian theory.
Subsequently, we present additional NMR OTOC datasets not included in the main text: additional datasets exploring the reproducibility of the method, and datasets with an on-site field added (that were used in the construction of Fig.~2(d) in the main text), and datasets showing learning of the two components of the Saupe tensor (see Eq.~\ref{eq:RDC_Simplified}). We then detail the HETCOR experiment performed to determine chemical shift terms in DMBP, which are used in learning experiment. The MQC experiments are subsequently presented, which determine the reference interaction parameters of toluene and validate the DMBP results.
Finally, we continue the Magnus expansion of the TARDIS sequence to third order, and show how higher order corrections affect the toluene OTOC.

\subsection{Notation}

To clarify notation, we summarize the small differences between NMR and QIS. In NMR, it is conventional to use spin angular momentum operators, while in QIS, the Pauli convention is typically preferred. The corresponding associations are as follows:
\begin{equation}
    \begin{aligned}
        I_{x}^{j}= \frac{1}{2} X_{j}, 
        \quad I_{y}^{j}=\frac{1}{2} Y_{j},
        \quad I_{z}^{j}=\frac{1}{2} Z_{j}.
    \end{aligned}
\end{equation}
Additionally, in NMR $I$ is typically reserved for proton spins, and $S$ denotes the heteronuclear species (${}^{13}$C in our case). QIS, on the other hand, adopts a linear indexing scheme (e.g. qubit 1, qubit 2, etc.), irrespective of nuclear species (where applicable). The conversion is as follows:
\begin{equation}
    \begin{aligned}
      \{I^{1}_{x},\dots,I^{N_{I}}_{x},S^{1}_{x},\dots,S^{N_{S}}_{x}\}
      \rightarrow
      \{X_{1},\dots,X_{N_{I}},X_{N_{I}+1},\dots,X_{N_{I}+N_{S}}\},
    \end{aligned}
\end{equation}
and similarly for $Y$ and $Z$. For NMR it is further convenient to introduce collective spin angular momentum operators
\begin{equation}
    \begin{aligned}
I_{\mu}=\sum_{j=1}^{N_{I}}I^{j}_{\mu},
\quad
S_{\mu}=\sum_{j=1}^{N_{S}}S^{j}_{\mu}.
    \end{aligned}
\end{equation}

Both QIS and NMR have a commonly used pictorial representation of their experiments, which are similar but slightly different.
In quantum information the representation is the circuit diagram~\cite{nielsen2010quantum}, where each qubit is given a line, and symbols are placed across lines to represent quantum gates that typically act on 1 or 2 qubits.
However, in NMR one does not have the ability to target single spins.
Instead, pulses are typically applied on all spins of the same species, with some possibility for frequency selectivity via pulse schemes such as BLEW12.
Thus, a pulse diagram in NMR has one line per species: pulses are represented by boxes sitting on this line, with the understanding that this is applied to all spins of this species.

\subsection{NMR Hamiltonian Overview}\label{sec:NMR_Hamiltonian}

\begin{figure}
    \includegraphics[width=0.5\textwidth]{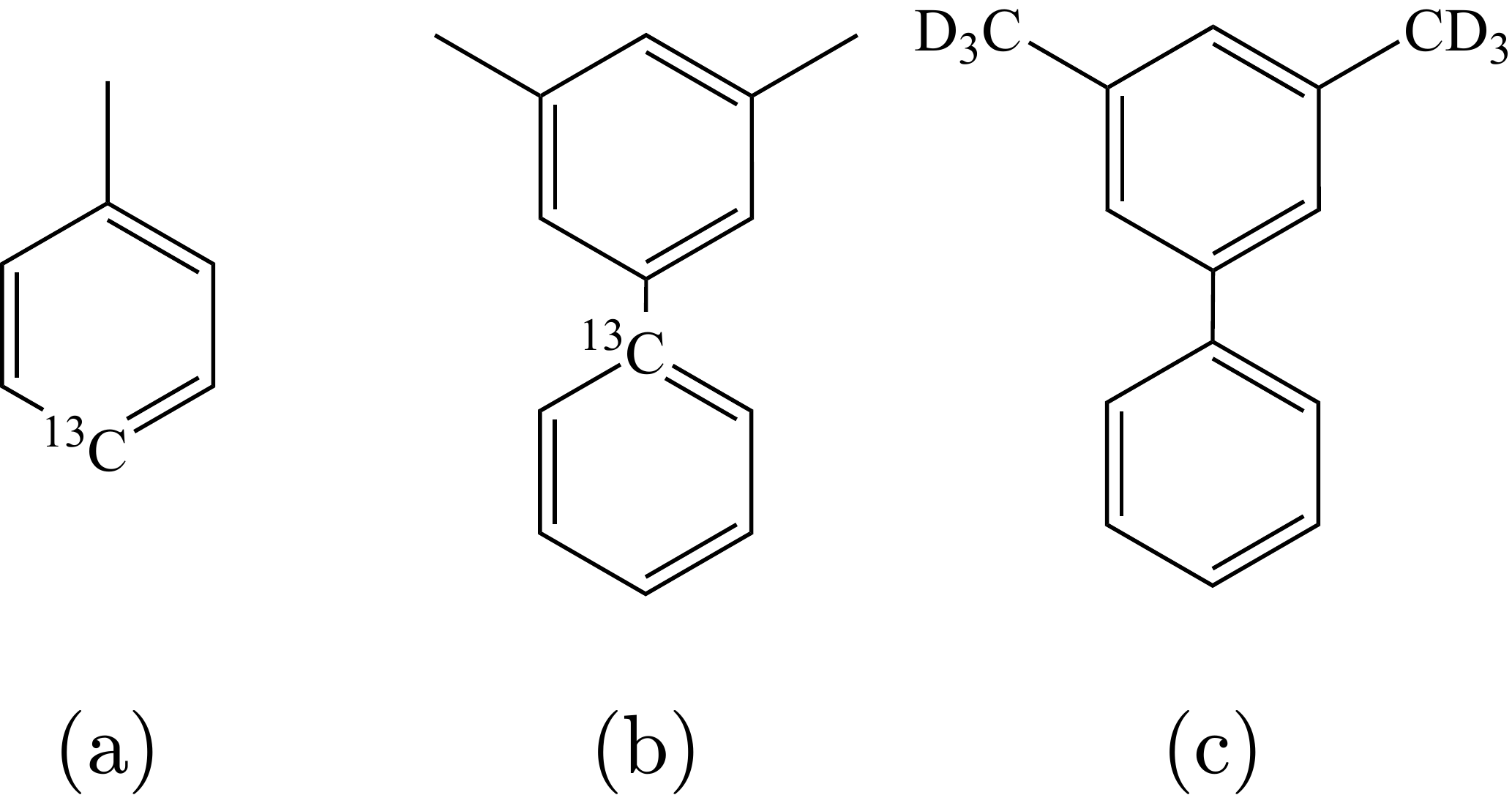}
    \caption{\label{fig:molecule_formula} Compounds used in OTOC and MQC experiments. (a)\sampleA, (b) Compound 2 or~\sampleB, (c) Compound 3 or~\sampleC.}
\end{figure}

NMR experiments are performed on the molecular systems shown in Fig.~\ref{fig:molecule_formula}, which we refer to \sampleA (or 9Q), \sampleB (or 15Q) and \sampleC, respectively. All experiments are performed in a nematic liquid crystal solvent leading to a partial alignment. Due to partial alignment, molecular tumbling is not fully isotropic resulting in the presence of residual intramolecular dipolar interactions~\cite{levitt2008spin}. The internal system Hamiltonian in rotating frame is then given by
\begin{equation}
    H_{\rm int}=H_{\rm CS}+H_{\rm J}+H_{\rm DD}.
    \label{eq:sys Hamiltonian}
\end{equation}
The chemical shift term $H_{\rm CS}$ describes the Zeeman interaction with the external magnetic field
\begin{equation}
    H_{\rm CS}=\sum_{i\in \proton}\omega_{i}I_{z}^{i},
\end{equation}
where $\omega_{i}$ is the chemically shifted resonance frequency relative to the $\proton$ channel carrier frequency. For the carbon nucleus, we assume the carrier frequency to be matched to its resonance frequency. Within the secular approximation the homo- and heteronuclear scalar coupling term $H_{J}$ is given by
\begin{equation}
    H_{\rm J}=2\pi\sum_{i,j\in \proton,i<j} J_{ij}(I_{x}^{i}I_{x}^{j}+I_{y}^{i}I_{y}^{j}+I_{z}^{i}I_{z}^{j})+2\pi\sum_{i\in \proton} J_{Ci}I_{z}^{i}S_{z},
\end{equation}
where $J_{ij}$ represents the coupling constant between the \textit{i}'th and \textit{j}'th proton spin, whereas $J_{Ci}$ represents the coupling constant between the $^{13}\text{C}$ and the \textit{i}'th proton spin. The dipolar interaction term $H_{\rm DD}$ represents through-space coupling of two spins. Within the secular approximation the dipolar interactions for homo- and heteronuclear spin pairs are given by
\begin{equation}
    H_{\rm DD}=2\pi\sum_{i,j\in \proton,i<j} D_{ij}\cdot\left(2I_{z}^{i}I_{z}^{j}-I_{x}^{i}I_{x}^{j}-I_{y}^{i}I_{y}^{j}\right)+2\pi\sum_{i\in\proton} D_{Ci}\cdot2I_{z}^{i}S_{z}
\end{equation}
Here, $D_{ij}$ denote the dipolar coupling between the \textit{i}'th and the \textit{j}'th proton spin, $D_{Ci}$ denotes the dipolar coupling between the $^{13}\text{C}$ and the \textit{i}'th proton spin. The effect of partial averaging is taken into account as follows
\begin{equation}
\begin{aligned} 
    &D_{ij}=-\frac{\mu_{0}h}{4\pi}\gamma_{i}\gamma_{j}\left\langle\frac{\frac{3}{2}(\hat{r}_{ij}\cdot\hat{n})^{2}-\frac{1}{2}}{r_{ij}^{3}}\right\rangle,
\label{eq:DD defination}
\end{aligned}
\end{equation}
Here, 
$\mu_{0}$ is the vacuum magnetic permeability and $h$ is Planck's constant. The unit vector $\hat{n}$ is aligned with the direction of the static external field, whereas $\hat{r}_{ij}$ points from spin $i$ to spin $j$, $r_{ij}$ represents the distance between the two spins. The angular brackets indicate averaging over the external and internal molecular motion. 

\begin{figure}
    \includegraphics[width=0.5\textwidth]{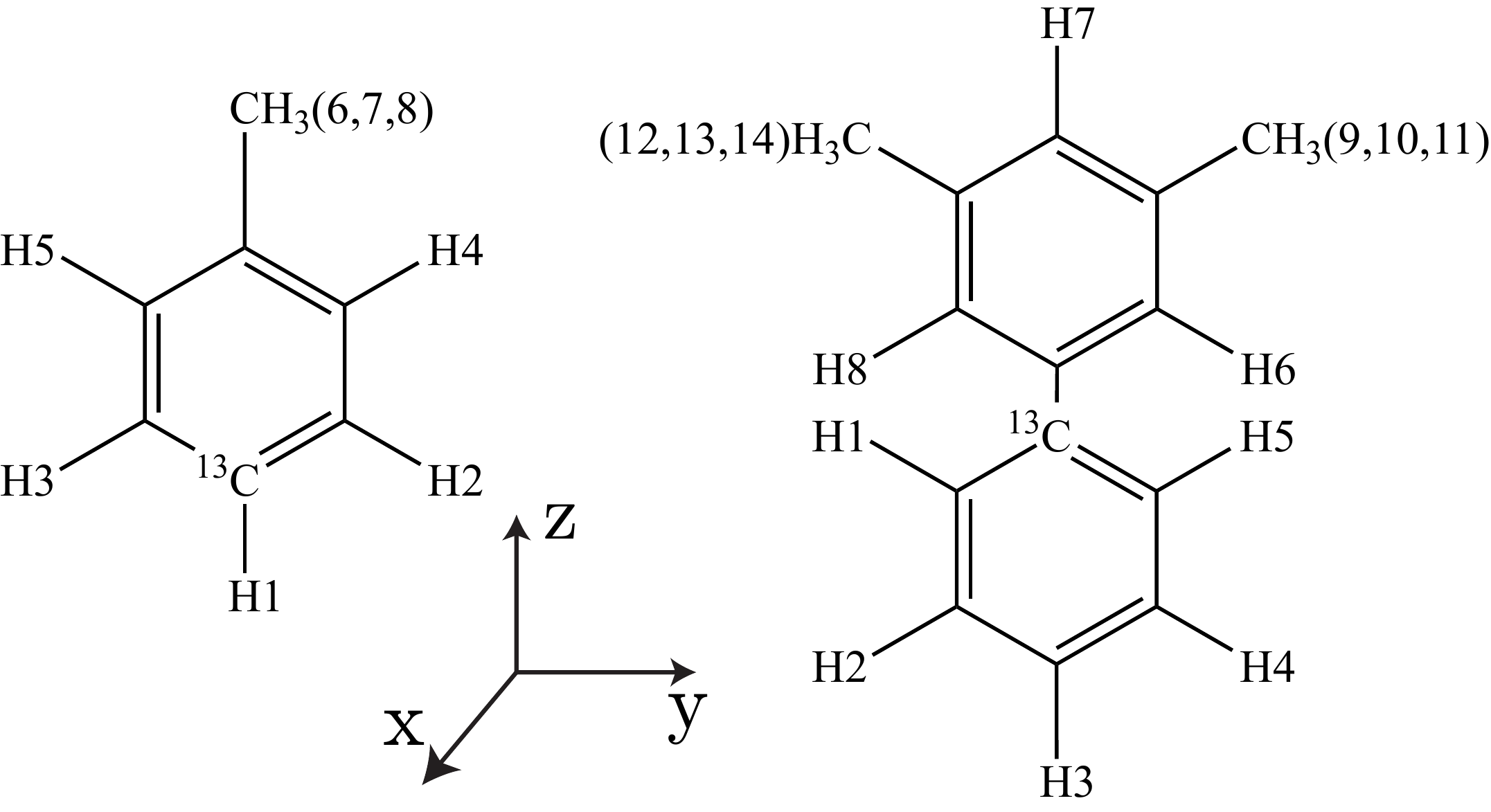}
    \caption{\label{fig:spin_indexing} Enumeration scheme for \sampleA (left) and \sampleB (right). In both molecules, the $z$-axis of the molecular frame is defined as the long axis of the molecule. For \sampleA, the $x$-axis is taken normal to the benzene ring plane. For \sampleB, the $x$-axis is taken normal to the unsubstituted phenyl ring.}.
\end{figure}

For toluene, we assume that the aromatic ring tumbles rigidly about the liquid crystal director, while the methyl group rotates freely. The flat potential energy surface leads to rapid and isotropic methyl rotation, making the three protons equivalent and imposing approximate $C_{2v}$ symmetry. With the ring rigidity and rapid methyl rotation decoupling external tumbling from internal motion, the residual dipolar couplings can be expressed in terms of the Saupe order tensor ($S_{ab}$)~\cite{saupe1964kernresonanzen}

\begin{equation}
\label{eq:RDC_Simplified}
\begin{aligned}
D_{ij} &= -\frac{\mu_{0} h}{4 \pi} \gamma_{i} \gamma_{j}
\left\langle
    \frac{
        \sum_{a,b = x,y,z}
        (\hat{r}_{ij} \cdot \hat{e}_{a})
        \left[
            \frac{3}{2} 
            (\hat{n} \cdot \hat{e}_{a})
            (\hat{n} \cdot \hat{e}_{b})
            - \frac{1}{2} \delta_{ab}
        \right]
        (\hat{r}_{ij} \cdot \hat{e}_{b})
    }{
        r_{ij}^{3}
    }
\right\rangle \\
&= -\frac{\mu_{0} h}{4 \pi} \gamma_{i} \gamma_{j}
\sum_{a,b = x,y,z}
\left\langle
    \frac{
        (\hat{r}_{ij} \cdot \hat{e}_{a})
        (\hat{r}_{ij} \cdot \hat{e}_{b})
    }{
        r_{ij}^{3}
    }
\right\rangle_{\rm internal}
\left\langle
    \frac{3}{2} 
    (\hat{n} \cdot \hat{e}_{a})
    (\hat{n} \cdot \hat{e}_{b})
    - \frac{1}{2} \delta_{ab}
\right\rangle_{\rm external} \\
&= -\frac{\mu_{0} h}{4 \pi} \gamma_{i} \gamma_{j}
\sum_{a,b = x,y,z} R_{ij,ab} S_{ab},
\end{aligned}
\end{equation}
The Saupe order captures the alignment of the molecular frame ${[}\hat{e}_{x},\hat{e}_{y},\hat{e}_{z}{]}$ with respect to the magnetic field direction (see Fig.~\ref{fig:spin_indexing}), whereas the quantities $R_{ij,ab}$ represent the internal ordering of the molecule. Owing to the molecular symmetry, the off-diagonal elements of $S_{ab}$ vanish. Given the tensor’s traceless nature, $S_{ab}$ is thus completely defined by its two independent diagonal components, $S_{yy}$ and $S_{zz}$.

The tensor $R_{ij,ab}$ for different spin pairs can be determined directly from the 
toluene structure. Between two spins on the aromatic ring, $\mathbf{r}_{ij}$ is fixed in 
the molecular frame, making $R_{ij,ab}$ straightforward to compute. 
For two methyl $^1\mathrm{H}$ spins, the effect of isotropic rotation is included by averaging over all methyl orientations, simplifying the expression to
\begin{equation}
    R_{ij,ab}=\frac{1}{r_{ij}^3}\begin{pmatrix}
\frac{1}{2} & 0& 0 \\
0 & \frac{1}{2} & 0 \\
0 & 0 & 0
\end{pmatrix}
\end{equation} 
Here, $r_{ij}$ represents the distance between two methyl $^1\mathrm{H}$ spins. 
Similarly, for the dipolar coupling between a methyl $^1\mathrm{H}$ spin and a spin on the ring, 
$R_{ij,ab}$ can be computed numerically as
\begin{equation}
    R_{ij,ab}=\left\langle
    \frac{
        (\hat{r}_{ij} \cdot \hat{e}_{a})
        (\hat{r}_{ij} \cdot \hat{e}_{b})
    }{
        r_{ij}^{3}
    }
\right\rangle_{\rm rotation}
\end{equation}
where the angular brackets denote averaging over uniform methyl group orientations. 

\subsection{OTOC Experiments}

\begin{figure}
    \includegraphics[width=0.7\textwidth]{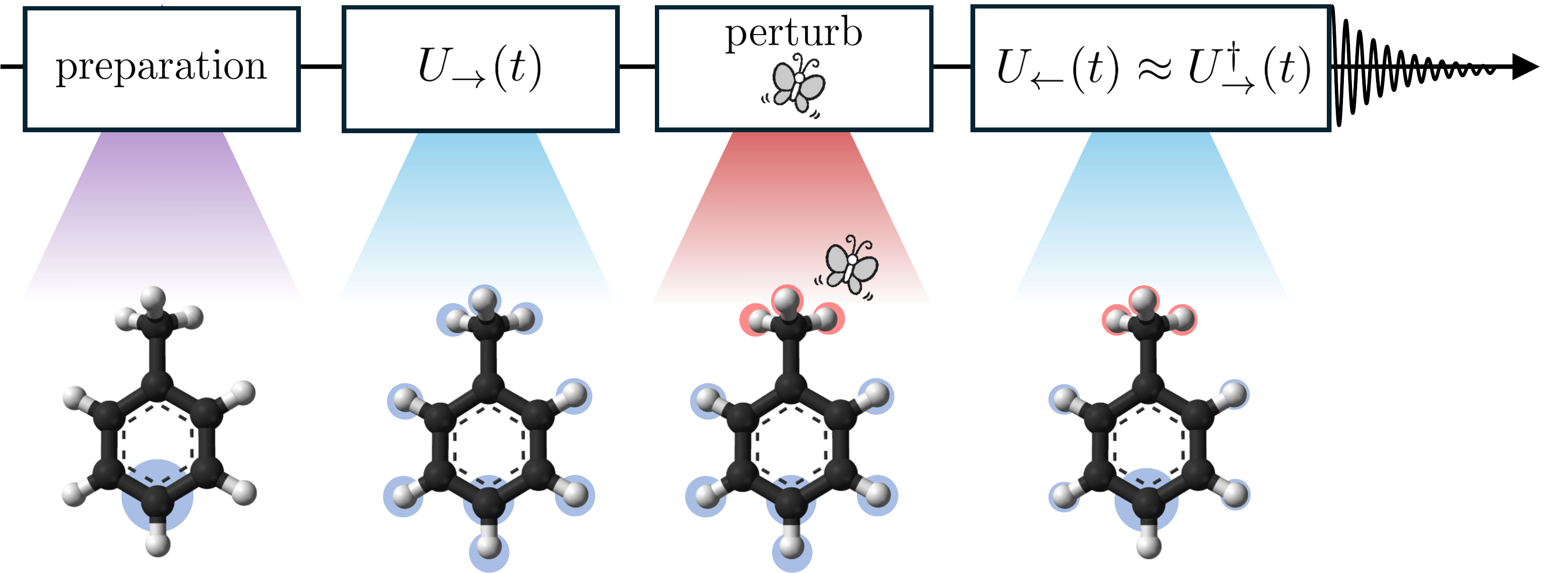}
    \caption{\label{fig:OTOC_general_scheme}
    Schematic protocol for OTOC-NMR illustrated for \sampleA. 0) a preparation period initializes polarization on a carbon spin (measurement spin). 1) An evolution period spreads the initial polarization across the carbon and proton network. 2) A selective perturbation of a subset of the protons acts as a butterly. 3) The evolution period is approximately reversed aiming to return all polarization back to the initial carbon spin. 4) Measure the resulting carbon FID.
    }.
\end{figure}

The formal definition of an out-of-time-order correlator (in the high-temperature limit) is given by~\cite{li2017measuring}
\begin{equation}
\begin{aligned}
    C(t)={\rm Tr}\{B^{\dagger}(t)M^{\dagger}B(t)M\}.
\end{aligned}
\end{equation}
With $M$ representing a local operator for the ``measurement" spin and $B(t)$ representing a perturbation on the ``butterfly" spins expressed in the Heisenberg picture ($B(t)=U^{\dagger}(t)BU(t)$). If we identify $M=S_{x}$ with the carbon spin angular momentum along $x$ and $B=V_{\rm H}$ with a selective proton pulse sequence, we may explicitly connect the OTOC to a typical NMR experiment 
\begin{equation}
\begin{aligned}
    C(t)={\rm Tr}\{S_{x}U^{\dagger}(t)V_{\rm H} U(t)S_{x}U^{\dagger}(t) V^{\dagger}_{\rm H} U(t)\}.
\end{aligned}
\end{equation}
This equation now says the following: evolve the carbon magnetisation forward in time, apply a proton selective pulse sequence, reverse the initial evolution period, and measure the carbon FID (see Fig.~\ref{fig:OTOC_general_scheme}). An outline of our experimental strategy is given below, explicit details of individual elements are given in the following:

\begin{enumerate}
\item  
Initialisation: After thermalization we selectively saturate all $^{1}{\rm H}$ spins and initialize the state of the system in $\rho_{0}\propto S_{x}$ by applying a $(\pi/2)_{y}$ pulse. The system is further subjected to a RF filter selecting a sub-ensemble with a tighter RF distribution. The second step is important to mitigate pulse error effects throughout the OTOC measurement.  
\item 
Forward evolution: To this end, we engineer periodic evolution under a dipolar double-quantum Hamiltonian utilizing a novel proton-carbon synchronized pulse element, denoted TARDIS described in detail below. To zeroth order the dynamics are governed by, $U_{\rightarrow}(t_{n})\approx \exp(-i\bar{H}^{(0)}_{\rightarrow}t_{n})$, where $t_{n}=nt_c$, $n$ is the cycles of the pulse sequence element and $t_c$ the cycle time. 
\item 
Local perturbation (butterfly): To locally perturb the system we devise a pulse sequence element that selectively rotates a subset of the $\proton$ spins (we choose the methyl protons throughout):
\begin{equation}
    V_{\rm H}=\exp\left(-i\theta\vec{n}\cdot\sum_{i\in\text{methyl}}\vec{I^{i}}\right)
    \label{eq:butterfly_approx}
\end{equation}
where $\Vec{n}$ is the rotation axis of the perturbation, $\vec{I^{i}}=\left(I^{i}_x,I^i_y,I^i_z\right)$ is the angular momentum vector.
\item 
Backward evolution: To achieve backward evolution we engineer periodic evolution under a double-quantum Hamiltonian that to zeroth order satisfies
\begin{equation}
    \bar{H}^{(0)}_{\leftarrow}=-\bar{H}^{(0)}_{\rightarrow},
\end{equation} 
so that $U_{\leftarrow}(t)\approx U_{\rightarrow}^{\dagger}(t)$.
\item 
Readout: The OTOC is obtained as a modulation of the signal amplitude of the resulting $\carbon$ FID and can be directly obtained from the intensity of spectrum.
\end{enumerate}




\subsection{The TARDIS pulse sequences}

\begin{figure}
    \includegraphics[width=1\textwidth]{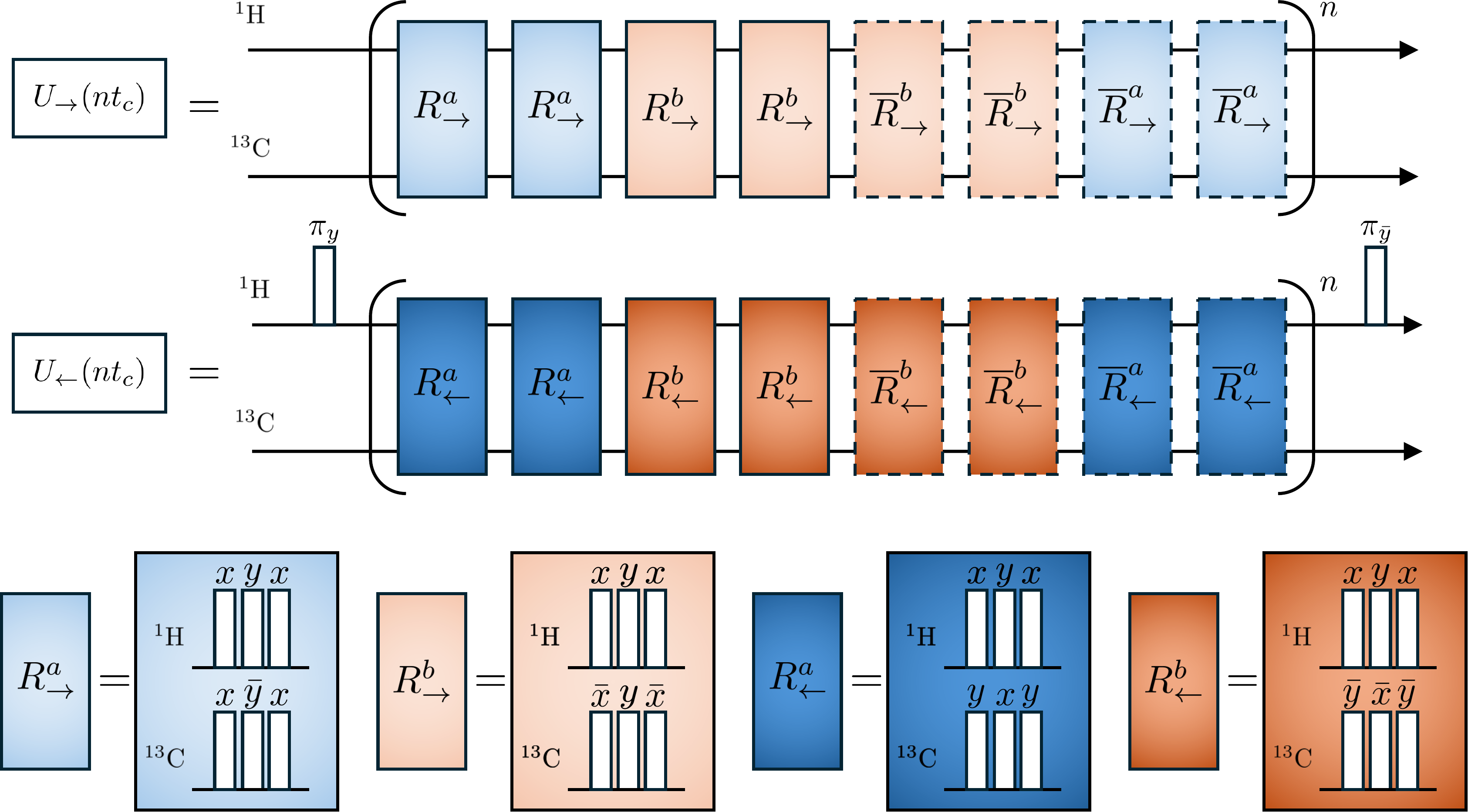}
    \caption{\label{fig:DREAM1} TARDIS-1 pulse sequence. (a) TARDIS-1 forward- and backward-evolution pulse sequences. Each forward-evolution cycle consists of 8 synchronized composite double-resonance pulses applied on the $\proton$ and the $\carbon$ channel: $U_{\rightarrow}(t_c)=R_{\rightarrow}^a R_{\rightarrow}^a R_{\rightarrow}^b R_{\rightarrow}^b \bar R_{\rightarrow}^b \bar R_{\rightarrow}^b \bar R_{\rightarrow}^a \bar R_{\rightarrow}^a$. Reversal of TARDIS-1 is achieved by a $\frac{\pi}{2}$ phase shift on the $\carbon$ pulses, combined with $\proton$ channel $\pi_{y}$ and $\pi_{\bar{y}}$ pulses applied before and after the reverse cycles. (b) Definition of composite pulses $R_{\rightarrow}^{a}$, $R_{\rightarrow}^{b}$, $R_{\leftarrow}^{a}$, 
        and $R_{\leftarrow}^{b}$. Each composite element consists of three synchronized $\pi/2$ sub-pulses with alternating 
        $x/y$ phases on both channels, with matched nutation amplitudes $\omega_{\mathrm{rf}}^{I} = \omega_{\mathrm{rf}}^{S}$ to ensure double-resonance conditions.}
\end{figure}

We first describe the generation of dipolar double-quantum evolution blocks, which are vital components of the OTOC-NMR scheme. These blocks are suitable targets for OTOC experiments because they encode spatially meaningful information and offer time-reversal capabilities~\cite{baum1985multiple}. To simultaneously reverse both homo- and heteronuclear dipolar interactions—while also providing favorable compensation for finite pulse widths and pulse errors—we devised two novel double-resonance Floquet driving protocols, denoted as TARDIS-1 and TARDIS-2. The interaction between the spin system and the externally applied RF pulses is described by
\begin{equation}
    H_{\rm rf}(t)=\omega_{\rm \text{rf}}^{I}(t)\left(I_{x}\cos\phi_{I}(t)-I_{y}\sin\phi_{I}(t)\right)+\omega_{\rm \text{rf}}^{S}(t)\left(S_{x}\cos\phi_{S}(t)-S_{y}\sin\phi_{S}(t)\right)
\end{equation}
Here $\omega_{\rm rf}^{I}$, $\omega_{\rm rf}^{S}$ are the nutation angular frequencies of the $\proton$, $\carbon$ spins under the RF field. While $\phi_{I}$, $\phi_{S}$ denote the phase of pulses in $\proton$, $\carbon$ channel respectively. Experimentally, the RF pulses are approximated as rectangular pulses, with a $\pi/2$ pulse width of $t_p=10.5\mu s$. The effect of the pulse sequence on the system is analyzed through an interaction picture decomposition
\begin{equation}
    U\left(t\right)=U_{\rm rf}\left(t\right)U_{\rm int}\left(t\right)
\end{equation}
Where
\begin{align}
    U_{\rm rf}(t)&=\mathcal T\exp\left[-i\int_{0}^{t}H_{\rm rf}(t^{\prime})dt^{\prime}\right],
    \\
    \tilde{U}_{\rm int}(t)&=\mathcal T\exp\left[-i\int_{0}^{t}\tilde{H}_{\rm int}(t^{\prime})dt^{\prime}\right].
\end{align}
$\tilde{H}_{\rm int}(t)$ represents the spin interactions in toggling frame, $\tilde{H}_{\rm int}(t)=U^{\dagger}_{\rm rf}(t)H_{\rm int}U_{\rm rf}(t)$. Under cyclicity we further have $U_{\rm rf}(nt_{c})=\mathbf{1}$, with $t_c$ being the cycle period, and $U(nt_{c})=\tilde{U}_{\rm int}^{n}(t_{c})$. According to Average Hamiltonian Theory (AHT)~\cite{haeberlen1968coherent}, the evolution operator over one period can be written as $U(t_{c})=\exp\left[-i\bar{H}t_{c}\right]$, with $\bar{H}$ being the effective or average Hamiltonian. Approximate expressions for $\bar{H}$ may be generated via the Magnus expansion: $\bar{H}=\bar{H}^{\left(0\right)}+\bar{H}^{\left(1\right)}+\bar{H}^{\left(2\right)}+\dots$. The first three terms are given by:
\begin{equation}
\label{eq:Magnus_high_order}
\begin{aligned}
    \bar{H}^{\left(0\right)}&=\frac{1}{t_{c}}\int_{0}^{t_{c}}\tilde{H}_{\rm int}\left(t^{\prime}\right)dt^{\prime},
    \\
    \bar{H}^{\left(1\right)}&=\frac{-i}{2t_{c}}\int_{0}^{t_{c}}dt_{2}\int_{0}^{t_{2}}dt_{1}\left[\tilde{H}_{\rm int}\left(t_{2}\right),\tilde{H}_{\rm int}\left(t_{1}\right)\right],
    \\
    \bar{H}^{\left(2\right)}&=\frac{-1}{6t_{c}}\int_{0}^{t_{c}}dt_{3}\int_{0}^{t_{3}}dt_{2}\int_{0}^{t_{2}}dt_{1}
    \left\{\left[\tilde{H}_{\rm int}\left(t_{3}\right),\left[\tilde{H}_{\rm int}\left(t_{2}\right),\tilde{H}_{\rm int}\left(t_{1}\right)\right] \right]\right.\\
    &\left.\quad+\left[\tilde{H}_{\rm int}\left(t_{1}\right),\left[\tilde{H}_{\rm int}\left(t_{2}\right),\tilde{H}_{\rm int}\left(t_{3}\right)\right]
    \right]
    \right\}.
\end{aligned}
\end{equation}



\subsubsection{TARDIS-1}
Fig.~\ref{fig:DREAM1} gives an overview of the TARDIS-1 sequence and its building blocks. Each forward-evolution cycle consists of 8 composite pulses applied in synchrony on the proton and carbon channel: 
\begin{align}
U_{\rightarrow}(t_c)=R_{\rightarrow}^a R_{\rightarrow}^a R_{\rightarrow}^b R_{\rightarrow}^b \bar R_{\rightarrow}^b \bar R_{\rightarrow}^b \bar R_{\rightarrow}^a \bar R_{\rightarrow}^a
\end{align}
As shown in Fig.~\ref{fig:DREAM1}(b), the building blocks $R_{\rightarrow}^a$ and $R_{\rightarrow}^b$ are given by the composition: $(\pi/2)_{x}(\pi/2)_{y/\bar{y}}(\pi/2)_{x}$. The result is a $\pi$ rotation along $(\mp 1, 1 ,0)$ affecting all Cartesian components of the spin Hamiltonian simultaneously. ``Barred" elements $\bar R_{\rightarrow}^b$, $\bar R_{\rightarrow}^a$ refer to $\pi$-phase-shifted counterparts (phase shift occurs on both channels). As indicated in Fig.~\ref{fig:DREAM1}(b), the $\frac{\pi}{2}$ pulses for protons and carbons are tuned to the same length. In practice this is achieved by matching the nutation amplitudes ($\omega_{\rm rf}^{I}=\omega_{\rm rf}^{S}$). A standard AHT calculation shows that the zeroth order average Hamiltonian is given by
\begin{equation}\label{eq:zeroth_order_magnus_tardis-1}
    \bar{H}^{\left(0\right)}_{\rightarrow}/2\pi=\frac{2}{3}\sum_{i\in ^{1}\rm{H}}(D_{Ci}+\frac{1}{2}J_{Ci})\cdot I^{i}_{z}S_{z}+\frac{1}{\pi}\sum_{i,j\in ^{1}\rm{H},i< j}D_{ij}\cdot(I^{i}_{x}I^{j}_{y}+I^{i}_{y}I^{j}_{x}),
\end{equation} 
indicating homonuclear double-quantum and heteronuclear $ZZ$ evolution. Time-reversal is achieved by applying a $\frac{\pi}{2}$ phase shift to all pulses on the carbon channel of $U_{\rightarrow}$, combined with $\pi_{y}$ and $\pi_{\bar{y}}$ pulses on the proton channel applied before and after the reverse cycles:
\begin{align}
U_{\leftarrow}(t_c)=\pi^{I}_{y}{[}R_{\leftarrow}^a R_{\leftarrow}^a R_{\leftarrow}^b R_{\leftarrow}^b \bar R_{\leftarrow}^b \bar R_{\leftarrow}^b \bar R_{\leftarrow}^a \bar R_{\leftarrow}^a{]}\pi^{I}_{\bar{y}}
\end{align}
Again, a standard AHT calculation shows that to zeroth order $\bar{H}^{\left(0\right)}_{\rightarrow}=-\bar{H}^{\left(0\right)}_{\leftarrow}$.

\begin{figure}
    \includegraphics[width=1\textwidth]{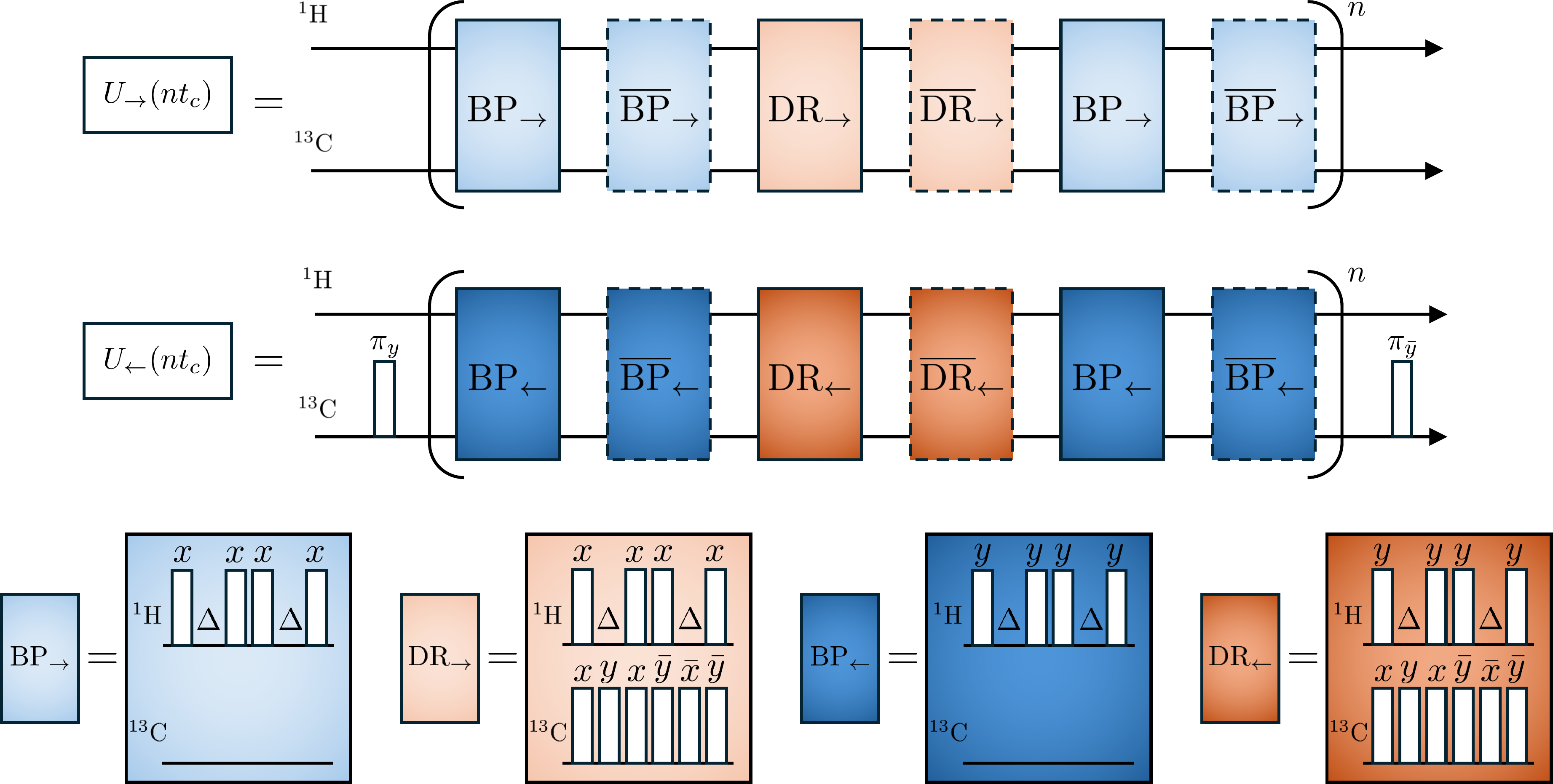}
    \caption{\label{fig:DREAM2} TARDIS-2 pulse sequence. (a) TARDIS-2 forward- and backward-evolution pulse sequences. 
        One forward-evolution cycle consists of six composite pulse blocks on the 
        $^{1}\mathrm{H}$ and $^{13}\mathrm{C}$ channels: 
        $U_{\rightarrow}(t_c) = \mathrm{BP}_{\rightarrow}\,\overline{\mathrm{BP}}_{\rightarrow}\,
        \mathrm{DR}_{\rightarrow}\,\overline{\mathrm{DR}}_{\rightarrow}\,
        \mathrm{BP}_{\rightarrow}\,\overline{\mathrm{BP}}_{\rightarrow}$. 
        The BP block is adapted from the Baum–Pines sequence and acts only on $^{1}\mathrm{H}$, 
        while the DR block is a double-resonance element acting on both channels. 
        Reversal of TARDIS-2 is obtained by a $\pi/2$ phase shift on all 
        $^{1}\mathrm{H}$ pulses and by applying $^{13}\mathrm{C}$ channel $\pi_{y}$ and $\pi_{\bar{y}}$ pulses before and after the reverse cycles. 
        (b) Definition of the pulse blocks $\mathrm{BP}_{\rightarrow}$, 
        $\mathrm{DR}_{\rightarrow}$, $\mathrm{BP}_{\leftarrow}$, and 
        $\mathrm{DR}_{\leftarrow}$. Each BP block consists of four synchronized $\pi/2$ 
        sub-pulses on the $^{1}\mathrm{H}$ channel with interleaved delays of the same length of a $\pi/2$ pulse ($\Delta=t_p$), while each DR 
        block contains alternating $\pi/2$ sub-pulses on both channels with $x/y$ phase patterns. 
        The nutation amplitudes $\omega_{\mathrm{rf}}^{I}$ and $\omega_{\mathrm{rf}}^{S}$ 
        are matched to maintain double-resonance conditions.
}
\end{figure}

\subsubsection{TARDIS-2}
Fig.~\ref{fig:DREAM2} gives an overview of the TARDIS-2 sequence and its building blocks. The forward evolution is composed of the concatenation of 6 pulse elements:
\begin{align}
U_{\rightarrow}(t_c)=\text{BP}_{\rightarrow}\overline{\text{BP}}_{\rightarrow}\text{DR}_{\rightarrow}\overline{\text{DR}}_{\rightarrow}\text{BP}_{\rightarrow}\overline{\text{BP}}_{\rightarrow}
\end{align}
The $\text{BP}_{\rightarrow}$ element is adapted from the Baum-Pines sequence~\cite{baum1985multiple}. The $\text{DR}_{\rightarrow}$ double-resonance block is summarized in Fig~\ref{fig:DREAM2}(b). Both the $\text{BP}_{\rightarrow}\overline{\text{BP}}$ block and the $\text{DR}_{\rightarrow}\overline{\text{DR}}$ block are cyclic and generate a zeroth order average Hamiltonian of the type:
\begin{equation}
    \bar{H}^{\left(0\right)}/2\pi=\eta_{\rm heter}\sum_{i\in ^{1}\rm{H}}(D_{Ci}+\frac{1}{2}J_{Ci})\cdot I^{i}_{z}S_{z}+\eta_{\rm homo}\sum_{i,j\in ^{1}\rm{H},i< j}D_{ij}\cdot(I^{i}_{y}I^{j}_{y}-I^{i}_{x}I^{j}_{x}).
\end{equation} 
Similar to TARDIS-1, the TARDIS-2 average Hamiltonian generates homonuclear double-quantum and heteronuclear $ZZ$ evolution. The respective scaling factor are $\{\eta_{\rm heter},\eta_{\rm homo}\}=\{0,1\}$ for the $\text{BP}_{\rightarrow}\overline{\text{BP}}$ block and $\{\eta_{\rm heter},\eta_{\rm homo}\}=\{\frac{2}{3},1\}$ for $\text{DR}_{\rightarrow}\overline{\text{DR}}$ block. By interleaving these two blocks, the scaling factors are averaged, leading to the overall scaling factor for one TARDIS-2 forward cycle: $\{\eta_{\rm heter},\eta_{\rm homo}\}=\{\frac{2}{9},1\}$ (This may be seen by utilizing the methods developed in reference~\cite{hohwy1998systematic}). Compared to TARDIS-1, TARDIS-2 has favorable properties in systems where the $\carbon$ is directly bonded to a $\proton$ (for example, \sampleA). In such cases a strong $\carbon-\proton$ coupling can partially truncate the homonuclear dipolar coupling, leading to an undesirable OTOC slowdown. 

Time-reversal is achieved by applying a $\frac{\pi}{2}$ phase shift to all pulses on the proton channel of $U_{\rightarrow}$, combined with $\pi_{y}$ and $\pi_{\bar{y}}$ pulses on the carbon channel applied before and after the reverse cycles:
\begin{align}
U_{\leftarrow}(t_c)=\pi^{S}_{y}{[}\text{BP}_{\leftarrow}\overline{\text{BP}}_{\leftarrow}\text{DR}_{\leftarrow}\overline{\text{DR}}_{\leftarrow}\text{BP}_{\leftarrow}\overline{\text{BP}}_{\leftarrow}{]}\pi^{S}_{\bar{y}}
\end{align}
A straightforward calculation shows that the zeroth order average Hamiltonians are related by: $\bar{H}^{\left(0\right)}_{\rightarrow}=-\bar{H}^{\left(0\right)}_{\leftarrow}$.

\subsubsection{TARDIS-2 sequences with on-site field}\label{sec:dream_transverse}

A uniform on-site field can be introduced for the $\proton$ by the following phase shifting scheme (adapted from \cite{wei2018exploring}): In the $m$-th cycle of the forward TARDIS-2 sequence, the earlier $\text{BP}_{\rightarrow}\overline{\text{BP}}_{\rightarrow}$ block is phase shifted by $(3m-3)\phi$ on the $\proton$ channel, the $\text{DR}_{\rightarrow}\overline{\text{DR}}_{\rightarrow}$ block by $(3m-2)\phi$, and the later $\text{BP}_{\rightarrow}\overline{\text{BP}}_{\rightarrow}$ block by $(3m-1)\phi$. The evolution operator for each cycle is given by:
\begin{equation}
\begin{aligned}
    U(t_{c},0)=&\exp\!\left(i\cdot2\phi I_{z}\right)\exp\!\left(-i\bar{H}^{\left(0\right)}_{\text{BP}_{\rightarrow}\overline{\text{BP}}}\cdot t_{c}/3\right)\exp\!\left(-i\cdot 2\phi I_{z}\right)\\
    &\quad\quad\cdot
    \exp\!\left(i\phi I_{z}\right)\exp\!\left(-i\bar{H}^{\left(0\right)}_{\text{BP}_{\rightarrow}\overline{\text{BP}}}\cdot t_{c}/3\right)\exp\!\left(-i\phi I_{z}\right)\\
    &\quad\quad\cdot
    \exp\!\left(-i\bar{H}^{\left(0\right)}_{\text{BP}_{\rightarrow}\overline{\text{BP}}}\cdot t_{c}/3\right)\\
    \approx& \exp\!\left(i\cdot3\phi I_{z}\right)\exp\!\left[-i\left(\bar{H}^{\left(0\right)}_{\rightarrow}+\frac{3\phi}{t_c}I_{z}\right)t_c\right]\\
    U(2t_{c},t_c)=&\exp\!\left(i\cdot3\phi I_{z}\right)U(t_{c},0)\exp\!\left(-i\cdot3\phi I_{z}\right)\\
    &\quad\quad ......\\
    U(nt_{c},(n-1)t_c)=&\exp\!\left(i\cdot3(n-1)\phi I_{z}\right)U(t_c,0)\exp\!\left(-i\cdot3(n-1)\phi I_{z}\right)
\end{aligned}
\end{equation}
Hence the $n$-cycle propagator is:
\begin{align}
    U(nt_{c},0)=U(nt_{c},(n-1)t_c)...U(2t_{c},t_c)U(t_{c},0)\approx &\exp\!\left(i\cdot3n\phi I_{z}\right)\left[\exp\!\left[-i\left(\bar{H}^{\left(0\right)}_{\rightarrow}+\frac{3\phi}{t_c}I_{z}\right)t_c\right]\right]^n\\
    =&\exp\!\left(i\cdot3n\phi I_{z}\right)\exp\!\left(-i\bar{H}^{(0)\prime}_{\rightarrow}\cdot nt_c\right)
\end{align}
with the effective zero-order average Hamiltonian $\bar{H}^{(0)\prime}=\bar{H}^{(0)}+\frac{3\phi}{t_c}I_{z}$. The prefactor $\exp\!\left(i\cdot3n\phi I_{z}\right)$ can be compensated by phase shifting the perturbation sequence by $3n\phi$. An inverse on-site field can be introduced for the backward evolution using an analogous phase-shifting scheme. Finally, the zero-order average Hamiltonian resulting from the on-site field TARDIS-2 sequence is given by
\begin{equation}
    \bar{H}^{\left(0\right)\prime}_{\rightarrow}/2\pi=\frac{2}{9}\sum_{i\in ^{1}\rm{H}}(D_{Ci}+\frac{1}{2}J_{Ci})\cdot I^{i}_{z}S_{z}+\sum_{i,j\in ^{1}\rm{H},i< j}D_{ij}\cdot(I^{i}_{y}I^{j}_{y}-I^{i}_{x}I^{j}_{x})+hI_{z}
\end{equation} 
With $h=\frac{3\phi}{2\pi t_c}$.

\subsection{Initial State Preparation and RF inhomogeneity}

\begin{figure}
    \includegraphics[width=1\textwidth]{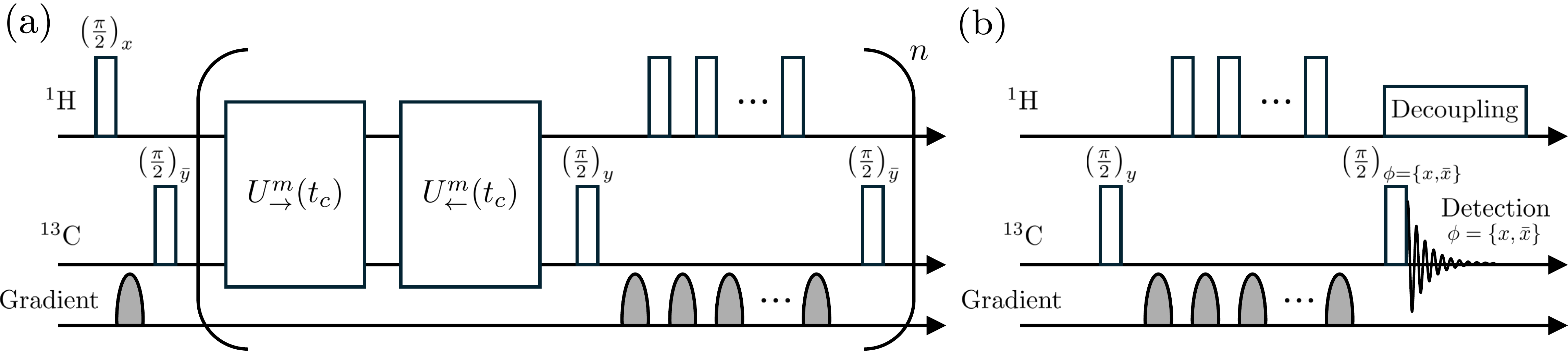}
    \caption{\label{fig:Initialization_measurement}Pulse sequences for (a)Initial state preparation and (b)Readout. 
(a) The sequence prepares the system in an initial state $\rho_{0} \propto S_{x}$ by rotating the $^{13}\mathrm{C}$ thermal magnetization and destroying the $^{1}\mathrm{H}$ magnetization. The $^{13}\mathrm{C}$ magnetization is then subjected to an RF pre-selection filter consisting of $m = 10$ TARDIS forward/backward cycles. This filtering step selects ensemble members with small RF inhomogeneity deviations and enhances homogeneity across the sample. Subsequently, a series of gradient and random proton pulses selects pure $^{13}\mathrm{C}$ $x$-magnetization. The process is repeated for $n = 2$. 
(b) Before detection, the density matrix is filtered to retain the $S_{x}$ component using the same scheme as the second half of the initialization filter. The $^{13}\mathrm{C}$ FID is then recorded under proton decoupling.
}
\end{figure}

The system initially starts out at thermal equilibrium, $\rho_{\text{eq}}=\epsilon_{\proton}I_{z}+\epsilon_{\carbon}S_{z}$, where $\epsilon_{\proton}$, $\epsilon_{\carbon}$ is the polarization of $\proton$ and $\carbon$ respectively. 
From here, we want to prepare the system in $\rho_{0}\propto M=S_{x}$.
Moreover, there is a relatively large RF inhomogeneity across our sample (i.e. molecules at different physical locations experience a different strength of the applied RF pulses, which results in miscalibrated gates).
To account for this, we want to further pre-select a smaller sample sub-region with a relatively consistent RF strength, by decohering the spin population outside this region to the identity.

We achieve both of the above goals via the pulse sequence scheme shown in Fig.~\ref{fig:Initialization_measurement}(a).
An initial $\pi/2$ pulse on the proton channel followed by a gradient pulse spoils the proton magnetization.
Afterwards the $\carbon$ magnetization is rotated onto the $x$-axis and the whole system subjected to an RF pre-selection filter.
The filter consists of $m=10$ cycles of TARDIS (TARDIS-1 for \sampleB experiments and TARDIS-2 for \sampleA experiments) forward evolution followed by $m=10$ cycles of backward evolution.
This echo-based filter selectively saturates ensemble members with large RF deviations from the nominal reference.
Subsequently, a string of gradient and random proton pulses selects pure $\carbon$ $x$-magnetization.
The combination of gradient and proton pulses is used to ensure destruction of all $\proton$ coherences after the TARDIS filter.
The process is repeated twice [$n=2$ in Fig.~\ref{fig:Initialization_measurement}(a)] to further enhance RF homogeneity.

\begin{figure}
    \includegraphics[width=0.5\textwidth]{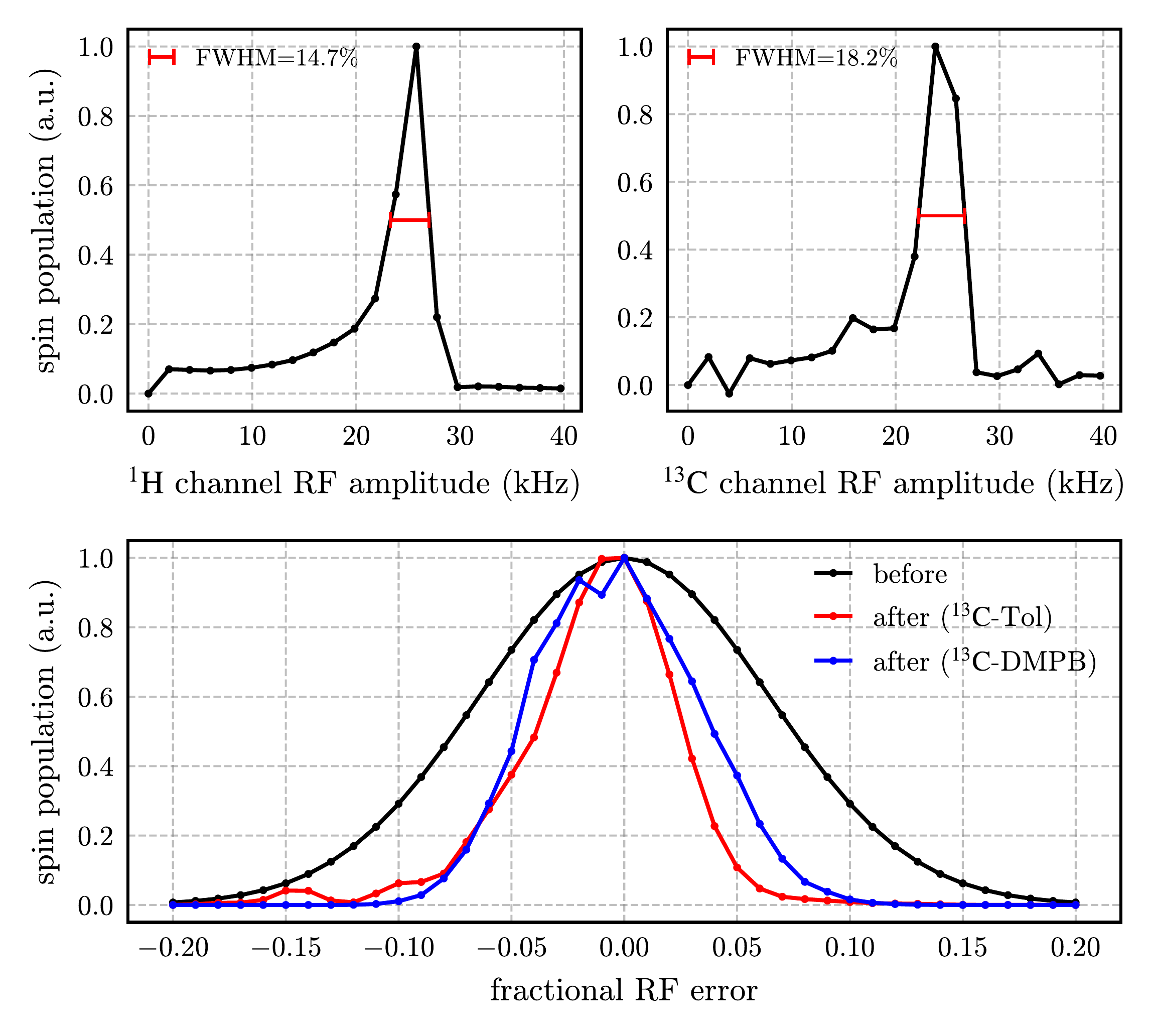}
    \caption{\label{fig:RF_error} Upper left: RF inhomogeneity profile of $\proton$ coil. Upper right: RF inhomogeneity profile of $\carbon$ coil. Bottom: fractional RF error for \sampleA and \sampleB before and after the pre-selection sequence. For \sampleB, the filtered RF error distribution is calculated using the dipolar couplings of the $u_{4}$ PMF surface.}
\end{figure}

To model the filtered RF inhomogeneity in the OTOC experiment, we explicitly measure the RF distribution profile produced by the NMR coil, and model the pre-selection sequence.
To determine the RF amplitude distribution of the $\proton$ coil and the $\carbon$ coil, we prepare $12.5 \%$ v/v toluene-1-$\carbon$ in benzene-$d_6$ of the same volume as other samples used in OTOC experiment.
The RF amplitude inhomogeneity was obtained by Fourier transform of the nutation curve, following the method in Ref.~\cite{guenneugues1999method}.
For both the $^{13}\text{C}$ channel and the $^1\text{H}$ channel, we observe that the profile peaks at around ~25kHz, with a FWHM around $15\%$ [Fig.~\ref{fig:RF_error}(top)].
We thus approximate the pre-filtered fractional RF error $\epsilon$ for both $^{13}$C and $^{1}$H by a Gaussian with standard deviation $0.064$, i.e. $p_{\text{coil}}(\epsilon)=\mathcal{N}(\mu=0,\sigma=0.064)$.
To model the effect of filtering, we calculate an $\epsilon$ dependent damping factor $F(\epsilon)$, corresponding to the return probability of the $X$ polarization on the $^{13}$C spin after a single block of the sequence in Fig.~\ref{fig:Initialization_measurement}(a)
\begin{align}
    F(\epsilon)=\text{tr}\left[S_{x}(U_{\leftarrow}(\epsilon))^{10}(U_{\rightarrow}(\epsilon))^{10}S_{x}(U_{\rightarrow}^{\dagger}(\epsilon))^{10}(U_{\leftarrow}^{\dagger}(\epsilon))^{10}\right]
\end{align}
Here $U_{\rightarrow}(\epsilon)$/$U_{\leftarrow}(\epsilon)$ represent TARDIS forward/backward evolution with RF error of $\epsilon$. The 'filtered' RF error distribution $p(\epsilon)$ can then be written as:
\begin{align}
    p(\epsilon)&=p_{\text{coil}}(\epsilon)\times F^{2}(\epsilon)
\end{align}
Note that $F(\epsilon)$ is squared since the filtering cycle is executed twice. The `filtered' RF error distribution is shown in Fig.~\ref{fig:RF_error}(bottom).

\subsection{Perturbation}

\begin{figure}
    \includegraphics[width=0.5\textwidth]{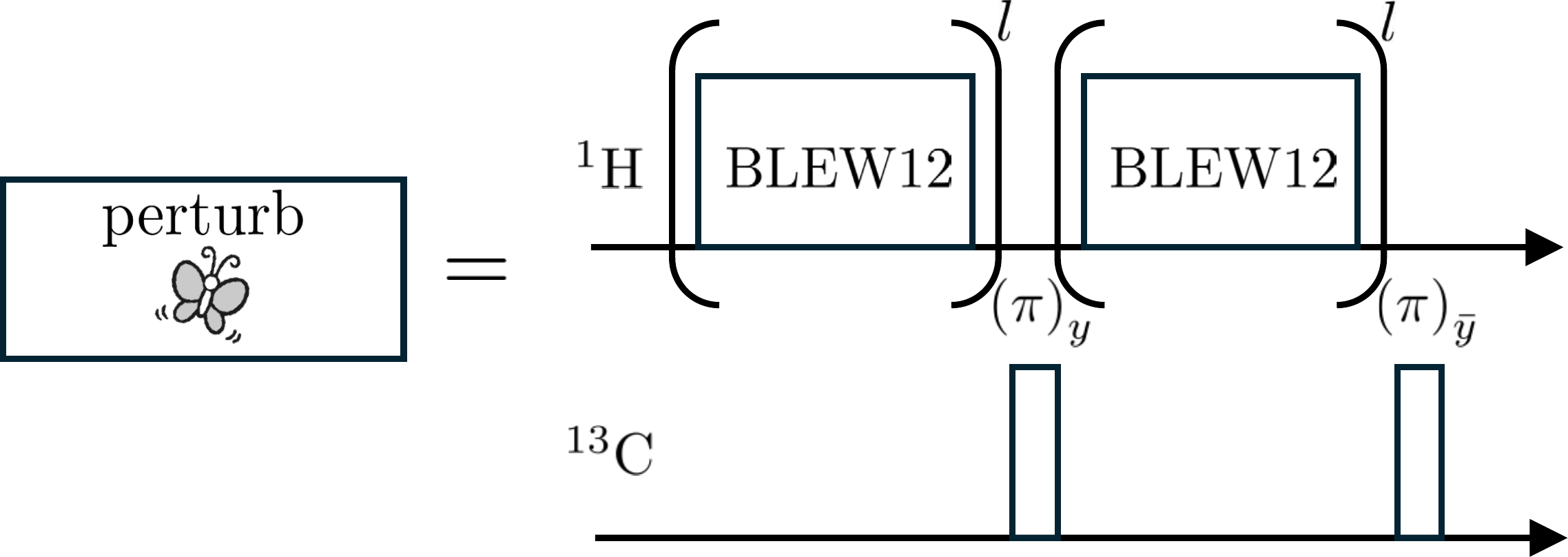}
    \caption{\label{fig:perturbation} Perturbation pulse sequence. The perturbation sequence is based on the homonuclear dipolar decoupled pulse sequence BLEW12. Two $\carbon$ $\pi$ pulses before and after the later BLEW12 block serves to refocus the heteronuclear dipolar coupling. $l=2$ across all experiments.}.
\end{figure}


For both \sampleA and \sampleB we choose the methyl protons as the butterfly spins. To selectively address the methyl group we utilize their chemical shift difference with respect to the remaining ring protons. As perturbation we choose a local rotation of the methyl protons, which may be achieved through the application of a BLEW12~\cite{burum1981low} pulse sequence element (see Fig.~\ref{fig:perturbation}). The BLEW12 elements on the $\proton$ channel removes their dipolar couplings and generates the following zeroth order average Hamiltonian:
\begin{align}
\bar{H}^{(0)}_{\text{BLEW12}}&=\eta\left[2\pi\sum_{i\in \proton} \left(D_{Ci}+\frac{1}{2}J_{Ci}\right) \left(\vec{n}\cdot \vec{I}^{i}\right)S_{z} + \sum_{i\in \proton}\omega_{i}(\vec{n}\cdot \vec{I}^{i})\right]
\end{align}
Where $\eta=\frac{2\sqrt{5}}{3\pi}$ is the scaling factor and $\vec{n}=\frac{1}{\sqrt{5}}(0,2,1)$ is the effective rotation axis. The sandwich of $\carbon$ $\pi$ pulses before and after the second $\text{BLEW12}$ block refocus the heteronuclear dipolar couplings. In total we then find
\begin{align}
V_{\rm H}=\exp\left(-i\eta\times 2l\sum_{i\in \proton}\omega_{i}(\vec{n}\cdot \vec{\sigma}_{i})\cdot12t_{p}\right),
\end{align}
where $t_{p}$ is the duration of a single pulse, and 12$t_p$ is the length of one BLEW12 cycle. The factor $2l$ corresponds to the total number of BLEW12 cycles applied, and $l$ is set to 2 in all experiments. If the carrier frequency is set to approximately match the ring proton resonances, equation \ref{eq:butterfly_approx} follows.

\subsection{Readout}

Before detection, the density matrix is once again filtered to isolate its $S_{x}$ component. The filtering scheme is identical to the second half of the initialization filter. The carbon FID is recorded under proton decoupling (see Fig.~\ref{fig:Initialization_measurement}(b)). Since each molecule contains only one $\carbon$, proton decoupling leads to a single peak in the carbon spectrum. After phase-correction, the magnitude of the carbon magnetization is extracted by fitting a Lorentzian line shape function to the spectrum.

\subsection{Additional OTOC and learning experiments}

\begin{figure}
    \centering
    \includegraphics[width=\linewidth]{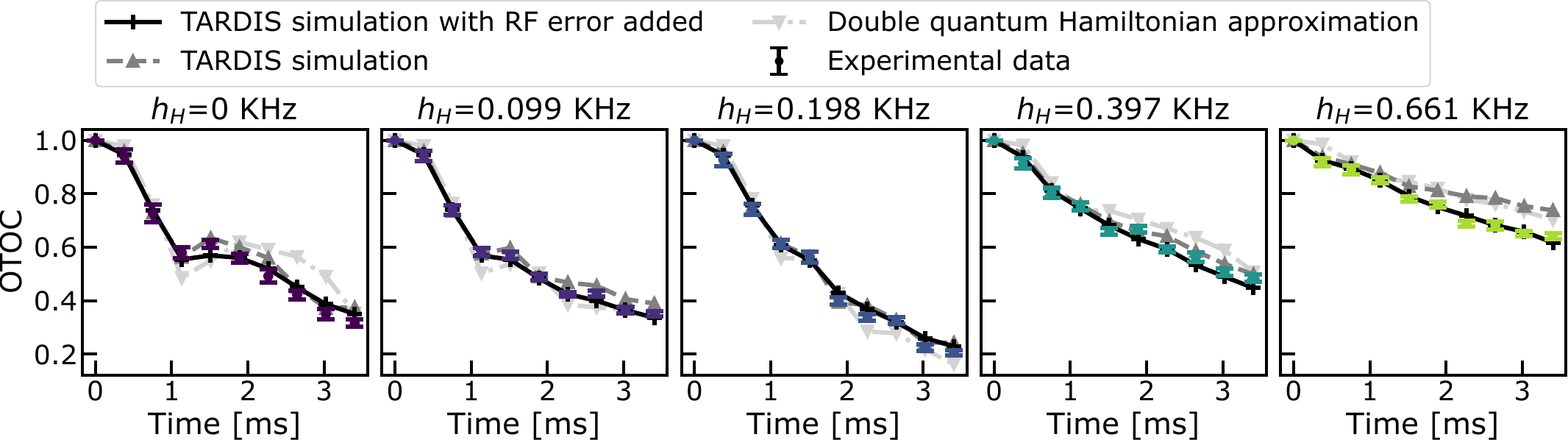}
    \caption{Extended data showing OTOCs taken with five different on-site fields, using the experimental protocol defined in Sec.~\ref{sec:dream_transverse}.
    This data was used in the generation of Fig.~2(e) of the main text.
    Colored points denote experimental data (error bars are $2\sigma$ confidence intervals).
    Lines denote three levels of simulation: the double quantum Hamiltonian approximation (i.e. zeroth-order Magnus expansion), a full simulation of the TARDIS-2 sequence (and accurate simulation of BLEW12 butterfly pulse) without RF inhomogeneity, and the same, averaged over the externally-determined RF inhomogeneity given in Fig.~\ref{fig:RF_error}(bottom, blue).}
    \label{fig:all_toluene_otocs}
\end{figure}

\begin{figure}
    \centering
    \includegraphics[width=0.5\linewidth]{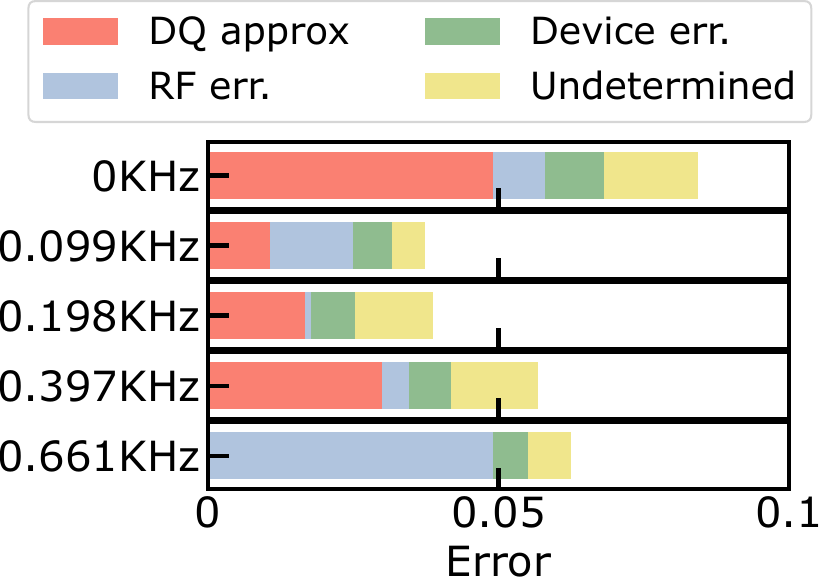}
    \caption{Budgeting the error in the simulations of Fig.~\ref{fig:all_toluene_otocs}.
    For each of the five experiments, we split the observed error between the DQ approximation and the experimental data into components (calculated by measuring the improved RMSE as the accuracy of the simulation increases).
    After subtracting the observed error bars from the experimental data, the residual error is marked as undetermined.}
    \label{fig:toluene_error_budget}
\end{figure}

\begin{figure}
    \centering
    \includegraphics[width=0.7\linewidth]{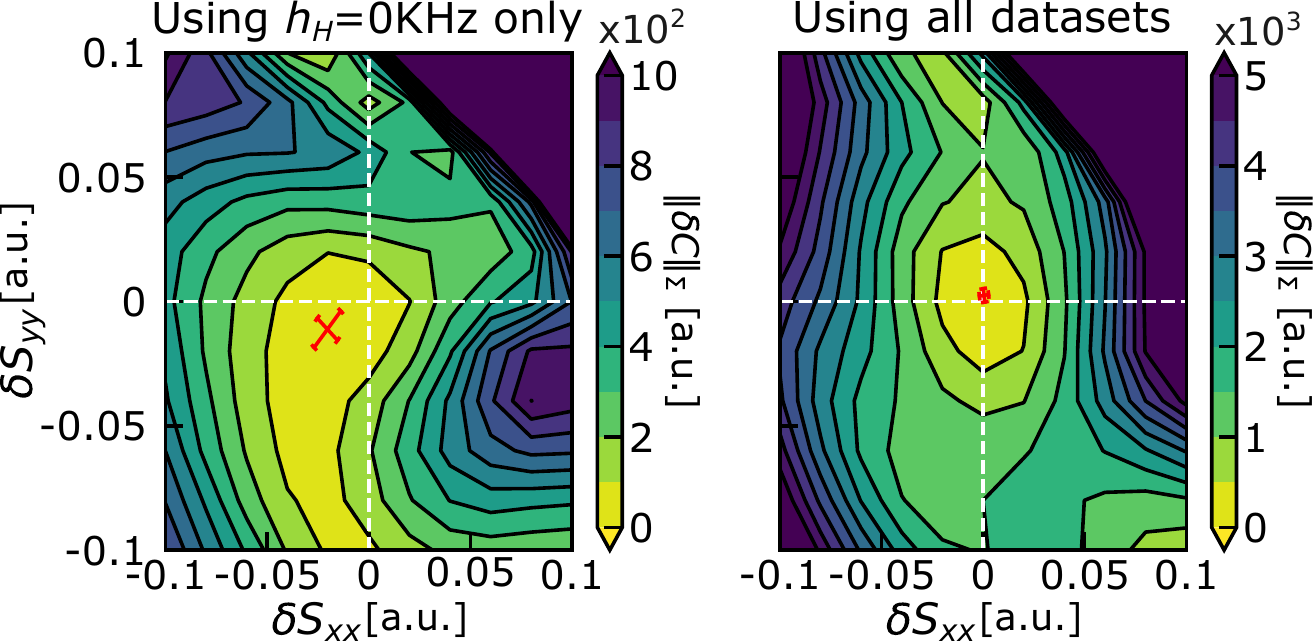}
    \caption{Learning the two free components of the Saupe tensor (Eq.~\eqref{eq:RDC_Simplified}) from either (left) a single dataset, or (right) all data shown in Fig.~\ref{fig:all_toluene_otocs}.
    We show contour plots of the covariance-weighted squared error as a function of the two free components of Eq.~\eqref{eq:RDC_Simplified}, $S_{xx}$ and $S_{yy}$, in terms of their shifts from the reference values given in Tab.~\ref{tab:9Q_old_sample} (white dashed liens).
    The minima of the two surfaces is found by fitting a quadratic function and marked in red.
    The size of the red bars give $2\sigma$ intervals along the principal components of the covariance matrix, calculated by resampling the experimental data and reperforming the fit.
    }
    \label{fig:saupe_tensor_learning}
\end{figure}

Here we give additional experimental datasets that were not added to the main text due to space constraints.
In Fig.~\ref{fig:all_toluene_otocs} we show an extended set of OTOCs with different onsite potentials, and plot behind this classical approximations of these curves at different levels of theory.
In Fig.~\ref{fig:toluene_error_budget}, we expand on this by showing an error budget for the curves in Fig.~\ref{fig:all_toluene_otocs}, by dividing the error into different levels of theory.
We observe that our classical modelling reproduces the experimental data well across the entire curve; indeed the dataset shown in the main text has the largest residual error of the datasets shown here.
Finally, in Fig.~\ref{fig:saupe_tensor_learning}, we demonstrate using either a single or five OTOC curves to learn the two free components of the Saupe tensor (Eq.~\eqref{eq:RDC_Simplified}), which were otherwise determined as part of the MQC fitting procedure described in Sec.~\ref{sec:MQC}.
We observe that when using a single dataset our cost function is relatively insensitive to a correlated change between the two components, which indicates that the data is low in quantity.
However, this disappears when we use all datasets, and we obtain 95$\%$ confidence intervals that are only a factor $2$ larger than the reference data.

\begin{figure}
    \centering
    \includegraphics[width=0.8\linewidth]{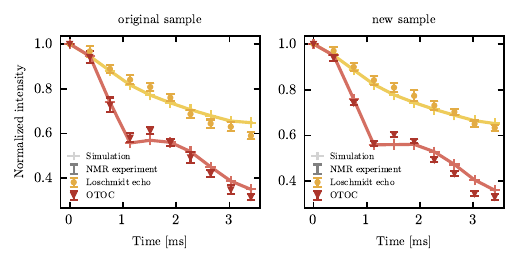}
    \caption{Retaken dataset shown in Fig. 2(c) of the main text on a new sample (right), compared with the dataset from the original sample (left, replotted from the main text), demonstrating reproducibility. Experimental NMR data points are shown with $2\sigma$ confidence intervals (CIs) and compared to numerical simulations (lines).
    }
    \label{fig:reproducibility}
\end{figure}

In addition, to test the reproducibility of our experiments, we prepared a new \sampleA sample and repeated the HMQC experiment. Fitting of the new HMQC spectrum shows that, compared to the original sample, the components of the Saupe order tensor differ by less than 0.005, while the chemical shift terms and $J_{C1}$ vary by less than 30 Hz. We attribute these small differences to variations in the toluene concentration: since toluene is highly volatile and the sample preparation involves mixing toluene and EBBA at 358 K, different degrees of toluene loss may occur during the preparation of separate samples. Subsequently, we remeasured the dataset shown in Fig. 2(c) of the main text using the new sample, and the comparison is presented in Fig.~\ref{fig:reproducibility}. Although the two datasets were acquired a year apart—during which drift in the magnetic field, amplifier performance, and other experimental conditions could have occurred—the agreement between experiment and simulation remains consistent after a simple recalibration of the spectrometer. These results demonstrate the high reproducibility of our OTOC experiment.

\subsection{HETCOR Experiment}
\begin{figure}
    \includegraphics[width=0.5\textwidth]{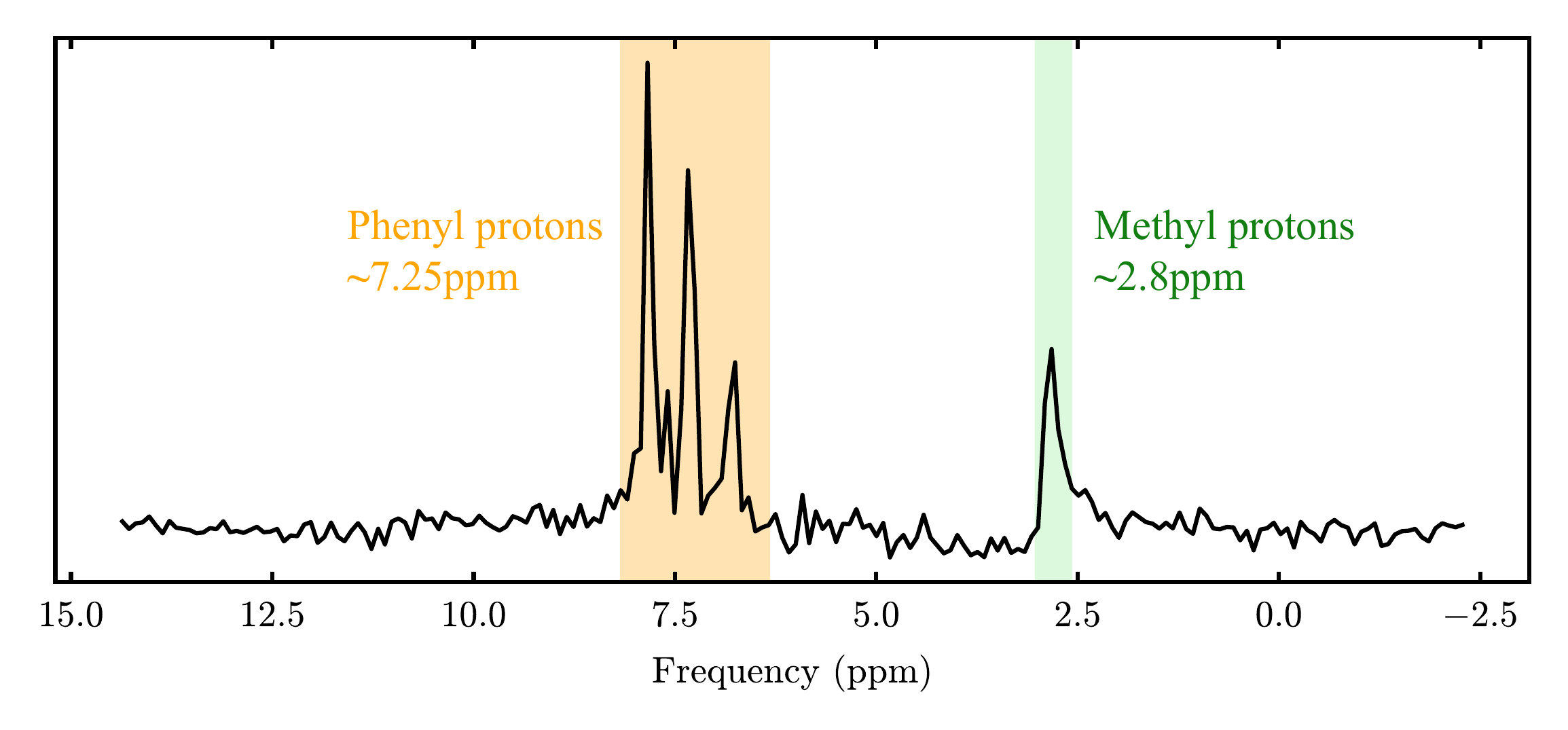}
    \caption{\label{fig:15Q_HETCOR} 
    HETCOR spectrum of \sampleB. CP time = 2.5~ms; 100~$t_1$ increments, one scan each. $\pi/2$ pulse = 10.5~$\mu$s, giving an effective $F_1$ width of 8363~Hz (BLEW12-scaled). $F_2$ width = 15~kHz. The 1D projection was obtained by integrating the $F_2$ peak, zero-filling to 200 points, and Fourier transforming along $F_1$. The peak at 2.8~ppm (shaded in green) is assigned to the methyl protons, whereas the peaks around 7.25~ppm (shaded in yellow) are attributed to the ring protons.
}
\end{figure}

$^{1}\text{H}$ chemical shift information is a prerequisite for \sampleB OTOC-learning experiment. Because of large number of transitions, no resolved peak is observed in $^{1}\text{H}$ single quantum spectrum, all the readout needs to be performed in $^{1}\text{H}$ decoupled $^{13}\text{C}$ single quantum spectrum. Thus, we performed heteronuclear correlation (HETCOR) experiment, with a sequence adapted from \cite{hong1996measurement}, shown in Fig.~\ref{fig:MQC_pulse_sequence}(c). The obtained spectrum is shown in Fig.~\ref{fig:15Q_HETCOR}.

The peak at 2.8 ppm was assigned to methyl protons, while the three resolved peaks around 7.25 ppm correspond to phenyl protons. Assignment of these peaks to individual protons was not successful. We attempted HETCOR experiment with varying cross polarization buildup times to obtained additional information for peak assignment. However, since only the enriched carbon is observable in the $^{13}\text{C}$ spectrum, insufficient information was obtained for reliable peak assignment. Therefore, we approximate that all the phenyl protons have chemical shift of 7.25 ppm. In OTOC experiment, We chose the carrier frequency to be on resonance with the phenyl protons. The relative chemical shift for \sampleB is summarized in Table \ref{tab:15Q_chemical_shift}.

\begin{table}[h!]
  \centering
  \caption{Relative chemical shift (in Hz) for \sampleB, determined from the HECTOR experiment, }
  \label{tab:15Q_chemical_shift}
  \begin{tabular}{p{5cm} p{2cm}}
    \hline
    Parameter & Value \\
    \hline
    $\omega_{1}/2\pi,\omega_{2}/2\pi,...,\omega_{8}/2\pi$ & 0 \\
    $\omega_{9}/2\pi,\omega_{10}/2\pi,...,\omega_{14}/2\pi$ & -2225 \\
    \hline
  \end{tabular}
\end{table}

\subsection{MQC Experiments}\label{sec:MQC}

\begin{figure}
    \includegraphics[width=\textwidth]{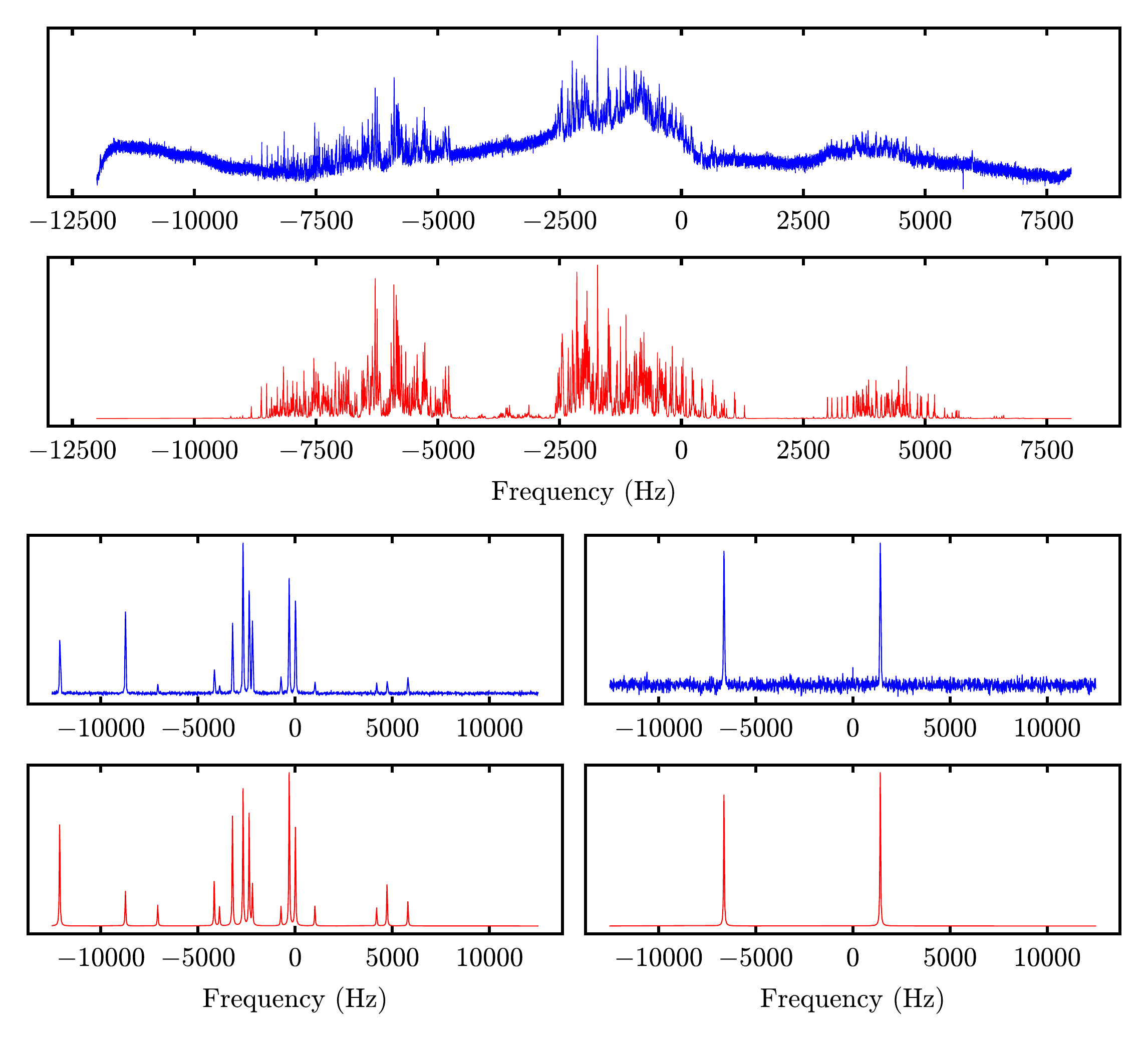}
    \caption{\label{fig:9Q_HMQC_fitting} Experimental (blue) and simulated (red) single-quantum spectrum (top), and HMQC spectra for $\Delta m_{S}=0$, $\Delta m_{I}=7$ (bottom left), 8 (bottom right). Single-quantum spectrum was taken to be one-shot. 32k data points are acquired and the spectral width is 20kHz. For HMQC experiments, $\tau=10\,\text{ms}$, 2500 $t_1$ increments, 6 scans per increment. Spectrum width is 25kHz for $F_{1}$ and 20kHz for $F_{2}$. The magnitude spectrum of $F_{1}$ is obtained by summing the absolute values of the Fourier transforms across different points in $F_{2}$. The average linewidth is 60Hz.}.
\end{figure}

To cross-validate the parameters inferred from OTOC-NMR Hamiltonian learning and to establish full or partial ground-truth reference data we employ a spectral estimation strategy based on multiple-quantum-coherence (MQC) NMR~\cite{field2003multiple,weitekamp1982determination}. In general spectral estimation via a standard pulse-acquire spectrum is difficult due to the large number of observable NMR transitions fulfilling the selection rule $\Delta M_{z}=1$. To simplify spectral complexity and reduce spectral crowding multiple-quantum techniques aim to indirectly observe higher-order NMR transitions with $\Delta m_{I}\neq 1$. 

\subsubsection{MQC spectra for Toluene}
\begin{figure}
    \includegraphics[width=1\textwidth]{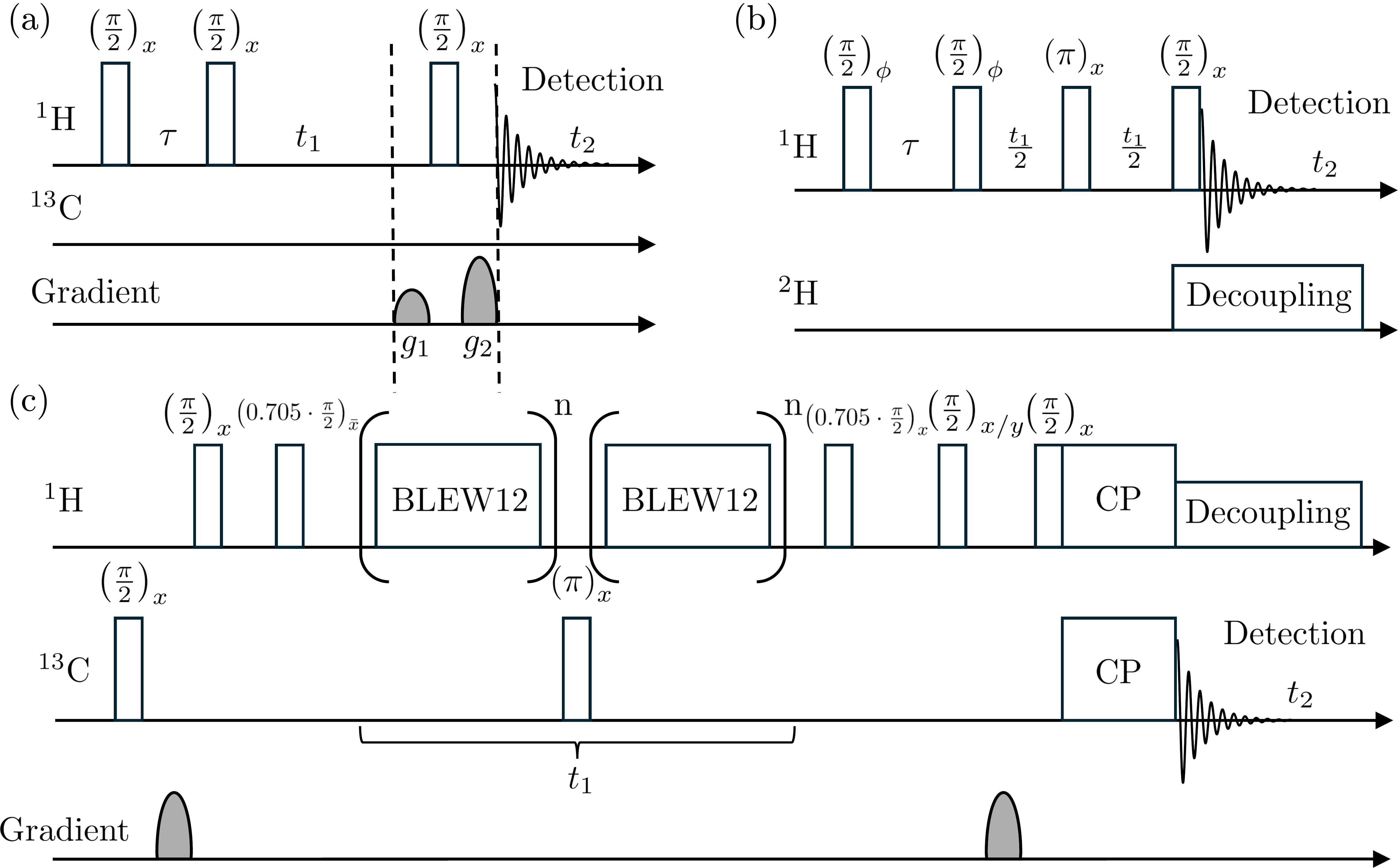}
    \caption{\label{fig:MQC_pulse_sequence} Pulse sequences for (a) HMQC experiment performed on \sampleA. (b)  Proton MQC experiment performed on \sampleC. (c) HETCOR experiment performed on \sampleB. (a) High coherence orders are generated using two $\pi/2$ pulse with a free evolution in between. Free evolution during the $t_{1}$ period leads to phase-encoding onto individual quantum coherence terms. A subsequent $\pi/2$ pulse converts coherence order operators back into observable 1Q terms which are read out during the $t_{2}$ period. Two gradient pulses are applied to filter out signal components originating from a specific coherence order operator. The gradient strength may be adjusted according to the selection rule $g_{2}:g_{1}=\Delta m_{I}+\frac{\gamma_{S}}{\gamma_{I}}\Delta m_{S}$. (b) Similar to (a), high coherence orders are generated using two $\pi/2$ pulse seperated by a free-evolution period. The method of time proportional phase incrementation (TPPI)~\cite{drobny1979faraday} is employed, in which the phase $\phi$ is incremented proportionally with $t_{1}$, introducing an artificial offset term into the $t_1$ evolution period, enabling separation of spectrum from different coherence orders. The central $\pi$ pulse within $t_1$ refocuses both $\proton$ chemical shifts and the $\proton-\deuterium$ dipolar couplings evolution. To enhance the $\proton$ signal energy during detection, double-quantum decoupling is applied via continuous-wave irradiation at the center of $\deuterium$ spectrum, effectively suppressing heteronuclear dipolar couplings.  (c) During the $t_1$ evolution period, the $\proton$ magnetization evolves under chemical shift interactions. The BLEW-12 multiple pulse sequence removes $\proton$-$\proton$ homonuclear coupling, while a $\carbon$ $\pi$ pulse refocuses the $\carbon$-$\proton$ heteronuclear dipolar coupling. Two $0.705\cdot\frac{\pi}{2}$ pulses sandwiching the BLEW-12 multiple-pulse train effectively rotate the precession plane into the $xy$ plane. Following a gradient z-filter that seperately selects the cosine- and sine-modulated component, the magnetization is transferred to $\carbon$ via cross-polarization and detected during $t_2$.}
\end{figure}

Spectral estimation of the residual dipolar coupling parameters and chemical shifts for \sampleA was performed using the heteronuclear MQC (HMQC) experiments shown in Fig.~\ref{fig:MQC_pulse_sequence}(a). Using the analysis in section~\ref{sec:NMR_Hamiltonian} and taking the molecular equilibrium positions from ~\cite{field2003multiple} (summarized in Table~\ref{tab:atom coordinate}), the residual dipolar coupling are completely determined by the molecular Saupe order tensor (see equation~\ref{eq:RDC_Simplified}). Except for the strong heteronuclear J coupling $D_{C1}$, all other J couplings are taken from solution-state NMR data ignoring small J anisotropies ~\cite{field2003multiple,schaefer1983spin}. As a result only 6 free parameters remain for spetral estimation.

It may be shown that the combined information contained in the $\Delta m_{I}=7$, $\Delta m_{S}=0$ and $\Delta m_{I}=8$, $\Delta m_{S}=0$ HMQC spectra is sufficient to determine all free parameters. We then fit the free parameters to minimize the RMS deviation between the experimental and simulated peak positions. The estimated interaction parameters are summarized in Table~\ref{tab:9Q_old_sample}. A comparison of numerically simulated HMQC spectra (red), generated using the estimated parameters, with the experimental data (blue) is shown in Fig.~\ref{fig:9Q_HMQC_fitting}. 

\begin{table}[h!]
  \centering
  \caption{Atom coordinate for Toluene-4-$^{13}\rm{C}$.}
  \label{tab:atom coordinate}
  \begin{tabular}{p{3cm} p{3cm} p{3cm} p{3cm}}
    \hline
    Nucleus & x/\text{\AA} & y/\text{\AA} & z/\text{\AA} \\
    \hline
    H1 & 0.0 & 0.0 & 0.0 \\
    H2 & 0.0 & -2.1486 & 1.2417$\pm$0.005 \\
    H3 & 0.0 & 2.1486 & 1.2417$\pm$0.005 \\
    H4 & 0.0 & -2.1290$\pm$0.001 & 3.7051$\pm$0.009 \\
    H5 & 0.0 & 2.1290$\pm$0.001 & 3.7051$\pm$0.009 \\
    H6 & -0.5150$\pm$0.0005 & -0.8919$\pm$0.0005 & 5.7510$\pm$0.008 \\
    H7 & -0.5150$\pm$0.0005 & 0.8919$\pm$0.0005 & 5.7510$\pm$0.008 \\
    H8 & 1.0299$\pm$0.0005 & 0.0 & 5.7510$\pm$0.008 \\
    $^{13}\mathrm{C}^{a}$ & 0.0 & 0.0 & 1.0866 \\
    \hline
  \end{tabular}
  \begin{flushleft}
  \footnotesize
    $^{a}$ Proton coordinates are taken from ~\cite{field2003multiple}, whereas the $\carbon$ coordinate obtained from PubChem database~\cite{kim2016pubchem} is less precise and may bias $D_{C1}$. This bias is absorbed by the free parameter $J_{C1}$, since only the combination $D_{C1} + \tfrac{1}{2}J_{C1}$ is relevant.\\
    \end{flushleft}
\end{table}

\begin{table}
  \centering
  \caption{HMQC-estimated interaction parameters (in Hz) and Saupe order tensor of~\sampleA. Dipolar couplings are calculated from the estimated Saupe order tensor and the molecular structure (Table~\ref{tab:atom coordinate})$^a$}
  \label{tab:fitting result}
  \begin{tabular}{p{3cm} p{4cm} p{6cm} p{3cm}}
    \hline
    Parameter&Value&Parameter&Value\\
    \hline
    $S_{yy}$ & -0.0176$\pm$0.0005&$J^{c}_{45}$&1.97\\
    $S_{zz}$ & 0.1758$\pm$0.0003&$D_{12},D_{13}$&-242.25\\
    $\omega_{1}/2\pi$ & 19$\pm$9 &$D_{14},D_{15}$&-196.73\\
    $\omega_{2}/2\pi$, $\omega_{3}/2\pi$ & -72$\pm$14&$D_{16},D_{17},D_{18}$&-100.95\\
    $\omega_{4}/2\pi$, $\omega_{5}/2\pi$ & -111$\pm$14&$D_{23}$&26.64 \\
    $\omega_{6}/2\pi$, $\omega_{7}/2\pi$,$\omega_{8}/2\pi$ & -2348$\pm$6&$D_{45}$&27.38\\
    $J_{C1}^{b}$& 559$\pm$16&$D_{24},D_{35}$&-1412.40 \\
    $J_{C2}^{c},J_{C3}$ & 1.07&$D_{25},D_{34}$&-30.53\\
    $J_{C4}^{c},J_{C5}$ & 7.54&$D_{26},D_{27},D_{28},D_{36},D_{37},D_{38}$&-126.91\\
    $J_{C6}^{c},J_{C7},J_{C8}$ & 0.84&$D_{46},D_{47},D_{48},D_{56},D_{57},D_{58}$&-405.22\\
    $J_{12}^{c},J_{13}$ & 7.48&$D_{67},D_{68},D_{78}$&1860.23\\
    $J_{14}^{c},J_{15}$ & 1.27&$D_{C1}$&-4139.45\\
    $J_{23}^{c}$ & 1.49&$D_{C2},D_{C3}$&50.16\\
    $J_{24}^{c},J_{35}$ & 7.68&$D_{C4},D_{C5}$&-77.68\\
    $J_{25}^{c},J_{34}$ & 0.62&$D_{C6},D_{C7},D_{C8}$&-45.33\\
    \hline
  \end{tabular}
  \begin{flushleft}
  \footnotesize
    $^{a}$ All 18 peaks in $\Delta m_{I}=7$, $\Delta m_{S}=0$ and $\Delta m_{I}=8$, $\Delta m_{S}=0$ HMQC spectra were used in fitting, yielding in an r.m.s error of 7.2Hz\\
    $^{b}$ In the fitting, $J_{C1}$ appears larger than its typical isotropic value. This reflects compensation for the imprecise $^{13}$C coordinate rather than a physical deviation, and does not affect the analysis since only $D_{C1} + \tfrac{1}{2}J_{C1}$ enters the Hamiltonian.\\
    $^{c}$ Non-direct bonded $J$-couplings were fixed to their isotropic values from the literature~\cite{field2003multiple,schaefer1983spin}.\\
    \end{flushleft}
  \label{tab:9Q_old_sample}
\end{table}

\subsubsection{MQC experiment of DMBP}
\begin{figure}
    \includegraphics[width=\textwidth]{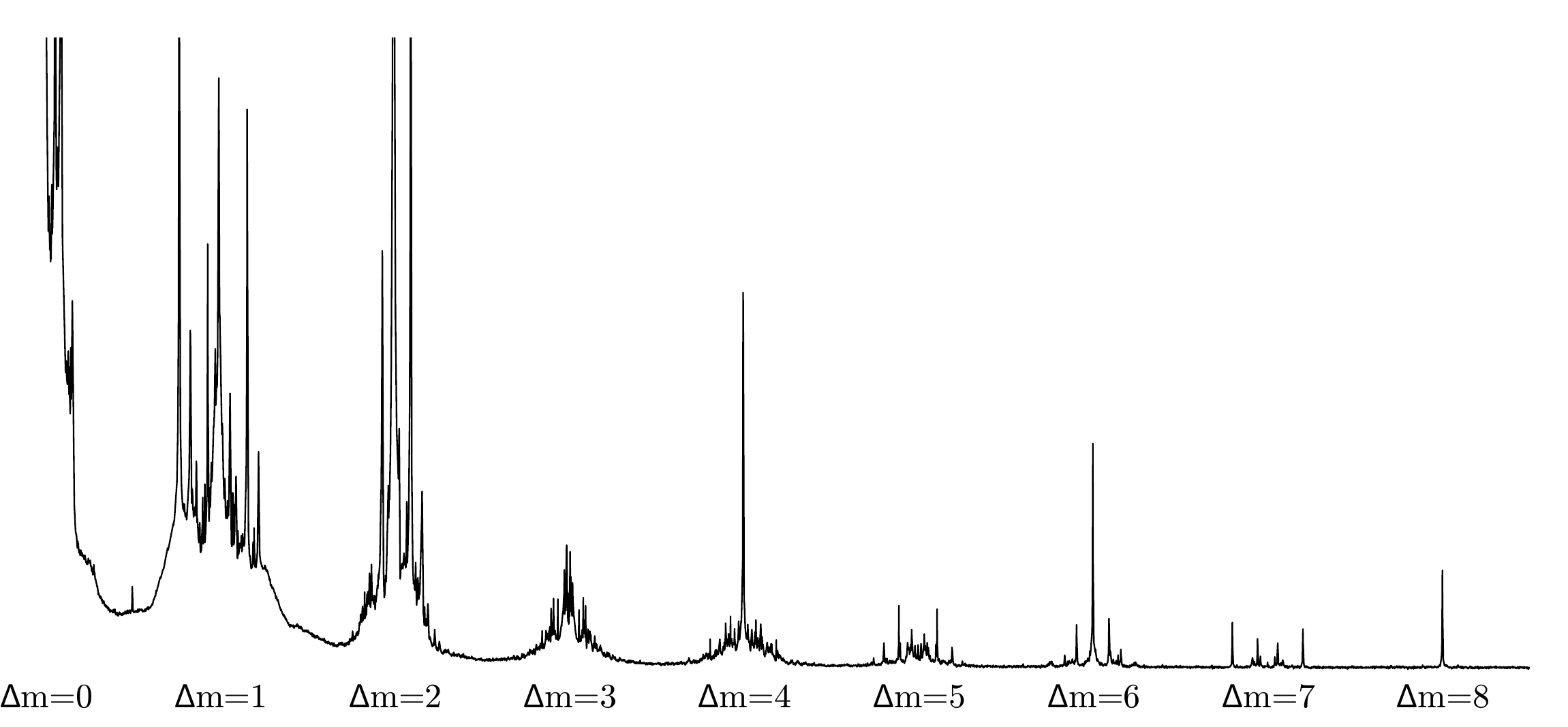}
    \caption{\label{fig:15Q_MQC_full} Proton MQC spectrum of \sampleC. The individual multiple-quantum coherence orders were separated by time proportional phase instrumentation (TPPI). The spectrum is obtained by magnitude-averaging of three spectra with $\tau = 0.8$, 1.0, and 1.2 ms. For all experiments, 16k $t_{1}$ increments, 2 scans per increment. The offset introduced by TPPI is 25kHz. Spectrum width is 450kHz for $F_{1}$ and 10kHz for $F_2$. The magnitude of deuterium decoupling rf field is about 3kHz in frequency unit.}.
\end{figure}

\begin{figure}
    \includegraphics[width=\textwidth]{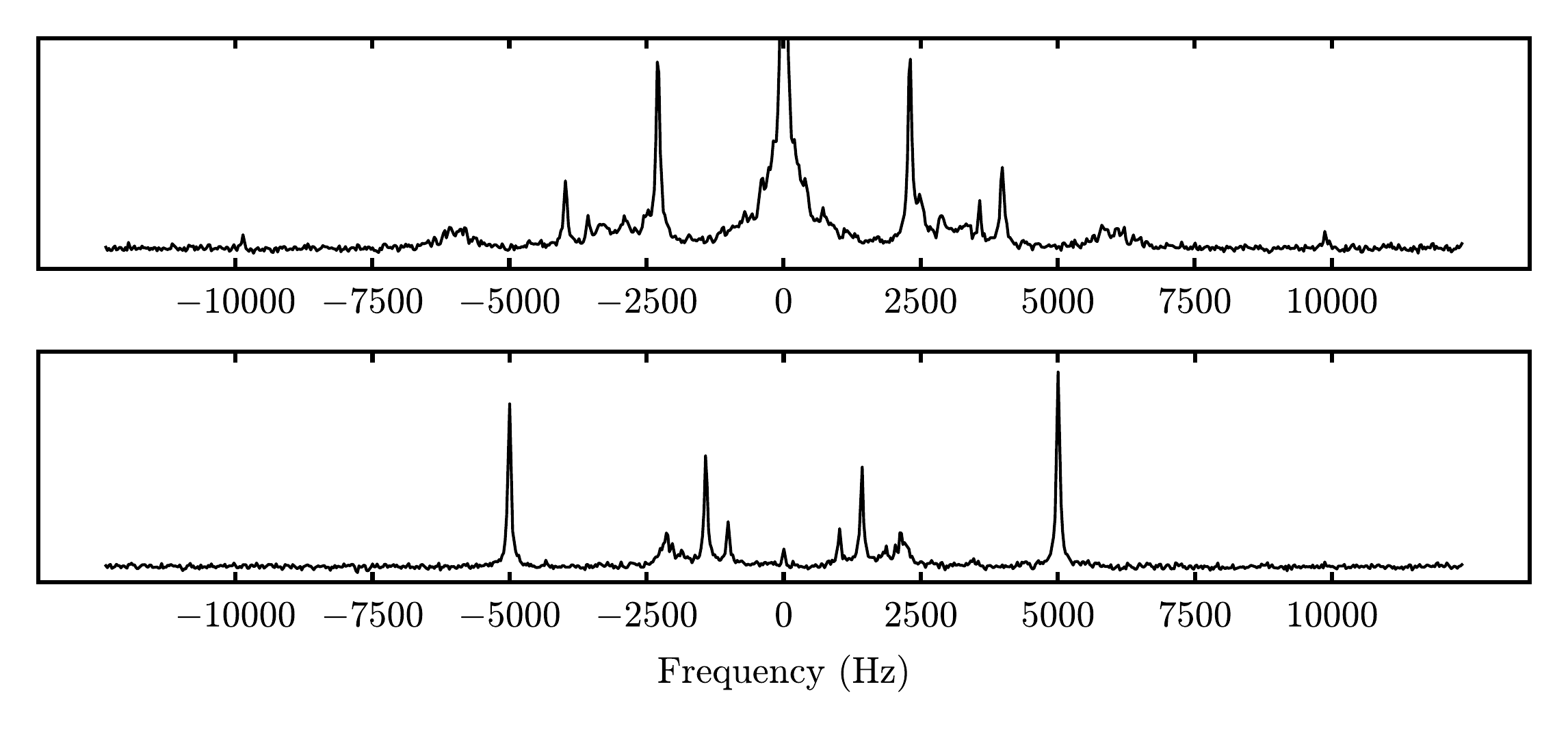}
    \caption{\label{fig:15Q_MQC} $\Delta m=6$ (top) and $\Delta m=7$ (bottom) proton MQC spectrum of \sampleC. Peaks are symmetrically distributed around 0Hz. 11 pairs of peaks are identified and assigned. The average linewidth is about 70Hz.}.
\end{figure}

Since MQC intensity decays rapidly with increasing coherence order~\cite{bengs2025fundamental}, direct MQC experiments on \sampleB are impractical due to the large number of $\proton$ spins (15). To address this, we synthesized \sampleC, in which the methyl $\proton$ are replaced by $\deuterium$, and performed proton MQC experiments using the experiment scheme reported in~\cite{sinton1984multiple}(Fig.~\ref{fig:MQC_pulse_sequence}(b)). In this sequence, a central $\pi$ pulse in the $t_1$ evolution refocuses both $\proton$ chemical shifts and $\proton-\deuterium$ dipolar couplings. Three spectra were acquired with $\tau = 0.8$, 1.0, and 1.2 ms and subsequently magnitude-averaged. The resulting full spectrum containing all coherence orders is shown in Fig.~\ref{fig:15Q_MQC_full}. The $\Delta m=6$ and $\Delta m=7$ spectra (Fig.~\ref{fig:15Q_MQC}) exhibit peaks symmetrically distributed around 0Hz, which were used for spectral estimation. We observed significant peak broadening, possibly due to the large quadrupolar splitting of $\deuterium$ and strong heteronuclear dipolar couplings, which lead to imperfect refocusing during the $t_1$ evolution period. In addition, the relatively low solute concentration and limited deuterium decoupling power of our system result in a poor signal-to-noise ratio. Consequently, only 11 pairs of peaks in  $\Delta m=6$ and $\Delta m=7$ spectra could be reliably identified and assigned, which are insufficient to determine all 14 unique dipolar couplings. Therefore, we directly evaluated the RMS deviation between the experimental peak positions and those predicted from the dipolar couplings obtained by MD using various candidate PMF surfaces, to identify the best-matching surface (result shown in Fig. 4(d) of the main text). Since J-couplings are much smaller then the spetra linewidth, they are ignored in simulations.

\subsection{Higher order Magnus expansion terms}\label{sec:magnus}
Our goal of this section is to define a family of average Hamiltonians that closely approximates the full pulse sequence used in the experiment. The goal of this approximation is to make emulation of the NMR experiment feasible with reasonable depth quantum circuits.
The NMR Hamiltonian, including coupling of the proton and carbon spins to the external field, is given by
\begin{equation}
    H(t)=H^{\rm H}_{\rm rf}(t)+H^{\rm C}_{\rm rf}(t)+H_{\rm int}.
\end{equation}
The external field Hamiltonians take the form 
\begin{equation}
    \begin{aligned}
        &H^{\rm H}_{\rm rf}(t)=(1+\epsilon_{\rm H})H^{\rm H}_{\rm nom}(t),
        \\
        &H^{\rm C}_{\rm rf}(t)=(1+\epsilon_{\rm C})H^{\rm C}_{\rm nom}(t),
    \end{aligned}
\end{equation}
where $\epsilon_{\rm H}$ and $\epsilon_{\rm C}$ account for pulse amplitude imperfections due to $B_{1}$-inhomogeneity, and the nominal radio-frequency Hamiltonians are defined by
\begin{equation}
    \begin{aligned}
    &H^{\rm H}_{\rm nom}(t)=\omega^{\rm H}_{\rm nom}\big\{\cos(\phi_{\rm H}(t))I_{x}-\sin(\phi_{\rm H}(t))I_{y}\big\},
    \\
    &H^{\rm C}_{\rm nom}(t)=\omega^{\rm C}_{\rm nom}\big\{\cos(\phi_{\rm C}(t))S_{x}-\sin(\phi_{\rm C}(t))S_{y}\big\}.
    \end{aligned}
\end{equation}
The nominal pulse amplitude $\omega_{\rm nom}$ is defined such that $\omega_{\rm nom}\tau_{\pi/2}=\pi/2$, the phase $\phi(t)$ is pulse sequence and channel dependent. Both $H^{\rm H}_{\rm rf}(t)$ and $H^{\rm C}_{\rm rf}(t)$ produce simple rotation operations
\begin{equation}
    \begin{aligned}
    &U^{\rm H}_{\rm rf}(t)=\mathcal T\exp(-i \int_{0}^{t}H^{\rm H}_{\rm rf}(t^{\prime})dt^{\prime})\in {\rm SU}(2),
    \\
    &U^{\rm C}_{\rm rf}(t)=\mathcal T\exp(-i \int_{0}^{t}H^{\rm C}_{\rm rf}(t^{\prime})dt^{\prime})\in {\rm SU}(2),
    \end{aligned}
\end{equation}
and admit a decomposition in terms of Euler rotations
\begin{equation}
\label{eq:Urf_euler}
    \begin{aligned}
    &U^{\rm H}_{\rm rf}(t)=R(\Omega^{\rm H}(t))=R_{z}(\alpha^{\rm H}(t))R_{y}(\beta^{\rm H}(t))R_{z}(\gamma^{\rm H}(t)),
    \\
    &U^{\rm C}_{\rm rf}(t)=R(\Omega^{\rm C}(t))=R_{z}(\alpha^{\rm C}(t))R_{y}(\beta^{\rm C}(t))R_{z}(\gamma^{\rm C}(t)).
    \end{aligned}
\end{equation}

The propagator for the full TARDIS-1 OTOC experiment may be represented as a concatenation of 6 basic elements
\begin{equation}
    \begin{aligned}
U_{\rm TARDIS-1,OTOC}= \{U_{\leftarrow}(nt_c)\}\{\Pi^{\rm H}_{y}\Pi^{\rm C}_{-y}\}\{U_{\rm BLEW12}^2\}\{\Pi^{\rm C}_{y}\}\{U_{\rm BLEW12}^2\}\{U_{\rightarrow}(nt_c)\}
    \end{aligned}
\end{equation}
Where $\Pi$ denotes the propagator of the $\pi$ pulse. By design, $U_{\leftarrow}$, $U_{\rightarrow}$ and $U_{\rm BLEW12}$ are so-called {\em cyclic} pulse sequence elements. As a consequence, the {\em nominal} part of the radio-frequency propagator, when considered in isolation, produces no net rotation upon completion of a period
\begin{equation}
    \begin{aligned}
U_{\rightarrow}^{\rm nom}= \pm \mathbbm{1},
\quad 
U_{\leftarrow}^{\rm nom}= \pm \mathbbm{1},
\quad 
U_{\rm BLEW12}^{\rm nom}= \pm \mathbbm{1}.
    \end{aligned}
\end{equation}
For the non-trivial unitaries we then utilise an interaction frame decomposition factoring out the nominal part of the radio-frequency Hamiltonian
\begin{equation}
    \begin{aligned}
&U_{\rightarrow}(nt_c)=U^{\rm nom}_{\rightarrow}\tilde{U}_{\rightarrow}(nt_c)=\pm\tilde{U}_{\rightarrow}(nt_c),
\\
&U_{\leftarrow}(nt_c)=U^{\rm nom}_{\leftarrow}(nt_c)\tilde{U}_{\leftarrow}=\pm\tilde{U}_{\leftarrow}(nt_c),
\\
&U_{\rm BLEW12}=U^{\rm nom}_{\rm BLEW12}\tilde{U}_{\rm BLEW12}=\pm\tilde{U}_{\rm BLEW12},
    \end{aligned}
\end{equation}
The corresponding interaction frame propagator $\tilde{U}$ is generally given by
\begin{equation}
    \begin{aligned}
\tilde{U}(t_c)&=\mathcal T\exp(-i \int_{0}^{t_c}\tilde{H}(t^{\prime})dt^{\prime}),
\end{aligned}
\end{equation}
with
\begin{equation}
\begin{aligned}
\tilde{H}(t^{\prime})&=U^{\rm nom\dagger}(t^{\prime})\{H(t^{\prime})-(H^{\rm H}_{\rm nom}(t^{\prime})+H^{\rm C}_{\rm nom}(t^{\prime})\}U^{\rm nom}(t^{\prime})
\\
&=\epsilon_{\rm H}\tilde{H}^{\rm H}_{\rm nom}(t^{\prime})+\epsilon_{\rm C}\tilde{H}^{\rm C}_{\rm nom}(t^{\prime})+\tilde{H}_{\rm int}(t^{\prime}),
    \end{aligned}
\end{equation}
and
\begin{equation}
    \begin{aligned}
U^{\rm nom}(t^{\prime})=\mathcal T\exp(-i \int_{0}^{t^{\prime}}H_{\rm nom}(s)ds). 
    \end{aligned}
\end{equation}

The propagator for the full TARDIS-1 sequence is then alternatively given by
\begin{equation}\label{eq:U_DREAM1}
    \begin{aligned}
U_{\rm TARDIS-1, OTOC}= \{\tilde{U}_{\leftarrow}(nt_c)\}\{\Pi^{\rm H}_{y}\Pi^{\rm C}_{-y}\}\{\tilde{U}_{\rm BLEW12}^2\}\{\Pi^{\rm C}_{y}\}\{\tilde{U}_{\rm BLEW12}^2\}\{\tilde{U}_{\rightarrow}(nt_c)\}.
    \end{aligned}
\end{equation}
\begin{equation}
    \begin{aligned}
U_{\rm TARDIS-1, LE}= \{\tilde{U}_{\leftarrow}(nt_c)\}\{\Pi^{\rm H}_{y}\}\{\tilde{U}_{\rightarrow}(nt_c)\}.
    \end{aligned}
\end{equation}
This representation is suitable for average Hamiltonian theory (AHT) approximation schemes, so that the interaction frame propagators $\tilde{U}$ may be estimated by
\begin{equation}
    \begin{aligned}
&\tilde{U}_{\rightarrow}(nt_c)=\exp(-i \bar{H}_{\rightarrow} nt_c)\simeq \exp(-i (\bar{H}^{(0)}_{\rightarrow}+\bar{H}^{(1)}_{\rightarrow}+\dots) nt_c),
\\
&\tilde{U}_{\leftarrow}(nt_c)=\exp(-i \bar{H}_{\leftarrow} nt_c)\simeq \exp(-i (\bar{H}^{(0)}_{\leftarrow}+\bar{H}^{(1)}_{\leftarrow}+\dots) nt_c),
\\
&\tilde{U}_{\rm BLEW12}=\exp(-i \bar{H}_{\rm BLEW12} T_{\rm BLEW12})\simeq \exp(-i (\bar{H}^{0}_{\rm BLEW12}+\bar{H}^{1}_{\rm BLEW12}+\dots) T_{\rm BLEW12}).
    \end{aligned}
\end{equation}
The first few terms of this approximation may be calculated via the Magnus expansion (Eq.~\ref{eq:Magnus_high_order}).
The Euler decomposition given by Eq.~\ref{eq:Urf_euler} provides a convenient method for the calculation of the average Hamiltonian terms. To this end, the spin Hamiltonian is expanded in terms of irreducible tensor operators $Q_{lm}$ of the rotation group ${\rm SO}(3)$
\begin{equation}
    \begin{aligned}
   R(\Omega)Q_{lm}R^{\dagger}(\Omega)=\sum_{n=-l}^{+l}Q_{ln}D^{l}_{nm}(\Omega),
    \end{aligned}
\end{equation}
where $D^{l}_{nm}(\Omega)$ are Wigner matrix elements of spherical rank $l$. In the absence of $B_{1}$-inhomogeneities ($\epsilon=0$), the spin Hamiltonian in the interaction frame of the nominal rf part may be decomposed as follows
\begin{equation}
    \begin{aligned}
    \tilde{H}(t)=\tilde{H}^{\rm HH}_{\rm DD}+\tilde{H}^{\rm H}_{\rm CS}+\tilde{H}^{\rm HC}_{\rm DD},
    \end{aligned}
\end{equation}
\begin{equation}
    \begin{aligned}
    &\tilde{H}^{\rm HH}_{\rm DD}\leftarrow \sum_{m=-2}^{+2} Q_{2m}D^{2}_{m0}(\Omega^{\rm H}(t)),
    \\
    &\tilde{H}^{\rm H}_{\rm CS}\leftarrow \sum_{m=-1}^{+1} Q_{1m}D^{1}_{m0}(\Omega^{\rm H}(t)),
    \\
    &\tilde{H}^{\rm HC}_{\rm DD}\leftarrow \sum_{m,n=-1}^{+1} Q_{1m}K_{1n}D^{1}_{m0}(\Omega^{\rm H}(t))D^{1}_{n0}(\Omega^{\rm C}(t)).
    \end{aligned}
\end{equation}
The decomposition for $H^{\rm HC}_{\rm DD}$ follows from that fact that pulses may be applied individually to the protons and the carbons, so technically speaking these operators span representations for ${\rm SO}(3)\otimes {\rm SO}(3)$. The basis tensors are explicitly given by
\begin{align}
{[}Q{]}_{2m}=\sum_{i < j, i,j \in \proton}D_{ij}
\begin{pmatrix}
\sqrt{\frac{3}{2}}\left( I_{x}^iI_{x}^j - i\left(I_{x}^iI_{y}^j + I_{y}^iI_{x}^j - i\left( I_{y}^iI_{y}^j \right) \right) \right) \\
\sqrt{\frac{3}{2}}\left(  I_{x}^iI_{z}^j - i\left(I_{y}^iI_{z}^j + I_{z}^iI_{y}^j + i\left( I_{z}^iI_{x}^j \right) \right)   \right) \\
-  I_{x}^iI_{x}^j - I_{y}^iI_{y}^j + 2I_{z}^iI_{z}^j \\
-\sqrt{\frac{3}{2}}\left(  I_{x}^iI_{z}^j + i\left(I_{y}^iI_{z}^j + I_{z}^iI_{y}^j - i\left( I_{z}^iI_{x}^j \right) \right)   \right) \\
\sqrt{\frac{3}{2}}\left( I_{x}^iI_{x}^j + i\left(I_{x}^iI_{y}^j + I_{y}^iI_{x}^j + i\left( I_{y}^iI_{y}^j \right) \right) \right) \\
\end{pmatrix}
\end{align}
\begin{align}
{[}Q{]}_{1m}=\sum_{i\in \proton}\omega_{i}
\begin{pmatrix}
\frac{1}{\sqrt{2}}\left( I_{x}^i - i I_{y}^i \right) \\
I_{z}^i \\
-\frac{1}{\sqrt{2}}\left( I_{x}^i + i I_{y}^i \right)
\end{pmatrix}
\end{align}
\begin{align}
{[}Q{]}_{1m}{[}K{]}_{1n}=\sum_{i \in \proton}\left(D_{Ci} + J_{Ci}/2 \right)
\begin{pmatrix}
I_{x}^iS_{x} - iI_{x}^iS_{y} -i I_{y}^iS_{x} - I_{y}^iS_{y} \\
\sqrt{2} \left( I_{x}^iS_{z} - i I_{y}^iS_{z} \right) \\
-I_{x}^iS_{x} - iI_{x}^iS_{y} +i I_{y}^iS_{x} - I_{y}^iS_{y} \\
\sqrt{2} \left( I_{z}^iS_{x} - i I_{z}^iS_{y} \right) \\
2I_{z}^iS_{z} \\
-\sqrt{2} \left( I_{z}^iS_{x} + i I_{z}^iS_{y} \right) \\
-I_{x}^iS_{x} + iI_{x}^iS_{y} -i I_{y}^iS_{x} - I_{y}^iS_{y} \\
-\sqrt{2} \left( I_{x}^iS_{z} + i I_{y}^iS_{z} \right) \\
I_{x}^iS_{x} + iI_{x}^iS_{y} +i I_{y}^iS_{x} - I_{y}^iS_{y} \\
\end{pmatrix} 
\end{align}
If we move to a linear indexing scheme
\begin{equation}
\begin{aligned} 
({[}Q{]}_{2m},{[}Q{]}_{1m},{[}Q{]}_{1m}{[}K{]}_{1n})\rightarrow (C_{1},C_{2},\dots),
\end{aligned}
\end{equation}
the interaction frame Hamiltonian takes the simple form
\begin{equation}
    \begin{aligned}
    \tilde{H}(t)=\sum_{j=1}^{N}c_{j}(t)C_{j},
    \end{aligned}
\end{equation}
where $c_{j}(t)$ are the appropriate Wigner matrix elements. If we further define a commutator basis
\begin{equation}
    \begin{aligned}
    C_{ij}=&{[}C_{i},C_{j}{]},
    \\
    C_{ijk}=&{[}C_{i},{[}C_{j},C_{k}{]}{]},
    \end{aligned}
\end{equation}
the expansion terms of the Magnus series are given by
\begin{equation}
    \begin{aligned}
    &\bar H^{(0)}=\frac{1}{t_c}\sum_{i=1}^{N}C_{i}w_{i},
    \\
    &\bar H^{(1)}=\frac{1}{2it_c}\sum_{i,j=1}^{N}C_{ij}w_{ij},
    \\
    &\bar H^{(2)}=-\frac{1}{6t_c}\sum_{i,j,k=1}^{N}C_{ijk}(w_{ijk}+w_{kji}),
    \end{aligned}
\end{equation}
with the weights
\begin{equation}
    \begin{aligned}
    &w_{i}=\int_{0}^{t_c}c_{i}(t_{1})dt_{1},
    \\
    &w_{ij}=\int_{0}^{t_c}\int_{0}^{t_{1}}c_{i}(t_{1})c_{j}(t_{2})dt_{2}dt_{1}=\int_{0}^{t_c}c_{i}(t_{1})w_{j}(t_{1})dt_{1},
    \\
    &w_{ijk}=\int_{0}^{t_c}\int_{0}^{t_{1}}\int_{0}^{t_{2}}c_{i}(t_{1})c_{j}(t_{2})c_{k}(t_{3})dt_{3}dt_{2}dt_{1}=\int_{0}^{t_c}c_{i}(t_{1})w_{jk}(t_{1})dt_{1}.
    \end{aligned}
\end{equation}
To include rf errors it is convenient to augment the basis tensors by
\begin{align}
{[}F^{\rm H}{]}_{1m}=\omega^{\rm H}_{\rm nom}
\begin{pmatrix}
\frac{1}{\sqrt{2}}(I_{x}-i I_{y})
\\
I_{z}
\\
-\frac{1}{\sqrt{2}}(I_{x}+i I_{y})
\end{pmatrix}
\end{align}
\begin{align}
{[}F^{\rm C}{]}_{1m}=\omega^{\rm C}_{\rm nom}
\begin{pmatrix}
\frac{1}{\sqrt{2}}(S_{x}-i S_{y})
\\
S_{z}
\\
-\frac{1}{\sqrt{2}}(S_{x}+i S_{y})
\end{pmatrix}
\end{align}
The rf Hamiltonian may then be written as
\begin{equation}
    \begin{aligned}
        &H^{\rm H}_{\rm rf}(t)=-(1+\epsilon_{\rm H})\frac{\omega^{\rm H}_{\rm nom}}{\sqrt{2}}\sum_{m=-1}^{+1}m F^{\rm H}_{1m}e^{-i m \phi_{\rm H}(t)},
        \\
        &H^{\rm C}_{\rm rf}(t)=-(1+\epsilon_{\rm C})\frac{\omega^{\rm C}_{\rm nom}}{\sqrt{2}}\sum_{m=-1}^{+1}m F^{\rm C}_{1m}e^{-i m \phi_{\rm C}(t)},
    \end{aligned}
\end{equation}
The rf error term in the interaction frame of the nominal pulse propagator is then given by
\begin{equation}
    \begin{aligned}
        &\tilde{H}^{\rm H}_{\epsilon}(t)
        =-\epsilon_{\rm H}\frac{\omega^{\rm H}_{\rm nom}}{\sqrt{2}}\sum_{m,n=-1}^{+1}mF^{\rm H}_{1n}D_{nm}(\Omega^{\rm H}(t))e^{-i m \phi_{\rm H}(t)}
        =-\epsilon_{\rm H}\frac{\omega^{\rm H}_{\rm nom}}{\sqrt{2}}\sum_{n=-1}^{+1}F^{\rm H}_{1n}f^{\rm H}_{n}(t),
        \\
        &\tilde{H}^{\rm C}_{\epsilon}(t)
        =-\epsilon_{\rm C}\frac{\omega^{\rm C}_{\rm nom}}{\sqrt{2}}\sum_{m,n=-1}^{+1}mF^{\rm C}_{1n}D_{nm}(\Omega^{\rm C}(t))e^{-i m \phi_{\rm C}(t)}
        =-\epsilon_{\rm C}\frac{\omega^{\rm C}_{\rm nom}}{\sqrt{2}}\sum_{n=-1}^{+1}F^{\rm C}_{1n}f^{\rm C}_{n}(t),
    \end{aligned}
\end{equation}
where the time-dependent part is independent of the fractional error and may be pre-computed once.

\begin{figure}
    \centering
    \includegraphics[width=0.32\linewidth]{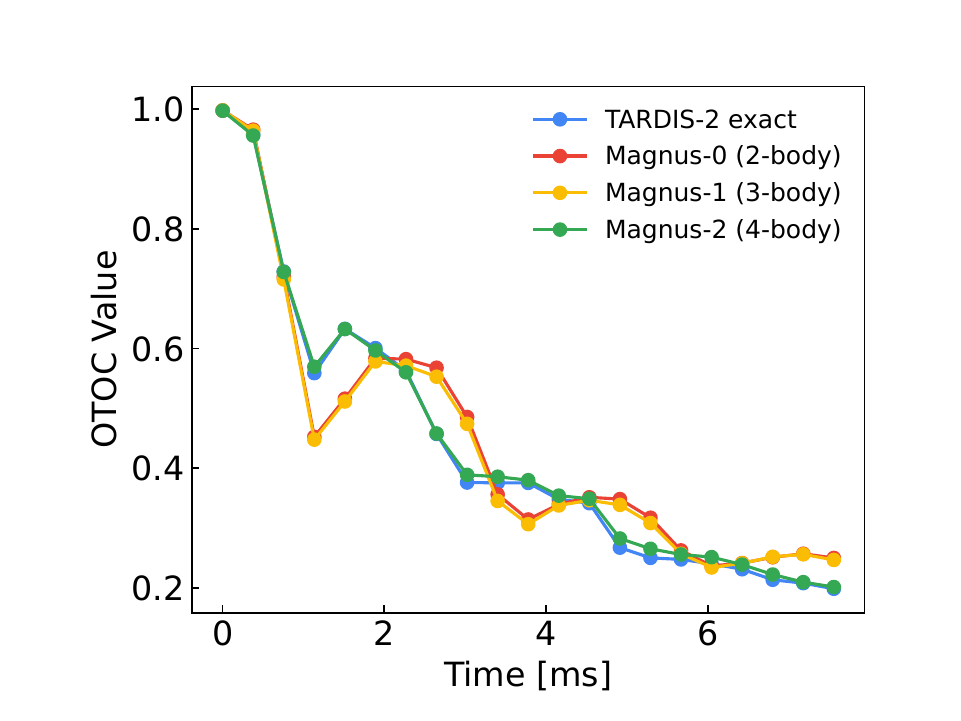}
    \includegraphics[width=0.32\linewidth]{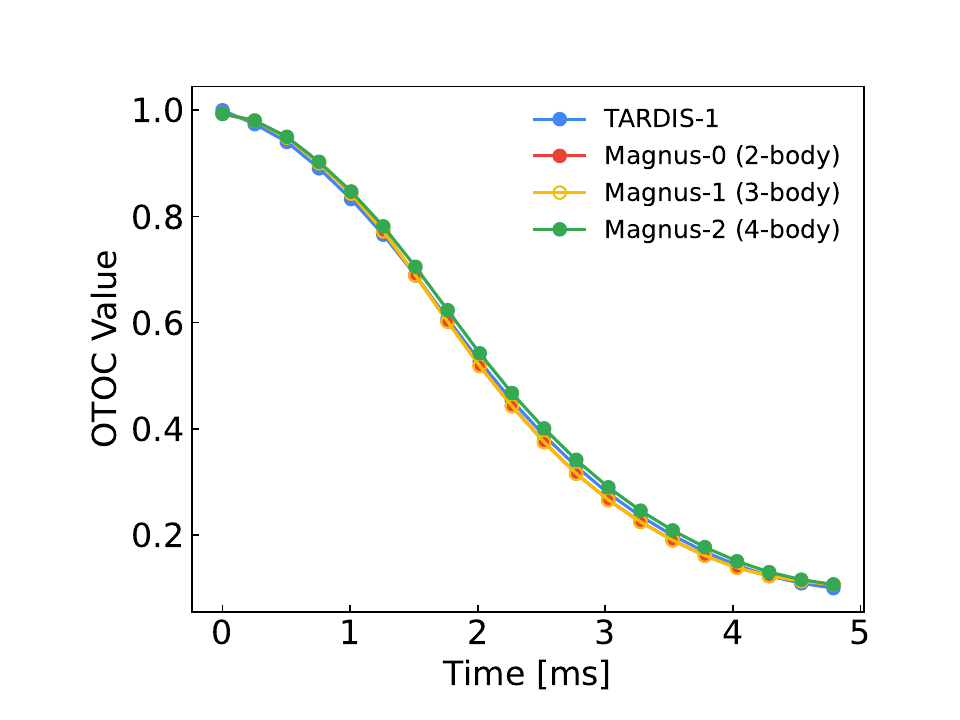}
        \includegraphics[width=0.32\linewidth]{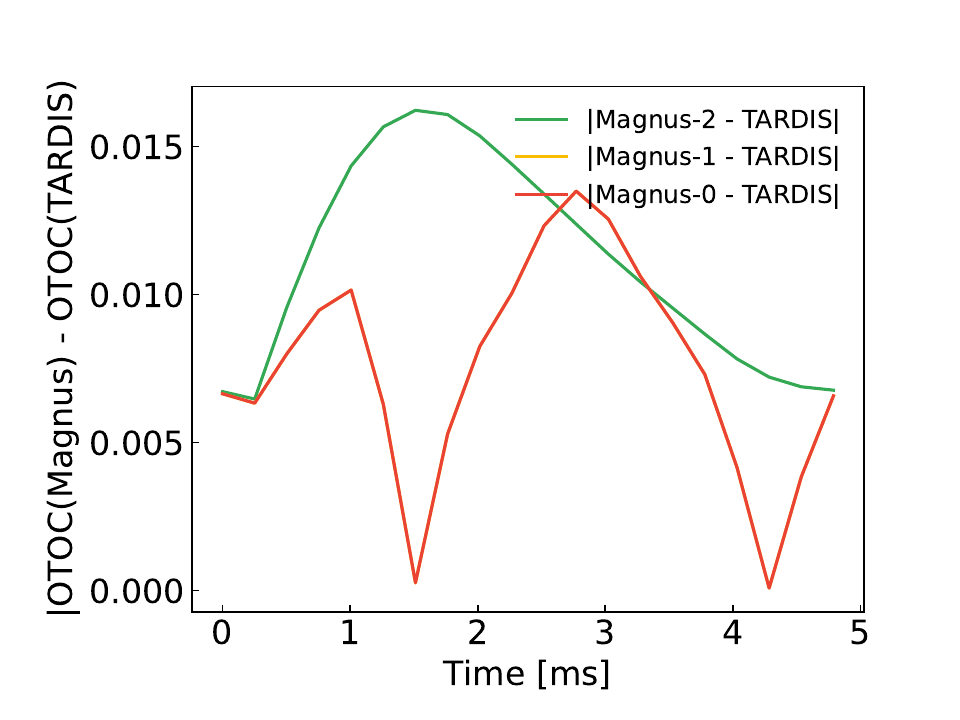}
    \caption{(\textit{left}) OTOC versus experiment time demonstrating the convergence of the Magnus expansion unitaries (Eq.~\eqref{eq:Magnus_high_order}) up to second order. For each order in the Magnus expansion correction we indicate the many-body order of the correction Hamitlonian. The similarity of green (Magnus-2) and blue (full TARDIS-2 sequence) indicates convergence of the average Hamiltonian expansion. Average Hamiltonian theory is applied to the forward, backwards, and BLEW12 sequence. The $\pi$ and $-\pi$ pulses on the $\carbon$ channel, which contain the full dipolar Hamiltonian background evolution for  $2t_p=21\mu s$, is simulated explicitly. (\textit{center}) OTOC versus time demonstrating the convergence of the Magnus expansion unitaries towards the full TARDIS-1 unitary for the 15-spin system. (\textit{right}) Difference in OTOC values for the 15-spin system at different orders in Magnus expansion and the TARDIS-1 OTOC value. Magnus-1 is indistinguishable from Magus-0, by design.}
    \label{fig:magnus_DREAM2}
\end{figure}

\subsection{Understanding features of the toluene OTOC decay}

In Fig.~2(c) of the main text we showed the observed NMR OTOC for toluene, which in addition to the typical decay profile of an OTOC contained a notable bump around 1.5~ms.
We further observed in Fig.~2(d) that aroud this bump the OTOC had an inversion feature; increasing the ortho-meta HH coupling (which should typically cause faster decay) caused the OTOC to increase, rather than decreasing.
In this section, we outline our understanding of the physical mechanism causing this phenomenon, and provide further numerical simulations to back up our theory.

\begin{figure}
    \centering
    \includegraphics[width=\linewidth]{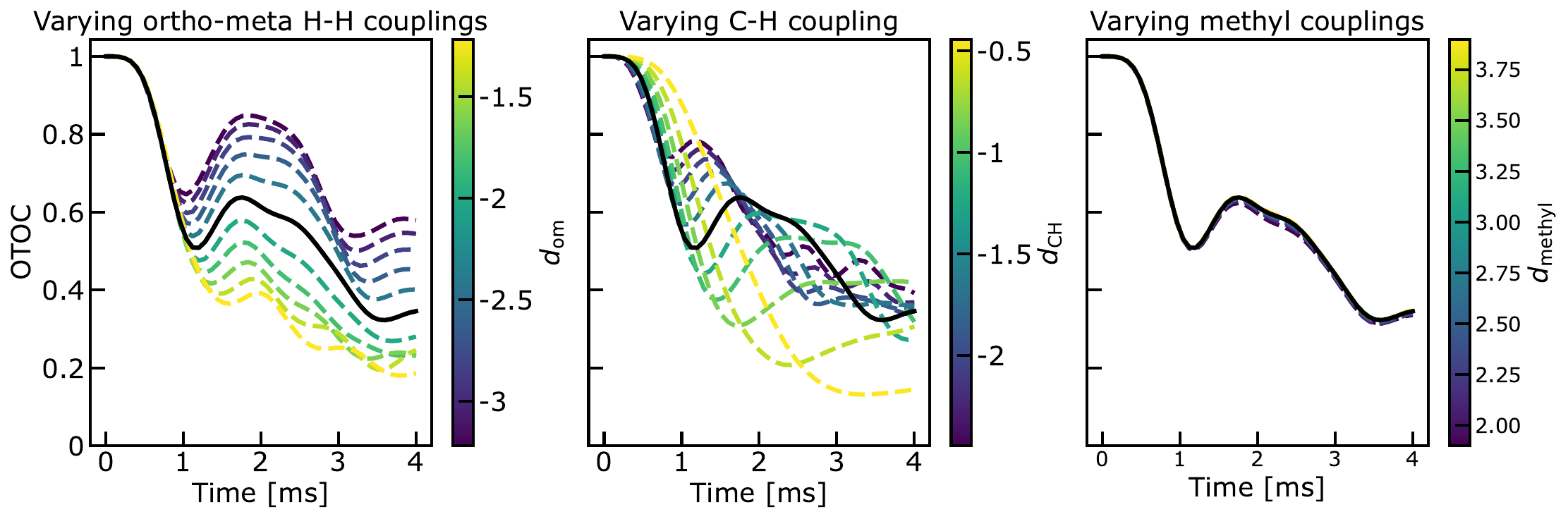}
    \caption{Plots showing the change in the toluene OTOC as different couplings in the molecule are changed.
    All simulations are done to lowest order in average Hamiltonian theory (i.e. using the DQ Hamiltonian directly).
    For each plot, the black line shows the expected OTOC decay using the reference parameters described in the text, while the coloured dashed lines show the effect of changing a single coupling (as described in title).
    }
    \label{fig:coupling_comparison}
\end{figure}

To understand this feature, we first note that we can separate the toluene spins into three clusters:
\begin{enumerate}
    \item The $^{13}$C and para-$^{1}$H spin.
    \item The ortho- and meta-$^{1}$H spins
    \item The methyl-$^{1}$H spins.
\end{enumerate}
The intra-cluster dipolar couplings are significantly larger ($3-10\times$) than the inter-cluster couplings, and so they form bound states.
We hypothesize that these bound states cause two effects to the OTOC:
\begin{enumerate}
    \item The operators within the C-H bound state oscillate quickly, which directly causes the OTOC oscillations.
    \item The H-H bound state acts as a barrier, preventing the decay of the C-H bound state into the rest of the system.
\end{enumerate}
We evidence this hypothesis in Fig.~\ref{fig:coupling_comparison}, where we vary the large couplings within each cluster simultaneously\footnote{Note that changing the ortho-meta coupling achieves a similar result to the stretching performed in Fig.~2(d) of the main text. The results here are not identical because a) here we do not change couplings other than the single H-H coupling, and b) here we take the zeroth order of the Magnus expansion.}.
The crucial observation here is that changing the ortho-meta H-H coupling changes the height of the OTOC fluctuations but not the frequency, whilst changing the C-H coupling changes the frequency but has a lesser effect on the height, which qualitatively matches what we would expect from the above hypothesis.
By contrast, we observe that the OTOC decay is almost completely independent of the intra-methyl couplings, which serve as a baseline.

In order to observe the feature detailed above, we needed the presence of bound states, but we also required they be isolated - i.e.~cut off from the rest of the molecule save for coupling through another bound state.
In a larger molecule we expect to continue to have bound states, however we would expect cases of isolated bound states to be more rare due to the increased number of spins in the system.
Thus, we expect that the features observed here will be less frequent in larger system sizes, and the only characteristic feature of large-scale OTOCs will be their decay.

\subsection{Predicting ultimate decoherence distance limits in measuring OTOCs}

In this section, we estimate the ultimate length limit of an OTOC in a nuclear magnetic spin system due to the decoherence alone.
Though decoherence was not the primary source of Loschmidt echo decay observed in Fig.~2(c) of the main text, the sources that were observed (higher-order corrections to the Magnus expansion and RF inhomogeneity) can be fixed in turn by going to higher pulse power spectrometers and by using a smaller sample holder respectively.
Thus, we argue that the ultimate limit we should consider comes from the decoherence of the large spin cluster formed by forward evolution.
We note here that the size of this spin cluster can in principle be reduced by selective deuteration, and that decoherence times in NMR systems can be reduced by lowering temperature.
However, this additional overhead increases the complexity of the OTOC experiment, so we do not consider these (or other) experimental techniques for this simple calculation.

Our calculation proceeds by imagining the growth of a cluster starting from a single spin, assuming that all spins within the cluster decohere at a rate $\tau_{\mathrm{d}}$.
We can perform an OTOC using any spin (and observe a signal) in this cluster as a butterfly.
We assume that the experiment lasts a time $t_{\mathrm{exp}}$; in this time the cluster grows to a volume $V_{\mathrm{s}}$, which entangles $N_{\mathrm{s}}$ spins depending on the density of the protein: $\rho_{\mathrm{s}}=N_{\mathrm{s}}/V_{\mathrm{s}}$
As a rough back-of-the-envelope calculation, we can estimate typical values for our physical constants as
\begin{equation}\label{eq:nuclear_system_constants}
    \tau_{\mathrm{d}}\approx 1~\mathrm{s},\hspace{1cm} t_{\mathrm{exp}}\approx 1~\mathrm{ms},\hspace{1cm} \rho_{\mathrm{s}}\approx 1.38\mathrm{g\; protein\; cm}^{-3}\approx 10^{-25}~\mathrm{g\; H\; \text{\AA}^{-3}\approx 0.1~\mathrm{spin\;\text{\AA}^{-3}}},
\end{equation}
where $1.38$~g\, cm$^{-3}$ is the mean density of protein, of which about $6-10\%$ is hydrogen (and we recall that the mass of a hydrogen atom is around $10^{-24}$~g), and our approximation for $t_{\mathrm{exp}}$ is based on previous numerical simulations of OTOCs in solid state proteins (which yielded $t_{\mathrm{exp}}\in [0.5~\mathrm{ms},2~\mathrm{ms}]$ at distances of $10-12$~\AA.

\begin{figure}
    \centering
    \includegraphics[width=0.5\linewidth]{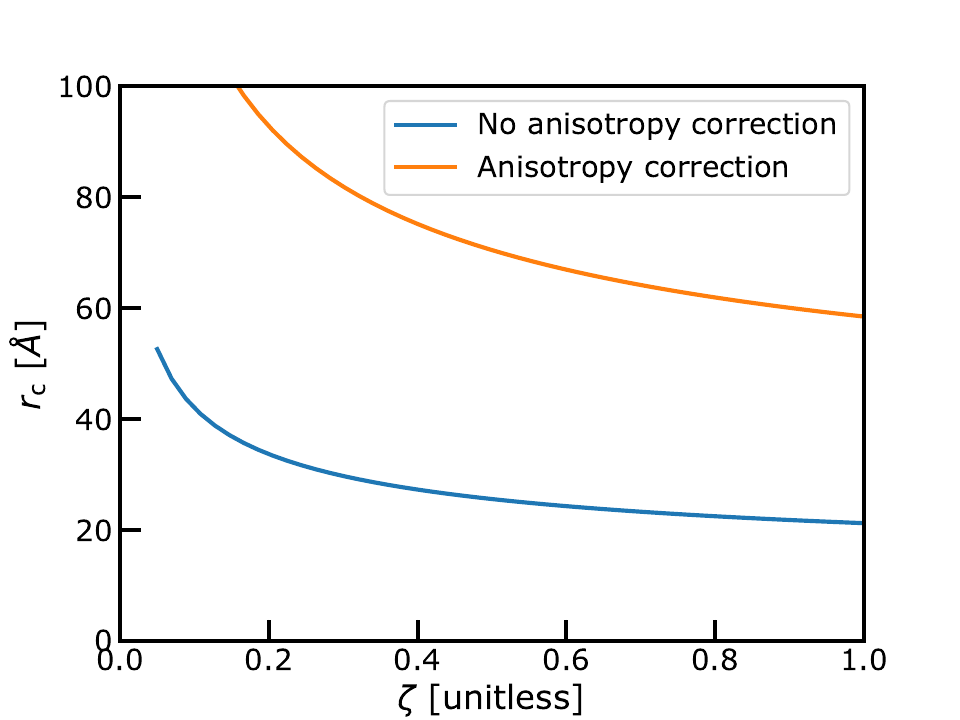}
    \caption{Plot of Eq.~\eqref{eq:rc_fixed_1} (blue) and Eq.~\eqref{eq:rc_fixed_2} (orange)}
    \label{fig:rc_fixed_1}
\end{figure}

We need to consider two complications in developing an accurate model of spin cluster growth:
\begin{enumerate}
    \item Not every spin stays within the cluster for the same time.
    \item The growth of the spin cluster is not spherically symmetric, due to the distance dependence of the dipolar interaction.
\end{enumerate}
Let us first assume spherically symmetric growth, and attempt to capture the time dependence accurately.
We can assign each spin $s$ a time $t_s$ that corresponds to how long the spin is active in the OTOC.
Then, our decoherence criteria is
\begin{equation}
    \sum_s t_s=\tau_{\mathrm{d}}.
\end{equation}
Without the directional dependence on the dipolar coupling, Ref.~\cite{zhou20operator} suggests that the operator light cone for a $1/r^3$ coupling in a $3$-dimensional system should grow as $e^{\sqrt{\ln(x)}}$.
However, the directional dependence will likely reduce the effective dimension below $3$, in which case we expect a linear or superlinear growth of the operator.
This implies that spins at distance $x_{\mathrm{s}}$ from the measurement spin will become entangled in a time
\begin{equation}
    t(x_{\mathrm{s}}) = ax^{\zeta}\rightarrow t_{\mathrm{s}}=t_{\mathrm{exp}}-t(x_{\mathrm{s}})=a(r_{\mathrm{c}}^{\zeta}-x^{\zeta}_{\mathrm{s}}),
\end{equation}
for some $\zeta\in [0,1]$.
We can thus write our decoherence criteria as
\begin{equation}
    \tau_{\mathrm{d}} = \sum_s a(r_{\mathrm{c}}^{\zeta}-x^{\zeta}_{\mathrm{s}})=\int_0^{r_{\mathrm{c}}} dx A(r_{\mathrm{c}}^{\zeta}-x^{\zeta})n(x),
\end{equation}
where $n(x)$ counts the number of spins in a layer $dx$ thick around the radius $x$.
Ignoring the directional dependence for now, we have $n(x)=4\pi x^2 \rho_{\mathrm{s}}$, and so
\begin{equation}
    \tau_{\mathrm{d}} = \int_0^{r_{\mathrm{c}}} dx\; 4\pi\rho_{\mathrm{s}} A(r_{\mathrm{c}}^{\zeta}x^2 -x^{2+\zeta})=4\pi a\rho_{\mathrm{s}}\Big[\tfrac{1}{3}r_{\mathrm{c}}^{\zeta}x^3-\tfrac{1}{3+\zeta}x^{3+\zeta}\Big]_0^{r_{\mathrm{c}}}=4\pi a\rho_{\mathrm{s}}r_{\mathrm{c}}^{3+\zeta}\frac{\zeta}{9+3\zeta}=4\pi\rho_{\mathrm{s}}t_{\mathrm{exp}}r_{\mathrm{c}}^3\frac{\zeta}{9+3\zeta}
\end{equation}
This can be rearranged to give
\begin{equation}\label{eq:rc_fixed_1}
    r_{\mathrm{c}}=\Big[\tfrac{10000}{4\times \pi} \tfrac{9+3\zeta}{\zeta}\Big]^{1/3}\mathrm{\text{\AA}}.
\end{equation}
We plot this as the blue line in Fig.~\ref{fig:rc_fixed_1}, and observe that this yields a result between $20$~\AA~and $40$~\AA~depending on the value of $\zeta$.

\begin{figure}
    \centering
    \includegraphics[width=0.5\linewidth]{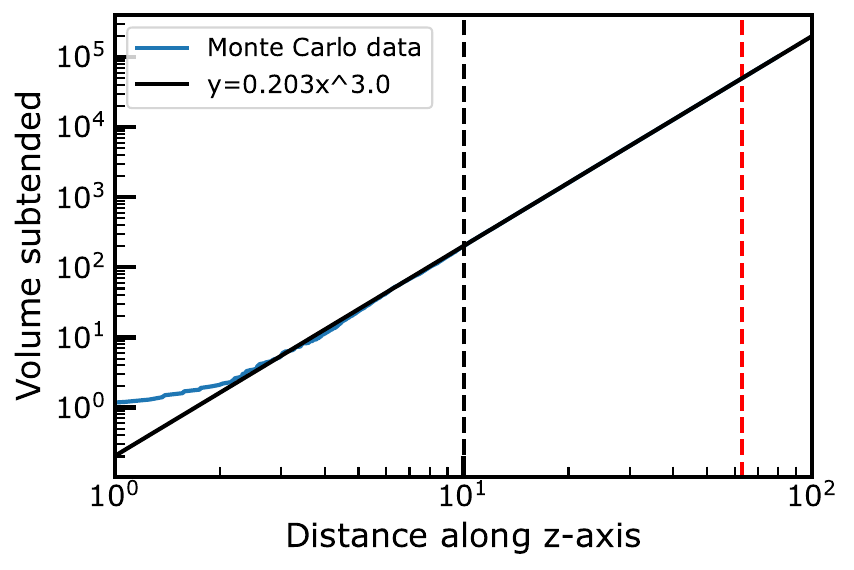}
    \caption{Determination of the volume enclosed by the (absolute) equipotential lines of the dipolar interaction, as a function of the direction along the $z$-axis (the axis of quickest coupling).}
    \label{fig:numerical_integration}
\end{figure}

Next, let us try to consider the effect of directional dependence.
As a simplest first option, we can fix our measurement spin $M$ at some co-ordinate in space, and assume that the spin cluster evolves following the equipotential lines of the $1/r^3(1-3\cos^2(\theta))$ dipolar interaction.
In Fig.~\ref{fig:numerical_integration} we determine this volume via numerical integration and find it to scale as $V=0.2 r_{\mathrm{c}}^3$, which implies that along the $z$-direction, we have $n(x)=0.6x^2\rho_{\mathrm{s}}$.
Substituting this into the above yields
\begin{equation}\label{eq:rc_fixed_2}
    r_{\mathrm{c}}=\Big[\tfrac{10000}{0.6} \tfrac{9+3\zeta}{\zeta}\Big]^{1/3}\mathrm{\text{\AA}}.
\end{equation}
We plot this as the orange line in Fig.~\ref{fig:rc_fixed_1}, and observe that distances of over $60$~\AA~could be measured in an OTOC under the correct circumstances.
Note that in this expression we would expect the exponent of the OTOC growth $\zeta$ to likely be $1$ due to the dimensional reduction in the system.
The above is furthermore a vast simplification of the operator growth dynamics in dipolar-coupled spin systems; understanding this in more detail is a clear target for future work.

%% file: md.tex
To simulate the structure of the solute in the liquid crystal environment, we ran molecular dynamics with a molecular mechanics force field. We use non-bonded parameters of GAFF version 2.11~\cite{wang2004development}, but we replace the bonded parameters with Grappa version 1.3.1~\cite{seute2025grappa}. Although Grappa was designed for biomolecules, the training set includes a very large dataset of small organic molecules and we have observed empirically that the grappa force field makes the aliphatic chains in the liquid crystal more flexible which is consistent with modifications made to GAFF for liquid crystal systems~\cite{zhang2011atomistic}. For reference, we provide the GAFF force-field  structure
\begin{align}
E_{\text{pair}} =& \sum_{\text{bonds}}K_{r}\left(r - r_{\text{eq}}\right)^{2} +  \sum_{\text{angles}}K_{\theta}\left(\theta - \theta_{\text{eq}}\right)^{2} \\
+& \sum_{\text{dihedrals}}\frac{V_{n}}{2}\left[1 + \cos\left(n \phi - \gamma\right)\right] \nonumber \\
+& \sum_{i<j}\left[\frac{A_{ij}}{R_{ij}^{12}} - \frac{B_{ij}}{R_{ij}^{6}} + \frac{q_{i}q_{j}}{\epsilon R_{ij}} \right] \nonumber
\end{align}
where non-bonding parameters are $A_{ij}$ and $B_{ij}$. In the following section we outline the construction and parameterization of the molecular dynamics model and the dihedral learning problem described in the main text.
\subsection{15Q Simulation details}
\label{sec:MD:details}
Starting structures were generated with Packmol version 21.0.0~\cite{martinez2009packmol} at a concentration of 2.05\% $\rm{v/v}$ 1,3-dimethyl-5-(phenyl-$^{13}\rm{C}$)benzene (DMPB) and a starting density of 1.01 g/ml. The liquid crystal molecules were approximately aligned along the x-axis of the simulation box with dimensions [65\AA, 50\AA, 50\AA] by providing angle constraints in the Packmol input file. The initial structures were relaxed by minimizing the system at constant volume and then equilibrated with 500 ps of NPT dynamics. All simulations use a 2 fs time step with hydrogen bond constraints and a hydrogen mass of 4 amu. NPT calculations used a Monte Carlo barostat with updates every 25 steps. All ensemble averages were evaluated with respect to 8 independent trajectories with production times of 1 $\mu$s each. All molecular dynamics simulations were performed using OpenMM \cite{eastman2023openmm}.

\subsection{15Q \texorpdfstring{$P_{2}$}{P2} vs temperature Calibration}
The second-rank orientational order parameter of a molecule, $P_{2}$, characterizes the degree of molecular alignment and is defined as
\begin{align}
    P_{2}=\left\langle\frac{3\cos^{2}\theta-1}{2}\right\rangle
\end{align}
Here, the angle bracket denotes an ensemble average and $\theta$ is the angle between the molecular reference (long) axis and the liquid crystal director. For flexible molecules such as 5CB, the reference axis is chosen as the principal axis of its inertia tensor.

As reported in Ref.~\citenum{pizzirusso2014order}, for biphenyl dissolved in nematic liquid crystal 5CB, the biphenyl $P_{2}$ is directly proportional to $P_{2}$ of 5CB ($P_{2}^{\text{5CB}}$). Therefore, for a similar system like ours, it is essential to first match $P_{2}^{\text{5CB}}$ in both experiment and MD in order for MD to accurately reproduce the experimental orientational order of the 15Q molecule. The $P_{2}^{\text{5CB}}$ from MD can differ from experiment in two ways: (1) the critical temperature can be different and (2) the behavior of $P_2$ relative to the critical temperature can be different. As pointed out in Ref.~\citenum{pizzirusso2014order}, even with a well-optimized force field, $P_{2}^{\text{5CB}}$ can still differ by $\sim10\%$ between MD and experiment at the same reduced temperature. Furthermore, even determining the critical temperature accurately from MD can be difficult since $P_{2}^{\text{5CB}}$ becomes hard to converge near the critical point. However, since we are not concerned with the temperature dependence of orientational order itself, rather than attempting to further optimize MD, we adopted a simpler strategy: we experimentally measured $P_{2}^{\text{5CB}}$ as a function of temperature, and used this calibration curve to determine the experimental temperature that reproduces the same $P_{2}^{\text{5CB}}$ value as in MD ($T_{\text{match}}$). $T_{\text{match}}$ was determined to be 289K, and was used for subsequent OTOC experiments. This approach was later validated by MQC experiments. Though the MD value of $P_{2}^{\text{5CB}}$ at 300K is likely not sufficiently converged, the calculations at 294K are well converged and the predicted $P{2}$ value thus serves as the reference point for calibration relative to experiment. 

\begin{figure}[H]
    \centering
    \includegraphics[width=0.5\linewidth]{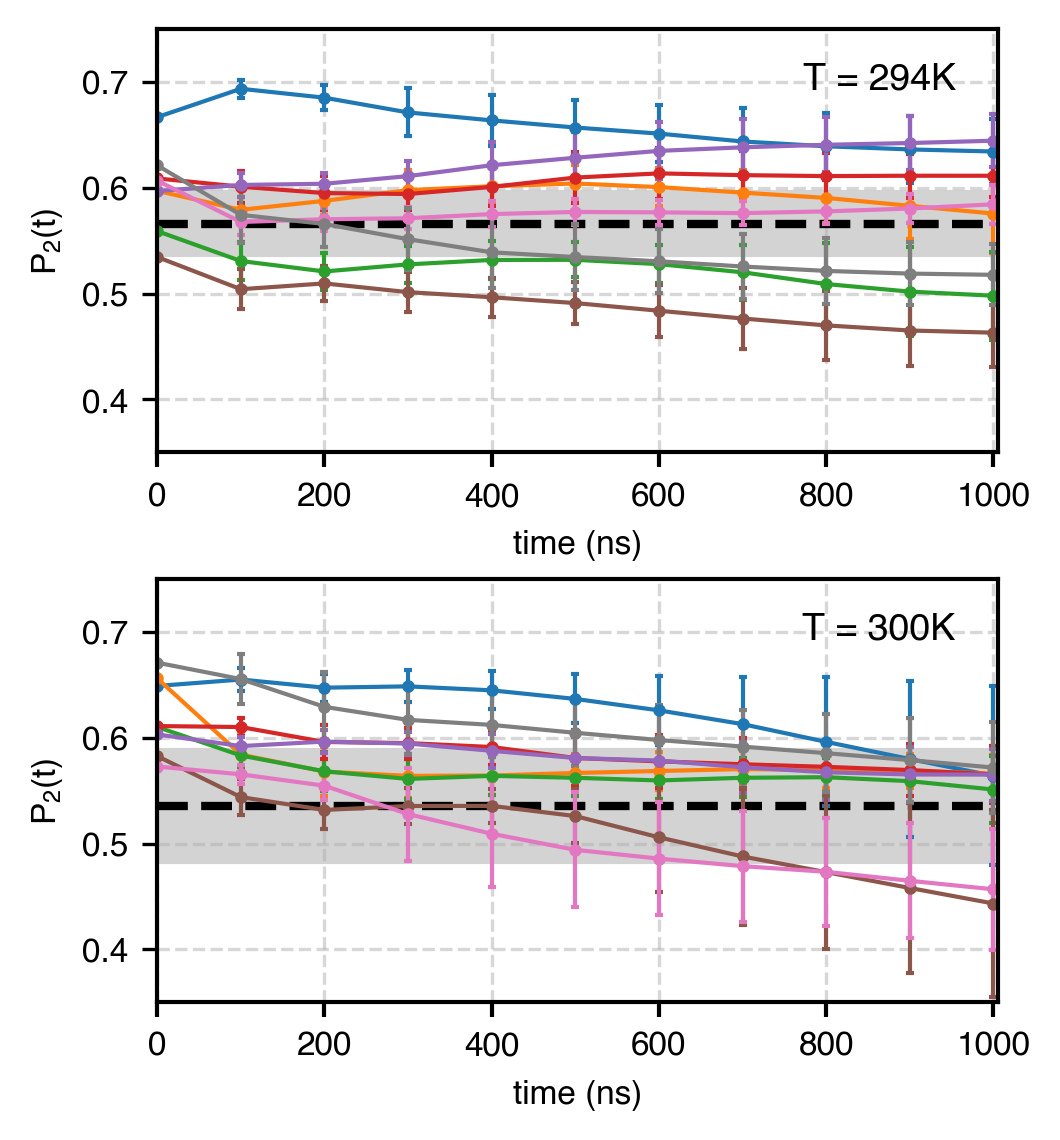}
    \caption{Cumulative average and standard deviation of $P_2$ along each trajectory for the simulations run at $T=294$ K and $T=300$ K. The black dashed line and shaded gray regions correspond to the overall mean and error for the full ensemble. The ensemble error is evaluated as the square root of the mean variance.}
    \label{fig:P2_Error_Analysis}
\end{figure}

The simulated $P_2$ values at each temperature are obtained by averaging expectation values over 8 independent 1 $\mu$s trajectories, as described in Section~\ref{sec:MD:details}. In Figure~\ref{fig:P2_Error_Analysis}, we present the cumulative average and standard deviation of $P_2$ value along each trajectory. The cumulative average for each trajectory was calculated according to 
\begin{equation}\label{eq:cumulative_p2}
P_2(t) = \frac{1}{t}\int_0^t dt' \  p_2(t'),
\end{equation}
where $p_2(t')$ is the instantaneous orientational order parameter.

While the cumulative averages for each trajectory run at $T=294$ K have largely stabilized toward the end of the runtime, several of the trajectories run at $T=300$ K are not converged, as can be seen by the relatively large changes in the cumulative average combined with the increasing standard deviations over the second half of the trajectory.

Experimentally, $P_{2}^{\text{5CB}}$ was determined by measuring the $^{13}\text{C}$ chemical shifts of 5CB~\cite{guo1991determination}. It has been shown that there exists a simple semi-empirical relationship between the $^{13}\text{C}$ chemical shifts and the order parameter:
\begin{equation}
    \Delta \sigma=aP_{2}^{\text{5CB}}+b
\end{equation}
where $\Delta \sigma=\sigma_{\text{n}}-\sigma_{\text{iso}}$, with $\sigma_{\text{n}}$, $\sigma_{\text{iso}}$ being the $^{13}\text{C}$ chemical shifts in nematic and isotropic phases. The empirical coefficients $a$ and $b$ are listed in~\cite{guo1991determination}\footnote{The empirical coefficients $a$, $b$ for $\text{C}-3^{\prime}$ listed in ~\cite{guo1991determination} are incorrect, as pointed out in ~\cite{sandstrom1996orientational}. Therefore, all carbons in the 5CB biphenyl cores except $\text{C}-3^{\prime}$ were used to calculate $P_{2}^{\text{5CB}}$. ~\cite{sandstrom1996orientational} also identified several additional errors. }

\begin{figure}[H]
    \centering
    \includegraphics[width=8.5cm]{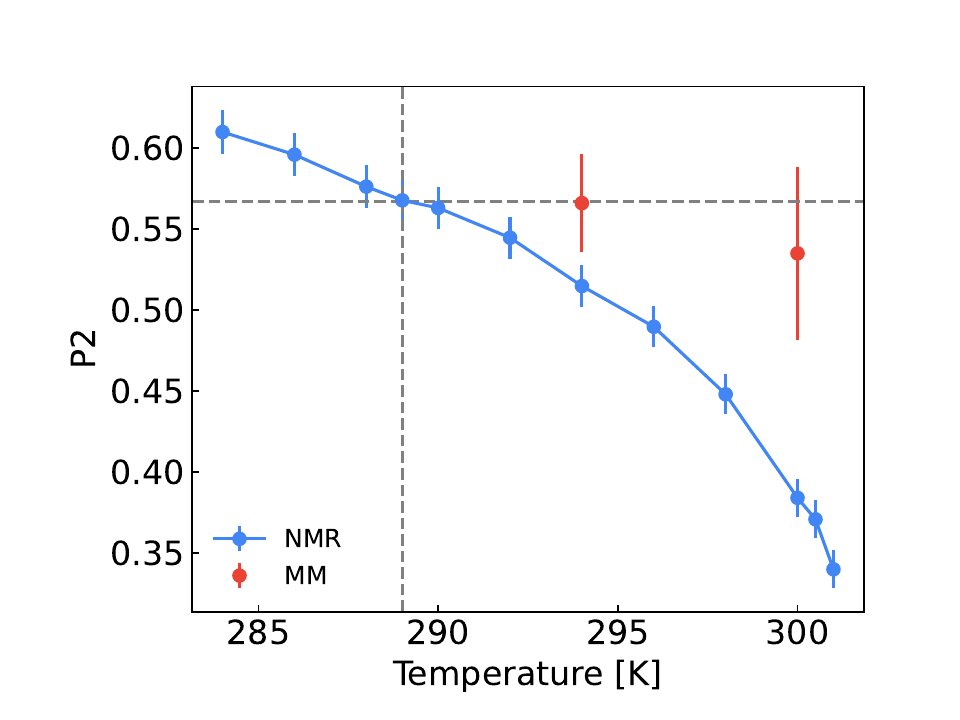}
    \caption{Calibration curve of experimental $P_{2}^{\text{5CB}}$ vs. temperature, overlaid with MD-simulated $P_{2}^{\text{5CB}}$. MD simulation at both 294K and 300K yields $P_{2}^{\text{5CB}}\sim0.566$, which corresponds to $T_{\text{match}}=289K$. Error bars on experimental $P_{2}^{\text{5CB}}$ are obtained by propagating the reported uncertainties in the empirical coefficients $a$ and $b$~\cite{guo1991determination}}
    \label{fig:P2_NMR_MM}
\end{figure}

\subsection{Analysis of timescales for the 15Q system}\label{sec:MD:timscales}
\begin{figure}
    \includegraphics[width=0.4\textwidth]{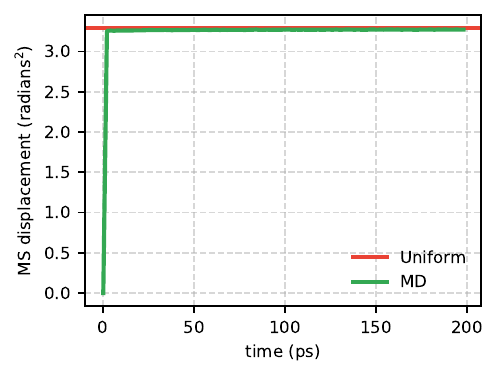}
    \includegraphics[width=0.4\textwidth]{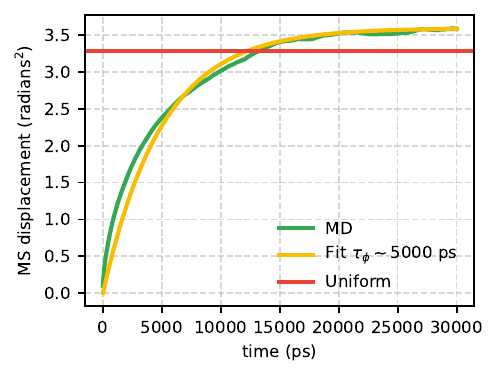}
    \includegraphics[width=0.4\textwidth]{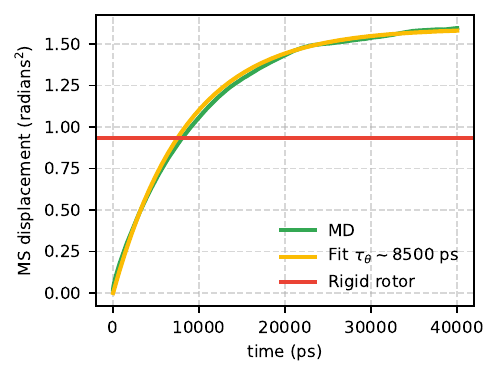}
    \includegraphics[width=0.4\textwidth]{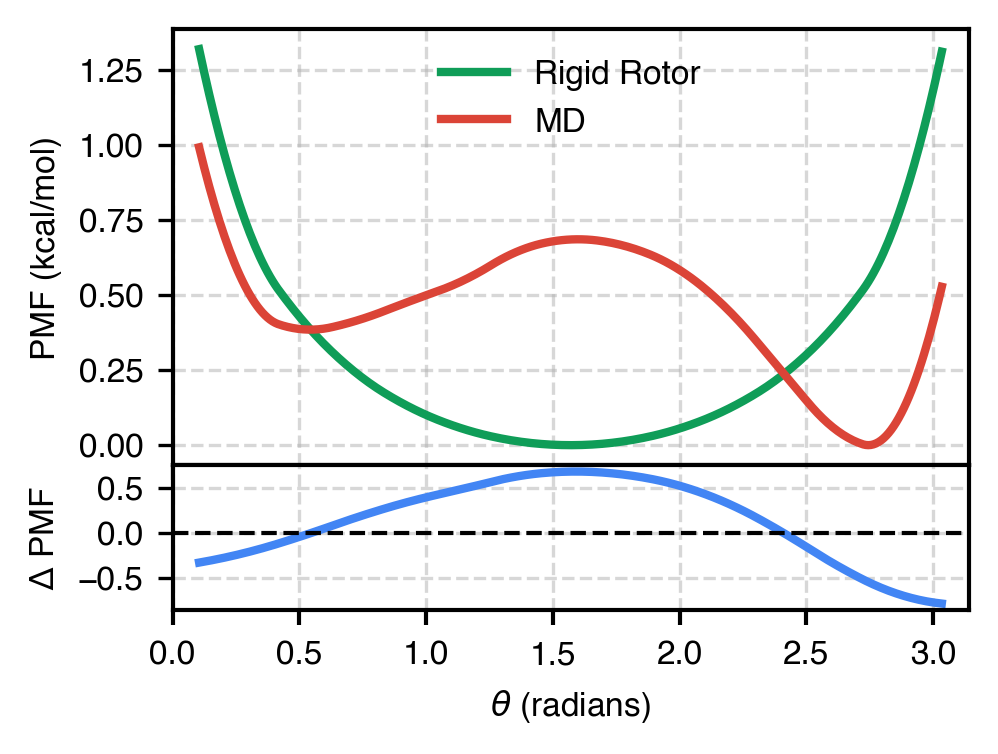}
    \caption{\label{fig:dmbp_md_analysis} The mean-square displacement as a function of simulation time for the methyl-rotation ($\chi$), the inter-ring torsion ($\phi$) and the overall alignment with the liquid crystal director ($\theta$). The diffusion constants are $D_{\chi}^{-1} <$ 1 ps, $D_{\phi}^{-1} =$ 900 ps and $D_{\theta}^{-1} =$ 4 ns. We also show the PMF along the alignment with the liquid crystal director from MD, assuming a rigid rotor, and the difference between the two which reveals that interactions with the liquid crystal destabilize perpendicular orientations and stabilize parallel and antiparallel orientations.}
\end{figure}

For the 15Q system we verify that the dominant time-scales impacting the OTOC indicating there is no observable low-frequency mode on the scale of the experiment. We analyze the timescales associated with the methyl rotation angle, $\chi$, the inter-phenyl torsion angle ($\phi$) and the overall alignment with the liquid crystal director, ($\theta$). We can analyze the timescale associated with these angular quantities by computing the mean-squared displacement in each coordinate as a function of time. For example, we compute,
\begin{equation}
    \langle (\phi(t_0) - \phi(t_0 + t))^2\rangle = \left\langle\frac{1}{N}\sum_{i = 1}^N (\phi_i(t_0) - \phi_i(t_0 + t))^2\right\rangle_{t_0},
\end{equation}
where $N$ is the number of solute molecules in the simulation. Here, the expectation values are evaluated by averaging over time origins, $t_0$. In the long-time limit, the two angles should become uncorrelated so that
\begin{equation}
    \lim_{t\rightarrow \infty}\langle (\phi(t_0) - \phi(t_0 + t))^2\rangle = 2 \langle \phi^2\rangle - 2 \langle \phi\rangle^2 = 2\text{Var}[\phi].
\end{equation}
For torsions, such as $\phi$ and $\chi$, which would be uniformly distributed on $[0, 2\pi)$ in the absence of interactions, this means that
\begin{equation}
    \lim_{t\rightarrow \infty}\langle (\phi(t_0) - \phi(t_0 + t))^2\rangle \approx \frac{(2\pi)^2}{6}
\end{equation}
The same is true for the polar angle, $\theta$, except that it is not uniformly distributed because of the non-uniform density of states. In the absence of any correlations, the distribution will be $p(\theta) = 1/2 \sin\theta $. This means that, in the long time limit, the mean squared displacement should be approximately
\begin{equation}
     \lim_{t\rightarrow \infty}\langle (\theta(t_0) - \theta(t_0 + t))^2\rangle \approx 2\text{Var}[\theta] = \pi^2/2 - 4.
\end{equation}
Given these expressions for the limiting behavior at long times, we can fit the mean-squared displacement to a function of the form
\begin{equation}
    \text{MSD}[X](t) = C_X\left[1 - e^{-D_Xt}\right] \qquad [X = \phi, \theta],
\end{equation}
where $C_X$ is fixed according to the empirically observed long-time limit, which differs from the uncorrelated case, and $D$ is like a diffusion constant with units of inverse time so that we can define a relaxation time, $\tau_X \equiv 1 / D_X$. Plots of the mean-square displacement are shown in Figure~\ref{fig:dmbp_md_analysis}. The inverses of the fitted constants, $D_X$, provide estimates of the timescale associated with motion in these coordinates.

\subsection{15Q Analysis of the effect of the non-isotropic environment}
From molecular dynamics simulations, we can extract the free energy surface, or potential of mean force (PMF), along the coordinate that represents the alignment of solute molecules with the liquid crystal director, $\theta$. Specifically, the alignment is obtained by evaluating the angle between the para axis of the biphenyl core of the DMBP molecule and the instantaneous liquid crystal director. As discussed in Section~\ref{sec:MD:timscales}, the density of states is not uniform along this coordinate, so the non-isotropic behavior is most easily seen by subtracting the PMF of a non-interacting rigid rotor as shown in Figure~\ref{fig:dmbp_md_analysis}. As expected, we observe that configurations that are aligned (either parallel or antiparallel) with the director of the liquid crystal are stabilized by the interaction with the liquid crystal environment (negative $\Delta$PMF), while perpendicular orientations are destabilized.

\subsection{15Q Analysis of the dipolar coupling matrix elements}
\begin{figure}
    \centering
    \includegraphics[width=0.4\linewidth]{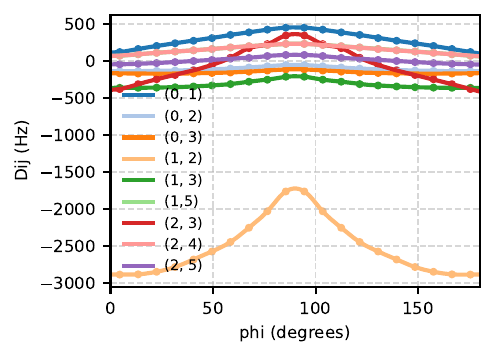}
    \includegraphics[width=0.4\linewidth]{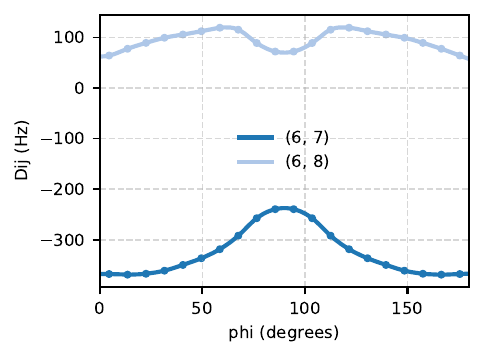}
    \includegraphics[width=0.4\linewidth]{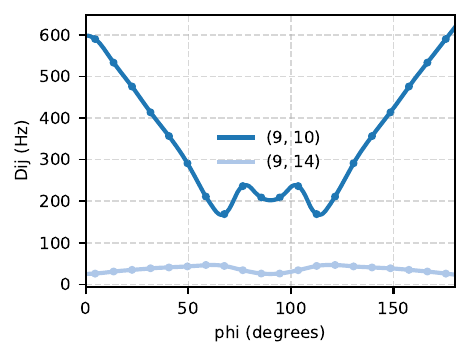}
    \includegraphics[width=0.4\linewidth]{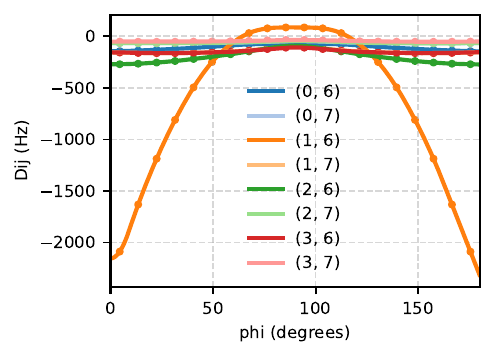}
    \includegraphics[width=0.4\linewidth]{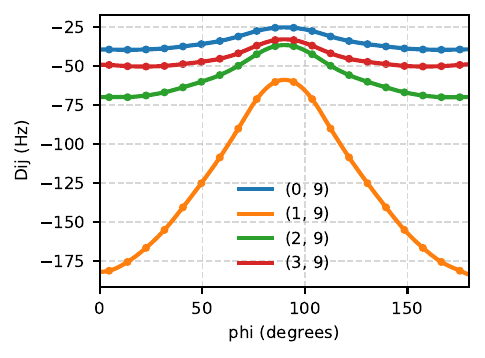}
    \includegraphics[width=0.4\linewidth]{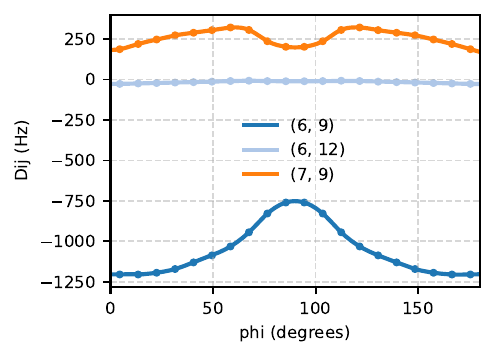}
    \caption{All unique elements of $D_{ij}$ as a function of the torsion angle. The first row shows the intra-ring elements, the second row shows the intra-methyl and inter-ring elements, and the third row shows the ring-methyl elements.}
    \label{fig:Dij}
\end{figure}

To further analyze the dipolar couplings in particular, we plot the matrix elements of the $D_{ij}$ matrix as a function of the torsion angle, $\phi$, in Figure~\ref{fig:Dij}. The indexing is consistent with the atom labels in Figure~\ref{fig:spin_indexing} with the labeled carbon as index zero. As expected, the couplings between hydrogens 1 and 6 (or any other equivalent pair) are most dramatically correlated with $\phi$. For other elements of $D_{ij}$ the effect is more subtle, but still significant.

To assess the uncertainty in predicted dipolar coupling matrix elements, we perform a block averaging analysis on the dipolar coupling time series data for each solute molecule. We note that the dipolar coupling matrix elements are largely determined by the inter-ring torsion, $\phi$, and the alignment with the liquid crystal director, $\theta$. We define the blocks to be proportional to the $\theta$ relaxation time, and choose $20$ nanosecond blocks (where the associated mean squared displacement is roughly representative of no correlation), resulting in 50 blocks per solute molecule per trajectory. We subsequently define the standard error as $S = \frac{\sigma_{b}}{\sqrt{M_b}}$ where $\sigma_b$ is the standard deviation of block averages, and the $M_b$ is the total number of blocks ($M_b= 20\times50\times8=8000$). The resulting bounds on the 95\% confidence interval for a given dipolar coupling matrix element is then evaluated as $D_{ij}\pm1.96S_{ij}$.

\begin{table}[h!]
  \centering
  \caption{Computed dipolar coupling values (in Hz) and their 95\% confidence intervals for the 15Q system.}
  \label{tab:fitting result_15q}
  \begin{tabular}{p{2cm} p{3cm} p{2cm} p{3cm}}
    \hline
    Parameter&Value&Parameter&Value\\
    \hline
    $D_{C,1}$  & $237.27 \pm 22.49$   & $D_{C,2}$ & $-129.40 \pm 2.81$ \\
    $D_{C,3}$  & $-170.46 \pm 4.85$   & $D_{C,6}$ & $-121.01 \pm 4.81$ \\
    $D_{C,7}$  & $-67.43 \pm 1.92 $   & $D_{C,9}$ & $-38.30 \pm 0.85$ \\
    $D_{1,2}$  & $-2753.91 \pm 78.43$ & $D_{1,3}$ & $-352.21 \pm 10.13$ \\
    $D_{1,4}$  & $-21.41 \pm 7.90$    & $D_{1,5}$ & $125.77 \pm 11.66$ \\
    $D_{1,6}$  & $-844.83 \pm 30.18$  & $D_{1,7}$ & $-160.86 \pm 4.31$ \\
    $D_{1,9}$  & $-152.64 \pm 3.97$   & $D_{2,3}$ & $-184.78 \pm 41.98$ \\
    $D_{2,6}$  & $-235.79 \pm 6.27$   & $D_{2,9}$ & $-66.03 \pm 1.75$ \\
    $D_{3,6}$  & $-161.29 \pm 4.25$   & $D_{3,7}$ & $-52.18 \pm 1.48$ \\
    $D_{3,9}$  & $-49.42 \pm 1.32$    & $D_{6,7}$ & $-357.45 \pm 8.46$ \\
    $D_{6,8}$  & $97.92 \pm 12.61$    & $D_{6,9}$ & $-1158.80 \pm 31.29$ \\
    $D_{6,12}$ & $-19.38 \pm 2.65$   & $D_{7,9}$ & $268.22 \pm 34.05$ \\
    $D_{9,10}$ & $411.47 \pm 49.54$  & $D_{9,12}$ & $ 38.31 \pm 4.89$ \\
    \hline
  \end{tabular}
\end{table}

\subsection{QM correction to the PMF}
A simple correction to the PMF along the DMBP dihedral angle can be evaluated by approximating the expectation value of Boltzmann reweighting factor. In order to evaluate this correction, we consider each DMBP solute molecule, whose coordinates are denoted by $x$, as a QM region. We then consider the solvent and all other solute molecules as an MM region, and their coordinates are denoted by $R$. The QM/MM energy is then given by
\begin{equation} \label{eq:qmmm_energy}
    V_{QM/MM}(x, R) = V_{MM}(x, R) + V_{QM}(x) - V_{MM}(x),
\end{equation}
such that the PMF as a function of $\phi$ for each solute molecule can be expressed as
\begin{equation}\label{eq:qmmm_pmf_phi}
\begin{split}
    F_{QM/MM}(\phi) &= -\beta^{-1} \log\left(\int dx dR \ e^{-\beta V_{QM/MM}(x, R)} \delta(\phi'(x) - \phi)\right) + C\\
    & = -\beta^{-1} \log\left( \left\langle e^{-\beta(V_{QM}(x) - V_{MM}(x))}\delta(\phi'(x) - \phi) \right\rangle_{W} \right) + C'.
\end{split}
\end{equation}
Here, we have used Eq. \eqref{eq:qmmm_energy} for the energy and have defined the solvation free energy surface of the solute in the liquid crystal as
\begin{equation}
    W(x) =  -\beta^{-1} \log\left(\int dR \ \exp(-\beta V_{MM}(x, R))\right).
\end{equation}
In this simple scheme, the QM/MM PMF can then be obtained by evaluating the conditional expectation value of a Boltzmann reweighting factor that is proportional to the difference between the vacuum QM and vacuum MM energies using solute configurations from the ensemble generated by a molecular mechanics potential.

\begin{figure}
    \centering
    \includegraphics[width=0.45\linewidth]{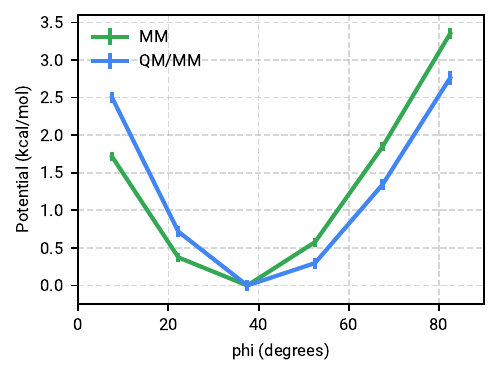}
    \includegraphics[width=0.45\linewidth]{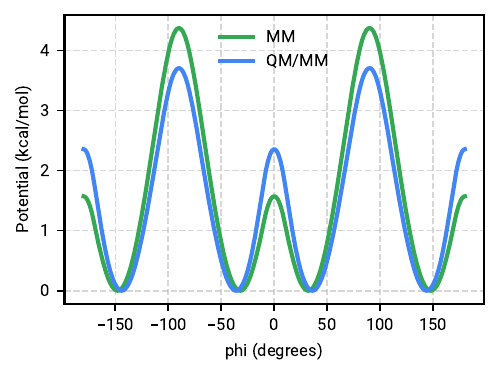}
    \caption{The average MM and estimated QM/MM potential energy as a function of the torsion angle computed in the ensemble generated from MM trajectories (left). The QM-corrected PMF (all data is mapped to (-180, 180] and interpolated) (right).}
    \label{fig:PMF_correction}
\end{figure}

To simplify the analysis and circumvent statistical error, we then approximate the exponential to first order. After expanding the logarithm, we find that the QM/MM PMF can be approximated by a correction to the MM PMF,
\begin{equation}
    F_{\mathrm{QM/MM}}(\phi) \approx F_{\mathrm{MM}}(\phi) + \Delta E(\phi),
\end{equation}
where the correction is given by
\begin{equation}
    \Delta E(\phi) = \frac{\int dR dx e^{-\beta V_{MM}(x, R)}(V_{QM}(x) - V_{MM}(x))\delta(\tilde{\phi}(x) - \phi) }
    {\int dR dx e^{-\beta V_{MM}(x, R)} \delta(\tilde{\phi}(x) - \phi)}.
\end{equation}
The correction is computed by binning the structures from all MM trajectories as a function of $\phi$ and then computing the average of $V_{QM} - V_{MM}$ within in each bin. The results of this analysis are shown in Figure~\ref{fig:PMF_correction}. The QM correction has the effect of lowering the barrier at 90 (and -90) degrees and raising the barrier at 0 (and 180) degrees.
\subsection{15Q learning experiment candidate PMFs}
\begin{figure}[H]
    \centering
    \includegraphics[width=0.45\linewidth]{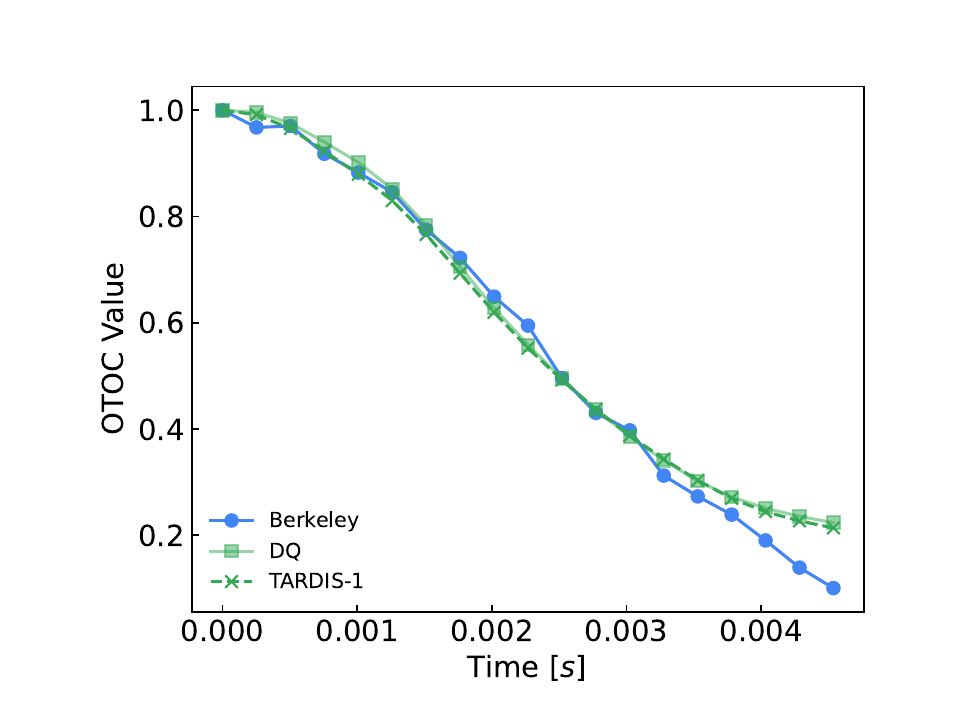}
    \includegraphics[width=0.45\linewidth]{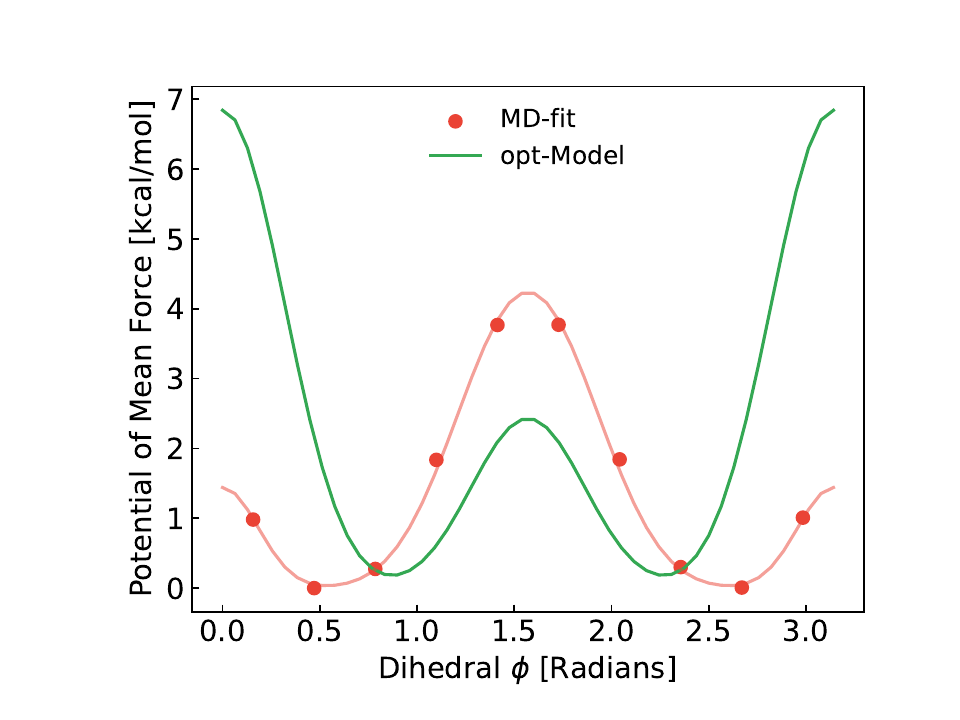}
    \caption{(\textit{left}) Comparison of OTOC decay curves simulated using a double-quantum Hamiltonian (green squares) and the full TARDIS-1 pulse sequence (green cross) and the NMR OTOC decay curve generated using the TARDIS-1 pulse sequence (blue). (\textit{right}) PMFs for MD/MM (red) compared to the PMF determined by fitting the double-quantum OTOC to the 300K NMR data (green).}
    \label{fig:Q15_model_fit_to300K_data}
\end{figure}

The 15Q learning experiment involves generating 9 candidate PMFs and then using NMR data to confirm which PMF generates $D_{ij}$ coupling coefficients that give the lowest RMSE or generates MQC spectra with the lowest RMSE in peak position. Our strategy was to identify possible PMF curves and interpolate between them as a more efficient substitute to gradient descent on the parameter space of the model for the PMFs. This greatly simplifies the learning experiment to identifying which PMF matches the experiment the best. To generate the 9 candidate PMFs we interpolate between the PMF obtained from MD/MM, a vacuum torsion scan of the DMBP with the wB97X-D3BJ functional using a def2-tzvp basis, and an artificial shifted double well potential. The artificial double well potential is obtained by fitting the parameters of a double-quantum 15Q Hamiltonian Eq.~\eqref{eq:zeroth_order_magnus_tardis-1} to an OTOC measured using the TARDIS-1 gate sequence on a sample of DMBP in 5CB liquid crystal at 300 Kelvin. The model used to fit the NMR data (Fig.~\ref{fig:Q15_model_fit_to300K_data}(left) blue) is
\begin{align}\label{eq:ML_model_fit}
U'(c, a, m, n) = c\left((1-a) \cos(x)^{n} + a \sin(x)^{m}\right)
\end{align}
where the optimal parameters are found to be $c=9.271505070720735$, $a=0.2620267312627955$, $m=12$, $n=10$. The optimal parameters were determined by Gaussian Process optimization implemented in the Bayesian Optimization Python Package~\cite{bayesoptpython}. The optimization compared the L2-norm of the difference in OTOC decay points computed at times $[2, 4, 6, 8, 9, 10, 12, 13]$ (index starting at zero) of the model (Eq.~\eqref{eq:ML_model_fit} and NMR data.  Shown in Fig.~\ref{fig:Q15_model_fit_to300K_data} along with the NMR TARDIS-1 OTOC decay curve is the simulated double-quantum OTOC decay curve and the simulated TARDIS-1 OTOC decay curve using the optimal parameters.  The right side of Fig.~\ref{fig:Q15_model_fit_to300K_data} plots Eq.~\eqref{eq:ML_model_fit} at the optimal parameters (green) compared to the MD/MM PMF.  

\begin{figure}[H]
    \centering
    \includegraphics[width=0.5\linewidth]{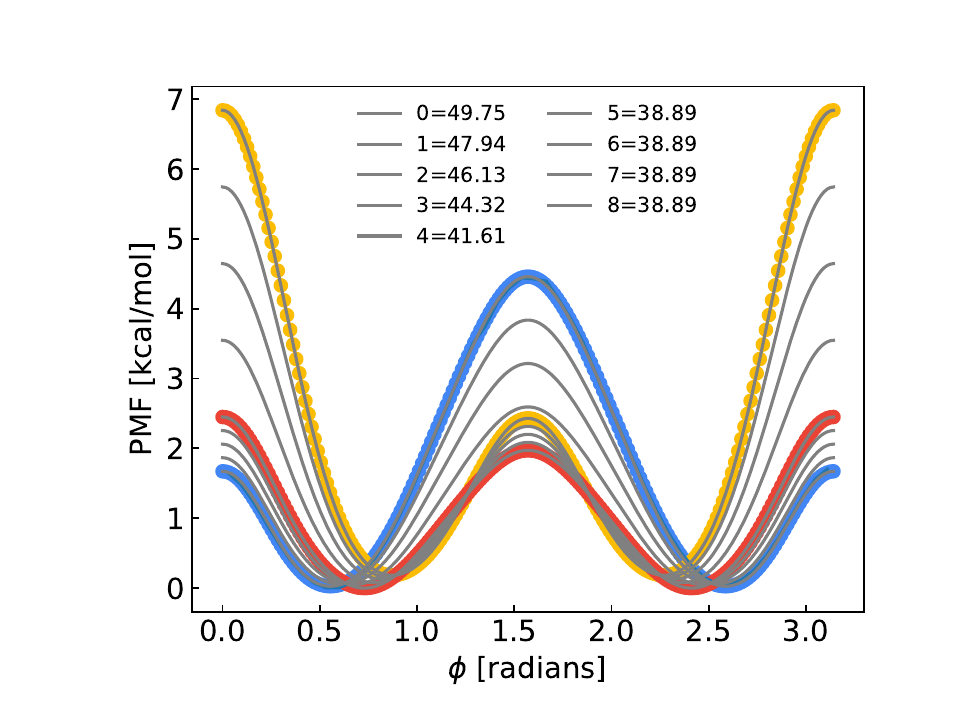}
    \caption{9 candidate PMFs (grey) interpolated between the artificial double well (yellow), vacuum DFT curve (red), and MD/MM (blue).  The legend indicates the minimum of each PMF. The $U_{0}$ PMF is the artificial double well, the $U_{4}$ PMF is the vacuum DFT curve, and $U_{8}$ PMF is MD/MM. }
    \label{fig:pmfs_candidate_with_minimum}
\end{figure}

Following a similar double well fitting protocol as Ref.~\citenum{pizzirusso2014order} the first seven even Fourier coefficients can be fit to approximate a double well between $[0, \pi]$ for the dihedral angle of DMBP. We first fit the coefficients of Eq.~\eqref{eq:pmf_even_fourier_coeffs} for MD/MM PMF, vacuum DFT torsion scan, and the aforementioned artificially fit at higher temperature PMF. The Fourier coefficients for these three curves labeled as candidate PMF $8,4,0$ in Table~\ref{tab:even_fourier_coefficients_of_candidate_PMFS}, respectively.  We then linearly interpolate between the three curves producing curves $[1, 2, 3]$ and curves $[5, 6, 7]$.  All Fourier coefficients for all 9 curves are shown in Table~\ref{tab:even_fourier_coefficients_of_candidate_PMFS}. Figure~\ref{fig:pmfs_candidate_with_minimum} shows all interpolated PMFs along with the dihedral angle with the minimum free energy.
\begin{align}\label{eq:pmf_even_fourier_coeffs}
U(\phi) = \sum_{n}u_{n}\cos(2 n \phi)
\end{align}
\begin{table}[H]
    \centering
    \begin{tabular}{|c|c|c|c|c|c|c|c|c|c|}
    \hline \hline
Candidate PMF Index & 0 & 1 & 2 & 3 & 4 & 5 & 6 & 7 & 8 \\
\hline 
$u_{0}$ & 2.2318 & 1.9306 & 1.6294 & 1.3282 & 1.0270 & 1.1817 & 1.3364 & 1.4910 & 1.6457 \\
$u_{2}$ & 1.8669 & 1.4057 & 0.9445 & 0.4834 & 0.0222 & -0.3806 & -0.7834 & -1.1861 & -1.5889 \\
$u_{4}$ & 2.1908 & 1.9118 & 1.6329 & 1.3539 & 1.0750 & 1.1241 & 1.1732 & 1.2224 & 1.2715 \\
$u_{6}$ & 0.3404 & 0.3026 & 0.2647 & 0.2269 & 0.1890 & 0.1828 & 0.1766 & 0.1704 & 0.1642 \\
$u_{8}$ & 0.2119 & 0.1809 & 0.1499 & 0.1189 & 0.0879 & 0.0962 & 0.1045 & 0.1127 & 0.1210 \\
$u_{10}$ & -0.0009 & 0.0063 & 0.0134 & 0.0205 & 0.0276 & 0.0283 & 0.0290 & 0.0297 & 0.0304 \\
$u_{12}$ & 0.0012 & 0.0059 & 0.0105 & 0.0152 & 0.0199 & 0.0216 & 0.0233 & 0.0250 & 0.0267 \\
\hline \hline
    \end{tabular}
    \caption{Coefficients for each candidate PMF interpolated from the yellow to red to blue curves of Fig~\ref{fig:pmfs_candidate_with_minimum} where the interpolating function is defined in Eq.~\eqref{eq:pmf_even_fourier_coeffs}.}
    \label{tab:even_fourier_coefficients_of_candidate_PMFS}
\end{table}

Finally, we note that though one of these candidate PMFs curves are obtained from a vacuum DFT torsion scan of DMBP the selection of this curve giving the lowest RMSE compared to NMR TARDIS-1 is purely coincidence. This coincidence does not mean vacuum DFT is sufficient for accurately modeling free energy surfaces in liquid crystal environments. In this work, the vacuum DFT curve is merely used as a candidate interpolation point. A full PMF computed with density functional theory would require \textit{ab initio} MD with an appropriate density functional energy for forces. Though \textit{ab initio} MD on the systems used in this work may be possible it would be a heroic computational effort and difficult to justify experimental expense.
\subsection{9Q Simulation details and results}
Simulations of the 9Q (Toluene) system in EBBA liquid crystal were performed using molecular mechanics parameters from GAFF version 2.11~\cite{wang2004development}. The initial structures were generated with Packmol version 21.0.0~\cite{martinez2009packmol} at a concentration of 0.33\% $\rm{wt\%}$ Toluene and a starting density of 1.00 g/ml. The EBBA molecules were approximately aligned along the x-axis of the simulation box with dimensions [52\AA, 30\AA, 30\AA] by providing angle constraints in the Packmol input file. The initial structures were relaxed by minimizing the system at constant volume and then equilibrated with 500 ps of NPT dynamics at a temperature of 300 K. All simulations use a 2 fs time step with hydrogen bond constraints and a hydrogen mass of 4 amu. NPT calculations used a Monte Carlo barostat with updates every 25 steps. After equilibration, expectation values were evaluated from 8 independent from 1 $\mu$s production runs.

\begin{figure}
    \includegraphics[width=0.4\textwidth]{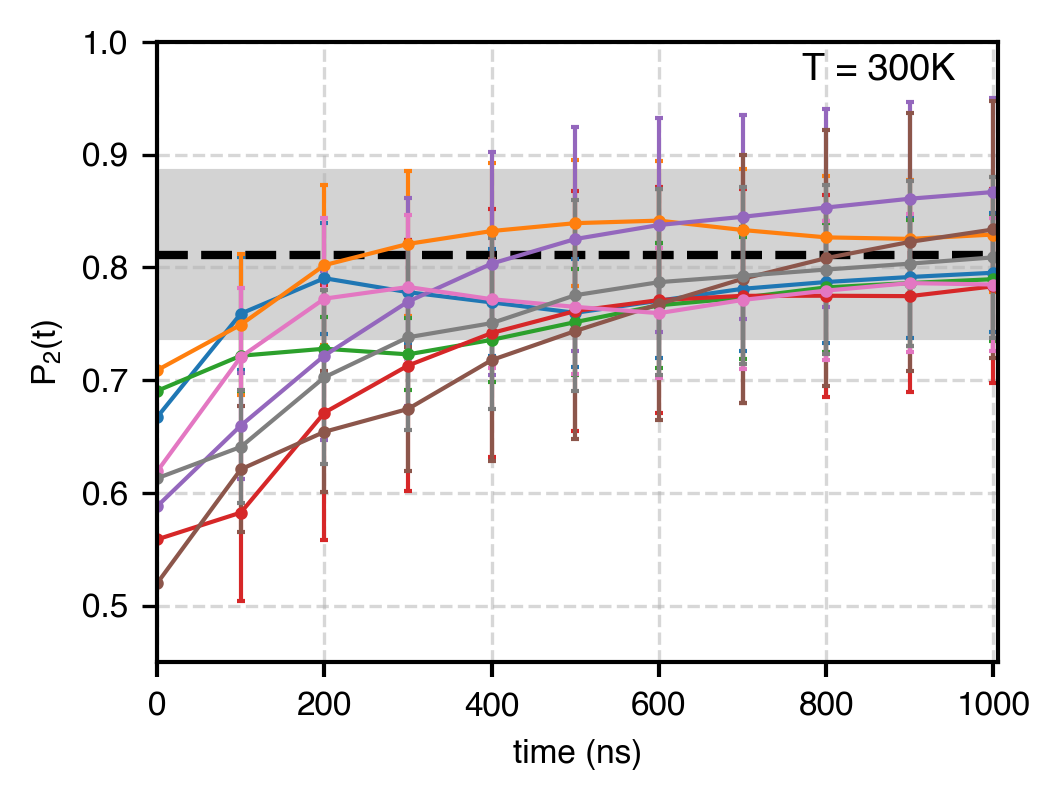}
    \includegraphics[width=0.4\textwidth]{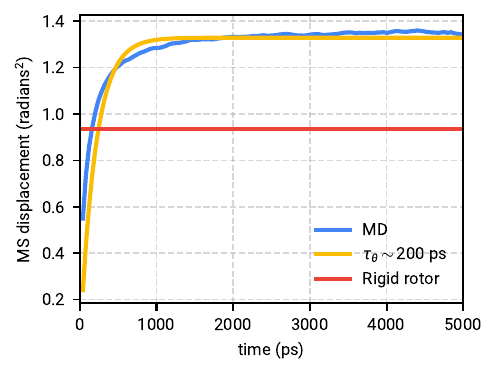}
    \includegraphics[width=0.4\textwidth]{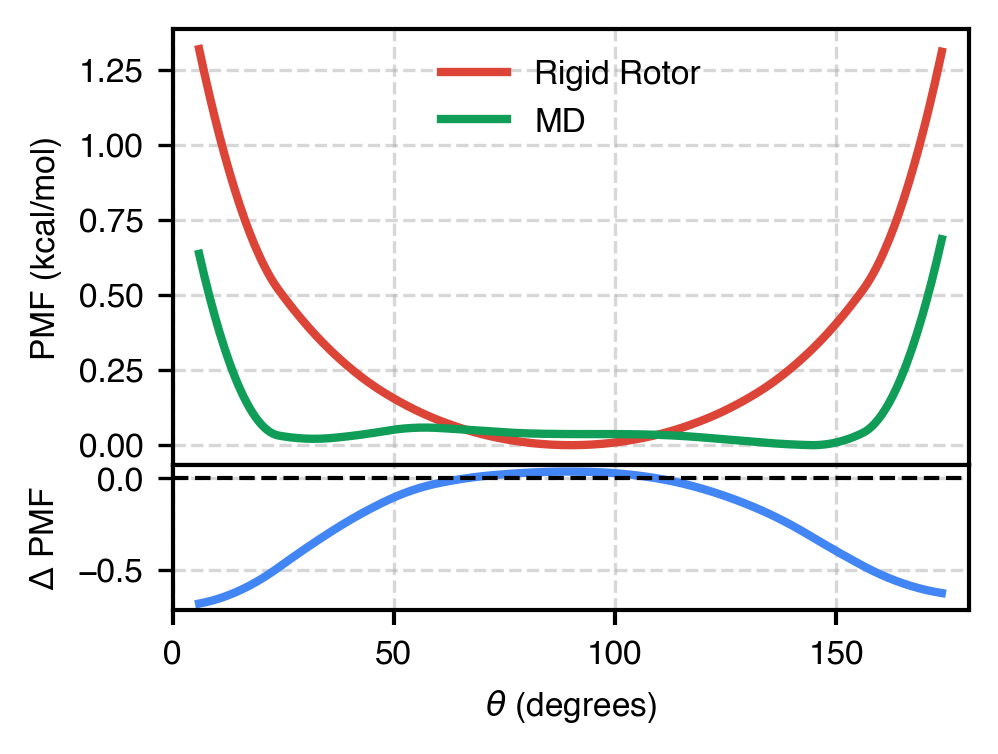}
    \caption{\label{fig:9Q_results} The cumulative average of the orientational order parameter for an EBBA liquid crystal for 8 independent 1 $\mu$s trajectories simulated at T=300 K (top-left). The mean-square displacement as a function of simulation time for the overall alignment of Toluene with the liquid crystal director (top-left). The potential of mean force along the liquid crystal alignment angle, $\theta$ from the simulation (green) and assuming a rigid rotor (red). The difference between these (blue) reveals that interactions with the liquid crystal stabilize the parallel and anti-parallel configurations (bottom). 
    }
\end{figure}

At a temperature of 300K, the EBBA liquid crystal becomes highly ordered, and multiple trajectories were observed to have (at least transiently) crystallized. This is reflected by the high $P_2$ value and the large variance associated with each trajectory in Figure~\ref{fig:9Q_results}.

%% file: trotter.tex
We will now show that we can can use low order Trotter formulas to simulate the time evolution generated by the average Hamiltonians described in Section~\ref{sec:magnus} to obtain circuits of reasonable depth that are capable of simulating the OTOC experiment to sufficiently high accuracy. Given the error bars on the experiment, we consider an error in the OTOC value that is $< 0.02$ to be acceptable. Unless otherwise noted, all simulations reported in this section are noiseless with the dipolar Hamiltonian derived from the $u = 4$ PMF of the 15Q system.

\subsection{Analysis of Trotter accuracy}\label{sec:trotter_analysis}
When using a product formula to simulate the time evolution for short times, the error in the OTOC will be one order higher in general, and two orders higher for the first-order Trotter.
To show this result, the error due to using an order-$p$ product formula can be expressed using the effective
Hamiltonian
\begin{equation}
    H_{\rm eff} = H + \delta H t^{p} + \cO(t^{p+1}),
\end{equation}
where $H$ is the true Hamiltonian, and $\delta H$ is an error term in the form of multi-commutators of the terms in $H$.
Now let $M(t)$ be $M$ after time evolution under $H$, so
\begin{equation}
    M(t) = e^{-iHt} M e^{iHt} \, .
\end{equation}
Then, if $M_{\rm prod}(t)$ is $M$ with the time evolution under the product formula, the error can be given as
\begin{equation}
    M_{\rm prod}(t) - M(t) = - i[\delta H,M] t^{p+1} + \cO(t^{p+2}).
\end{equation}
The OTOC value is given by
\begin{equation}
    C(t) = \Tr(B^\dag M(t) B M(t)) \, ,
\end{equation}
so the error from using the product formula is
\begin{align}
\Delta C(t) &= -i t^{p+1} \Tr(B^\dag[\delta H,M] B M(t)) -i t^{p+1} \Tr(B^\dag M(t) B [\delta H,M])  + \cO(t^{p+2})\nn
&=-i t^{p+1} \Tr(B^\dag[\delta H,M] B M) -i t^{p+1} \Tr(B^\dag M B [\delta H,M])  + \cO(t^{p+2}) .
\end{align}
In the second line we have used the fact that $M$ only differs by $M(t)$ by $\cO(t)$, and so the difference can be bundled into $\cO(t^{p+2})$.
Next, we can simplify the error to
\begin{align}\label{eq:firstcase}
   \Delta C(t) &= -i t^{p+1} \Tr([\delta H,M] B M B^\dag)-i t^{p+1} \Tr([\delta H,M] B^\dag M B) + \cO(t^{p+2}) \nn
    &=-2i t^{p+1} \Tr([\delta H,M] M) + \cO(t^{p+2})\nn
    &=-2i t^{p+1}\left(  \Tr(\delta H M M) -  \Tr(M\delta H M) \right) + \cO(t^{p+2})\nn
    &=-2i t^{p+1}\left(  \Tr(\delta H M M) -  \Tr(\delta H M M) \right) + \cO(t^{p+2})\nn
    &= \cO(t^{p+2}) \, .
\end{align}
The first line uses the cyclic permutation property, the second uses the fact that $M$ and $B$ commute, the third expands the commutator, and the fourth uses cyclic permutation again.

This increase in the order is for the total evolution time, rather than for later time steps.
For approximation of time evolution using product formulae, the total time $t$ is broken up into $r$ steps, and there is error $\cO((t/r)^{p+1})$ for each step, for a total of $\cO(r(t/r)^{p+1})$.
The above reasoning is not exact for each of the $r$ time steps (so these do not give error order $\cO((t/r)^{p+2})$).
However, because the total error is $\cO(t^{p+1})$ in the total evolution time, for short overall time the above reasoning shows that the error is reduced.

In the case where the product formula is first order we can go further and show that one more term in the error cancels, giving the error term of order $t^4$.
The conditions to be used in this derivation are that
\begin{enumerate}
    \item $B$ and $M$ commute,
    \item $MM=I$,
    \item $M H M = -H_{\rm het} + H_{\rm hom}$,
    \item $H_{\rm het}$ is a sum of commuting terms,
    \item the terms in $H_{\rm hom}$ have Pauli operators on pairs of qubits excluding qubit 1,
    \item the terms in $H_{\rm het}$ have Pauli operators on pairs of qubits \emph{including} qubit 1, and
    \item $B$ is a tensor product of single-qubit operations.
\end{enumerate}
Here, $H_{\rm het}$ and $H_{\rm hom}$ are the heterogeneous and homogeneous terms in the Hamiltonian, respectively.
The first two conditions listed are those already used above to show that the error is one order higher.

The effective Hamiltonian for the first-order product formula can be written in the form
\begin{equation}
    H_{\rm eff} = H + H_1 t + H_2 t^2 + \cO(t^3) \, ,
\end{equation}
where $H_1$ and $H_2$ are in the form of commutators.
Then when we consider the time-evolved $M$, we obtain error up to third order
\begin{align}
    M_{\rm prod}(t) - M(t) &= -i[H_1 t + H_2 t^2,M] t - [H,[H_1,M]] t^3/2 - [H_1,[H,M]] t^3/2 + \cO(t^4) \nn
    &=-i[H_1,M] t^2 -i[H_2,M] t^3 - [H,[H_1,M]] t^3/2 - [H_1,[H,M]] t^3/2 + \cO(t^4) \, .\label{fourterms}
\end{align}
Here, the first two terms come from $n=1$ in the time evolution, whereas the other two come from $n=2$.
Substituting that into the OTOC gives error
\begin{align}\label{eq:secondcase}
    \Delta C(t) &= \Tr(B^\dag (-i[H_1,M] t^2) B M) + \Tr(B^\dag M B (-i[H_1,M]t^2))\nn
    & \quad +\Tr(B^\dag (-i[H_2,M] t^3) B M) + \Tr(B^\dag M B (-i[H_2,M]t^3))\nn
    & \quad + \Tr(B^\dag (-i[H_1,M] t^2) B (- i[H,M] t))  + \Tr(B^\dag (-i[H,M] t) B (- i[H_1,M] t^2)) \nn
    & \quad + \Tr(B^\dag (- \tfrac 12 [H,[H_1,M]] t^3) B M) + \Tr(B^\dag M B (- \tfrac 12 [H,[H_1,M]] t^3)) \nn
    & \quad + \Tr(B^\dag (- \tfrac 12 [H_1,[H,M]] t^3) B M) + \Tr(B^\dag M B (- \tfrac 12 [H_1,[H,M]] t^3))+\cO(t^4) \, .
\end{align}
These lines correspond to the four terms in Eq.~\eqref{fourterms}, except the third line corresponds to cross terms between the first term in Eq.~\eqref{fourterms} and the first-order term for the time-evolution in $M$.
The first and second terms in the third line correspond to considering error in the first $M(t)$ versus the second $M(t)$ in the OTOC.

The first two lines of Eq.~\eqref{eq:secondcase} are zero using exactly the same reasoning as above in Eq.~\eqref{eq:firstcase}.
To show the remaining terms are zero, we simplify using cyclic permutation within the trace, and condition 1 that $M$ and $B$ commute, to give $BMB^\dagger=M$.
This gives
\begin{align}
    \Delta C(t)&=- \Tr([H_1,M] B [H,M] B^\dag)t^3  - \Tr([H_1,M] B^\dag [H,M] B)t^3 \nn
    & \quad - \Tr([H,[H_1,M]] M)t^3 \nn
    & \quad - \Tr([H_1,[H,M]] M)t^3+\cO(t^4) \, .
\end{align}
For line one we have used a cyclic permutation.
For lines two and three we have used $BMB^\dagger=M$ and cyclic permutations, as well as combining the two factors of $1/2$.

Expanding these commutators then gives
\begin{align}
    \Delta C(t) &= - \Tr(H_1 M B H M B^\dag-M H_1 B H M B^\dag-H_1 M B M H B^\dag+ M H_1 B M H B^\dag)t^3 \nn
    & \quad - \Tr(H_1 M B^\dag H M B-M H_1 B^\dag H M B-H_1 M B^\dag M H B+ M H_1 B^\dag M H B)t^3 \nn
    & \quad - \Tr(H H_1 M M - H M H_1 M - H_1 M H M + M H_1 H M)t^3 \nn
    & \quad - \Tr(H_1 H M M - H_1 M H M - H M H_1 M + M H H_1 M)t^3+\cO(t^4)  \nn
    &=- 2\Tr(H_1 B M H M B^\dag-H_1 B H B^\dag)t^3 - 2\Tr(H_1 B^\dag M H M B-H_1 B^\dag H B)t^3 \nn
    &\quad - 4\Tr(H_1 H - H_1 M H M)t^3+\cO(t^4) \, .
\end{align}
We have made several simplifications using the condition that $B$ and $M$ commute, as well as that $MM=I$ (condition 2).
Now conditions 3 means that $M H M$ flips the sign of the heterogeneous part, and leaves the homogeneous part unchanged, so gives
\begin{align}
    H - M H M &= 2H_{\rm het}\, .
\end{align}
The error therefore simplifies to
\begin{align}\label{eq:traces}
    \Delta C(t)&=4\Tr(H_1 B H_{\rm het} B^\dag)t^3 + 4\Tr(H_1 B^\dag H_{\rm het} B)t^3 - 8\Tr(H_1 H_{\rm het})t^3+\cO(t^4)  \, .
\end{align}

We can prove the traces are zero by expanding the expressions inside the trace as sums of Pauli strings.
The only Pauli string that can give nonzero trace is the identity.
The term $H_1$ in the error corresponds to commutators of terms in the Hamiltonian.
Because of condition 4, $H_1$ will not contain any commutators between terms in $H_{\rm het}$.
We can then consider commutators between terms within $H_{\rm hom}$, or between those in $H_{\rm hom}$ and those in $H_{\rm het}$.
Condition 5 means that commutators between terms in $H_{\rm hom}$ must have the identity on qubit 1.
Condition 6 means that $H_{\rm het}$ has a non-identity Pauli operator on qubit 1, and so any product between terms in $H_{\rm het}$ and commutators between terms in $H_{\rm hom}$ also has a non-identity Pauli on qubit 1.
As a result its trace is zero.

This means the remaining terms in $H_1$ we need to consider are commutators of terms in $H_{\rm hom}$ and $H_{\rm het}$.
If a term in $H_{\rm het}$ does not commute with a term in $H_{\rm hom}$, then the two terms must have Pauli operators at a matching location.
(They cannot have two matching locations because of conditions 5 and 6.)
As a result, the commutator must not be equal to the identity at \emph{three} locations; qubit 1 and the two qubits where the term from $H_{\rm hom}$ has Pauli operators.
Because terms from $H_{\rm het}$ have Pauli operators at only two locations, the product of these terms and the commutator must again have at least one non-identity Pauli operator.
This reasoning therefore implies that $\Tr(H_1 H_{\rm het})=0$.

We also have terms in Eq.~\eqref{eq:traces} with $B H_{\rm het} B^\dag$ and $B^\dag H_{\rm het} B$.
Condition 7 is that $B$ only acts locally on qubits, and therefore leaves identity operators in the Pauli strings unchanged.
As a result, $B H_{\rm het} B^\dag$ and $B^\dag H_{\rm het} B$ are still sums of terms with non-identity operators at only two locations.
Hence, the same reasoning as for $\Tr(H_1 H_{\rm het})$ holds, and the remaining traces in Eq.~\eqref{eq:traces} are equal to zero.
This shows that the order $t^3$ terms all cancel, and the OTOC error must be at least order $t^4$ when using the first-order product formula.

Again, the reasoning is not exact for time steps that do not start at time $0$, so we do not obtain error $\cO(r(t/r)^4)$ when using $r$ time steps.
The primary condition that is violated for steps with a nonzero initial time is that $B$ and $M$ commute, because $M$ is replaced with $M(t_0)$.
The time-evolved $M$ no longer commutes with $B$, though for short times it will approximately commute.
For first-order product formulae, condition 3 is violated as well, but again will be approximately true for short times.
These considerations mean that the error is still significantly reduced when using first-order product formulae over multiple time steps.

\subsection{Numerical tests of Trotter accuracy}\label{sec:numerical_trotter_accuracy}
\begin{figure}
    \centering
    \includegraphics[width=0.3\linewidth]{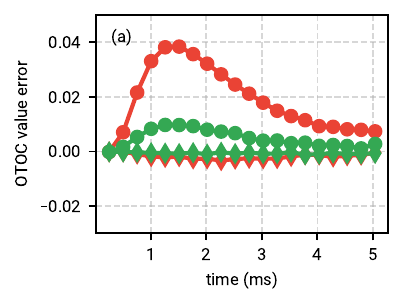}
    \includegraphics[width=0.3\linewidth]{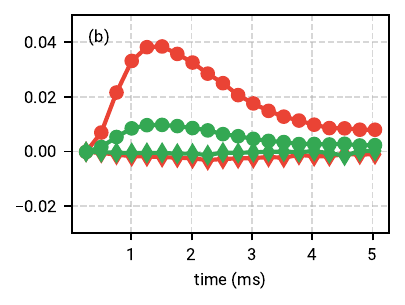}
    \includegraphics[width=0.3\linewidth]{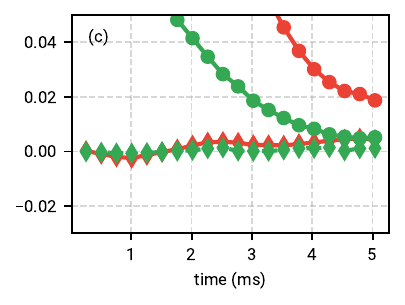}
    \includegraphics[width=0.3\linewidth]{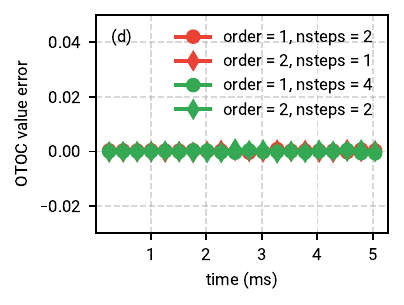}
    \includegraphics[width=0.3\linewidth]{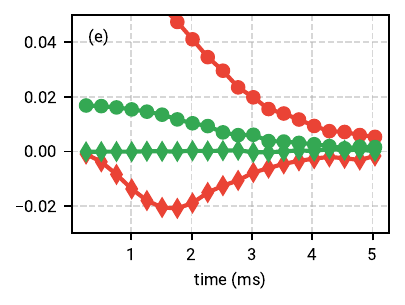}
    \caption{The error in the OTOC value as a result of individually approximating each unitary with 1st or 2nd order Trotter. In particular, we show the effect of Trotterizing the forward evolution ($\tilde{U}_{\rightarrow}$)(a), the backward evolution, $\tilde{U}_{\leftarrow}$, (b), the BLEW12 term, $\tilde{U}_{\mathrm{BL12}}$, (c), the initial $\pi$ pulse, $\Pi_{y}^C$, (d) and the final $-\pi$ pulse, $\Pi_y^H\Pi_{-y}^C$,  (e). Though these terms have different error profiles, we find that in all cases 2nd order Trotter outperforms the 1st order formula for roughly the same circuit depth and only a very small number of steps are needed for an error that is small relative to a target accuracy of 0.02 in the OTOC curve.}
    \label{fig:trotter_error}
\end{figure}

\begin{figure}
    \centering
    \includegraphics[width=0.5\linewidth]{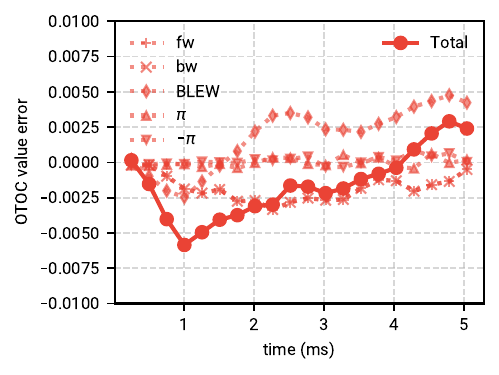}
    \caption{The error in the final OTOC value using 2nd order Trotter for all unitaries. A single time step is used for each unitary except for the $-\pi$ pulse which uses two Trotter steps. This scheme results in an error that is significantly less than our target accuracy of 0.02.}
    \label{fig:trotter_error_total}
\end{figure}

Within the AHT framework, we simulate the TARDIS-1 pulse sequence given in Eq.~\eqref{eq:U_DREAM1} using a low-order Magnus expansion for each unitary. For the 15Q system, we use only the first term (Magnus-0), and we order terms within each Trotter step by grouping them into commuting partitions which are then applied in series. 
The Trotter error for each term is shown in Figure~\ref{fig:trotter_error}. Given the discussion in Section~\ref{sec:numerical_trotter_accuracy}, it is not surprising to see that for 1st and 2nd order Trotter, only a very small number of time steps (1 or 2) is needed for accurate propagation. In Figure~\ref{fig:trotter_error_total} we show the Trotter error using one 2nd order Trotter step for each unitary with the exception of $-\pi$-pulse for which we use 2 steps. The numerical error for this scheme is shown in Figure~\ref{fig:trotter_error_total}.

\subsubsection{Precision and resources}
\begin{figure}
    \centering
    \includegraphics[width=0.5\linewidth]{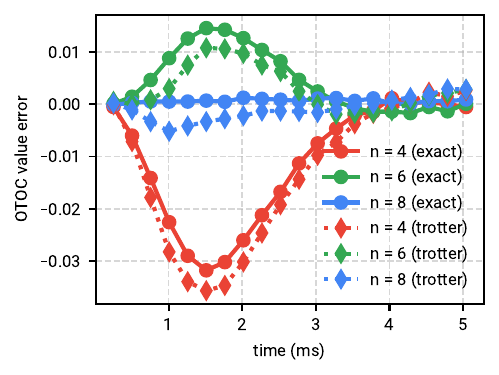}
    \caption{The error in the simulated OTOC curves for different truncations of the rotation angles. Here $n$ is the number of bits of decimal precision and the results are shown for both exact evolution and for the Trotter scheme described in Section~\ref{sec:numerical_trotter_accuracy}.}
    \label{fig:trotter_angle_prec}
\end{figure}

\begin{figure}
    \centering
    \includegraphics[width=0.45\linewidth]{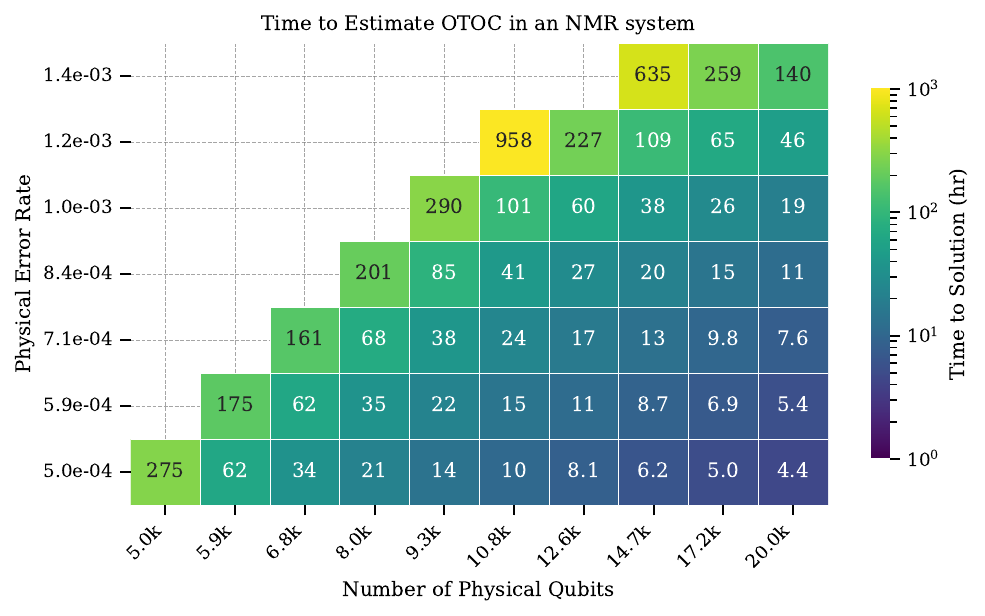}
    \includegraphics[width=0.45\linewidth]{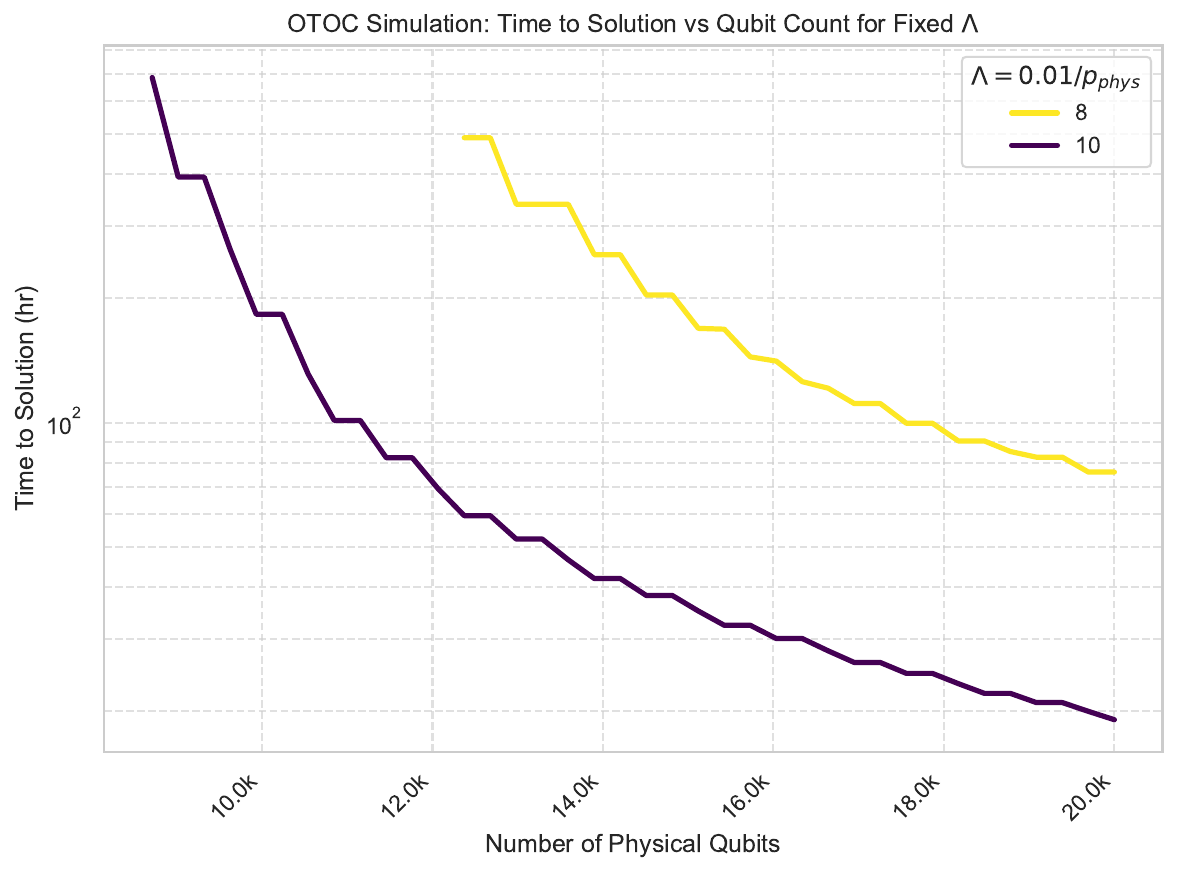}
    \caption{Resource estimates for the longest circuit (the final time point) needed to simulate the 15Q system. (left) Total runtime in the FLASQ model for different error rates and numbers of physical qubits. (right) Time to solution as a function of the number of physical qubits for different fixed values of $\Lambda$.}
    \label{fig:15Q_flasq}
\end{figure}

In Figure~\ref{fig:trotter_angle_prec} we show that truncating the angles in the Trotterized propagators to 6 fractional bits of precision (this roughly corresponds to an allowed error of $2^{-7} \approx 0.0078$ in the rotation angles) provides sufficient accuracy in the simulated OTOC curve.
Given these precision requirements, we compute the resources needed to simulate the OTOC curves using the FLuid Allocation of Surface code Qubits (FLASQ) cost model for early fault-tolerant quantum computing (manuscript in preparation). The results are shown in Figure~\ref{fig:15Q_flasq}.
These resource estimates indicate that these simulations are feasible on realistic models of early fault-tolerant devices. Note that we have not performed a careful optimization of these circuits and the FLASQ model inherently neglects certain optimizations, so we expect that it is possible to improve on these cost estimates.

\subsection{Extrapolating the cost of Trotterization in dipolar spin systems}

\begin{figure}
    \centering
    \includegraphics[width=0.7\linewidth]{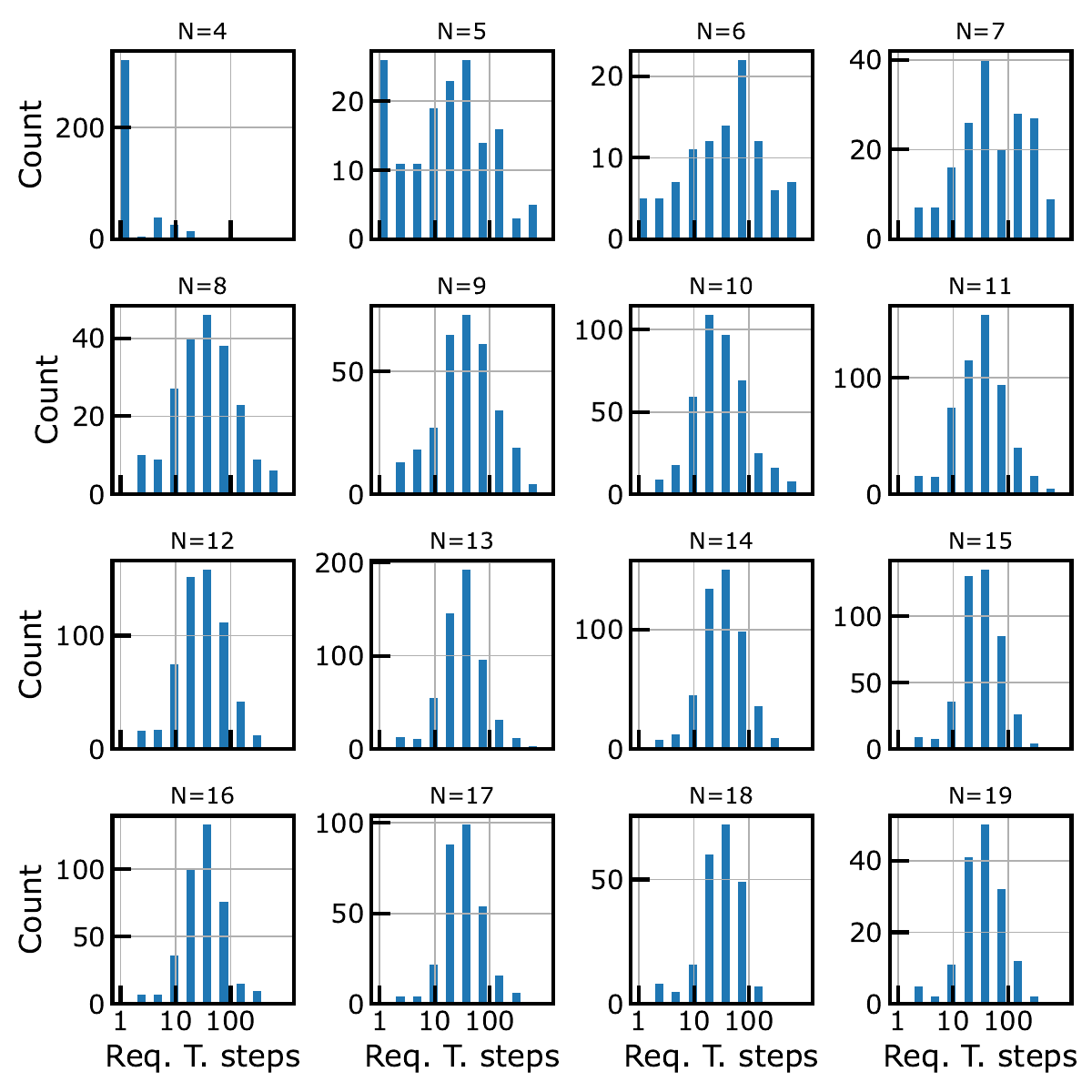}
    \caption{Histograms of the number of Trotter steps required to converge OTOCs on artificial dipolar spin systems (see text) to an error of $0.02$ at times when $C(t)\in [0.4, 0.5]$.
    The x-axis is plotted in log space.}
    \label{fig:trotter_histograms}
\end{figure}

\begin{figure}
    \centering
    \includegraphics[width=0.5\linewidth]{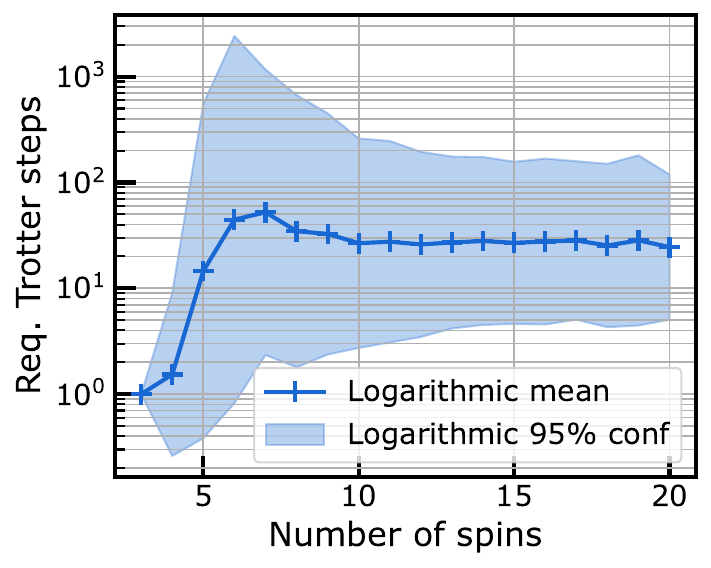}
    \caption{Plot of the the logarithmic mean and $95\%$ confidence interval of the data in Fig.~\ref{fig:trotter_histograms}.}
    \label{fig:trotter_convergence_generic_dipolar_system}
\end{figure}

One of the most important questions left open in our work is the quantum resource requirements to extend these methods to a practical demonstration of quantum advantage.
Here, we make a brief attempt at an order-of-magnitude estimate using a naive first-order Trotterization with no additional optimization.
The key challenge lies in estimating the number $K$ of all-to-all-coupled Trotter steps required to converge the OTOC.
Without further optimization, executing a Trotter step in both the forward and backward pass on $N$ qubits using a swap network requires $N(N-1)$ interactions, or $G(N)=3N(N-1)$ CZ gates.
For $N=50$ ($N=60$), we have $G(50)=7350$ ($G(60)=10620$) CZ gates per Trotter step; the total number of gates per circuit is $K\times G$.

The required number $K$ of Trotter steps is highly system and quantity specific.
Here we define a simplified model system of the nuclear spin systems studied in this work, and then outline a protocol to estimate an ensemble of Trotterized OTOCs throughout this system.
We define a model system by scattering spin positions $\bs{r}_i$ randomly in a large $3D$ box at a fixed density $\rho_s=0.1~\text{\AA}^{-3}$, and with a minimum distance between spins of $1.5$~\AA.
This density is chosen to match the density of spins in a protein (Eq.~\eqref{eq:nuclear_system_constants}), and the minimum distance is chosen to be slightly smaller than the $1.7$~\AA~distance between methyl spins (larger distances created packing issues due to the random scattering of spins).
From this, we generate a set of dipolar couplings $d_{ij}$ using Eq.~\eqref{eq:DD defination}, assuming the spins are protons fixed in space and the magnetic field points in the $\hat{z}$ direction.

Next, we choose pairs of measurement and butterfly spins $(m, b)$ at random separated by between $4$~\AA~and $12$~\AA, sufficiently within the edges of the box to prevent observing the boundaries.
(For efficiency reasons, we generate $20$ unique pairs within cubic regions of width $20$~\AA~plus an additional gap to the system boundary; we expect this volume to be large enough to prevent significant correlations between different problem instances.)
For efficiency, we desire to simulate OTOCs between these spin pairs including only relevant spectator spins $j$ (i.e. spins that contribute to the OTOC).
To achieve this, we construct a classical estimate of the time-of-flight $t^{(\mathrm{c})}_{j}$ for information to spread from the measurement spin through a spectator spin to the butterfly.
This estimate $t^{(\mathrm{c})}_j$ is calculated by treating the spins as a weighted graph with adjacency matrix $s_{ij}=d_{ij}^{-1}$, which we interpret as the time taken for information to propagate from spin $i$ to spin $j$.
Under this interpretation, the time taken for information to propagate from the measurement spin $m$ through spin $j$ to the butterfly spin $b$ is equal to the minimum weight of a path $m-\rightarrow j\rightarrow b$, which can be efficiently calculated using Dijkstra's algorithm.

Once the above time-of-flight calculation is complete, we order spins $j$ by their time-of-flight, and added them from the shortest $t^{(\mathrm{c})}_j$ to the largest until the OTOC converges to a value below $C(t)=0.5$.
(We approximate the infinite temperature OTOC by averaging over initial states until the standard error drops below $0.004$, which we judge to be sufficiently small to be irrelevant.)
As we add each spin, we first converge the Trotterization of the $n$-spin system by repeatedly doubling the number of Trotter steps $k$ and estimating $c_{n,k}(t)$ until we find $|c_{n,k}-c_{n,k}|<0.02$, yielding an estimate $K(N)=k/2$, $C_{n}(t)=c_{n,k}(t)$.
We then test whether $C_{n}(t)=C_{n-1}(t)$; if so we report the pair $(N,K)=(n-1, K(n-1))$ as the last steps were used to confirm convergence.
When we have not converged, we update the time $t$ to ensure the OTOC is near but below $0.5$, add a new spin, and restart the process.
(Due to a large chosen time gradation, this results in a final OTOC value $C(t)\in[0.4,0.5]$; we reject outlying experiments that fall below $0.4$.)

In Fig.~\ref{fig:trotter_histograms}, we plot histograms of the number of steps to converge, on a logarithmic axis, separated for different values of $N$.
We observe that the distribution is relatively symmetric within this log space, with a mode around $K=32$ for all $N>11$.
Beyond this surprisingly small value the distribution appears to be converging; in Fig.~\ref{fig:trotter_convergence_generic_dipolar_system} we plot the mean and $2\sigma$ interval of our obtained estimates of $K$ as a function of $N$.
We observe that the resulting mean appears to have converged quickly to a number around $K=30$.
This would correspond to needing $2.2\times 10^5$ ($3.2\times 10^5$) CZ gates for a $50$ ($60$) qubit simulation.
The standard deviation trend is slower: at $N=20$ we observe the population spread is over roughly an order of magnitude.
Extrapolating this over three additional decades is highly unreliable, and significantly more work needs to be done to determine the cost of Trotterization at larger system sizes.
However, we are optimistic that these results may suggest instances of OTOC estimation at $50-60$ qubits requiring $10^5-10^6$ two-qubit gates, and this work forms our justification for this claim in the main text.

\subsection{Correlation between different measures of Trotter error}

\begin{figure}
    \centering
    \includegraphics[width=\linewidth]{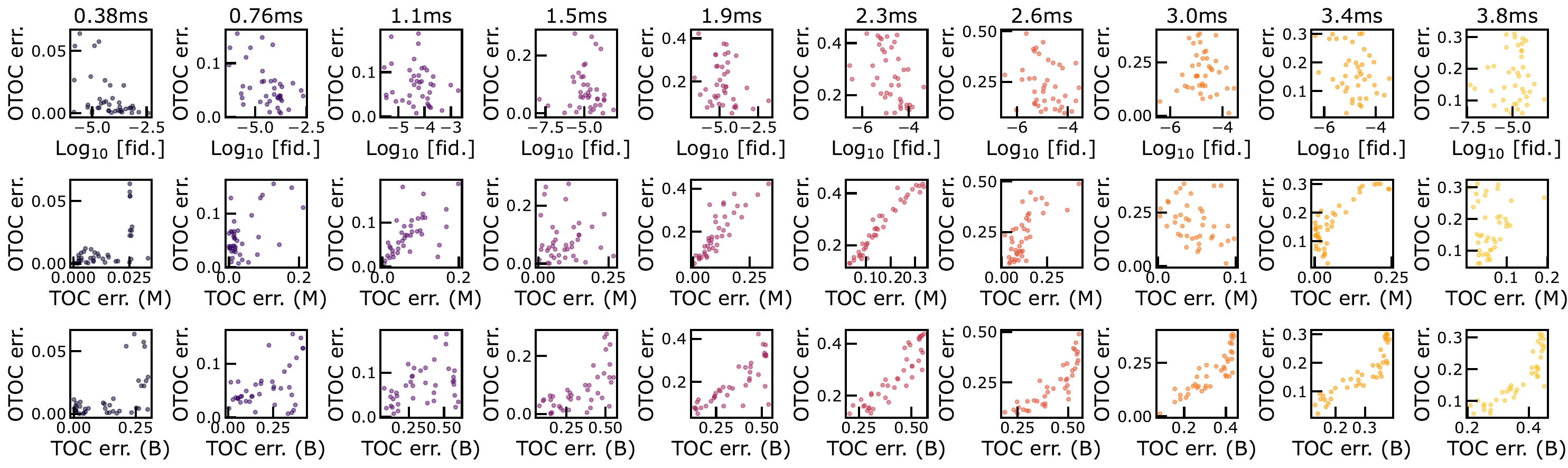}
    \caption{Plots of Trotter error as measured via OTOC error, unitary fidelity, or TOC error, for a range of different Trotterizations of the toluene DQ Hamiltonian differing by time, number of Trotter steps per timestep, and the ordering within the swap network. Points are coloured to separate different time instances.}
    \label{fig:otoc_fidelity_toc}
\end{figure}

Meaningfully capturing the error in Trotter approximations is a challenging task.
Traditionally, this has been estimated by holistic metrics such as unitary fidelity, or via matrix norms on the multiplicative or additive error term~\cite{childs2021theory}, which provide upper bounds on other desirable quantities.
However, ultimately one is only concerned with the accuracy of the data extracted at the end of the computation.
As we demonstrated in Sec.~\ref{sec:trotter_analysis}, this can be even asymptotically better than unitary fidelity metrics, due to error terms cancelling out.
Given the vast freedom in designing product formulas, we would like to optimize even further, but this runs into two problems.
The first is a spoofing problem: we can achieve arbitrary single expectation values using short-depth unitaries that bear no physical resemblance to the original Hamiltonian, so the error in such an expectation value is a completely unreliable product formula metric.
We have attempted to mitigate this concern in our AlphaEvolve-optimized circuits by testing across a landscape of times and Hamiltonians, and by inspecting the output product formula code (see the main text and Sec.~\ref{sec:alphaevolve} for further details); further investigation here is needed.
The second is a prediction problem: we desire the ability to estimate the error in an expectation value without being able to estimate the expectation value itself, which requires finding correlated error metrics.

In this section, we briefly investigate the challenge in solving the prediction problem, by comparing different error metrics on a range of product formula approximations to the evolution of toluene under the double quantum Hamiltonian approximation (Eq.~\eqref{eq:double-quantum}).
We generate approximate unitaries $U_{\mathrm{pf}}(t)\approx e^{-iHt}$ using swap networks with randomly chosen qubit orderings and a range of Trotter step sizes $dt$.
In Fig.~\ref{fig:otoc_fidelity_toc}, we measure the error in these approximations with four metrics:
\begin{enumerate}
    \item The OTOC error 
    \begin{equation}
        \mathrm{OTOC\;err.}=\tfrac{1}{2^N}\big|\mathrm{Trace}[MU^{\dagger}_{\mathrm{pf}}(t)B U_{\mathrm{pf}}(t)MU^{\dagger}_{\mathrm{pf}}(t)B^{\dagger} U_{\mathrm{pf}}(t)]-\mathrm{Trace}[Me^{iHt}B e^{-iHt}Me^{iHt}B^{\dagger} e^{-iHt}]\big|,
    \end{equation}
    where $M=X_{{}^{13}\mathrm{C}}$, and $B$ is the single-qubit approximate butterfly pulse to the BLEW-12 sequence (Eq.~\eqref{eq:butterfly_approx}) that acts on the methyl protons.
    \item The unitary fidelity, 
    \begin{equation}
        \mathrm{fid.}=\tfrac{1}{2^N}\mathrm{Trace}[U^{\dagger}_{\mathrm{pf}}e^{iHt}].
    \end{equation}
    \item The TOC error on the measurement operator (using the same definitions as above),
    \begin{equation}\mathrm{TOC\;err.\;(M)}=\tfrac{1}{2^N}\big|\mathrm{Trace}[MU^{\dagger}_{\mathrm{pf}}(t)M U_{\mathrm{pf}}(t)]-\mathrm{Trace}[Me^{iHt}M e^{-iHt}]\big|.
    \end{equation}
    \item The TOC error on the butterfly operator (using the same definitions as above),
    \begin{equation}\mathrm{TOC\;err.\;(B)}=\tfrac{1}{2^N}\big|\mathrm{Trace}[B^{\dagger}U^{\dagger}_{\mathrm{pf}}(t)B U_{\mathrm{pf}}(t)]-\mathrm{Trace}[B^{\dagger}e^{iHt}B e^{-iHt}]\big|.
    \end{equation}
\end{enumerate}
Overall, we observe that the predictive power of any of the other three metrics for the error in an OTOC is somewhat limited.
The unitary fidelity is a particularly bad metric: despite the OTOC being often well approximated by various product formulas shown here, this never rises above $10^{-2}$ (optimally the unitary fidelity should be $1$), and we observe no correlation between this metric and the OTOC.
This should not be surprising given the results in Sec.~\ref{sec:trotter_analysis}, as the leading contributions to the error in the unitary fidelity cancel out in the OTOC error.
The result here implies that there are is little correlation between the contributions that cancel and those that remain.
We observe some correlation between the two TOC errors and the OTOC error, though this correlation is curiously inconsistent across different times, and not terribly strong for the most part (with a notable exception at $2.3$~ms).
This demonstrates the need for further investigation into the mechanisms behind Trotter error to improve our predictive power.

%% file: alphaevolve.tex

\begin{figure}[t]
    \centering 

        \centering
        \includegraphics[height=7.5cm]{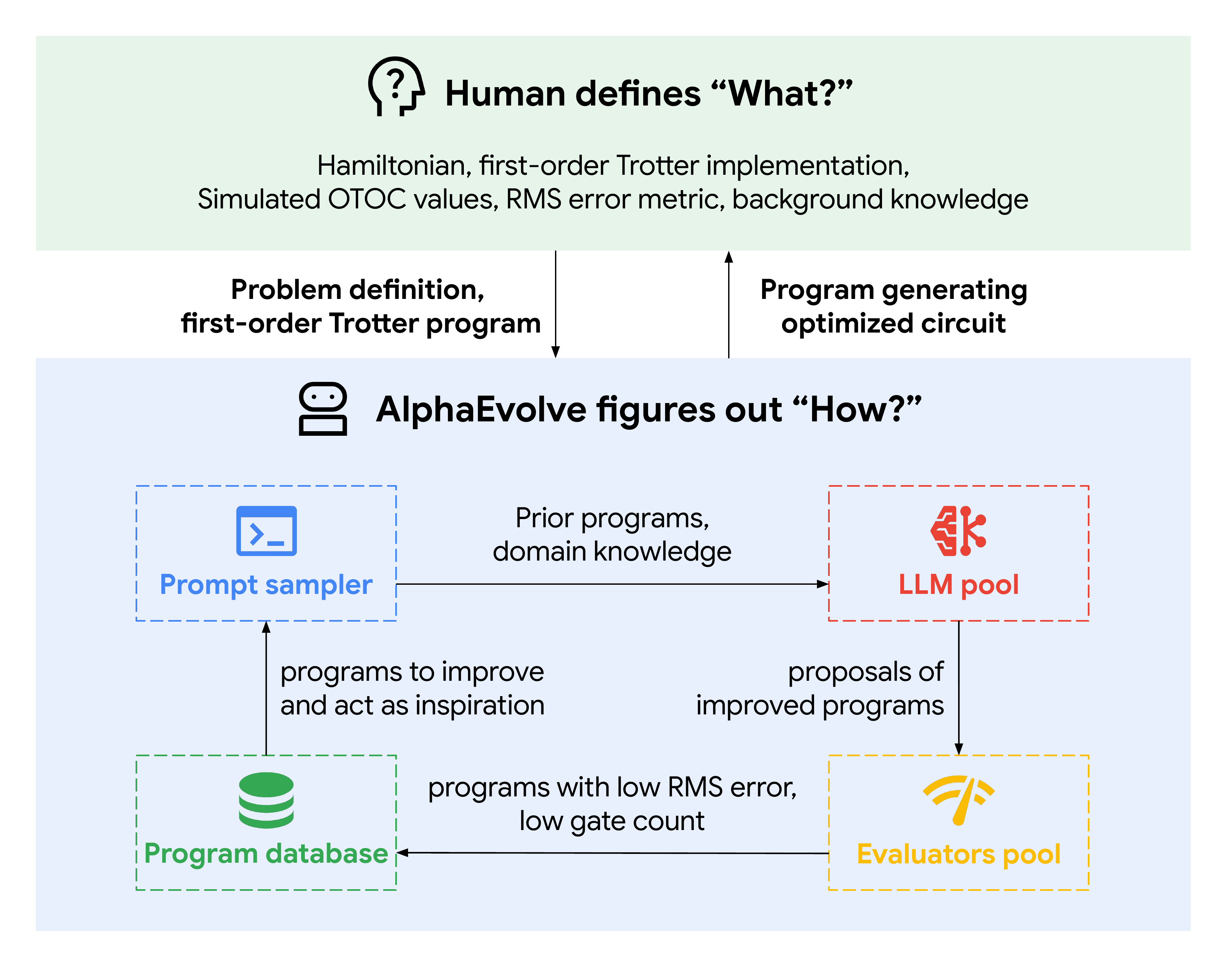}
    \caption{Overview of AlphaEvolve optimization cycle as presented in ~\cite{novikov2025alphaevolvecodingagentscientific}}
    \label{fig:ae_high_level}
\end{figure}

We employed AlphaEvolve, an evolutionary coding agent that uses LLMs, to discover Python functions which generate quantum circuits for approximating Hamiltonian evolution over a landscape of time and Hamiltonian parameters. The evolutionary process was seeded with a human-written, first-order Trotter formula for a 15-qubit chain and iteratively improved programs through an LLM-driven mutation, evaluation, and selection loop, as shown in Fig~\ref{fig:ae_high_level}. The optimization targeted the root mean square error (RMSE) of the OTOC averaged across the entire parameter landscape, with the LLM's mutation prompts guided by a mixture of domain-specific instructions and a tabular error matrix feedback to perform in-context learning on high-error regions.

To accelerate convergence and enhance search space exploration, we ran multiple independent AlphaEvolve instances that were periodically synchronized with the best-performing solution. We also overcame a key computational bottleneck in the evaluation step by implementing a highly parallelized KAK decomposition for efficient CZ gate counting, a tool which the final evolved programs learned to utilize for more efficient circuit generation.

In order to assess the efficacy of prompting AlphaEvolve with problem- and domain-specific knowledge, we analysed the learning curve of the program evolution. To get a high-level overview, we group prompt instructions given to AlphaEvolve into five general categories presented in Table~\ref{tab:prompts_table}. The generic prompts (GENERIC) are simply general guidelines to improve the program. The prompts grouped under domain knowledge, topology, and gate count span quantum computing knowledge (THRIFT \cite{Bosse2025-ha}, higher order Trotter formulas), molecular structure, and circuit compilation respectively. Parameter tuning then explicitly instructs AlphaEvolve to adjust coupling strengths and time steps based on the given parameters.
\newcommand{\colorsquare}[1]{\raisebox{3pt}{\fcolorbox{black}{#1}{\null}}}

\begin{table}[b]
\centering 
\caption{ Additional prompt instructions by category}
\label{tab:prompts_table}
\renewcommand{\arraystretch}{1.3} 
\begin{tabular}{
  p{0.19\textwidth} 
  p{0.19\textwidth} 
  p{0.19\textwidth} 
  p{0.19\textwidth} 
  p{0.19\textwidth} 
}
\hline
\textbf{Domain Knowledge} & \textbf{Topology} & \textbf{Parameter tuning} & \textbf{Gate count} & \textbf{Generic} \\
\hline
\colorsquare{red} \textbf{DOMAIN} & \colorsquare{cyan} \textbf{TOPO} & \colorsquare{lightgray} \textbf{PARAM} & \colorsquare{yellow} \textbf{GATES} & \colorsquare{white} \textbf{GENERIC} \\
\hline

Implement THRIFT \cite{Bosse2025-ha} by identifying a strongly and weakly coupled subset of atoms 
& Adjust couplings based on vicinity to “unimportant” atoms
& Adjust step dt based on parameter $p$
& Reduce the circuit depth based on $t$ and $p$
& Suggest an idea to improve the code \\
\hline

Implement a 2nd order Suzuki-Trotter formula
& Adjust frequency of some couplings based on shift parameter $p$
& Adjust effective time step with an envelope that may depend on atom, $p$, butterfly/measurement indices
& Increase the circuit depth based on t and $p$ 
& Suggest an idea to optimize the code \\
\hline

Implement a 4th order Suzuki-Trotter formula
& Handle components of interaction graph independently
& Adjust couplings based on $p$
& Build a multi-pass approach, filtering out unimportant gates/adding further steps 
& \\
\hline

Invent new gates based on the Weyl piece (use KAK/Cartan decompositions of $SU(4)$)
& Lightcone based on butterfly and measurement indices by tracking spatial operator support
& Derive scaling exponents from $p$ and exponentiate couplings by it
& Implement a 4\textsuperscript{th} order Suzuki-Trotter formula
& \\
\hline

Implement fractional Trotter steps/partial layers
& Keep the ${}^{13}$C atom stationary
& 
& 
& \\

\hline
\end{tabular}
\end{table}

\begin{figure}[t]
    \centering 
        \includegraphics[height=6.5cm]{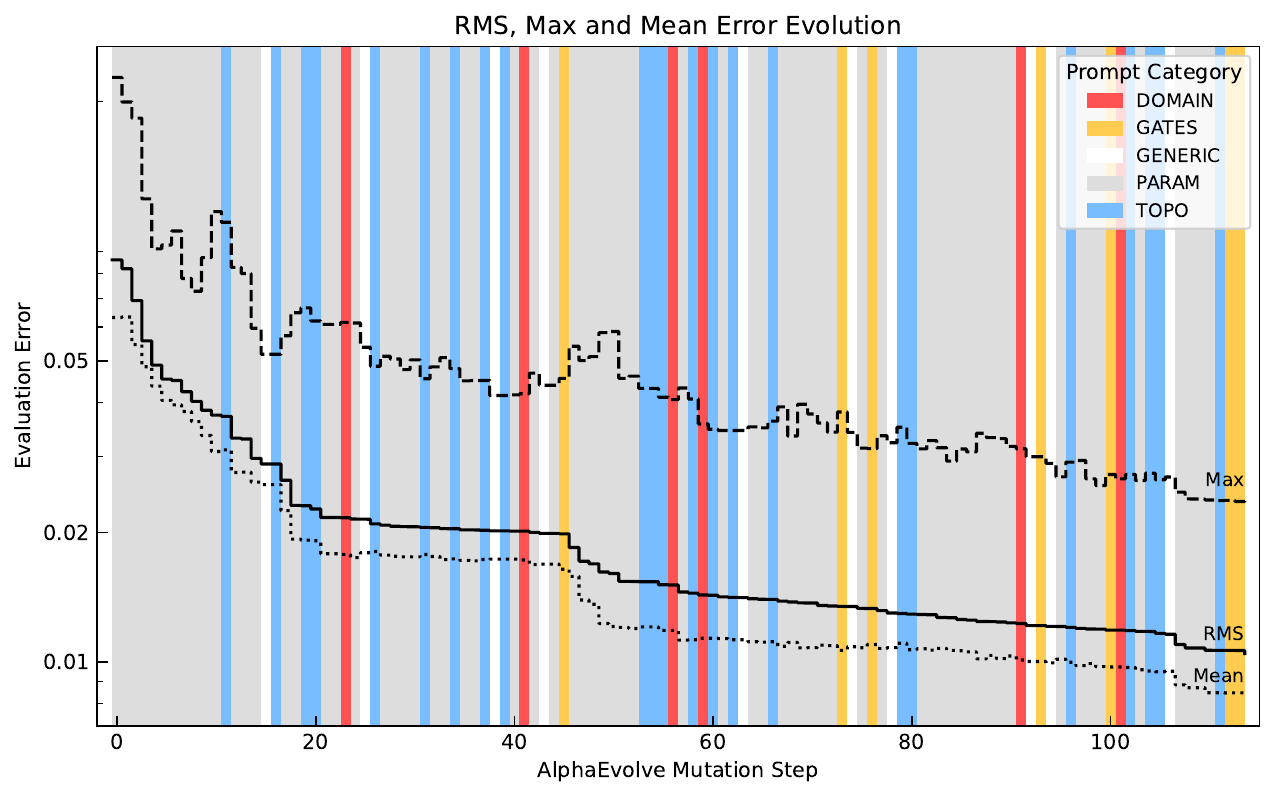}
        \hspace{2cm}
        \includegraphics[height=6.5cm]{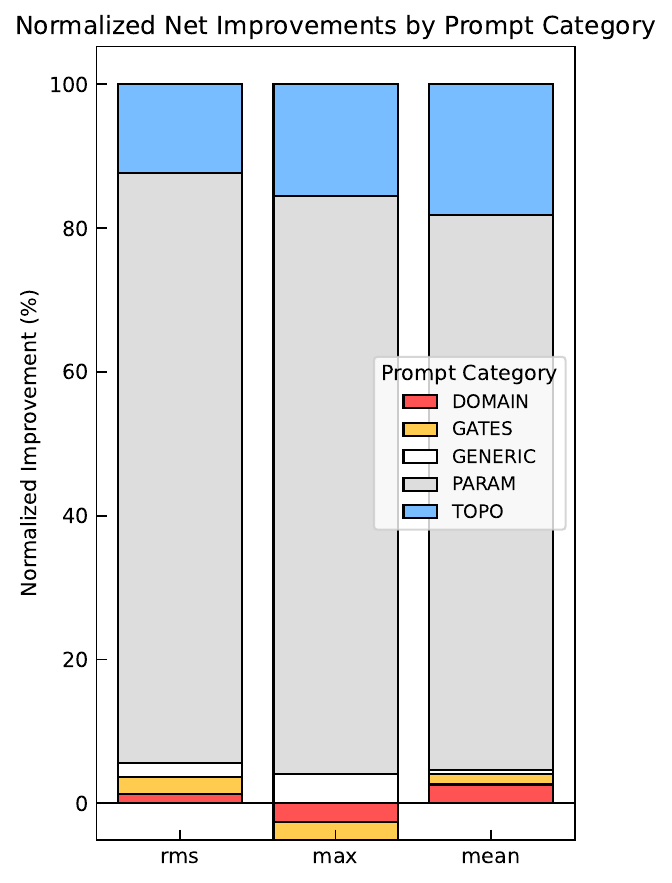}
        \caption{Prompt category contributions. (\textbf{left}) Evaluation error as function of mutation number. Bold line represents optimized metric (rms error), dashed lines correspond auxiliary tracked metrics (mean, min). The colored stripes indicate the type of prompt category used that resulted in this score change, as listed in Table~\ref{tab:prompts_table}. 
        (\textbf{right}) Histogram of a cumulative prompt category importance. The importance is calculated as a weighted sum over all score improvements for a given category.}
    \label{fig:prompt_study}
\end{figure}

To understand how AlphaEvolve proceeds and get a sense of which of the instruction prompts are most beneficial, Figure~\ref{fig:prompt_study}(left) shows an optimization curve with the improvement in RMSE against evolution steps. The colored background annotates which category of prompt was used to obtain the improvements. In addition, in Figure~\ref{fig:prompt_study}(right) we provide a histogram of cumulative importance of the prompt categories used. We note that we see a lot of variation from run to run, so the lineage of this particular program is not a prescription of how to proceed optimally; repeated runs will produce a completely different result. We do note however that the performance (in terms of RMSE reached by the end of the optimization) can be reliably achieved.

The parameter tuning prompts yield most of the improvements, as seen in \Cref{fig:prompt_study}(left), followed by instructions to utilize the molecular topology. Domain knowledge, generic and gates-specific prompts less frequently yield improvements. When summing the net contributions of each prompt towards the overall aggregated min, mean, and RMSE scores, we find a similar picture: the overwhelming improvement comes from parameter-specific instructions, followed by topology. With respect to the max error, domain- and gate-specific prompts seem detrimental; however, we note that many of the parameter- and topology-specific improvements only make sense after the domain-specific improvements have been applied; the order in which mutations take place naturally matters a lot.

\begin{figure}[t]
\caption{Pseudocode describing the AlphaEvolve-generated algorithm to generate Trotter circuits.}
\label{fig:ae_pseudocode}
\begin{minipage}{0.83\textwidth}
\begin{lstlisting}[style=mystyle, backgroundcolor=\color{backcolour}]
Function GenerateTrotterNetwork_Improved(
        Time t
        CZ budget MaxCZ
        Hamiltonian couplings H
        Parameter p
        Butterfly/Measurement indices I
    ):
    // Concept a: Adaptive Hamiltonian Scaling
    H_scaled := ApplyAdaptiveCouplingScaling(H, p)
    // Concept b: Hamiltonian Decomposition
    Components := DecomposeHamiltonian(H_scaled, I)
    // Concept c: Dynamic Budgeting (Two-Pass Approach)
    TotalCostPerStep := 0
    For C in Components Do
        DummyCircuit := TrotterStep(C, t, 1, p)
        TotalCostPerStep += CountCZGates(DummyCircuit)
    End For
    N := CalculateTrotterSteps(t, p, MaxCZ, TotalCostPerStep)
    OverallGateSequence := []
    For C in Components Do
        ComponentCircuit := TrotterStep(C, t, N, p)
        OverallGateSequence += ComponentCircuit
    End For
    Return OverallGateSequence
End Function

Function TrotterStep(
        ComponentData C
        Time t
        Steps N
        Parameter p
    ):
    ComponentGateSequence := []
    For k in 0..N-1 Do
        // Concept d: Time Step Modulation.
        dt := CalculateEffectiveTimeSteps(t, N, p, k)
        CurrentStepGates := []
        For LayerIdx in 0..C.LayersPerStep-1 Do
            // Concept e: Adaptive Symmetric Second-Order Trotterization
            CurrentStepGates += Adaptive2ndOrderLayer(dt, p, t, C)
            CurrentStepGates += SwapNetwork(LayerIdx, C)
        End For
    // Concept f: Lightcone Pruning
    ComponentGateSequence += LightconePruning(CurrentStepGates, C)
    End For
    Return ComponentGateSequence
End Function

\end{lstlisting}
\end{minipage}
\centering
\end{figure}

To further give a high-level intuition of what kinds of modifications AlphaEvolve suggests, we present an translation of the full, optimized Trotter generator function into pseudocode in Figure~\ref{fig:ae_pseudocode}. As the full evolved program has a length of roughly 1300 lines of code and thus is hard to interpret directly, we prompt an LLM to convert the final AlphaEvolve optimized program into pseudocode as a starting point, and then manually try to match the extracted concepts with the given code. We highlight that the overall program structure and concepts used are highly specific to the simulation target at hand, and that some of the concepts invoked may be implemented in an incomplete, wrong or obscure way; if implemented from first principles, these concepts might individually work quite differently, and their efficacy could depend on the order in which the successive mutations were applied.

The concepts AlphaEvolve used in the final program in \Cref{fig:ae_pseudocode} are outlined below. We note that all of them find their origin in at least one of the prompt instructions from Table~\ref{tab:prompts_table}, such as THRIFT scaling, or utilizing the time and landscape parameters to attenuate the parameters in a nonlinear fashion. The two-pass approach was also translated into a major component, as well as light cone filtering as a final pruning step.

\begin{itemize}
\item \emph{Adaptive Hamiltonian Scaling.} The Hamiltonian couplings are scaled, by modulating them with an exponent dependent on the parameter $p$, via $J \mapsto \mathrm{sign}(J) |J|^{a(p)}$ for some polynomial $a(p)$.

\item \emph{Hamiltonian Decomposition.} The Hamiltonian is decomposed based on its interaction graph.

\item \emph{Dynamic Budgeting.} In a two-step approach we first estimate the CZ cost of a single, pruned Trotter step across all components. The number of Trotter steps is then the minimum of a heuristic desired $N$ (which depends on $t$ and $p$) and the maximum $N$ allowed by the CZ budget.

\item \emph{Time Step Modulation.} Scale time steps via $t \mapsto \mathrm{sign}(t) |t|^{e(p, t)}$ for some polynomial $e(p, t)$. This implements a time-dependent envelope, global THRIFT scaling, and scaling dependent on the importance of the qubits (vicinity to measurement indices, vicinity to important atoms, e.g.\ the carbon atom).

\item \emph{Adaptive Symmetric Second-Order Trotterization.} Uses a symmetric decomposition within each layer based on the formula $\exp(-\mathrm i(A+B)\Delta t) \approx \exp(-\mathrm i A x \Delta t)\exp(-i B \Delta t) \exp(-i A(1-x)\Delta t)$ for some $x = x(p, t)$; the Hamiltonian is split into $A$ (background) and $B$ (perturbation) in a THRIFT-inspired way.

\item \emph{Lightcone Pruning.} Gates that do not causally affect the final measured observables are removed.
\end{itemize}

Overall, while it remains unclear to what extent the above instructions were correctly implemented as intended and whether the concepts could function well independently, we see this experiment as a promising first step towards automatically translating scientific knowledge into optimized domain-specific tools.

%% file: classical.tex
As a preliminary assessment of the classical resources required to simulate out-of-time-ordered correlators (OTOCs) for nuclear spin ensembles, we consider a spin-1/2 double-quantum Hamiltonian with long-ranged interactions on an $N$-site chain
\begin{align}
    H = \sum_{i<j} \frac{J}{4 |i-j|} \left( X_i X_j - Y_i Y_j \right)
\end{align}
where $X_i$, $Y_i$ are Pauli operators on site $i$.
This gives a simplified model with similar behavior to our 3 dimensional dipolar Hamiltonian with a $1/r^3$ coupling, that we can use for scaling analyses of different classical methods.

\subsection{MPO analysis}

Matrix product operator (MPO) evolution is a powerful method for simulating the operator dynamics and out-of-time-ordered correlators of such one-dimensional quantum systems. MPOs provide an efficient representation of operators that are local or have low operator entanglement. Starting from an MPO representation of the butterfly operators $A$ and $B$, the computation of out-of-time-ordered correlators can be approximated via computing the Heisenberg evolution $A(t)$ at truncated maximum bond dimension for the time-evolved MPO. Prior work has demonstrated that such MPO evolution can capture the early-onset behavior and wave front dynamics of OTOCs in one-dimensional systems, while appreciable OTOC growth and late-time regimes remain out of reach even for local Hamiltonians unless operator entanglement is bounded \cite{PhysRevB.100.104303,xu2020accessing,PhysRevB.104.104307,PRXQuantum.5.010201}.

To estimate MPO simulation resources for OTOCs for the long-ranged double quantum Hamiltonian, we study a Trotterized time evolution using the ``$U^{\rm II}$'' approximation \cite{PhysRevB.91.165112} of the time evolution operator $e^{i\hat{H} \delta t}$ at fixed time step $\delta t = 0.1/J$, and monitor the convergence of OTOCs (within this fixed time-step Trotter evolution) as a function of maximum bond dimension for the butterfly operator. As the required MPO bond dimension for representing the long-ranged double quantum Hamiltonian itself diverges polynomially with system size, we employ the MPO compression algorithm of Ref. \cite{PhysRevB.102.035147} and truncate the Hamiltonian bond dimension at 42. At this fixed Trotter step and Hamiltonian bond dimension, the problem is representative of the complexity of simulating OTOCs for long-ranged interactions, but permits isolating the resource requirements (bond dimension) needed to represent the time-evolved butterfly operators.

We choose a pair of butterfly $\hat{X}$ operators at positions $n=N/3$ and $m=2N/3$ of the $N$-site chain. As the overall system is invariant under time translations (by the Trotter step $\delta t$), the OTOC can be computed using either one-sided (1S) evolution of a single butterfly operator to time $t$, or two-sided (2S) time evolution of both butterfly operators forward/backward in time
\begin{align}
    \mathcal{O}_{1S}(t) &= \frac{1}{\mathcal{N}} \langle X_n(t) X_m(0) X_n(t) X_m(0) \rangle~, \\
    \mathcal{O}_{2S}(t) &= \frac{1}{\mathcal{N}} \langle X_n(\tfrac{t}{2}) X_m(-\tfrac{t}{2}) X_n(\tfrac{t}{2}) X_m(-\tfrac{t}{2}) \rangle ~,
\end{align}
where $\mathcal{N}$ is the Hilbert space dimension.
These quantities are equivalent in the exact limit. To assess the convergence of MPO simulations at truncated bond dimension, this therefore motivates a simple error metric
\begin{align}
    \epsilon_{\rm rel}(t) = 2\frac{|\mathcal{O}_{1S}(t) - \mathcal{O}_{2S}(t)|}{|\mathcal{O}_{1S}(t) + \mathcal{O}_{2S}(t)|} ~.
\end{align}
that is expected to converge approximately monotonically at large bond dimension.

\begin{figure}[H]
    \centering
    \includegraphics[width=0.8\textwidth]{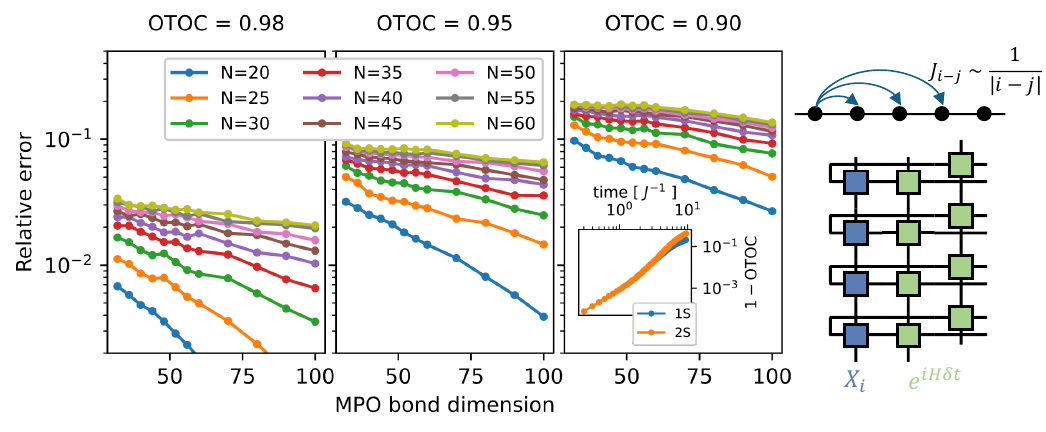}
    \caption{Scaling of classical resources for simulating out-of-time-ordered correlators (OTOCs) via matrix product operator (MPO) evolution, for a spin-1/2 double-quantum Hamiltonian with long-ranged $1/R$ interactions on an $N$-site chain. Relative simulation error as a function of MPO bond dimension is estimated by comparing the obtained OTOCs for one-sided (1S) MPO time evolution of a single butterfly operator [$\langle X_n(t) X_m(0) X_n(t) X_m(0) \rangle / \mathcal{N}$] to the two-sided (2S) MPO evolution of both butterfly operators forward/backward in time [$\langle X_n(\tfrac{t}{2}) X_m(-\tfrac{t}{2}) X_n(\tfrac{t}{2}) X_m(-\tfrac{t}{2}) \rangle / \mathcal{N}$]. The relative error between these approaches is plotted as a function of MPO bond dimension at fixed OTOC decay levels 0.98 (2\%), 0.95 (5\%), and 0.90 (10\%). Inset depicts the early-growth behavior of the commutator OTOC.}
    \label{fig:MPO_resource_scaling}
\end{figure}

Fig. \ref{fig:MPO_resource_scaling} depicts the relative error incurred at three fixed OTOC decay levels (0.98, 0.95, 0.90) as a function of maximum MPO bond dimension for system sizes $N=20$ to $N=60$ for the long-ranged double-quantum Hamiltonian. The relative error decreases with increasing bond dimension, however the rate of this decrease slows rapidly with growing number of spins. Extrapolating the observed scaling behavior to larger bond dimensions suggests that computing appreciable OTOC growth, accessible in experiments, remains out reach using standard MPO-based methods even for spin chains with non-local interactions. We anticipate that similar scaling behaviors will be observed for three-dimensional spin ensembles with dipolar interactions, as the long-range nature of the interactions is expected to lead to rapid growth of operator entanglement and large required bond dimensions.

\subsection{Monte Carlo analysis}

It is perhaps unsurprising that MPO techniques should find OTOC estimation challenging, due to the operator entanglement generated by OTOCs in even random circuits.
However, not all classical methods face an entanglement barrier, and Monte Carlo techniques have shown reasonable prowess in estimating OTOCs in random circuits~\cite{google25constructive}, due to the lack of large loop interference in the Pauli path picture.
However, in our case interference between Pauli paths is generated not only by the OTOC (which connects Pauli paths at either end), but by Hamiltonian dynamics, which can be expected to cause path interference even within a single unitary.

To test this hypothesis, we apply the tensor-network Monte Carlo (TNMC) algorithm in Pauli space~\cite{google25constructive} on small systems using the model defined at the start of this section, and compare the output to exact dynamics.
The TNMC algorithm works by ``cutting'' a circuit $U$ into sub-circuits $U_j$, and propagating the butterfly operator between these blocks via Monte Carlo.
(For Hamiltonian time-evolution, this requires that we choose a Trotterization or other simulation method to convert $e^{iHt}$ into a circuit.)
For each $U_j$, we calculate
\begin{equation}
    U^{\dag}_j B_{j-1}U_j=\sum_{P\in \mathbb{P}^{N}}b_{P}(j) P.
\end{equation}
We then sample from the distribution $b^2_P(j)$, which by unitarity of $U_j$ is normalized to $1$.
Repeating for each $j$, this generates an estimate $B_{\mathrm{MC}}(t)$ for the evolved butterfly operator $B(t)$, from which we calculate the OTOC as
\begin{equation}
    \mathrm{OTOC}_{\mathrm{MC}} = \mathrm{Trace}[\rho B_{\mathrm{MC}}(t)MB_{\mathrm{MC}}(t)M].
\end{equation}
The cost of classical estimation scales only with the cost of simulating the hardest $U_j$, which can be significantly smaller than the cost of simulating $U$ if sufficient cuts are made.

\begin{figure}
    \centering
    \includegraphics[width=\linewidth]{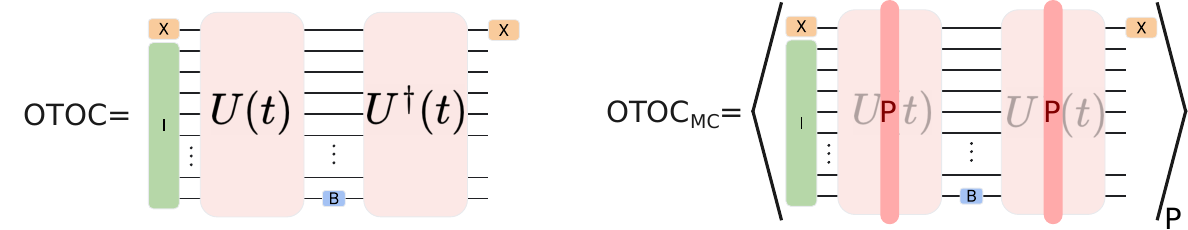}
    \caption{Explanation of the TNMC algorithm as implemented in this work.
    (left) a typical OTOC circuit generated by a unitary $U(t)$ (in our case $U(t)=e^{iHt}$).
    (right) The TNMC algorithm inserts a single layer of Pauli operators in the middle of a circuit decomposition of $U(t)$, and an identical layer of Pauli operators in $U^{\dag}(t)$.
    These are averaged over the inserted layer $P$ (drawing $I,X,Y,Z$ on each qubit with equal probability).
    The TNMC algorithm can estimate the resulting ensemble average, with a classical complexity scaling as the cost of simulating an OTOC using only half the circuit.}
    \label{fig:cutting_explanation}
\end{figure}

It turns out that this is equivalent to estimating the ensemble average of a set of circuits where, before each $U_j$, we insert randomly chosen Pauli operators on each qubit, and apply the same Pauli operator after $U_j^{\dagger}$.
(i.e. Pauli operators are chosen i.i.d for each qubit and each $j$, but are identical between the forward and inverse evolution.)
Intuitively, the inserted random Pauli operators ``dephase'' the operator $U_j^\dagger B_{j-1} U_j$ in the Pauli basis. This deletes constructive interference between the Pauli operators $P$ at the boundary of $U_j$, which has an identical effect to Monte Carlo sampling the $P$.
We show a cartoon of this process in Fig.~\ref{fig:cutting_explanation} for a single cut.
This connection gives a useful method to simulate the performance of the TNMC algorithm using circuit simulation alone.

\begin{figure}
    \centering
    \includegraphics[width=0.48\linewidth]{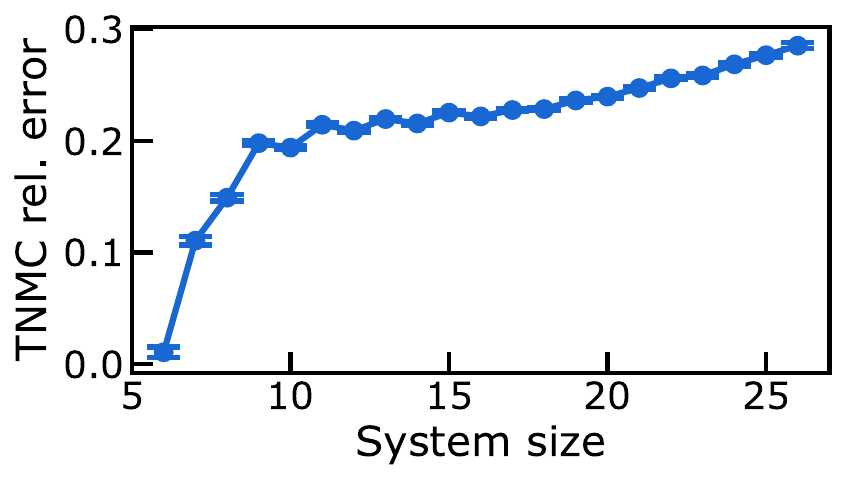}
    \includegraphics[width=0.48\linewidth]{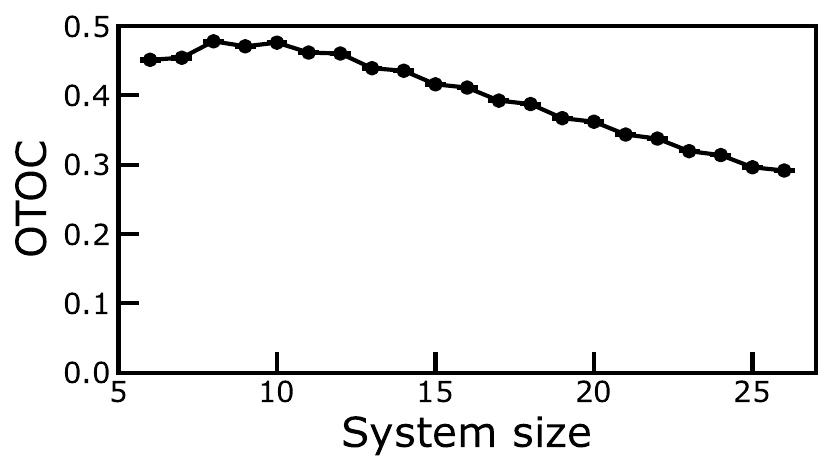}
    \caption{Results of TNMC estimation on systems of up to $26$ qubits with a single cut inserted in the middle of the circuit.
    (left) Relative error in TNMC estimation, normalized by the OTOC itself.
    (right) Target OTOC values used for (left).}
    \label{fig:tnmc_error}
\end{figure}

In practice, the all-to-all circuit simulations needed for long-range Hamiltonians are much deeper than they are wide.
Hence, we expect the TNMC algorithm to need to cut a $50-60$ qubit circuit multiple times in order to obtain practically simulable pieces.
In Fig.~\ref{fig:tnmc_error}, we observe that even a single cut in these circuits yields a significant OTOC error.
Here, we simulated the 1D chain described at the start of this section for a range of times, throughout which the OTOC is held to approximately $0.4$.
We observe that the TNMC relative error quickly increases to 20$\%$, and grows with the system size.
(This growth is in some part due to the OTOC decay with system size; the absolute error holds static at around $0.1$.)
This demonstrates that OTOC simulation methods that have simulated random circuit OTOCs to comparably high fidelity do not immediately extend to the Hamiltonian OTOCs considered here.
Testing other methods, and especially custom methods targeting Hamiltonian OTOCs is a clear target for future work.

%% file: sc_expt.tex
\subsection{Basic processor benchmarks}
Our quantum simulation is performed on a Willow quantum computer similar to that described in \cite{google25constructive} (see also \cite{google2025QEC_bellow_threshold}). The processors in this lineage are based on a 2D grid of flux tunable superconducting transmon qubits connected by flux tunable couplers between nearest neighbours. In this work, the mean frequency of qubit operation is $\sim6.2$~GHz and their anharmonicity is $\sim$210~MHz.

\begin{figure}[b!]
    \centering
    \includegraphics[width=\linewidth]{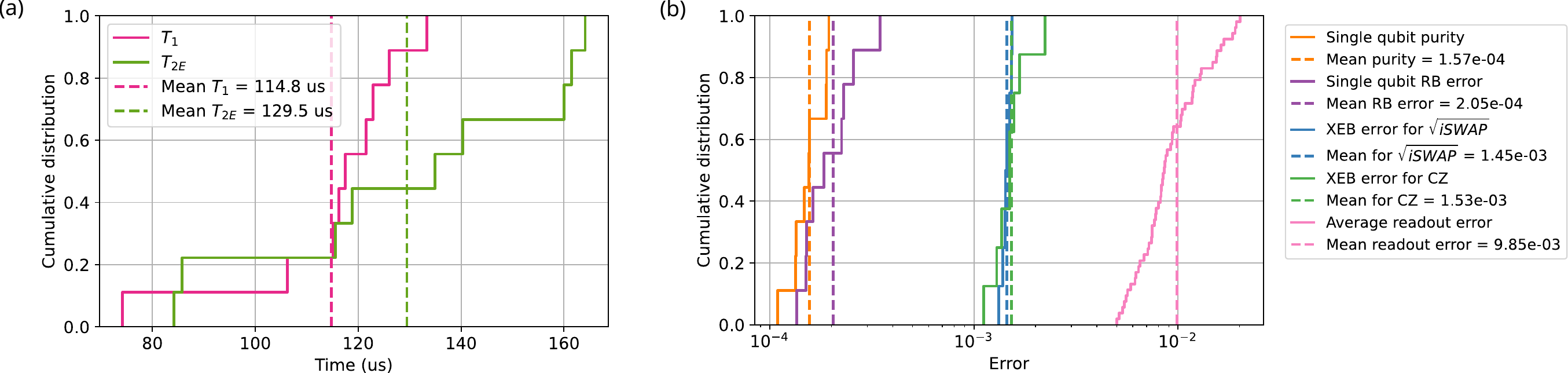}
    \caption{Base qubit and gate performance metrics. In panel (a) we show a cumulative distribution of single photon excitation lifetime $T_1$ for the qubits used in the quantum simulation. The qubits are arranged in a linear chain. The coherence time $T_{2E}$ of each qubit as measured by a Hahn-Echo sequence is also shown. In panel (b) we show a cumulative distribution of single and two qubit gate performance for each qubit and qubit pair involved in the quantum simulation. The single qubit performance is assessed by randomized benchmarking (RB), which is bounded from below by the purity of the state at the end of the sequence. The two-qubit gate performance is assessed here in terms of the CZ and $\sqrt{i\textrm{SWAP}}$ XEB cycle Pauli error. Finally, we also present a cumulative distribution of the average [$1-p(0|0)/2-p(1|1)/2$] readout error on the measurement qubit. The readout error distribution is over the quantum simulation time, about six hours.}
    \label{fig:T1_T2E_1Q_2Q_ro}
\end{figure}

Given that the quantum simulation of the two molecules addressed in this work require nine and fifteen qubits only, we can select a subsection of the 105 qubit Willow with performance better than average in the relevant benchmark metrics. The qubits are arranged in a line, and the unique measurement qubit in the quantum simulation is at one end of it. In this configuration, the all-to-all coupling is optimally achieved by a SWAP network as adapted from \cite{kivlichan2018quantum}. In \cref{fig:T1_T2E_1Q_2Q_ro} (a) we show the cumulative distribution of the lifetime of the single photon excitation in the qubits used for this work, which is $T_1=$114~\textmu s in average. In the same figure we also show the coherence time of these qubits as measured by a Hahn-Echo sequence and is in average $T_{2E}=$130~\textmu s. The performance of single qubit microwave gates, which we measure by Clifford randomized benchmarking (RB). In \cref{fig:T1_T2E_1Q_2Q_ro} (b) we show the total RB error and purity which are $\epsilon_{\textrm{RB}}$ = 0.00020 and $\epsilon_{\textrm{purity}}$ = 0.00015 in average, respectively. The same figure shows the performance of controlled-$Z$ (CZ) gates and square-root of $i$SWAP ($\sqrt{i\textrm{SWAP}}$) gates as measured by the cross entropy benchmarking (XEB) error, a previously established method \cite{Mi2021TimeCrystal, Arute2019Supremacy, mi21information}. Here gates are calibrated in parallel and therefore include errors arising from cross-talk effects in the parallelization policy for the linear chain experiment, whose moments are divided between two (``even" and ``odd") layers. Here we chose to inform the cycle Pauli error because it is a previously reported metric \cite{Mi2021TimeCrystal} allowing comparison of performance as the technology rapidly develops. This metric includes error contributions from two single qubit gates and a single two-qubit gate. This reflects well the structure of the OTOC-NMR quantum simulation where, as detailed below, two-qubit gates always appear accompanied by two single qubit gates in each layer of the Trotterization. The XEB cycle errors are respectively $\epsilon_{\textrm{CZ}}$=0.0015 and $\epsilon_{\sqrt{i\textrm{SWAP}}}=$0.0014, in average. In \cref{fig:T1_T2E_1Q_2Q_ro} (b) we also show the readout performance of the single measure qubit in the quantum simulation. The probabilities of preparing and measuring a given computational state [$p(0|0)$ and $p(1|1)$] on the measurement qubit are measured repeatedly during a typical experimental run of several hours. 
The corresponding distribution in \cref{fig:T1_T2E_1Q_2Q_ro} (b) shows the averaged readout error $\epsilon_{\textrm{ro}}=1-p(0|0)/2-p(1|1)/2$ during the quantum simulation and averages over six hours to 0.00985.  Although in general these metrics vary with a timescale of several hours and require recalibration during the data taking sessions, the figures here reflect the typical operating performance of our experiments.

\subsection{SWAP network: efficient all-to-all interaction from only nearest-neighbours coupling}
\label{sec:fSim_cal}

The Hamiltonian targeted by the superconducting quantum simulation,

\begin{equation}
\label{eq:double-quantum}
    \hat H = \sum d^{XY}_{ij}(X_iX_j-Y_iY_j) + \sum d^{ZZ}_{ij}Z_iZ_j,
\end{equation}

is known as the double-quantum Hamiltonian due to the term $X_iX_j-Y_iY_j=\sigma_i^+\sigma_j^++\sigma_i^-\sigma^-_j$. This is a squeezing term that creates and destroys coherently excitations in pairs. It is therefore not excitation number conserving and typical of driven systems. Our faculty to implement its Trotter decomposition of this Hamiltonian relies on the ability to implement high fidelity SWAP network to simulate an all-to-all spin interaction with hardware with only nearest neighbour connectivity \cite{kivlichan2018quantum}. 

\begin{figure}[b!]
    \centering
    \includegraphics[width=\linewidth]{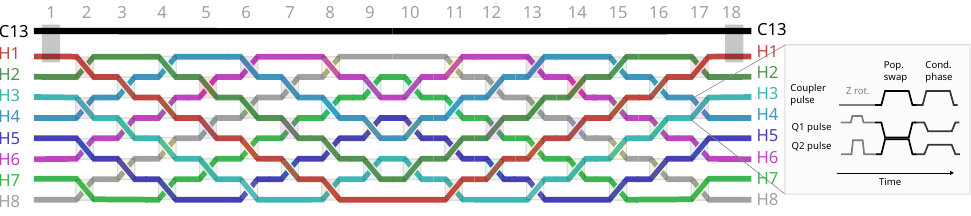}
    \caption{fSim SWAP network for the quantum simulation of Toluene. The circuit illustrates a single Trotter step of the double-quantum Hamiltonian simulation for Toluene on nine superconducting qubits arranged in a line. Horizontal rails represent superconducting transmon qubits, while meandering colored rails track the state of the nine molecular spins [one Carbon (C13) and eight Hydrogen protons(H1 to H8)] as their quantum states are swapped across the physical qubits. The colored boxes represent the tunable fSim$(\theta,\phi)$ two-qubit gates including single-qubit gate conjugations. The layers of non-overlapping gates are numbered from 1 to 18. The gates are implemented using a two-pulse composite sequence depicted in the right-most blow up. Note the exclusion of the Carbon spin from the main SWAP network, which reduces the gate count by 17\%.}
    \label{fig:SWAP_network}
\end{figure}

A circuit implementing a SWAP network Hamiltonian simulation of Toluene is shown in \cref{fig:SWAP_network}. Each two qubit interaction is depicted as a box connecting two adjacent qubit rails. The different colored boxes represent unique Fermionic simulation (fSim) gates \cite{Arute2019Supremacy, foxen2020demonstrating, mi21information,Mi2021TimeCrystal, Mi2022NoiseResilient, Morvan2022BoundStates} angle configuration encoding a spin-spin simulated interaction. The SWAP exchange is depicted as crossing wires. Superconducting qubits correspond to each of the horizontal rails and play the role of, i.e. simulate, different spins in the molecule. Each of the eight Hydrogen protons spins is assigned a color in this representation. The meandering color rails cross every other color rail in a single step of the network. 
This represents how each spin's quantum state is swapped across the superconducting qubits. So, the qubit-spin mapping changes in every step of the simulation. Importantly, this is optimal in the sense that the number of gates in a Trotter step is equal to the number of pairs \cite{kivlichan2018quantum}. That is, the number of gates is minimal and will not be reduced by an increased hardware connectivity: even if one would count with all-to-all connectivity, one would still require one gate per pair to simulate the all-to-all coupling. The technical challenge is to achieve state swapping efficiently and with sufficiently low error. This poses a metrological challenge over the implementation of tunable-angle fSim gates.

Due to the geometry of para-Carbon$^{13}$-Toluene addressed in this work, the Carbon spin, labelled ``C13" in \cref{fig:SWAP_network}, only interacts with the Hydrogen proton labelled ``H1". Or rather, its interaction with every other spin is two orders of magnitude smaller and therefore negligible in the timescale of the experiment.  Spin C13 can be therefore extracted from the SWAP network to reduce the gate count by 17\%, and therefore the error introduced by imperfect gates without otherwise affecting the learning experiment. We note also that the gates removed are full SWAP and therefore having the lowest performance. This technique is, in principle, general but does modify the ordinary Trotter formula. In particular, after one Trotter step the qubit-spin mapping is inverted leaving the qubits simulating C13 and H1 far apart and unable to interact (see mid layers 9 and 10 in \cref{fig:SWAP_network}, red and black rails). A second Trotter step needs to be applied to bring them next to each other before a new Trotterized interaction can take place between them (see layer 18 in \cref{fig:SWAP_network}), which corresponds to working in the interaction frame of the C-H coupling.

\subsection{Trotterization of a double-quantum Hamiltonian embedded in a SWAP network}

The Trotter approximation for the time evolution generated by the double-quantum Hamiltonian \cref{eq:double-quantum} is constructed by sequentially applying the unitaries corresponding to the individual terms of the Hamiltonian. For a time increment $\delta t$ these read
$$\textrm{SWAP}\times\mathrm{exp}[-id^{XY}_{ij}\delta t(X_iX_j-Y_iY_j)]=\begin{pmatrix}
\cos2d^{XY}_{ij}\delta t & 0 & 0 & -i\sin2d^{XY}_{ij}\delta t \\
0 & 0 & 1 & 0 \\
0 & 1 & 0 & 0 \\
-i\sin2d^{XY}_{ij}\delta t & 0 & 0 & \cos2d^{XY}_{ij}\delta t
\end{pmatrix},$$

and

$$\textrm{SWAP}\times\mathrm{exp}[-id^{ZZ}_{ij}\delta t Z_iZ_j]=\begin{pmatrix}
1 & 0 & 0 & 0 \\
0 & 0 & e^{i2d^{ZZ}_{ij}\delta t} & 0 \\
0 & e^{i2d^{ZZ}_{ij}\delta t} & 0 & 0 \\
0 & 0 & 0 & 1
\end{pmatrix},$$

As we will now discuss, these unitaries are equivalent, up to single qubit rotations, to the so-called (fSim) gates. These are the gates native to the Willow processor, and present a natural fit of this hardware to this application.

In a Willow processor, the effective two qubit gate dynamics can be modelled by the Hamiltonian

\begin{equation}
H(g, \Delta, \eta) = \begin{pmatrix}
0 & g & 0 & 0 & 0 \\
g & \Delta & 0 & 0 & 0 \\
0 & 0 & \Delta & \sqrt{2}g & \sqrt{2}g \\
0 & 0 & \sqrt{2}g & 2\Delta + \eta & 0 \\
0 & 0 & \sqrt{2}g & 0 & \eta
\end{pmatrix}.
\end{equation}

It describes the evolution of two neighbouring transmon qubits, detuned by an amount $\Delta$ and coupled with a single excitation resonant swap rate $g$, in the truncated oscillator basis $\{|01\rangle,|10\rangle,|11\rangle,|20\rangle,|02\rangle\}$. The ability to control both $\Delta$ and $g$ as a function of time allows the digital control of the interaction. The parameter $\eta$ is the transmon anharmonicity, approximated in this expression to be the same for both oscillator and constant.

The integration of the Schr\"odinger equation yields a fSim unitary with five effective parameters of the form

\begin{equation}
\text{fSim}(\theta, \phi, \Delta_+, \Delta_{-}^{\textrm{diag}}, \Delta_{-}^{\textrm{off diag}}) = 
\begin{pmatrix}
1 & 0 & 0 & 0 \\
0 & e^{i(\Delta_+ + \Delta_{-}^{\textrm{diag}})} \cos \theta & -i e^{i(\Delta_+ - \Delta_{-}^{\textrm{off diag}})} \sin \theta & 0 \\
0 & -i e^{i(\Delta_+ + \Delta_{-}^{\textrm{off diag}})} \sin \theta & e^{i(\Delta_+ - \Delta_{-}^{\textrm{diag}})} \cos \theta & 0 \\
0 & 0 & 0 & e^{i(2\Delta_+ + \phi)}
\end{pmatrix}.
\end{equation}

By applying $Z$-phase gates on each qubit before and after the interaction it is possible to gauge out the single qubit phases $\Delta_+, \;\Delta_{-}^{\textrm{diag}}, \;\Delta_{-}^{\textrm{off diag}}$ and bring the interaction unitary to the canonical fSim form

\begin{equation}
\text{fSim}(\theta, \phi) = 
\begin{pmatrix}
1 & 0 & 0 & 0 \\
0 &  \cos \theta & -i  \sin \theta & 0 \\
0 & -i  \sin \theta &  \cos \theta & 0 \\
0 & 0 & 0 & e^{i \phi}
\end{pmatrix}.
\end{equation}

That is 
$$
\text{fSim}(\theta, \phi) = (Z(\varphi_1)\otimes Z(\varphi_2)) \text{fSim}(\theta, \phi, \Delta_+, \Delta_{-}^{\textrm{diag}}, \Delta_{-}^{\textrm{off diag}}) (Z(\varphi_3)\otimes Z(\varphi_4)),$$
with $\varphi_1 = \Delta_{-}^{\textrm{off diag}} + \Delta_{-}^{\textrm{diag}} + 2\Delta_+$, $\varphi_2 = \Delta_{-}^{\textrm{off diag}} + \Delta_{-}^{\textrm{diag}} -2 \Delta_+$, $\varphi_3 =  \Delta_{-}^{\textrm{diag}} - \Delta_{-}^{\textrm{off diag}} $, and $\varphi_4 =  -\Delta_{-}^{\textrm{diag}} - \Delta_{-}^{\textrm{off diag}}$ and $Z(\varphi)=\textrm{diag}(1,e^{-i\varphi})$.

We note that it is possible to obtain with a single pulse a fSim gate with arbitrary two-qubit angles \cite{rosenberg2024dynamics,Morvan2022BoundStates}. While previous work was able to operate in this regime, it was exploited that given sets of two-qubit angles are more natural than other. For piecewise constant dependence of the parameters, one can approximate the swap angle during the gate changes as $\theta\approx t\sqrt{g^2 +\Delta^2/4}$ and the conditional phase changes as $\phi\approx t g^2/\Delta_{12}$ where $\Delta_{12}$ is the detuning between $|11\rangle$ and $|02\rangle$. The different scaling with $t$ and $g$ allows to tune both angles independently. When the experimental exploration at hand allows to only tune one of the knobs, say $g$, keeping the other one ($t$) fixed, this is the simple and most efficient way to proceed \cite{Morvan2022BoundStates,Neill2021QuantumRing,mi21information,Arute2020SeparatedDynamics,Mi2021TimeCrystal,Mi2022NoiseResilient}. On the other hand, imposing the single pulse fSim approach to produce arbitrary control of both $\theta$ and $\phi$ almost certainly forces too long gate times that degrades performance by decoherence, or too high coupling constant $g$ which enhances leakage probability outside of the computational manifold. In our case, both two-qubit angles are simultaneously specified by the NMR Hamiltonian and are required with high precision and high fidelity for the simulation. We therefore chose not to proceed with the single pulse fSim approach and instead adapted a two-pulse approach from earlier work \cite{foxen2020demonstrating}.

\subsection{Gate decomposition for composite fSim calibration}

To define the mapping between the Trotterization of the double-quantum Hamiltonian and the native gates on Willow it is convenient to introduce the composite gate unitary

$$U_{\textrm{comp}}(\varphi,\phi_{\textrm{cond}},\theta_{\textrm{fSim}},\phi_{\textrm{fSim}}) = Z_i(\varphi)Z_j(\varphi) \textrm{CPHASE}_{ij}(\phi_{\textrm{cond}})  \textrm{fSim}_{ij}(\theta_{\textrm{fSim}},\phi_{\textrm{fSim}})$$

Where $\textrm{CPHASE}$ is a form of fSim gate defined by convention as $\textrm{CPHASE}_{ij}(\phi_{\textrm{cond}}) = \textrm{fSim}_{ij}(\theta_{\textrm{fSim}}=0,\phi_{\textrm{fSim}}=-\phi_{\textrm{cond}}).$  The motivation to implement a composite gate from a CPHASE gate and a fSim gate is that it allows one to use independent calibrations to grossly target a population swapping gate with the desired $\theta_{\textrm{fSim}}$, unencumbered by the associated spurious $\phi_{\textrm{fSim}}$. Thereafter, independent control of $\phi_{\textrm{cond}}$ allows to obtain an unconstrained effective conditional phase for the composite unitary, therefore providing independent control over the effective two two-qubit angles of the composite gate. Experimental imperfections make all the effective angles of the composite gate differ by more than the target tolerance of 20 mrad from what one would predict from an ideal composition of the constituent gates. These errors are addressed on a second, fine tuned, calibration of the composite gate as a whole.

Experimental imperfection aside from now, this notation makes it is easy to see that 

$$\textrm{SWAP}\times\mathrm{exp}[-id^{XY}_{ij}\delta t(X_iX_j-Y_iY_j)]=X_i U_{\textrm{comp}} X_j,$$
where $U_{\textrm{comp}}$ is specified by $\phi_{\textrm{cond}} = \phi_{\textrm{fSim}}$, $\phi_{\textrm{cond}} = \pi - \phi_{\textrm{fSim}}$, $\theta_{\textrm{fSim}}=\pi/2 + 2 d^{XY}_{ij} \delta t$, $\varphi = -(\phi_{\textrm{cond}} + \phi_{\textrm{fSim}})/2$. While 

$$\textrm{SWAP}\times\mathrm{exp}[-id^{ZZ}_{ij}\delta t Z_iZ_j]=U_{\textrm{comp}},$$
where $U_{\textrm{comp}}$ is specified by $\phi_{\textrm{cond}} = \phi_{\textrm{fSim}}$, $\phi_{\textrm{cond}} = \pi - \phi_{\textrm{fSim}}+4d^{ZZ}_{ij}\delta t$, $\theta_{\textrm{fSim}}=\pi/2$, $\varphi = -(\phi_{\textrm{cond}} + \phi_{\textrm{fSim}})/2$.

Alternatively, considering together the $(X_iX_j-Y_iY_j)$ term and the $Z_iZ_j$ term, one can write 

$$U_{\textrm{comp}}(\varphi,\phi_{\textrm{cond}},\theta_{\textrm{fSim}},\phi_{\textrm{fSim}}) = \text{fSim}(\theta_{\textrm{comp}}, \phi_{\textrm{comp}}, \Delta_+, \Delta_{-}^{\textrm{diag}}, \Delta_{-}^{\textrm{off diag}}),$$ 

with the mapping

$$\theta_{\textrm{comp}}=\pi/2 + 2 d^{XY}_{ij} \delta t,\;\;\;\;\; \phi_{\textrm{comp}} = -\pi-4d^{ZZ}_{ij}\delta t,$$

$$\Delta_+=-\phi/2, \;\;\;\;\;\text{and}\;\;\;\;\;\Delta_{-}^{\textrm{diag}}=\Delta_{-}^{\textrm{off diag}}=0.$$

\subsection{Composite fSim gate calibration}

For a small Trotter step $\delta t$ (i.e., $d^{XY}_{ij} \delta t,d^{ZZ}_{ij} \delta t \ll\pi)$ every two-qubit gate in the Trotterization at hand is a perturbation of the SWAP gate. This requires only a precise calibration of both $\theta_{\textrm{comp}}$ and $\phi_{\textrm{comp}}$ around a small range around the SWAP values. However, such a decomposition would require far too many gates degrading the circuit fidelity below what can be recovered by current error mitigation techniques. A back of the envelope calculation can be used to estimate the gate cost of such an approach. With our implementation in mind, taking $d^{XY}_{ij}\delta t = 1/10$, the total simulated circuit extending to 50$\times\delta t$ (this corresponds to $d^{XY}_{ij}\approx 2$~kHz and total simulated time evolution of 2.5~ms forward, and 2.5~ms backwards) and the nine qubit experiment requiring 36 gates to implement a SWAP network, the gate count estimate yields $2\times50\times36 = 3600$. With current two-qubit gate fidelity of at best 0.998, the smallest circuit fidelity is estimated to be $0.998^{3600}\approx0.0007$. 

\begin{figure}[t!]
    \centering
    \includegraphics[width=\linewidth]{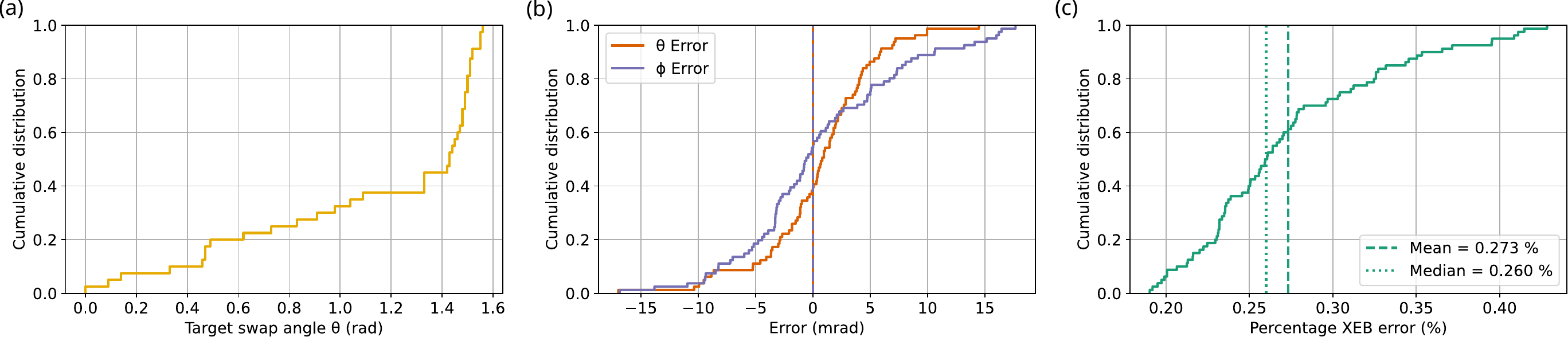}
    \caption{(a) Histogram showing the distribution of the 80 unique target swap angles $\theta$ required for the full learning quantum simulation of Toluene. They span the full range available $[0,\pi]$. (b) Distribution of two-qubit angle errors achieved by the iterative calibration protocol. All final errors are below the $20$~mrad by construction. (c) Distribution of the Cross-Entropy Benchmarking (XEB) error for the 80 calibrated, phase-matched fSim gates as they are deployed in the quantum simulation circuit. This metric includes the error contributions from both the two-qubit pulse and the required single-qubit Z gates used for phase matching.}
    \label{fig:target_angle_error_XEB}
\end{figure}

In order to perform our quantum simulation with sufficient fidelity we perform longer Trotter steps to bound the two-qubit gate count to only a few hundreds. To match the NMR data we chose $\delta t = 0.375$~ms for our simulation. In consequence, the gates calibrated do depart appreciably from the SWAP angles. Due to the rich scale of energies involved in Toluene the range of two-qubit angles that need to be calibrated span the entire domain from 0 to $\pi$. The conditional phase angles required for our SWAP network are all $\pi$ while the cognitional phase for the C13-H1 interaction are either 1.02~mrad for the gates at the beginning and end of the Trotterized evolution and doubles to 2.04 rad for gates in the middle of the circuit where the end gate of one step in compiled with the first gate of the next (these are the gates in layers 1 and 18 in \cref{fig:SWAP_network}). The swap angles involved in the full learning experiment are instead plotted in \cref{fig:target_angle_error_XEB} (a). The wide range of angles adds to the complexity of the calibration required for this experiments, as different calibration techniques are required for different angle ranges to achieve the required precision 20~mrad (see numerical section XI). 

The hardware gate pulses used to engineer these interactions are implemented with a series of base band trapezoidal pulses (see \cref{fig:SWAP_network}) with a fidelity-optimized pulse duration of 12~ns (8~ns hold time) for the physical phase-matching $Z$ gates, 47~ns (45~ns hold time) for the population swap pulse, and 33~ns for the conditional phase pulse (31~ns hold time). The coupler pulses have a total duration of 26~ns (18~ns hold time) for the population swap pulse and 22~ns (16~ns hold time). In practice, each gate in the composite gate is calibrated independently to provide an initial guess for the implementation. Then, the composite gate is calibrated as a whole to obtain the effective angles, which normally differ from the predicted ones from the independent calibration of the pulses by a few tens of miliradians. An iterative produce is then put in place to obtained the desired precision. 

The iterative calibration procedure develops as follows. Having calibrated the CPHASE pulse as in previous work \cite{Mi2021TimeCrystal} we move to calibrate the population swap pulse. The gate duration is optimized to implement a  $\sqrt{i\textrm{SWAP}}$ gate since with a swap angle of $\theta = \pi/4$ lies in the centre of the distribution of typical gates we will calibrate for. For fix pulse duration, we use fast two-qubit tomography to fine tune the qubit frequencies to maximize the population transfer totalling a complete ${i\textrm{SWAP}}$. At the calibrated qubit detuning we use fast two-qubit tomography again to determine both the swap angle and the conditional phase that the pulse imprints over the qubit pair as a function of the max coupling amplitude $g_{\textrm{max}}$ in the trapezoidal coupler pulse. In agreement with a simple model, we fit the swap angle $\theta_{\textrm{fsim}}$ by a linear dependence in $g_{\textrm{max}}$ and the conditional phase $\phi_{\textrm{fsim}}$ by a quadratic dependent in $g_{\textrm{max}}$. The variance of the fit errors is $<50$~mrad for both angles consistently. This fit provides a first good coarse guess to build the composite pulse.

A composite pulse is initially built by combining a fSim pulse with the swap angle dictated by the targetted Hamiltonian simulation $\theta_{\textrm{fsim}} = \pi/2 + 2 d^{XY}_{ij} \delta t$ folded in the [0,$\pi/2$] interval, and a CPHASE gate. The fSim pulse interaction strength $g_{\textrm{max}}$ is dictated by the linear fit obtained for $\theta_{\textrm{fsim}}$. The CHPASE pulse is set by an independently calibrated angle of $\phi_{\textrm{cond}} = -\pi-4d^{ZZ}_{ij}\delta t - \phi_{\textrm{fsim}}$, where $\phi_{\textrm{fsim}}$ is the spurious phase of the fSim gate (previously obtained from the quadratic fit against the interaction strength $g_{\textrm{max}}$). We note that the CPHASE, introduce to compensate the spurious conditional phase of the population swapping fSim pulse, has itself a small spurious swap angle of $\sim30$~mrad that will add to the final effective swap angle. These two swap rotations are in general not collinear and are taken care of by the iterative calibration protocol provided the following technicality. Due to the spurious swap angle in the CPHASE pulse ($\sim30$ mrad), the order in which the two interaction pulses in the composite gate is applied matters, if one aims for sub 20 mrad precision. The arbitrary angle between this consecutive swap rotations makes them non-commuting rotations. Furthermore, the single qubit physical $Z$ gates involved (see \cref{fig:SWAP_network}) are physical flux pulses (as opposed to virtual phase update on the microwave drives) as these $Z$ gates do not commute trough any of the swapping interaction. Even if it approximately commutes with the CPHASE, the finite spurious angle creates a non-commutativity error larger than our 20 mrad budget. These $Z$ gates have a phase that depends on the context. They depend on the single qubit gates that have been applied just before, since consecutive single qubit gates get merged together for efficiency in the compilation. Due to phase tracking, they even depend on idling time durations. Having a physical $Z$ gate between the two pulses in the composite gate, then, affects in an undesirable way the calibration and the final effective swap angle. It changes the relative angle of the two consecutive swap rotations of the gate as a function of context, as it changes therefore the relative angle with which these two rotations interfere. This is easily avoided by ensuring the physical $Z$ gates are not played between the interaction pulses but before them (or, alternatively, after).

With a first composite pulse composed, high precision Floquet calibrations, tailored to the target angle range, are used to measure the swap angle and the conditional phase angle of the composite pulse. For small swap angles ($<30$~mrad) we adapted to the two pulse fSim implementation the technique discussed in \cite{Mi2021TimeCrystal}, while for larger angles we adapted the technique described in \cite{Arute2020SeparatedDynamics}, and adjusted the number of gates used in the periodic calibration as well as the fitting procedure to the value of the expected angle. This first high precision measurements of both the swap and conditional phase angles of the composite gate are compared to the target angles. If any one of the two two-qubit angles differs from target by more that 20~mrad a correction proportional to the errors is applied to the control for compensation and a new high precision measurement of both angles is performed. That is, the error in the conditional phase feeds back into the control of the CPHASE pulse (i.e., the location of the detuning-coupling parameter along the minimum-leakage arc, see \cite{foxen2020demonstrating,Mi2021TimeCrystal}), while the error in the swap angle feeds back to the control of the fSim gate (i.e., $g_{\textrm{max}})$. Now with two high precision measurements for each, for the swap angle and the conditional phase angle, we perform a local linear extrapolation targeting the control parameters for zero error. The procedure of local linear extrapolation repeats taking into account all previous high precision measurement until the target precisions is attained in both angles. This procedure is done in parallel for all gates compatible with simultaneous calibration (at most 4, see \cref{fig:SWAP_network}), and always succeeds. It typically concludes after three rounds of simultaneous interpolation lasting $\sim$ 10 minutes total. The distribution of two-qubit angle errors is shown in \cref{fig:target_angle_error_XEB} (b) for our full Toluene learning experiment. For a given pulse shape and pair, periodic calibrations are able to reproducibly measure both angles with a shot-to-shot spread (precision) of $<5$~mrad.

For Toluene, 33 different fSim gates needed to be calibrated independently down to a precision of $<20$~mrad in order to simulate a single OTOC trajectory. Furthermore, for the learning experiment, nine different trajectories are needed. However, not all $9\times 33$ fsim gates are meaningfully different: rounding the interaction angles to 10~mrad we are left with only 80 unique gates to calibrate and still maintain an excellent approximation of the evolution. The unique-angle two-qubit gates appear in the quantum simulation in different combination, totalling 107 distinct moments. Calibrating this number of distinct moments is impractical, but we found that in-context calibration does not provide an advantage. While simultaneous calibration of different gates does degrade their XEB fidelity, this degradation is independent of the specific context. Importantly, due to the symmetries of the molecules, many of the distinct calibration moments differ by a single or only a few gates. We then shuffle the 80 distinct fSim gates needed for the full Hamiltonian learning experiment into calibration moments where each gate appears only once. This reduces the calibration overhead to only 28 moments with typical wall-clock time of 5hs of calibration.

Once the pulse parameter mapping to the target two-qubit angles is obtained, the gate's single qubit angles are measured by XEB fitting. The single qubit angles $\Delta_+, \;\Delta_{-}^{\textrm{diag}},$ and $\Delta_{-}^{\textrm{off diag}}$ from the XEB fit are then used to phase-match the gates and build the desired composite canonical $\text{fSim}(\theta, \phi)$ gate by single qubit $Z$-phase conjugation, as explained above. This produces the final composite fSim gates used in our simulation. The XEB fidelity of these gates is the measured once again to take into account the phase-matching correcting as a final metric assessing their performance. In \cref{fig:target_angle_error_XEB} (c) we show the XEB error of the phase-matched gates. The median cycle Pauli infidelity for the composite-phase matched gate is 0.0026.

%% file: modeling.tex
\subsection{Trajectory simulation of quantum circuits}

To validate our experimental results and accurately account for the impact of device noise, we performed comprehensive numerical simulations of the NMR OTOC protocol. These simulations were designed to faithfully reproduce the experimental conditions, incorporating a multi-layered noise model parameterized by the device benchmarks detailed in the main text (see Basic processor benchmarks). Our numerical simulations were performed using an in-house quantum trajectory simulator \texttt{kraus-sim}. Quantum channels are represented by Kraus-operators, defined by the respective action on the density matrix $\rho$ as:

$$\mathcal{E}(\rho) = \sum_{j \ge 1} K_j \rho K_j^\dagger, \quad \text{where} \quad \sum_{j \ge 1} K_j^\dagger K_j = I$$

\texttt{kraus-sim} simulates the evolution of a pure state vector $\ket{\psi}$ instead of the full density matrix $\rho$, offering a significant computational advantage at the cost of induced sampling error. Each trajectory represents one independent ``shot" of the experiment. The simulation begins with a pure initial state $|\psi\rangle$. For each channel $\mathcal{E}$ in the circuit, one of its Kraus operators $K_j$ is randomly sampled and applied to the state. The probability $p(j)$ of selecting a specific operator $K_j$ is given by the Born rule:$$p(j) = \langle\psi| K_j^\dagger K_j |\psi\rangle$$After sampling, the state is updated and re-normalized to

$$|\psi'\rangle = \frac{K_j |\psi\rangle}{\sqrt{p(j)}}$$

This process, which is equivalent to the state back-action from a POVM measurement with elements $M_j = K_j^\dagger K_j$, is repeated for every channel in the circuit. The final experimental statistics are recovered by averaging the outcomes of many independent trajectories.
A more detailed account of the quantum trajectories method can be found in Ref. \cite{isakov2021simulationsquantumcircuitsapproximate}

\subsection{Inhomogeneous noise sensitivity}

\begin{figure}
    \centering
    \includegraphics[width=\linewidth]{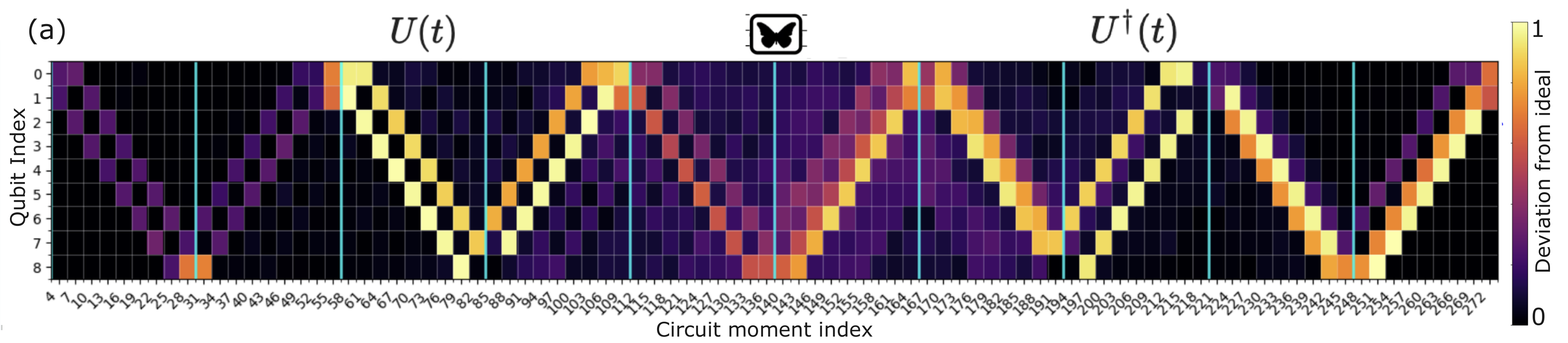}
    \caption{Spatio-temporal sensitivity map of the 9-qubit OTOC circuit to a single bit-flip error. A single Pauli-X gate was deterministically inserted at each point in the circuit, defined by the qubit index (vertical axis) and the circuit moment index (horizontal axis). The color at each point indicates the magnitude of the deviation of the final OTOC value from the ideal, noiseless result. The circuit is composed of a forward time evolution $U(t)$ followed by a backward time evolution $U^\dagger(t)$. The bright regions of high sensitivity clearly trace the path of the logical state of the Carbon spin (initially on Qubit 0) as it is swapped through the device via the SWAP network, illustrating the operator's "light cone."}
    \label{fig:9q_noise_sensitivity}
\end{figure}

To understand the circuit's susceptibility to errors, we performed a first-order sensitivity analysis by simulating the deterministic injection of a single error at all points in the circuit. This study allows us to map the spatio-temporal impact of a local error on the final OTOC observable, revealing the flow of information and the most sensitive regions of the quantum simulation. In particular, the methodology consists of running a series of ideal, noiseless simulations of the OTOC circuits. For each independent simulation, a single X gate was deterministically inserted at a unique location to emulate a bit flip error, defined by a specific qubit and a specific circuit moment (i.e., after a particular gate layer). We then computed the absolute deviation of the final measured OTOC value from the ideal, error-free case. The results of this analysis are visualized in the heatmap in Fig.~\ref{fig:9q_noise_sensitivity}.

The heatmap in Fig.~\ref{fig:9q_noise_sensitivity} plots this OTOC deviation as a function of the error's location. The vertical axis represents the physical qubit index while the horizontal axis represents the circuit moment index, or time. The circuit consists of a forward evolution $U(t)$ followed by the butterfly operator and the backward evolution $U^\dagger(t)$. The bright yellow and white regions indicate locations where a single bit-flip has a large impact on the final OTOC, while the dark purple regions indicate low sensitivity. The most striking feature of this analysis is the clear, structured pattern of error sensitivity inherited from the structured nature of the Hamiltonian. The regions of highest impact are not randomly distributed but instead form distinct diagonal and V-shaped paths through the circuit's space-time volume. This pattern is a direct visualization of the SWAP network's operation. As highlighted, the logical state of the Carbon spin begins on the physical qubit indexed as 0. The bright path originating from C13 qubit at the start of the circuit traces the trajectory of this logical spin interaction with the Hydrogen qubits as it is swapped from one physical qubit to the next during the forward evolution $U(t)$, and then back during the backward evolution $U^\dagger(t)$.

This result demonstrates that the observed circuit's infidelity is dominated by errors that occur within the effective ``light cone" of the operator being measured. In the context of the OTOC, the measurement is sensitive to the evolution of the initially localized operator (on the Carbon spin). An error occurring on or nearby a qubit that is currently holding the logical state of the Carbon spin directly corrupts this information, leading to a large deviation in the final measurement. Conversely, an error on a distant qubit, outside this light cone, has a much smaller, higher-order effect on the final OTOC value, as its influence must propagate through many two-qubit gates to reach the operator of interest. This has important ramifications on calibration -- fSim gates within the lightcone have more strigent precision requirements.

\subsection{Agreement of simulations to experiment}

 The specific Kraus operators for our error channels were parameterized using the calibrated benchmarks from the Willow processor. Incoherent errors were applied following the ideal gate execution, scaled by the physical gate time set by the waveform. In particular, qubit decay and dephasing were modeling as a composite amplitude and phase damping channels set by the median characterized values across all qubits in the grid, T1=114.8 \textmu s. and T2 =130 \textmu s respectively. Additionally, coherent errors were modeled as additional small angle fSim($\theta, \phi$) following each ideal gate to capture imprecision in calibration of the swap and cphase angles for each composite gate. Furthermore, to model the imprecision in estimating the required physical Z gates to compensate for the composite gate, we additionally add an unwanted $Z^{\alpha_{1,2}}$ gate on each qubit. Importantly, we assume these parameters are set by the calibration and thus are reproducible per unique calibrated gate, i.e. the same coherent error is applied on each successive gates. To this end, for each unique fSim gate we sampled $\theta,\phi,\alpha_i$ parameters from a Gaussian distribution with zero mean and standard deviation set by the calibration characterization above. For the 15 qubit experiments, the gateset was not arbitrary angled fSim but rather utilized CZ gates and arbitrary single qubit microwave gates. However, the underlying coherent error channels were modeled equivalently, i.e. by a small angled fSim gate and physical Z gates following each CZ. Unique CZ gates were in turn specified simply by the qubit pair on which it acts. All taken together, coherent errors, amplitude and phase damping channels could not fully explain the single qubit fidelities as measured by randomized benchmarking error or the two qubit XEB cycle Pauli infidelities, originating from features not modeled such as low frequency noise, or leakage out of the computational subspace. One and two qubit depolarizing channels were appended to the noise channels following each gate to match experimental error rates. Readout error was modeled as a classical confusion matrix applied to the final measurements.
 
\begin{figure}
    \centering
    \includegraphics[width=\linewidth]{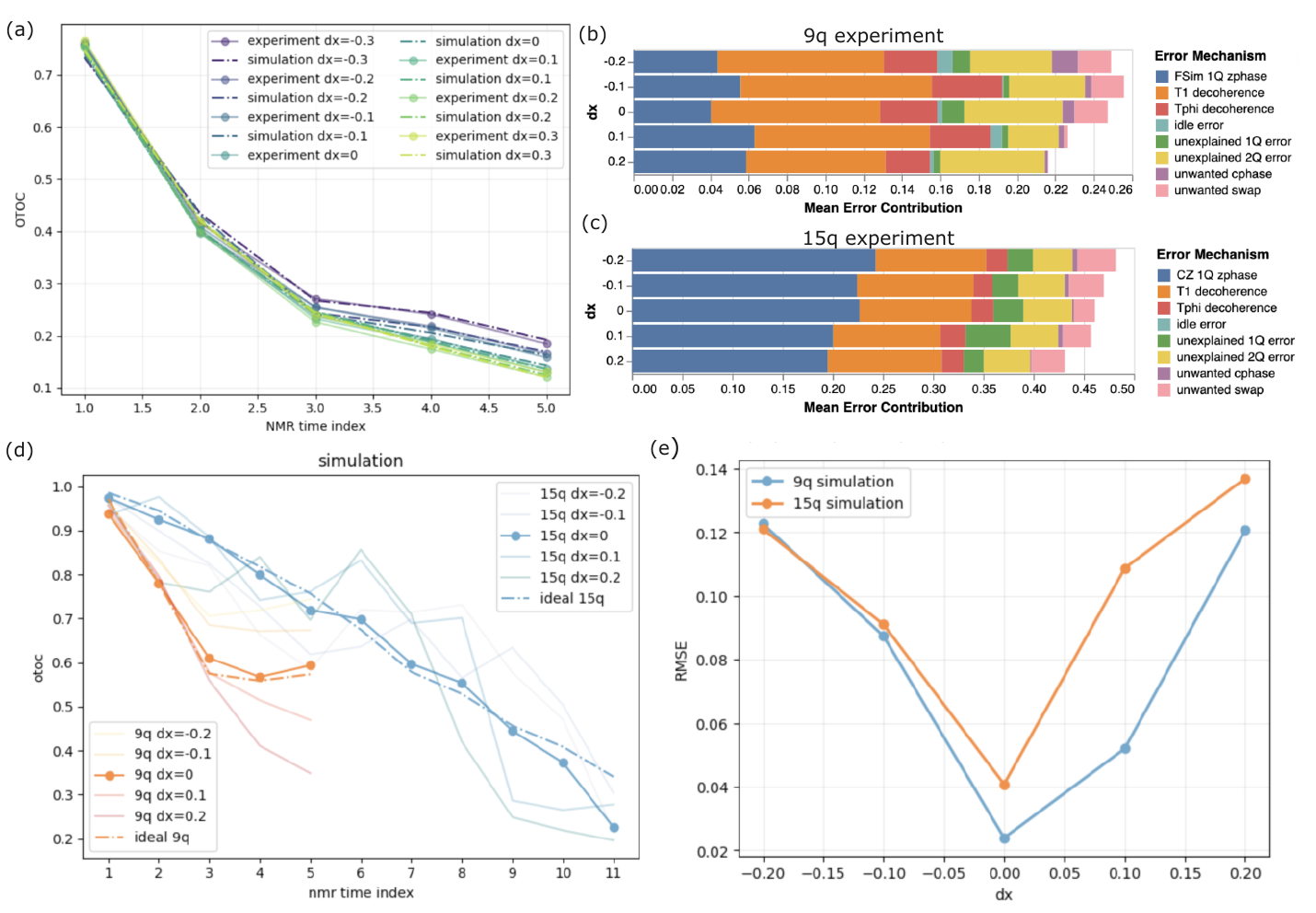}
    \caption{\textbf{Simulation validation and error analysis}. (a) Comparison of experimentally measured OTOC (solid lines) versus full noise-model simulation (dashed/dotted lines) for the 9-qubit experiment, showing strong agreement. The parameter $dx$ represents an additional, intentionally-added coherent error to test the model. (b, c) Mean error contribution per gate for the (b) 9q fSim-based experiment and (c) 15q CZ-based experiment, breaking down the sources of infidelity. (d, e) In-silico learning experiment for the (d) 9-qubit and (e) 15-qubit simulations. The calibrated noise model ($dx=0$, solid lines) is compared against mismatched models ($dx \neq 0$) and the ideal, noiseless evolution (dashed lines). (f) Root Mean Square Error (RMSE) between the simulated and ideal curves from (d) and (e), showing a distinct minimum at $dx=0$.}
    \label{fig:error_budget_analysis}
\end{figure}

The resulting noise model is validated in Fig.~\ref{fig:error_budget_analysis}(a), which shows the simulated OTOC decay plotted directly against the experimental data for the 9-qubit experiment. The solid lines represent the experimental data, while the dashed/dotted lines represent the output of our full noise-model simulation. The different colors correspond to experiments where a known, additional coherent error (parameterized by $dx$) was applied, testing the model's predictive power. The simulation accurately recapitulates the experimental results across all time steps and error parameters, confirming the validity of our model.

To better understand the dominant sources of infidelity, we performed a detailed error budget analysis, shown in Fig.~\ref{fig:error_budget_analysis}(b) and (c). These plots show the mean error contribution to the measured OTOC signal, averaged per gate, for the two different experimental sets of circuits. For the 9-qubit experiment (Fig. \ref{fig:error_budget_analysis}(b)), which uses the composite fSim gate set, the error budget is dominated by two-qubit errors. We note that a large contribution is the ``unexplained 2Q error", accounting for approximately 0.25 of the mean error. This is followed by coherent errors: ``unwanted swap" (approx. 0.10) and ``unwanted cphase" (approx. 0.08). We note that precision calibration discussed above helped mitigate the impact of these error channels. Decoherence during gates (``T1 decoherence," approx. 0.05) and ``unexplained 1Q error" (approx. 0.04)  are secondary, but non-negligible, contributions. For the 15-qubit experiment (Fig. \ref{fig:error_budget_analysis}(c)), which uses a CZ-based gate set, the error profile is qualitatively different. The coherent errors related to the fSim gate (``unwanted swap" and ``unwanted cphase") are negligible, but the uncalibrated Z-phase disproportionally impacts the OTOC fidelity. The error budget is also more evenly distributed. The ``unexplained 2Q error" is still a dominant contributor (approx. 0.022), but it is followed closely by "T1 decoherence" (approx. 0.018) and "unexplained 1Q error" (approx. 0.015). This indicates that for the CZ-based experiment, single-qubit errors and decoherence play a much larger relative role in the circuit fidelity than in the fSim-based experiment

Fig.~\ref{fig:error_budget_analysis}(d) and (e) demonstrate an in-silico learning experiment based on our validated simulation framework. In these plots, the dashed lines represent the ideal, noiseless evolution of the OTOC for the 9q and 15q systems, respectively. The solid blue and orange curves (dx=0) show the simulation using our calibrated noise model. The significant gap between the dx=0 curve and the ideal curve highlights the substantial impact of the characterized device noise on the quantum simulation.The lighter-colored curves show the simulated dynamics when the coherent error model is intentionally mismatched (e.g., $dx = -0.2, 0.1$, etc.). Qualitatively, the calibrated dx=0 simulation tracks the shape of the ideal evolution more closely than the mismatched models. This is quantified in Fig.~\ref{fig:error_budget_analysis}(f), which plots the Root Mean Square Error (RMSE) between each simulated $dx$ curve and the ideal, noiseless simulation. For both the 9q and 15q cases, the RMSE is clearly minimized at $dx=0.00$. This result is twofold: first, it demonstrates that our simulation of the experiment successfully recapitulates the results of the on-device learning experiment, and second, it confirms that the calibrated noise model is an accurate representation of the hardware

%% file: main.bbl
\begin{thebibliography}{108}%
\makeatletter
\providecommand \@ifxundefined [1]{%
 \@ifx{#1\undefined}
}%
\providecommand \@ifnum [1]{%
 \ifnum #1\expandafter \@firstoftwo
 \else \expandafter \@secondoftwo
 \fi
}%
\providecommand \@ifx [1]{%
 \ifx #1\expandafter \@firstoftwo
 \else \expandafter \@secondoftwo
 \fi
}%
\providecommand \natexlab [1]{#1}%
\providecommand \enquote  [1]{``#1''}%
\providecommand \bibnamefont  [1]{#1}%
\providecommand \bibfnamefont [1]{#1}%
\providecommand \citenamefont [1]{#1}%
\providecommand \href@noop [0]{\@secondoftwo}%
\providecommand \href [0]{\begingroup \@sanitize@url \@href}%
\providecommand \@href[1]{\@@startlink{#1}\@@href}%
\providecommand \@@href[1]{\endgroup#1\@@endlink}%
\providecommand \@sanitize@url [0]{\catcode `\\12\catcode `\$12\catcode `\&12\catcode `\#12\catcode `\^12\catcode `\_12\catcode `\%12\relax}%
\providecommand \@@startlink[1]{}%
\providecommand \@@endlink[0]{}%
\providecommand \url  [0]{\begingroup\@sanitize@url \@url }%
\providecommand \@url [1]{\endgroup\@href {#1}{\urlprefix }}%
\providecommand \urlprefix  [0]{URL }%
\providecommand \Eprint [0]{\href }%
\providecommand \doibase [0]{http://dx.doi.org/}%
\providecommand \selectlanguage [0]{\@gobble}%
\providecommand \bibinfo  [0]{\@secondoftwo}%
\providecommand \bibfield  [0]{\@secondoftwo}%
\providecommand \translation [1]{[#1]}%
\providecommand \BibitemOpen [0]{}%
\providecommand \bibitemStop [0]{}%
\providecommand \bibitemNoStop [0]{.\EOS\space}%
\providecommand \EOS [0]{\spacefactor3000\relax}%
\providecommand \BibitemShut  [1]{\csname bibitem#1\endcsname}%
\let\auto@bib@innerbib\@empty
\bibitem [{\citenamefont {Aleiner}\ \emph {et~al.}(2016)\citenamefont {Aleiner}, \citenamefont {Faoro},\ and\ \citenamefont {Ioffe}}]{aleiner16microscopic}%
  \BibitemOpen
  \bibfield  {author} {\bibinfo {author} {\bibfnamefont {Igor~L.}\ \bibnamefont {Aleiner}}, \bibinfo {author} {\bibfnamefont {Lara}\ \bibnamefont {Faoro}}, \ and\ \bibinfo {author} {\bibfnamefont {Lev~B.}\ \bibnamefont {Ioffe}},\ }\bibfield  {title} {\enquote {\bibinfo {title} {Microscopic model of quantum butterfly effect: Out-of-time-order correlators and traveling combustion waves},}\ }\href {https://arxiv.org/abs/1609.01251} {\bibfield  {journal} {\bibinfo  {journal} {Ann. Phys.}\ }\textbf {\bibinfo {volume} {375}},\ \bibinfo {pages} {378--406} (\bibinfo {year} {2016})}\BibitemShut {NoStop}%
\bibitem [{\citenamefont {Roberts}\ \emph {et~al.}(2015)\citenamefont {Roberts}, \citenamefont {Stanford},\ and\ \citenamefont {Susskind}}]{roberts15localized}%
  \BibitemOpen
  \bibfield  {author} {\bibinfo {author} {\bibfnamefont {Daniel~A.}\ \bibnamefont {Roberts}}, \bibinfo {author} {\bibfnamefont {Douglas}\ \bibnamefont {Stanford}}, \ and\ \bibinfo {author} {\bibfnamefont {Leonard}\ \bibnamefont {Susskind}},\ }\bibfield  {title} {\enquote {\bibinfo {title} {Localized shocks},}\ }\href {https://arxiv.org/abs/1409.8180} {\bibfield  {journal} {\bibinfo  {journal} {J. High Energy Phys.}\ }\textbf {\bibinfo {volume} {2015}},\ \bibinfo {pages} {51} (\bibinfo {year} {2015})}\BibitemShut {NoStop}%
\bibitem [{\citenamefont {Mi}\ \emph {et~al.}(2021)\citenamefont {Mi}, \citenamefont {Roushan}, \citenamefont {Quintana}, \citenamefont {Mandra}, \citenamefont {Marshall}, \citenamefont {Neill}, \citenamefont {Arute}, \citenamefont {Arya}, \citenamefont {Atalaya}, \citenamefont {Babbush}, \citenamefont {Bardin}, \citenamefont {Barends}, \citenamefont {Bengtsson}, \citenamefont {Boixo}, \citenamefont {Bourassa}, \citenamefont {Broughton}, \citenamefont {Buckley}, \citenamefont {Buell}, \citenamefont {Burkett}, \citenamefont {Bushnell}, \citenamefont {Chen}, \citenamefont {Chiaro}, \citenamefont {Collins}, \citenamefont {Courtney}, \citenamefont {Demura}, \citenamefont {Derk}, \citenamefont {Dunsworth}, \citenamefont {Eppens}, \citenamefont {Erickson}, \citenamefont {Farhi}, \citenamefont {Fowler}, \citenamefont {Foxen}, \citenamefont {Gidney}, \citenamefont {Giustina}, \citenamefont {Gross}, \citenamefont {Harrigan}, \citenamefont {Harrington}, \citenamefont {Hilton}, \citenamefont {Ho}, \citenamefont {Hong},
  \citenamefont {Huang}, \citenamefont {Huggins}, \citenamefont {Ioffe}, \citenamefont {Isakov}, \citenamefont {Jeffrey}, \citenamefont {Jiang}, \citenamefont {Jones}, \citenamefont {Kafri}, \citenamefont {Kelly}, \citenamefont {Kim}, \citenamefont {Kitaev}, \citenamefont {Klimov}, \citenamefont {Korotkov}, \citenamefont {Kostritsa}, \citenamefont {Landhuis}, \citenamefont {Laptev}, \citenamefont {Lucero}, \citenamefont {Martin}, \citenamefont {McClean}, \citenamefont {McCourt}, \citenamefont {McEwen}, \citenamefont {Megrant}, \citenamefont {Miao}, \citenamefont {Mohseni}, \citenamefont {Mruczkiewicz}, \citenamefont {Mutus}, \citenamefont {Naaman}, \citenamefont {Neeley}, \citenamefont {Newman}, \citenamefont {Niu}, \citenamefont {O'Brien}, \citenamefont {Opremcak}, \citenamefont {Ostby}, \citenamefont {Pato}, \citenamefont {Petukhov}, \citenamefont {Redd}, \citenamefont {Rubin}, \citenamefont {Sank}, \citenamefont {Satzinger}, \citenamefont {Shvarts}, \citenamefont {Strain}, \citenamefont {Szalay},
  \citenamefont {Trevithick}, \citenamefont {Villalonga}, \citenamefont {White}, \citenamefont {Yao}, \citenamefont {Yeh}, \citenamefont {Zalcman}, \citenamefont {Neven}, \citenamefont {Aleiner}, \citenamefont {Kechedzhi}, \citenamefont {Smelyanskiy},\ and\ \citenamefont {Chen}}]{mi21information}%
  \BibitemOpen
  \bibfield  {author} {\bibinfo {author} {\bibfnamefont {Xiao}\ \bibnamefont {Mi}}, \bibinfo {author} {\bibfnamefont {Pedram}\ \bibnamefont {Roushan}}, \bibinfo {author} {\bibfnamefont {Chris}\ \bibnamefont {Quintana}}, \bibinfo {author} {\bibfnamefont {Salvatore}\ \bibnamefont {Mandra}}, \bibinfo {author} {\bibfnamefont {Jeffrey}\ \bibnamefont {Marshall}}, \bibinfo {author} {\bibfnamefont {Charles}\ \bibnamefont {Neill}}, \bibinfo {author} {\bibfnamefont {Frank}\ \bibnamefont {Arute}}, \bibinfo {author} {\bibfnamefont {Kunal}\ \bibnamefont {Arya}}, \bibinfo {author} {\bibfnamefont {Juan}\ \bibnamefont {Atalaya}}, \bibinfo {author} {\bibfnamefont {Ryan}\ \bibnamefont {Babbush}}, \bibinfo {author} {\bibfnamefont {Joseph~C.}\ \bibnamefont {Bardin}}, \bibinfo {author} {\bibfnamefont {Rami}\ \bibnamefont {Barends}}, \bibinfo {author} {\bibfnamefont {Andreas}\ \bibnamefont {Bengtsson}}, \bibinfo {author} {\bibfnamefont {Sergio}\ \bibnamefont {Boixo}}, \bibinfo {author} {\bibfnamefont {Alexandre}\ \bibnamefont
  {Bourassa}}, \bibinfo {author} {\bibfnamefont {Michael}\ \bibnamefont {Broughton}}, \bibinfo {author} {\bibfnamefont {Bob~B.}\ \bibnamefont {Buckley}}, \bibinfo {author} {\bibfnamefont {David~A.}\ \bibnamefont {Buell}}, \bibinfo {author} {\bibfnamefont {Brian}\ \bibnamefont {Burkett}}, \bibinfo {author} {\bibfnamefont {Nicholas}\ \bibnamefont {Bushnell}}, \bibinfo {author} {\bibfnamefont {Zijun}\ \bibnamefont {Chen}}, \bibinfo {author} {\bibfnamefont {Benjamin}\ \bibnamefont {Chiaro}}, \bibinfo {author} {\bibfnamefont {Roberto}\ \bibnamefont {Collins}}, \bibinfo {author} {\bibfnamefont {William}\ \bibnamefont {Courtney}}, \bibinfo {author} {\bibfnamefont {Sean}\ \bibnamefont {Demura}}, \bibinfo {author} {\bibfnamefont {Alan~R.}\ \bibnamefont {Derk}}, \bibinfo {author} {\bibfnamefont {Andrew}\ \bibnamefont {Dunsworth}}, \bibinfo {author} {\bibfnamefont {Daniel}\ \bibnamefont {Eppens}}, \bibinfo {author} {\bibfnamefont {Catherine}\ \bibnamefont {Erickson}}, \bibinfo {author} {\bibfnamefont {Edward}\
  \bibnamefont {Farhi}}, \bibinfo {author} {\bibfnamefont {Austin~G.}\ \bibnamefont {Fowler}}, \bibinfo {author} {\bibfnamefont {Brooks}\ \bibnamefont {Foxen}}, \bibinfo {author} {\bibfnamefont {Craig}\ \bibnamefont {Gidney}}, \bibinfo {author} {\bibfnamefont {Marissa}\ \bibnamefont {Giustina}}, \bibinfo {author} {\bibfnamefont {Jonathan~A.}\ \bibnamefont {Gross}}, \bibinfo {author} {\bibfnamefont {Matthew~P.}\ \bibnamefont {Harrigan}}, \bibinfo {author} {\bibfnamefont {Sean~D.}\ \bibnamefont {Harrington}}, \bibinfo {author} {\bibfnamefont {Jeremy}\ \bibnamefont {Hilton}}, \bibinfo {author} {\bibfnamefont {Alan}\ \bibnamefont {Ho}}, \bibinfo {author} {\bibfnamefont {Sabrina}\ \bibnamefont {Hong}}, \bibinfo {author} {\bibfnamefont {Trent}\ \bibnamefont {Huang}}, \bibinfo {author} {\bibfnamefont {William~J.}\ \bibnamefont {Huggins}}, \bibinfo {author} {\bibfnamefont {L.~B.}\ \bibnamefont {Ioffe}}, \bibinfo {author} {\bibfnamefont {Sergei~V.}\ \bibnamefont {Isakov}}, \bibinfo {author} {\bibfnamefont {Evan}\
  \bibnamefont {Jeffrey}}, \bibinfo {author} {\bibfnamefont {Zhang}\ \bibnamefont {Jiang}}, \bibinfo {author} {\bibfnamefont {Cody}\ \bibnamefont {Jones}}, \bibinfo {author} {\bibfnamefont {Dvir}\ \bibnamefont {Kafri}}, \bibinfo {author} {\bibfnamefont {Julian}\ \bibnamefont {Kelly}}, \bibinfo {author} {\bibfnamefont {Seon}\ \bibnamefont {Kim}}, \bibinfo {author} {\bibfnamefont {Alexei}\ \bibnamefont {Kitaev}}, \bibinfo {author} {\bibfnamefont {Paul~V.}\ \bibnamefont {Klimov}}, \bibinfo {author} {\bibfnamefont {Alexander~N.}\ \bibnamefont {Korotkov}}, \bibinfo {author} {\bibfnamefont {Fedor}\ \bibnamefont {Kostritsa}}, \bibinfo {author} {\bibfnamefont {David}\ \bibnamefont {Landhuis}}, \bibinfo {author} {\bibfnamefont {Pavel}\ \bibnamefont {Laptev}}, \bibinfo {author} {\bibfnamefont {Erik}\ \bibnamefont {Lucero}}, \bibinfo {author} {\bibfnamefont {Orion}\ \bibnamefont {Martin}}, \bibinfo {author} {\bibfnamefont {Jarrod~R.}\ \bibnamefont {McClean}}, \bibinfo {author} {\bibfnamefont {Trevor}\ \bibnamefont
  {McCourt}}, \bibinfo {author} {\bibfnamefont {Matt}\ \bibnamefont {McEwen}}, \bibinfo {author} {\bibfnamefont {Anthony}\ \bibnamefont {Megrant}}, \bibinfo {author} {\bibfnamefont {Kevin~C.}\ \bibnamefont {Miao}}, \bibinfo {author} {\bibfnamefont {Masoud}\ \bibnamefont {Mohseni}}, \bibinfo {author} {\bibfnamefont {Wojciech}\ \bibnamefont {Mruczkiewicz}}, \bibinfo {author} {\bibfnamefont {Josh}\ \bibnamefont {Mutus}}, \bibinfo {author} {\bibfnamefont {Ofer}\ \bibnamefont {Naaman}}, \bibinfo {author} {\bibfnamefont {Matthew}\ \bibnamefont {Neeley}}, \bibinfo {author} {\bibfnamefont {Michael}\ \bibnamefont {Newman}}, \bibinfo {author} {\bibfnamefont {Murphy~Yuezhen}\ \bibnamefont {Niu}}, \bibinfo {author} {\bibfnamefont {Thomas~E.}\ \bibnamefont {O'Brien}}, \bibinfo {author} {\bibfnamefont {Alex}\ \bibnamefont {Opremcak}}, \bibinfo {author} {\bibfnamefont {Eric}\ \bibnamefont {Ostby}}, \bibinfo {author} {\bibfnamefont {Balint}\ \bibnamefont {Pato}}, \bibinfo {author} {\bibfnamefont {Andre}\ \bibnamefont
  {Petukhov}}, \bibinfo {author} {\bibfnamefont {Nicholas}\ \bibnamefont {Redd}}, \bibinfo {author} {\bibfnamefont {Nicholas~C.}\ \bibnamefont {Rubin}}, \bibinfo {author} {\bibfnamefont {Daniel}\ \bibnamefont {Sank}}, \bibinfo {author} {\bibfnamefont {Kevin~J.}\ \bibnamefont {Satzinger}}, \bibinfo {author} {\bibfnamefont {Vladimir}\ \bibnamefont {Shvarts}}, \bibinfo {author} {\bibfnamefont {Doug}\ \bibnamefont {Strain}}, \bibinfo {author} {\bibfnamefont {Marco}\ \bibnamefont {Szalay}}, \bibinfo {author} {\bibfnamefont {Matthew~D.}\ \bibnamefont {Trevithick}}, \bibinfo {author} {\bibfnamefont {Benjamin}\ \bibnamefont {Villalonga}}, \bibinfo {author} {\bibfnamefont {Theodore}\ \bibnamefont {White}}, \bibinfo {author} {\bibfnamefont {Z.~Jamie}\ \bibnamefont {Yao}}, \bibinfo {author} {\bibfnamefont {Ping}\ \bibnamefont {Yeh}}, \bibinfo {author} {\bibfnamefont {Adam}\ \bibnamefont {Zalcman}}, \bibinfo {author} {\bibfnamefont {Hartmut}\ \bibnamefont {Neven}}, \bibinfo {author} {\bibfnamefont {Igor}\ \bibnamefont
  {Aleiner}}, \bibinfo {author} {\bibfnamefont {Kostyantyn}\ \bibnamefont {Kechedzhi}}, \bibinfo {author} {\bibfnamefont {Vadim}\ \bibnamefont {Smelyanskiy}}, \ and\ \bibinfo {author} {\bibfnamefont {Yu}~\bibnamefont {Chen}},\ }\bibfield  {title} {\enquote {\bibinfo {title} {Information scrambling in computationally complex quantum circuits},}\ }\href {https://www.science.org/doi/10.1126/science.abg5029} {\bibfield  {journal} {\bibinfo  {journal} {Science}\ }\textbf {\bibinfo {volume} {374}},\ \bibinfo {pages} {1479} (\bibinfo {year} {2021})}\BibitemShut {NoStop}%
\bibitem [{\citenamefont {AI}\ and\ \citenamefont {Collaborators}(2025)}]{google25constructive}%
  \BibitemOpen
  \bibfield  {author} {\bibinfo {author} {\bibfnamefont {Google~Quantum}\ \bibnamefont {AI}}\ and\ \bibinfo {author} {\bibnamefont {Collaborators}},\ }\bibfield  {title} {\enquote {\bibinfo {title} {Constructive interference at the edge of quantum ergodic dynamics},}\ }\href {https://arxiv.org/pdf/2506.10191} {\bibfield  {journal} {\bibinfo  {journal} {arXiv:2506.10191}\ } (\bibinfo {year} {2025})}\BibitemShut {NoStop}%
\bibitem [{\citenamefont {Novikov}\ \emph {et~al.}(2025)\citenamefont {Novikov}, \citenamefont {Vũ}, \citenamefont {Eisenberger}, \citenamefont {Dupont}, \citenamefont {Huang}, \citenamefont {Wagner}, \citenamefont {Shirobokov}, \citenamefont {Kozlovskii}, \citenamefont {Ruiz}, \citenamefont {Mehrabian}, \citenamefont {Kumar}, \citenamefont {See}, \citenamefont {Chaudhuri}, \citenamefont {Holland}, \citenamefont {Davies}, \citenamefont {Nowozin}, \citenamefont {Kohli},\ and\ \citenamefont {Balog}}]{novikov2025alphaevolvecodingagentscientific}%
  \BibitemOpen
  \bibfield  {author} {\bibinfo {author} {\bibfnamefont {Alexander}\ \bibnamefont {Novikov}}, \bibinfo {author} {\bibfnamefont {Ngân}\ \bibnamefont {Vũ}}, \bibinfo {author} {\bibfnamefont {Marvin}\ \bibnamefont {Eisenberger}}, \bibinfo {author} {\bibfnamefont {Emilien}\ \bibnamefont {Dupont}}, \bibinfo {author} {\bibfnamefont {Po-Sen}\ \bibnamefont {Huang}}, \bibinfo {author} {\bibfnamefont {Adam~Zsolt}\ \bibnamefont {Wagner}}, \bibinfo {author} {\bibfnamefont {Sergey}\ \bibnamefont {Shirobokov}}, \bibinfo {author} {\bibfnamefont {Borislav}\ \bibnamefont {Kozlovskii}}, \bibinfo {author} {\bibfnamefont {Francisco J.~R.}\ \bibnamefont {Ruiz}}, \bibinfo {author} {\bibfnamefont {Abbas}\ \bibnamefont {Mehrabian}}, \bibinfo {author} {\bibfnamefont {M.~Pawan}\ \bibnamefont {Kumar}}, \bibinfo {author} {\bibfnamefont {Abigail}\ \bibnamefont {See}}, \bibinfo {author} {\bibfnamefont {Swarat}\ \bibnamefont {Chaudhuri}}, \bibinfo {author} {\bibfnamefont {George}\ \bibnamefont {Holland}}, \bibinfo {author} {\bibfnamefont
  {Alex}\ \bibnamefont {Davies}}, \bibinfo {author} {\bibfnamefont {Sebastian}\ \bibnamefont {Nowozin}}, \bibinfo {author} {\bibfnamefont {Pushmeet}\ \bibnamefont {Kohli}}, \ and\ \bibinfo {author} {\bibfnamefont {Matej}\ \bibnamefont {Balog}},\ }\href {https://arxiv.org/abs/2506.13131} {\enquote {\bibinfo {title} {Alphaevolve: A coding agent for scientific and algorithmic discovery},}\ } (\bibinfo {year} {2025}),\ \Eprint {http://arxiv.org/abs/2506.13131} {arXiv:2506.13131 [cs.AI]} \BibitemShut {NoStop}%
\bibitem [{\citenamefont {Levitt}(2008)}]{levitt2008spin}%
  \BibitemOpen
  \bibfield  {author} {\bibinfo {author} {\bibfnamefont {Malcolm~H}\ \bibnamefont {Levitt}},\ }\href@noop {} {\emph {\bibinfo {title} {Spin dynamics: basics of nuclear magnetic resonance}}}\ (\bibinfo  {publisher} {John Wiley \& Sons},\ \bibinfo {year} {2008})\BibitemShut {NoStop}%
\bibitem [{\citenamefont {Cho}\ \emph {et~al.}(2005)\citenamefont {Cho}, \citenamefont {Ladd}, \citenamefont {Baugh}, \citenamefont {Cory},\ and\ \citenamefont {Ramanathan}}]{cho2005multispin}%
  \BibitemOpen
  \bibfield  {author} {\bibinfo {author} {\bibfnamefont {Hyungjoon}\ \bibnamefont {Cho}}, \bibinfo {author} {\bibfnamefont {Thaddeus~D}\ \bibnamefont {Ladd}}, \bibinfo {author} {\bibfnamefont {Jonathan}\ \bibnamefont {Baugh}}, \bibinfo {author} {\bibfnamefont {David~G}\ \bibnamefont {Cory}}, \ and\ \bibinfo {author} {\bibfnamefont {Chandrasekhar}\ \bibnamefont {Ramanathan}},\ }\bibfield  {title} {\enquote {\bibinfo {title} {Multispin dynamics of the solid-state nmr free induction decay},}\ }\href {https://arxiv.org/pdf/cond-mat/0501578} {\bibfield  {journal} {\bibinfo  {journal} {Physical Review B—Condensed Matter and Materials Physics}\ }\textbf {\bibinfo {volume} {72}},\ \bibinfo {pages} {054427} (\bibinfo {year} {2005})}\BibitemShut {NoStop}%
\bibitem [{\citenamefont {Butler}\ \emph {et~al.}(2009)\citenamefont {Butler}, \citenamefont {Dumez},\ and\ \citenamefont {Emsley}}]{butler2009dynamics}%
  \BibitemOpen
  \bibfield  {author} {\bibinfo {author} {\bibfnamefont {Mark~C}\ \bibnamefont {Butler}}, \bibinfo {author} {\bibfnamefont {Jean-Nicolas}\ \bibnamefont {Dumez}}, \ and\ \bibinfo {author} {\bibfnamefont {Lyndon}\ \bibnamefont {Emsley}},\ }\bibfield  {title} {\enquote {\bibinfo {title} {Dynamics of large nuclear-spin systems from low-order correlations in liouville space},}\ }\href {https://www.sciencedirect.com/science/article/abs/pii/S0009261409008239?via%3Dihub} {\bibfield  {journal} {\bibinfo  {journal} {Chemical Physics Letters}\ }\textbf {\bibinfo {volume} {477}},\ \bibinfo {pages} {377--381} (\bibinfo {year} {2009})}\BibitemShut {NoStop}%
\bibitem [{\citenamefont {Dumez}\ \emph {et~al.}(2012)\citenamefont {Dumez}, \citenamefont {Halse}, \citenamefont {Butler},\ and\ \citenamefont {Emsley}}]{dumez2012first}%
  \BibitemOpen
  \bibfield  {author} {\bibinfo {author} {\bibfnamefont {Jean-Nicolas}\ \bibnamefont {Dumez}}, \bibinfo {author} {\bibfnamefont {Meghan~E}\ \bibnamefont {Halse}}, \bibinfo {author} {\bibfnamefont {Mark~C}\ \bibnamefont {Butler}}, \ and\ \bibinfo {author} {\bibfnamefont {Lyndon}\ \bibnamefont {Emsley}},\ }\bibfield  {title} {\enquote {\bibinfo {title} {A first-principles description of proton-driven spin diffusion},}\ }\href {https://pubs.rsc.org/en/content/articlehtml/2012/cp/c1cp22662b} {\bibfield  {journal} {\bibinfo  {journal} {Physical Chemistry Chemical Physics}\ }\textbf {\bibinfo {volume} {14}},\ \bibinfo {pages} {86--89} (\bibinfo {year} {2012})}\BibitemShut {NoStop}%
\bibitem [{\citenamefont {Baum}\ \emph {et~al.}(1985{\natexlab{a}})\citenamefont {Baum}, \citenamefont {Munowitz}, \citenamefont {Garroway},\ and\ \citenamefont {Pines}}]{baum_multiplequantum_1985}%
  \BibitemOpen
  \bibfield  {author} {\bibinfo {author} {\bibfnamefont {J.}~\bibnamefont {Baum}}, \bibinfo {author} {\bibfnamefont {M.}~\bibnamefont {Munowitz}}, \bibinfo {author} {\bibfnamefont {A.~N.}\ \bibnamefont {Garroway}}, \ and\ \bibinfo {author} {\bibfnamefont {A.}~\bibnamefont {Pines}},\ }\bibfield  {title} {\enquote {\bibinfo {title} {Multiple-quantum dynamics in solid state {{NMR}}},}\ }\href {\doibase 10.1063/1.449344} {\bibfield  {journal} {\bibinfo  {journal} {The Journal of Chemical Physics}\ }\textbf {\bibinfo {volume} {83}},\ \bibinfo {pages} {2015--2025} (\bibinfo {year} {1985}{\natexlab{a}})}\BibitemShut {NoStop}%
\bibitem [{\citenamefont {Gullion}\ and\ \citenamefont {Schaefer}(1989)}]{gullion1989rotational}%
  \BibitemOpen
  \bibfield  {author} {\bibinfo {author} {\bibfnamefont {Terry}\ \bibnamefont {Gullion}}\ and\ \bibinfo {author} {\bibfnamefont {Jacob}\ \bibnamefont {Schaefer}},\ }\bibfield  {title} {\enquote {\bibinfo {title} {Rotational-echo double-resonance nmr},}\ }\href {https://www.sciencedirect.com/science/article/abs/pii/0022236489902801} {\bibfield  {journal} {\bibinfo  {journal} {Journal of Magnetic Resonance (1969)}\ }\textbf {\bibinfo {volume} {81}},\ \bibinfo {pages} {196--200} (\bibinfo {year} {1989})}\BibitemShut {NoStop}%
\bibitem [{\citenamefont {Bennett}\ \emph {et~al.}(1998)\citenamefont {Bennett}, \citenamefont {Rienstra}, \citenamefont {Griffiths}, \citenamefont {Zhen}, \citenamefont {Lansbury~Jr},\ and\ \citenamefont {Griffin}}]{bennett1998homonuclear}%
  \BibitemOpen
  \bibfield  {author} {\bibinfo {author} {\bibfnamefont {Andrew~E}\ \bibnamefont {Bennett}}, \bibinfo {author} {\bibfnamefont {Chad~M}\ \bibnamefont {Rienstra}}, \bibinfo {author} {\bibfnamefont {Janet~M}\ \bibnamefont {Griffiths}}, \bibinfo {author} {\bibfnamefont {Weiguo}\ \bibnamefont {Zhen}}, \bibinfo {author} {\bibfnamefont {Peter~T}\ \bibnamefont {Lansbury~Jr}}, \ and\ \bibinfo {author} {\bibfnamefont {Robert~G}\ \bibnamefont {Griffin}},\ }\bibfield  {title} {\enquote {\bibinfo {title} {Homonuclear radio frequency-driven recoupling in rotating solids},}\ }\href {https://www.academia.edu/download/44888467/bennett.jchemphys108.pdf} {\bibfield  {journal} {\bibinfo  {journal} {The Journal of chemical physics}\ }\textbf {\bibinfo {volume} {108}},\ \bibinfo {pages} {9463--9479} (\bibinfo {year} {1998})}\BibitemShut {NoStop}%
\bibitem [{\citenamefont {Colombo}\ \emph {et~al.}(1988)\citenamefont {Colombo}, \citenamefont {Meier},\ and\ \citenamefont {Ernst}}]{COLOMBO1988189}%
  \BibitemOpen
  \bibfield  {author} {\bibinfo {author} {\bibfnamefont {M.G.}\ \bibnamefont {Colombo}}, \bibinfo {author} {\bibfnamefont {B.H.}\ \bibnamefont {Meier}}, \ and\ \bibinfo {author} {\bibfnamefont {R.R.}\ \bibnamefont {Ernst}},\ }\bibfield  {title} {\enquote {\bibinfo {title} {Rotor-driven spin diffusion in natural-abundance 13c spin systems},}\ }\href {\doibase https://doi.org/10.1016/0009-2614(88)87429-3} {\bibfield  {journal} {\bibinfo  {journal} {Chemical Physics Letters}\ }\textbf {\bibinfo {volume} {146}},\ \bibinfo {pages} {189--196} (\bibinfo {year} {1988})}\BibitemShut {NoStop}%
\bibitem [{\citenamefont {Grommek}\ \emph {et~al.}(2006)\citenamefont {Grommek}, \citenamefont {Meier},\ and\ \citenamefont {Ernst}}]{GROMMEK2006404}%
  \BibitemOpen
  \bibfield  {author} {\bibinfo {author} {\bibfnamefont {Andreas}\ \bibnamefont {Grommek}}, \bibinfo {author} {\bibfnamefont {Beat~H.}\ \bibnamefont {Meier}}, \ and\ \bibinfo {author} {\bibfnamefont {Matthias}\ \bibnamefont {Ernst}},\ }\bibfield  {title} {\enquote {\bibinfo {title} {Distance information from proton-driven spin diffusion under mas},}\ }\href {\doibase https://doi.org/10.1016/j.cplett.2006.07.005} {\bibfield  {journal} {\bibinfo  {journal} {Chemical Physics Letters}\ }\textbf {\bibinfo {volume} {427}},\ \bibinfo {pages} {404--409} (\bibinfo {year} {2006})}\BibitemShut {NoStop}%
\bibitem [{\citenamefont {Raleigh}\ \emph {et~al.}(1989)\citenamefont {Raleigh}, \citenamefont {Creuzet}, \citenamefont {Das~Gupta}, \citenamefont {Levitt},\ and\ \citenamefont {Griffin}}]{Raleigh1989}%
  \BibitemOpen
  \bibfield  {author} {\bibinfo {author} {\bibfnamefont {D.~P.}\ \bibnamefont {Raleigh}}, \bibinfo {author} {\bibfnamefont {F.}~\bibnamefont {Creuzet}}, \bibinfo {author} {\bibfnamefont {S.~K.}\ \bibnamefont {Das~Gupta}}, \bibinfo {author} {\bibfnamefont {M.~H.}\ \bibnamefont {Levitt}}, \ and\ \bibinfo {author} {\bibfnamefont {Robert~G.}\ \bibnamefont {Griffin}},\ }\bibfield  {title} {\enquote {\bibinfo {title} {Measurement of internuclear distances in polycrystalline solids. rotationally enhanced transfer of nuclear spin magnetization},}\ }\href {\doibase 10.1021/ja00194a057} {\bibfield  {journal} {\bibinfo  {journal} {Journal of the American Chemical Society}\ }\textbf {\bibinfo {volume} {111}},\ \bibinfo {pages} {4502--4503} (\bibinfo {year} {1989})},\ \Eprint {http://arxiv.org/abs/https://doi.org/10.1021/ja00194a057} {https://doi.org/10.1021/ja00194a057} \BibitemShut {NoStop}%
\bibitem [{\citenamefont {Creuzet}\ \emph {et~al.}(1991)\citenamefont {Creuzet}, \citenamefont {McDermott}, \citenamefont {Gebhard}, \citenamefont {van~der Hoef}, \citenamefont {Spijker-Assink}, \citenamefont {Herzfeld}, \citenamefont {Lugtenburg}, \citenamefont {Levitt},\ and\ \citenamefont {Griffin}}]{Creuzet1991}%
  \BibitemOpen
  \bibfield  {author} {\bibinfo {author} {\bibfnamefont {F.}~\bibnamefont {Creuzet}}, \bibinfo {author} {\bibfnamefont {A.}~\bibnamefont {McDermott}}, \bibinfo {author} {\bibfnamefont {R.}~\bibnamefont {Gebhard}}, \bibinfo {author} {\bibfnamefont {K.}~\bibnamefont {van~der Hoef}}, \bibinfo {author} {\bibfnamefont {M.~B.}\ \bibnamefont {Spijker-Assink}}, \bibinfo {author} {\bibfnamefont {J.}~\bibnamefont {Herzfeld}}, \bibinfo {author} {\bibfnamefont {J.}~\bibnamefont {Lugtenburg}}, \bibinfo {author} {\bibfnamefont {M.~H.}\ \bibnamefont {Levitt}}, \ and\ \bibinfo {author} {\bibfnamefont {R.~G.}\ \bibnamefont {Griffin}},\ }\bibfield  {title} {\enquote {\bibinfo {title} {Determination of membrane protein structure by rotational resonance nmr: Bacteriorhodopsin},}\ }\href {\doibase 10.1126/science.1990439} {\bibfield  {journal} {\bibinfo  {journal} {Science}\ }\textbf {\bibinfo {volume} {251}},\ \bibinfo {pages} {783--786} (\bibinfo {year} {1991})},\ \Eprint
  {http://arxiv.org/abs/https://www.science.org/doi/pdf/10.1126/science.1990439} {https://www.science.org/doi/pdf/10.1126/science.1990439} \BibitemShut {NoStop}%
\bibitem [{\citenamefont {Warren}\ and\ \citenamefont {Pines}(1981)}]{Warren1981}%
  \BibitemOpen
  \bibfield  {author} {\bibinfo {author} {\bibfnamefont {W.~S.}\ \bibnamefont {Warren}}\ and\ \bibinfo {author} {\bibfnamefont {A.}~\bibnamefont {Pines}},\ }\bibfield  {title} {\enquote {\bibinfo {title} {Analogy of multiple-quantum nmr to isotopic spin labeling},}\ }\href {\doibase 10.1021/ja00397a001} {\bibfield  {journal} {\bibinfo  {journal} {Journal of the American Chemical Society}\ }\textbf {\bibinfo {volume} {103}},\ \bibinfo {pages} {1613--1617} (\bibinfo {year} {1981})},\ \Eprint {http://arxiv.org/abs/https://doi.org/10.1021/ja00397a001} {https://doi.org/10.1021/ja00397a001} \BibitemShut {NoStop}%
\bibitem [{\citenamefont {Polson}\ and\ \citenamefont {Burnell}(1995)}]{Polson1995}%
  \BibitemOpen
  \bibfield  {author} {\bibinfo {author} {\bibfnamefont {James~M.}\ \bibnamefont {Polson}}\ and\ \bibinfo {author} {\bibfnamefont {E.~Elliott}\ \bibnamefont {Burnell}},\ }\bibfield  {title} {\enquote {\bibinfo {title} {Conformational equilibrium and orientational ordering: 1h‐nuclear magnetic resonance of butane in a nematic liquid crystal},}\ }\href {\doibase 10.1063/1.470367} {\bibfield  {journal} {\bibinfo  {journal} {The Journal of Chemical Physics}\ }\textbf {\bibinfo {volume} {103}},\ \bibinfo {pages} {6891--6902} (\bibinfo {year} {1995})}\BibitemShut {NoStop}%
\bibitem [{\citenamefont {Petkova}\ \emph {et~al.}(2002)\citenamefont {Petkova}, \citenamefont {Ishii}, \citenamefont {Balbach}, \citenamefont {Antzutkin}, \citenamefont {Leapman}, \citenamefont {Delaglio},\ and\ \citenamefont {Tycko}}]{petkova2002structural}%
  \BibitemOpen
  \bibfield  {author} {\bibinfo {author} {\bibfnamefont {Aneta~T}\ \bibnamefont {Petkova}}, \bibinfo {author} {\bibfnamefont {Yoshitaka}\ \bibnamefont {Ishii}}, \bibinfo {author} {\bibfnamefont {John~J}\ \bibnamefont {Balbach}}, \bibinfo {author} {\bibfnamefont {Oleg~N}\ \bibnamefont {Antzutkin}}, \bibinfo {author} {\bibfnamefont {Richard~D}\ \bibnamefont {Leapman}}, \bibinfo {author} {\bibfnamefont {Frank}\ \bibnamefont {Delaglio}}, \ and\ \bibinfo {author} {\bibfnamefont {Robert}\ \bibnamefont {Tycko}},\ }\bibfield  {title} {\enquote {\bibinfo {title} {A structural model for alzheimer's $\beta$-amyloid fibrils based on experimental constraints from solid state nmr},}\ }\href {https://www.pnas.org/doi/pdf/10.1073/pnas.262663499} {\bibfield  {journal} {\bibinfo  {journal} {Proceedings of the National Academy of Sciences}\ }\textbf {\bibinfo {volume} {99}},\ \bibinfo {pages} {16742--16747} (\bibinfo {year} {2002})}\BibitemShut {NoStop}%
\bibitem [{\citenamefont {Wasmer}\ \emph {et~al.}(2008)\citenamefont {Wasmer}, \citenamefont {Lange}, \citenamefont {Van~Melckebeke}, \citenamefont {Siemer}, \citenamefont {Riek},\ and\ \citenamefont {Meier}}]{wasmer2008amyloid}%
  \BibitemOpen
  \bibfield  {author} {\bibinfo {author} {\bibfnamefont {Christian}\ \bibnamefont {Wasmer}}, \bibinfo {author} {\bibfnamefont {Adam}\ \bibnamefont {Lange}}, \bibinfo {author} {\bibfnamefont {H{\'e}l{\`e}ne}\ \bibnamefont {Van~Melckebeke}}, \bibinfo {author} {\bibfnamefont {Ansgar~B}\ \bibnamefont {Siemer}}, \bibinfo {author} {\bibfnamefont {Roland}\ \bibnamefont {Riek}}, \ and\ \bibinfo {author} {\bibfnamefont {Beat~H}\ \bibnamefont {Meier}},\ }\bibfield  {title} {\enquote {\bibinfo {title} {Amyloid fibrils of the het-s (218--289) prion form a $\beta$ solenoid with a triangular hydrophobic core},}\ }\href@noop {} {\bibfield  {journal} {\bibinfo  {journal} {Science}\ }\textbf {\bibinfo {volume} {319}},\ \bibinfo {pages} {1523--1526} (\bibinfo {year} {2008})}\BibitemShut {NoStop}%
\bibitem [{\citenamefont {{Medeiros-Silva}}\ \emph {et~al.}(2023)\citenamefont {{Medeiros-Silva}}, \citenamefont {Dregni}, \citenamefont {Somberg}, \citenamefont {Duan},\ and\ \citenamefont {Hong}}]{medeiros-silva_atomic_2023}%
  \BibitemOpen
  \bibfield  {author} {\bibinfo {author} {\bibfnamefont {Jo{\~a}o}\ \bibnamefont {{Medeiros-Silva}}}, \bibinfo {author} {\bibfnamefont {Aurelio~J.}\ \bibnamefont {Dregni}}, \bibinfo {author} {\bibfnamefont {Noah~H.}\ \bibnamefont {Somberg}}, \bibinfo {author} {\bibfnamefont {Pu}~\bibnamefont {Duan}}, \ and\ \bibinfo {author} {\bibfnamefont {Mei}\ \bibnamefont {Hong}},\ }\bibfield  {title} {\enquote {\bibinfo {title} {Atomic structure of the open {{SARS-CoV-2 E}} viroporin},}\ }\href {\doibase 10.1126/sciadv.adi9007} {\bibfield  {journal} {\bibinfo  {journal} {Science Advances}\ }\textbf {\bibinfo {volume} {9}},\ \bibinfo {pages} {eadi9007} (\bibinfo {year} {2023})}\BibitemShut {NoStop}%
\bibitem [{\citenamefont {Shcherbakov}\ \emph {et~al.}(2022)\citenamefont {Shcherbakov}, \citenamefont {{Medeiros-Silva}}, \citenamefont {Tran}, \citenamefont {Gelenter},\ and\ \citenamefont {Hong}}]{shcherbakov_angstroms_2022}%
  \BibitemOpen
  \bibfield  {author} {\bibinfo {author} {\bibfnamefont {Alexander~A.}\ \bibnamefont {Shcherbakov}}, \bibinfo {author} {\bibfnamefont {Jo{\~a}o}\ \bibnamefont {{Medeiros-Silva}}}, \bibinfo {author} {\bibfnamefont {Nhi}\ \bibnamefont {Tran}}, \bibinfo {author} {\bibfnamefont {Martin~D.}\ \bibnamefont {Gelenter}}, \ and\ \bibinfo {author} {\bibfnamefont {Mei}\ \bibnamefont {Hong}},\ }\bibfield  {title} {\enquote {\bibinfo {title} {From {{Angstroms}} to {{Nanometers}}: {{Measuring Interatomic Distances}} by {{Solid-State NMR}}},}\ }\href {\doibase 10.1021/acs.chemrev.1c00662} {\bibfield  {journal} {\bibinfo  {journal} {Chemical Reviews}\ }\textbf {\bibinfo {volume} {122}},\ \bibinfo {pages} {9848--9879} (\bibinfo {year} {2022})}\BibitemShut {NoStop}%
\bibitem [{\citenamefont {Sels}\ \emph {et~al.}(2020)\citenamefont {Sels}, \citenamefont {Dashti}, \citenamefont {Mora}, \citenamefont {Demler},\ and\ \citenamefont {Demler}}]{sels2020quantum}%
  \BibitemOpen
  \bibfield  {author} {\bibinfo {author} {\bibfnamefont {Dries}\ \bibnamefont {Sels}}, \bibinfo {author} {\bibfnamefont {Hesam}\ \bibnamefont {Dashti}}, \bibinfo {author} {\bibfnamefont {Samia}\ \bibnamefont {Mora}}, \bibinfo {author} {\bibfnamefont {Olga}\ \bibnamefont {Demler}}, \ and\ \bibinfo {author} {\bibfnamefont {Eugene}\ \bibnamefont {Demler}},\ }\bibfield  {title} {\enquote {\bibinfo {title} {Quantum approximate bayesian computation for nmr model inference},}\ }\href {https://www.nature.com/articles/s42256-020-0198-x} {\bibfield  {journal} {\bibinfo  {journal} {Nature machine intelligence}\ }\textbf {\bibinfo {volume} {2}},\ \bibinfo {pages} {396--402} (\bibinfo {year} {2020})}\BibitemShut {NoStop}%
\bibitem [{\citenamefont {O’Brien}\ \emph {et~al.}(2022)\citenamefont {O’Brien}, \citenamefont {Ioffe}, \citenamefont {Su}, \citenamefont {Fushman}, \citenamefont {Neven}, \citenamefont {Babbush},\ and\ \citenamefont {Smelyanskiy}}]{obrien2022quantum}%
  \BibitemOpen
  \bibfield  {author} {\bibinfo {author} {\bibfnamefont {Thomas~E}\ \bibnamefont {O’Brien}}, \bibinfo {author} {\bibfnamefont {Lev~B}\ \bibnamefont {Ioffe}}, \bibinfo {author} {\bibfnamefont {Yuan}\ \bibnamefont {Su}}, \bibinfo {author} {\bibfnamefont {David}\ \bibnamefont {Fushman}}, \bibinfo {author} {\bibfnamefont {Hartmut}\ \bibnamefont {Neven}}, \bibinfo {author} {\bibfnamefont {Ryan}\ \bibnamefont {Babbush}}, \ and\ \bibinfo {author} {\bibfnamefont {Vadim}\ \bibnamefont {Smelyanskiy}},\ }\bibfield  {title} {\enquote {\bibinfo {title} {Quantum computation of molecular structure using data from challenging-to-classically-simulate nuclear magnetic resonance experiments},}\ }\href {https://link.aps.org/pdf/10.1103/PRXQuantum.3.030345} {\bibfield  {journal} {\bibinfo  {journal} {PRX Quantum}\ }\textbf {\bibinfo {volume} {3}},\ \bibinfo {pages} {030345} (\bibinfo {year} {2022})}\BibitemShut {NoStop}%
\bibitem [{\citenamefont {Algaba}\ \emph {et~al.}(2022)\citenamefont {Algaba}, \citenamefont {Ponce-Martinez}, \citenamefont {Munuera-Javaloy}, \citenamefont {Pina-Canelles}, \citenamefont {Thapa}, \citenamefont {Taketani}, \citenamefont {Leib}, \citenamefont {De~Vega}, \citenamefont {Casanova},\ and\ \citenamefont {Heimonen}}]{algaba2022co}%
  \BibitemOpen
  \bibfield  {author} {\bibinfo {author} {\bibfnamefont {Manuel~G}\ \bibnamefont {Algaba}}, \bibinfo {author} {\bibfnamefont {Mario}\ \bibnamefont {Ponce-Martinez}}, \bibinfo {author} {\bibfnamefont {Carlos}\ \bibnamefont {Munuera-Javaloy}}, \bibinfo {author} {\bibfnamefont {Vicente}\ \bibnamefont {Pina-Canelles}}, \bibinfo {author} {\bibfnamefont {Manish~J}\ \bibnamefont {Thapa}}, \bibinfo {author} {\bibfnamefont {Bruno~G}\ \bibnamefont {Taketani}}, \bibinfo {author} {\bibfnamefont {Martin}\ \bibnamefont {Leib}}, \bibinfo {author} {\bibfnamefont {In{\'e}s}\ \bibnamefont {De~Vega}}, \bibinfo {author} {\bibfnamefont {Jorge}\ \bibnamefont {Casanova}}, \ and\ \bibinfo {author} {\bibfnamefont {Hermanni}\ \bibnamefont {Heimonen}},\ }\bibfield  {title} {\enquote {\bibinfo {title} {Co-design quantum simulation of nanoscale nmr},}\ }\href {https://journals.aps.org/prresearch/abstract/10.1103/PhysRevResearch.4.043089} {\bibfield  {journal} {\bibinfo  {journal} {Physical Review Research}\ }\textbf {\bibinfo {volume}
  {4}},\ \bibinfo {pages} {043089} (\bibinfo {year} {2022})}\BibitemShut {NoStop}%
\bibitem [{\citenamefont {Schuster}\ \emph {et~al.}(2023)\citenamefont {Schuster}, \citenamefont {Niu}, \citenamefont {Cotler}, \citenamefont {O'Brien}, \citenamefont {McClean},\ and\ \citenamefont {Mohseni}}]{schuster2023learning}%
  \BibitemOpen
  \bibfield  {author} {\bibinfo {author} {\bibfnamefont {Thomas}\ \bibnamefont {Schuster}}, \bibinfo {author} {\bibfnamefont {Murphy}\ \bibnamefont {Niu}}, \bibinfo {author} {\bibfnamefont {Jordan}\ \bibnamefont {Cotler}}, \bibinfo {author} {\bibfnamefont {Thomas}\ \bibnamefont {O'Brien}}, \bibinfo {author} {\bibfnamefont {Jarrod~R}\ \bibnamefont {McClean}}, \ and\ \bibinfo {author} {\bibfnamefont {Masoud}\ \bibnamefont {Mohseni}},\ }\bibfield  {title} {\enquote {\bibinfo {title} {Learning quantum systems via out-of-time-order correlators},}\ }\href {https://link.aps.org/pdf/10.1103/PhysRevResearch.5.043284} {\bibfield  {journal} {\bibinfo  {journal} {Physical Review Research}\ }\textbf {\bibinfo {volume} {5}},\ \bibinfo {pages} {043284} (\bibinfo {year} {2023})}\BibitemShut {NoStop}%
\bibitem [{\citenamefont {Seetharam}\ \emph {et~al.}(2023)\citenamefont {Seetharam}, \citenamefont {Biswas}, \citenamefont {Noel}, \citenamefont {Risinger}, \citenamefont {Zhu}, \citenamefont {Katz}, \citenamefont {Chattopadhyay}, \citenamefont {Cetina}, \citenamefont {Monroe}, \citenamefont {Demler} \emph {et~al.}}]{seetharam2023digital}%
  \BibitemOpen
  \bibfield  {author} {\bibinfo {author} {\bibfnamefont {Kushal}\ \bibnamefont {Seetharam}}, \bibinfo {author} {\bibfnamefont {Debopriyo}\ \bibnamefont {Biswas}}, \bibinfo {author} {\bibfnamefont {Crystal}\ \bibnamefont {Noel}}, \bibinfo {author} {\bibfnamefont {Andrew}\ \bibnamefont {Risinger}}, \bibinfo {author} {\bibfnamefont {Daiwei}\ \bibnamefont {Zhu}}, \bibinfo {author} {\bibfnamefont {Or}~\bibnamefont {Katz}}, \bibinfo {author} {\bibfnamefont {Sambuddha}\ \bibnamefont {Chattopadhyay}}, \bibinfo {author} {\bibfnamefont {Marko}\ \bibnamefont {Cetina}}, \bibinfo {author} {\bibfnamefont {Christopher}\ \bibnamefont {Monroe}}, \bibinfo {author} {\bibfnamefont {Eugene}\ \bibnamefont {Demler}},  \emph {et~al.},\ }\bibfield  {title} {\enquote {\bibinfo {title} {Digital quantum simulation of nmr experiments},}\ }\href {https://www.science.org/doi/pdf/10.1126/sciadv.adh2594} {\bibfield  {journal} {\bibinfo  {journal} {Science Advances}\ }\textbf {\bibinfo {volume} {9}},\ \bibinfo {pages} {eadh2594} (\bibinfo
  {year} {2023})}\BibitemShut {NoStop}%
\bibitem [{\citenamefont {Fratus}\ \emph {et~al.}(2025)\citenamefont {Fratus}, \citenamefont {Enenkel}, \citenamefont {Zanker}, \citenamefont {Reiner}, \citenamefont {Marthaler},\ and\ \citenamefont {Schmitteckert}}]{fratus2025can}%
  \BibitemOpen
  \bibfield  {author} {\bibinfo {author} {\bibfnamefont {Keith~R}\ \bibnamefont {Fratus}}, \bibinfo {author} {\bibfnamefont {Nicklas}\ \bibnamefont {Enenkel}}, \bibinfo {author} {\bibfnamefont {Sebastian}\ \bibnamefont {Zanker}}, \bibinfo {author} {\bibfnamefont {Jan-Michael}\ \bibnamefont {Reiner}}, \bibinfo {author} {\bibfnamefont {Michael}\ \bibnamefont {Marthaler}}, \ and\ \bibinfo {author} {\bibfnamefont {Peter}\ \bibnamefont {Schmitteckert}},\ }\bibfield  {title} {\enquote {\bibinfo {title} {Can a quantum computer simulate nuclear magnetic resonance spectra better than a classical one?}}\ }\href {https://arxiv.org/pdf/2508.06448} {\bibfield  {journal} {\bibinfo  {journal} {arXiv preprint arXiv:2508.06448}\ } (\bibinfo {year} {2025})}\BibitemShut {NoStop}%
\bibitem [{\citenamefont {Marthaler}\ \emph {et~al.}(2025)\citenamefont {Marthaler}, \citenamefont {Pinski}, \citenamefont {Stadler}, \citenamefont {Rybkin},\ and\ \citenamefont {Walt}}]{marthaler2025good}%
  \BibitemOpen
  \bibfield  {author} {\bibinfo {author} {\bibfnamefont {Michael}\ \bibnamefont {Marthaler}}, \bibinfo {author} {\bibfnamefont {Peter}\ \bibnamefont {Pinski}}, \bibinfo {author} {\bibfnamefont {Pascal}\ \bibnamefont {Stadler}}, \bibinfo {author} {\bibfnamefont {Vladimir}\ \bibnamefont {Rybkin}}, \ and\ \bibinfo {author} {\bibfnamefont {Marina}\ \bibnamefont {Walt}},\ }\bibfield  {title} {\enquote {\bibinfo {title} {What is a good use case for quantum computers?}}\ }\href {https://arxiv.org/abs/2506.15426} {\bibfield  {journal} {\bibinfo  {journal} {arXiv preprint arXiv:2506.15426}\ } (\bibinfo {year} {2025})}\BibitemShut {NoStop}%
\bibitem [{\citenamefont {Angrisani}\ \emph {et~al.}(2024)\citenamefont {Angrisani}, \citenamefont {Schmidhuber}, \citenamefont {Rudolph}, \citenamefont {Cerezo}, \citenamefont {Holmes},\ and\ \citenamefont {Huang}}]{angrisani2024classically}%
  \BibitemOpen
  \bibfield  {author} {\bibinfo {author} {\bibfnamefont {Armando}\ \bibnamefont {Angrisani}}, \bibinfo {author} {\bibfnamefont {Alexander}\ \bibnamefont {Schmidhuber}}, \bibinfo {author} {\bibfnamefont {Manuel~S}\ \bibnamefont {Rudolph}}, \bibinfo {author} {\bibfnamefont {M}~\bibnamefont {Cerezo}}, \bibinfo {author} {\bibfnamefont {Zo{\"e}}\ \bibnamefont {Holmes}}, \ and\ \bibinfo {author} {\bibfnamefont {Hsin-Yuan}\ \bibnamefont {Huang}},\ }\bibfield  {title} {\enquote {\bibinfo {title} {Classically estimating observables of noiseless quantum circuits},}\ }\href@noop {} {\bibfield  {journal} {\bibinfo  {journal} {arXiv preprint arXiv:2409.01706}\ } (\bibinfo {year} {2024})}\BibitemShut {NoStop}%
\bibitem [{\citenamefont {Kechedzhi}\ \emph {et~al.}(2024)\citenamefont {Kechedzhi}, \citenamefont {Isakov}, \citenamefont {Mandr{\`a}}, \citenamefont {Villalonga}, \citenamefont {Mi}, \citenamefont {Boixo},\ and\ \citenamefont {Smelyanskiy}}]{kechedzhi2024effective}%
  \BibitemOpen
  \bibfield  {author} {\bibinfo {author} {\bibfnamefont {Kostyantyn}\ \bibnamefont {Kechedzhi}}, \bibinfo {author} {\bibfnamefont {Sergei~V}\ \bibnamefont {Isakov}}, \bibinfo {author} {\bibfnamefont {Salvatore}\ \bibnamefont {Mandr{\`a}}}, \bibinfo {author} {\bibfnamefont {Benjamin}\ \bibnamefont {Villalonga}}, \bibinfo {author} {\bibfnamefont {Xiao}\ \bibnamefont {Mi}}, \bibinfo {author} {\bibfnamefont {Sergio}\ \bibnamefont {Boixo}}, \ and\ \bibinfo {author} {\bibfnamefont {Vadim}\ \bibnamefont {Smelyanskiy}},\ }\bibfield  {title} {\enquote {\bibinfo {title} {Effective quantum volume, fidelity and computational cost of noisy quantum processing experiments},}\ }\href {https://www.sciencedirect.com/science/article/am/pii/S0167739X23004569} {\bibfield  {journal} {\bibinfo  {journal} {Future Generation Computer Systems}\ }\textbf {\bibinfo {volume} {153}},\ \bibinfo {pages} {431--441} (\bibinfo {year} {2024})}\BibitemShut {NoStop}%
\bibitem [{\citenamefont {Schuster}\ \emph {et~al.}(2024)\citenamefont {Schuster}, \citenamefont {Yin}, \citenamefont {Gao},\ and\ \citenamefont {Yao}}]{schuster2024polynomial}%
  \BibitemOpen
  \bibfield  {author} {\bibinfo {author} {\bibfnamefont {Thomas}\ \bibnamefont {Schuster}}, \bibinfo {author} {\bibfnamefont {Chao}\ \bibnamefont {Yin}}, \bibinfo {author} {\bibfnamefont {Xun}\ \bibnamefont {Gao}}, \ and\ \bibinfo {author} {\bibfnamefont {Norman~Y}\ \bibnamefont {Yao}},\ }\bibfield  {title} {\enquote {\bibinfo {title} {A polynomial-time classical algorithm for noisy quantum circuits},}\ }\href {https://arxiv.org/pdf/2407.12768} {\bibfield  {journal} {\bibinfo  {journal} {arXiv preprint arXiv:2407.12768}\ } (\bibinfo {year} {2024})}\BibitemShut {NoStop}%
\bibitem [{\citenamefont {Li}\ \emph {et~al.}(2017)\citenamefont {Li}, \citenamefont {Fan}, \citenamefont {Wang}, \citenamefont {Ye}, \citenamefont {Zeng}, \citenamefont {Zhai}, \citenamefont {Peng},\ and\ \citenamefont {Du}}]{li2017measuring}%
  \BibitemOpen
  \bibfield  {author} {\bibinfo {author} {\bibfnamefont {Jun}\ \bibnamefont {Li}}, \bibinfo {author} {\bibfnamefont {Ruihua}\ \bibnamefont {Fan}}, \bibinfo {author} {\bibfnamefont {Hengyan}\ \bibnamefont {Wang}}, \bibinfo {author} {\bibfnamefont {Bingtian}\ \bibnamefont {Ye}}, \bibinfo {author} {\bibfnamefont {Bei}\ \bibnamefont {Zeng}}, \bibinfo {author} {\bibfnamefont {Hui}\ \bibnamefont {Zhai}}, \bibinfo {author} {\bibfnamefont {Xinhua}\ \bibnamefont {Peng}}, \ and\ \bibinfo {author} {\bibfnamefont {Jiangfeng}\ \bibnamefont {Du}},\ }\bibfield  {title} {\enquote {\bibinfo {title} {Measuring out-of-time-order correlators on a nuclear magnetic resonance quantum simulator},}\ }\href {https://journals.aps.org/prx/pdf/10.1103/PhysRevX.7.031011} {\bibfield  {journal} {\bibinfo  {journal} {Physical Review X}\ }\textbf {\bibinfo {volume} {7}},\ \bibinfo {pages} {031011} (\bibinfo {year} {2017})}\BibitemShut {NoStop}%
\bibitem [{\citenamefont {Field}\ and\ \citenamefont {Ramadan}(2003)}]{field2003multiple}%
  \BibitemOpen
  \bibfield  {author} {\bibinfo {author} {\bibfnamefont {Leslie~D}\ \bibnamefont {Field}}\ and\ \bibinfo {author} {\bibfnamefont {Saadallah~A}\ \bibnamefont {Ramadan}},\ }\bibfield  {title} {\enquote {\bibinfo {title} {Multiple quantum nmr spectra of toluene and p-bromotoluene partially aligned in a nematic phase},}\ }\href@noop {} {\bibfield  {journal} {\bibinfo  {journal} {Magnetic Resonance in Chemistry}\ }\textbf {\bibinfo {volume} {41}},\ \bibinfo {pages} {933--938} (\bibinfo {year} {2003})}\BibitemShut {NoStop}%
\bibitem [{\citenamefont {Childs}\ \emph {et~al.}(2021)\citenamefont {Childs}, \citenamefont {Su}, \citenamefont {Tran}, \citenamefont {Wiebe},\ and\ \citenamefont {Zhu}}]{childs2021theory}%
  \BibitemOpen
  \bibfield  {author} {\bibinfo {author} {\bibfnamefont {Andrew~M}\ \bibnamefont {Childs}}, \bibinfo {author} {\bibfnamefont {Yuan}\ \bibnamefont {Su}}, \bibinfo {author} {\bibfnamefont {Minh~C}\ \bibnamefont {Tran}}, \bibinfo {author} {\bibfnamefont {Nathan}\ \bibnamefont {Wiebe}}, \ and\ \bibinfo {author} {\bibfnamefont {Shuchen}\ \bibnamefont {Zhu}},\ }\bibfield  {title} {\enquote {\bibinfo {title} {Theory of trotter error with commutator scaling},}\ }\href {https://link.aps.org/pdf/10.1103/PhysRevX.11.011020} {\bibfield  {journal} {\bibinfo  {journal} {Physical Review X}\ }\textbf {\bibinfo {volume} {11}},\ \bibinfo {pages} {011020} (\bibinfo {year} {2021})}\BibitemShut {NoStop}%
\bibitem [{\citenamefont {Low}\ and\ \citenamefont {Wiebe}(2018)}]{low2018hamiltonian}%
  \BibitemOpen
  \bibfield  {author} {\bibinfo {author} {\bibfnamefont {Guang~Hao}\ \bibnamefont {Low}}\ and\ \bibinfo {author} {\bibfnamefont {Nathan}\ \bibnamefont {Wiebe}},\ }\bibfield  {title} {\enquote {\bibinfo {title} {Hamiltonian simulation in the interaction picture},}\ }\href {https://arxiv.org/abs/1805.00675} {\bibfield  {journal} {\bibinfo  {journal} {arXiv preprint arXiv:1805.00675}\ } (\bibinfo {year} {2018})}\BibitemShut {NoStop}%
\bibitem [{\citenamefont {Kivlichan}\ \emph {et~al.}(2018)\citenamefont {Kivlichan}, \citenamefont {McClean}, \citenamefont {Wiebe}, \citenamefont {Gidney}, \citenamefont {Aspuru-Guzik}, \citenamefont {Chan},\ and\ \citenamefont {Babbush}}]{kivlichan2018quantum}%
  \BibitemOpen
  \bibfield  {author} {\bibinfo {author} {\bibfnamefont {Ian~D}\ \bibnamefont {Kivlichan}}, \bibinfo {author} {\bibfnamefont {Jarrod}\ \bibnamefont {McClean}}, \bibinfo {author} {\bibfnamefont {Nathan}\ \bibnamefont {Wiebe}}, \bibinfo {author} {\bibfnamefont {Craig}\ \bibnamefont {Gidney}}, \bibinfo {author} {\bibfnamefont {Al{\'a}n}\ \bibnamefont {Aspuru-Guzik}}, \bibinfo {author} {\bibfnamefont {Garnet Kin-Lic}\ \bibnamefont {Chan}}, \ and\ \bibinfo {author} {\bibfnamefont {Ryan}\ \bibnamefont {Babbush}},\ }\bibfield  {title} {\enquote {\bibinfo {title} {Quantum simulation of electronic structure with linear depth and connectivity},}\ }\href {https://arxiv.org/pdf/1711.04789} {\bibfield  {journal} {\bibinfo  {journal} {Physical review letters}\ }\textbf {\bibinfo {volume} {120}},\ \bibinfo {pages} {110501} (\bibinfo {year} {2018})}\BibitemShut {NoStop}%
\bibitem [{\citenamefont {Temme}\ \emph {et~al.}(2017)\citenamefont {Temme}, \citenamefont {Bravyi},\ and\ \citenamefont {Gambetta}}]{temme2017error}%
  \BibitemOpen
  \bibfield  {author} {\bibinfo {author} {\bibfnamefont {Kristan}\ \bibnamefont {Temme}}, \bibinfo {author} {\bibfnamefont {Sergey}\ \bibnamefont {Bravyi}}, \ and\ \bibinfo {author} {\bibfnamefont {Jay~M}\ \bibnamefont {Gambetta}},\ }\bibfield  {title} {\enquote {\bibinfo {title} {Error mitigation for short-depth quantum circuits},}\ }\href@noop {} {\bibfield  {journal} {\bibinfo  {journal} {Physical review letters}\ }\textbf {\bibinfo {volume} {119}},\ \bibinfo {pages} {180509} (\bibinfo {year} {2017})}\BibitemShut {NoStop}%
\bibitem [{\citenamefont {Li}\ and\ \citenamefont {Benjamin}(2017{\natexlab{a}})}]{li2017efficient}%
  \BibitemOpen
  \bibfield  {author} {\bibinfo {author} {\bibfnamefont {Ying}\ \bibnamefont {Li}}\ and\ \bibinfo {author} {\bibfnamefont {Simon~C}\ \bibnamefont {Benjamin}},\ }\bibfield  {title} {\enquote {\bibinfo {title} {Efficient variational quantum simulator incorporating active error minimization},}\ }\href@noop {} {\bibfield  {journal} {\bibinfo  {journal} {Physical Review X}\ }\textbf {\bibinfo {volume} {7}},\ \bibinfo {pages} {021050} (\bibinfo {year} {2017}{\natexlab{a}})}\BibitemShut {NoStop}%
\bibitem [{\citenamefont {McDermott}(2009)}]{mcdermott2009structure}%
  \BibitemOpen
  \bibfield  {author} {\bibinfo {author} {\bibfnamefont {Ann}\ \bibnamefont {McDermott}},\ }\bibfield  {title} {\enquote {\bibinfo {title} {Structure and dynamics of membrane proteins by magic angle spinning solid-state nmr},}\ }\href {https://www.annualreviews.org/content/journals/10.1146/annurev.biophys.050708.133719} {\bibfield  {journal} {\bibinfo  {journal} {Annual review of biophysics}\ }\textbf {\bibinfo {volume} {38}},\ \bibinfo {pages} {385--403} (\bibinfo {year} {2009})}\BibitemShut {NoStop}%
\bibitem [{\citenamefont {Han}\ \emph {et~al.}(2010)\citenamefont {Han}, \citenamefont {Ahn}, \citenamefont {Concel}, \citenamefont {Byeon}, \citenamefont {Gronenborn}, \citenamefont {Yang},\ and\ \citenamefont {Polenova}}]{han2010solid}%
  \BibitemOpen
  \bibfield  {author} {\bibinfo {author} {\bibfnamefont {Yun}\ \bibnamefont {Han}}, \bibinfo {author} {\bibfnamefont {Jinwoo}\ \bibnamefont {Ahn}}, \bibinfo {author} {\bibfnamefont {Jason}\ \bibnamefont {Concel}}, \bibinfo {author} {\bibfnamefont {In-Ja~L}\ \bibnamefont {Byeon}}, \bibinfo {author} {\bibfnamefont {Angela~M}\ \bibnamefont {Gronenborn}}, \bibinfo {author} {\bibfnamefont {Jun}\ \bibnamefont {Yang}}, \ and\ \bibinfo {author} {\bibfnamefont {Tatyana}\ \bibnamefont {Polenova}},\ }\bibfield  {title} {\enquote {\bibinfo {title} {Solid-state nmr studies of hiv-1 capsid protein assemblies},}\ }\href {https://pubs.acs.org/doi/abs/10.1021/ja908687k} {\bibfield  {journal} {\bibinfo  {journal} {Journal of the American Chemical Society}\ }\textbf {\bibinfo {volume} {132}},\ \bibinfo {pages} {1976--1987} (\bibinfo {year} {2010})}\BibitemShut {NoStop}%
\bibitem [{\citenamefont {Kong}\ \emph {et~al.}(2013)\citenamefont {Kong}, \citenamefont {Deng}, \citenamefont {Yan}, \citenamefont {Kim}, \citenamefont {Swisher}, \citenamefont {Smit}, \citenamefont {Yaghi},\ and\ \citenamefont {Reimer}}]{kong2013mapping}%
  \BibitemOpen
  \bibfield  {author} {\bibinfo {author} {\bibfnamefont {Xueqian}\ \bibnamefont {Kong}}, \bibinfo {author} {\bibfnamefont {Hexiang}\ \bibnamefont {Deng}}, \bibinfo {author} {\bibfnamefont {Fangyong}\ \bibnamefont {Yan}}, \bibinfo {author} {\bibfnamefont {Jihan}\ \bibnamefont {Kim}}, \bibinfo {author} {\bibfnamefont {Joseph~A}\ \bibnamefont {Swisher}}, \bibinfo {author} {\bibfnamefont {Berend}\ \bibnamefont {Smit}}, \bibinfo {author} {\bibfnamefont {Omar~M}\ \bibnamefont {Yaghi}}, \ and\ \bibinfo {author} {\bibfnamefont {Jeffrey~A}\ \bibnamefont {Reimer}},\ }\bibfield  {title} {\enquote {\bibinfo {title} {Mapping of functional groups in metal-organic frameworks},}\ }\href {https://www.science.org/doi/abs/10.1126/science.1238339} {\bibfield  {journal} {\bibinfo  {journal} {Science}\ }\textbf {\bibinfo {volume} {341}},\ \bibinfo {pages} {882--885} (\bibinfo {year} {2013})}\BibitemShut {NoStop}%
\bibitem [{\citenamefont {Karabanov}\ \emph {et~al.}(2011)\citenamefont {Karabanov}, \citenamefont {Kuprov}, \citenamefont {Charnock}, \citenamefont {van~der Drift}, \citenamefont {Edwards},\ and\ \citenamefont {Köckenberger}}]{Karabanov2011}%
  \BibitemOpen
  \bibfield  {author} {\bibinfo {author} {\bibfnamefont {Alexander}\ \bibnamefont {Karabanov}}, \bibinfo {author} {\bibfnamefont {Ilya}\ \bibnamefont {Kuprov}}, \bibinfo {author} {\bibfnamefont {G.~T.~P.}\ \bibnamefont {Charnock}}, \bibinfo {author} {\bibfnamefont {Anniek}\ \bibnamefont {van~der Drift}}, \bibinfo {author} {\bibfnamefont {Luke~J.}\ \bibnamefont {Edwards}}, \ and\ \bibinfo {author} {\bibfnamefont {Walter}\ \bibnamefont {Köckenberger}},\ }\bibfield  {title} {\enquote {\bibinfo {title} {On the accuracy of the state space restriction approximation for spin dynamics simulations},}\ }\href {\doibase 10.1063/1.3624564} {\bibfield  {journal} {\bibinfo  {journal} {The Journal of Chemical Physics}\ }\textbf {\bibinfo {volume} {135}},\ \bibinfo {pages} {084106} (\bibinfo {year} {2011})}\BibitemShut {NoStop}%
\bibitem [{\citenamefont {Stryer}\ and\ \citenamefont {Haugland}(1967)}]{Stryer1967}%
  \BibitemOpen
  \bibfield  {author} {\bibinfo {author} {\bibfnamefont {L}~\bibnamefont {Stryer}}\ and\ \bibinfo {author} {\bibfnamefont {R~P}\ \bibnamefont {Haugland}},\ }\bibfield  {title} {\enquote {\bibinfo {title} {Energy transfer: a spectroscopic ruler.}}\ }\href {\doibase 10.1073/pnas.58.2.719} {\bibfield  {journal} {\bibinfo  {journal} {Proceedings of the National Academy of Sciences}\ }\textbf {\bibinfo {volume} {58}},\ \bibinfo {pages} {719--726} (\bibinfo {year} {1967})},\ \Eprint {http://arxiv.org/abs/https://www.pnas.org/doi/pdf/10.1073/pnas.58.2.719} {https://www.pnas.org/doi/pdf/10.1073/pnas.58.2.719} \BibitemShut {NoStop}%
\bibitem [{\citenamefont {Krushelnitsky}\ \emph {et~al.}(2013)\citenamefont {Krushelnitsky}, \citenamefont {Reichert},\ and\ \citenamefont {Saalwachter}}]{krushelnitsky2013solid}%
  \BibitemOpen
  \bibfield  {author} {\bibinfo {author} {\bibfnamefont {Alexey}\ \bibnamefont {Krushelnitsky}}, \bibinfo {author} {\bibfnamefont {Detlef}\ \bibnamefont {Reichert}}, \ and\ \bibinfo {author} {\bibfnamefont {Kay}\ \bibnamefont {Saalwachter}},\ }\bibfield  {title} {\enquote {\bibinfo {title} {Solid-state nmr approaches to internal dynamics of proteins: from picoseconds to microseconds and seconds},}\ }\href {https://pubs.acs.org/doi/full/10.1021/ar300292p} {\bibfield  {journal} {\bibinfo  {journal} {Accounts of chemical research}\ }\textbf {\bibinfo {volume} {46}},\ \bibinfo {pages} {2028--2036} (\bibinfo {year} {2013})}\BibitemShut {NoStop}%
\bibitem [{\citenamefont {Childs}\ \emph {et~al.}(2019)\citenamefont {Childs}, \citenamefont {Ostrander},\ and\ \citenamefont {Su}}]{childs2019faster}%
  \BibitemOpen
  \bibfield  {author} {\bibinfo {author} {\bibfnamefont {Andrew~M}\ \bibnamefont {Childs}}, \bibinfo {author} {\bibfnamefont {Aaron}\ \bibnamefont {Ostrander}}, \ and\ \bibinfo {author} {\bibfnamefont {Yuan}\ \bibnamefont {Su}},\ }\bibfield  {title} {\enquote {\bibinfo {title} {Faster quantum simulation by randomization},}\ }\href {https://quantum-journal.org/papers/q-2019-09-02-182/} {\bibfield  {journal} {\bibinfo  {journal} {Quantum}\ }\textbf {\bibinfo {volume} {3}},\ \bibinfo {pages} {182} (\bibinfo {year} {2019})}\BibitemShut {NoStop}%
\bibitem [{\citenamefont {Campbell}(2019)}]{campbell2019random}%
  \BibitemOpen
  \bibfield  {author} {\bibinfo {author} {\bibfnamefont {Earl}\ \bibnamefont {Campbell}},\ }\bibfield  {title} {\enquote {\bibinfo {title} {Random compiler for fast hamiltonian simulation},}\ }\href {https://arxiv.org/pdf/1811.08017} {\bibfield  {journal} {\bibinfo  {journal} {Physical review letters}\ }\textbf {\bibinfo {volume} {123}},\ \bibinfo {pages} {070503} (\bibinfo {year} {2019})}\BibitemShut {NoStop}%
\bibitem [{\citenamefont {Zhao}\ \emph {et~al.}(2023)\citenamefont {Zhao}, \citenamefont {Bukov}, \citenamefont {Heyl},\ and\ \citenamefont {Moessner}}]{zhao2023making}%
  \BibitemOpen
  \bibfield  {author} {\bibinfo {author} {\bibfnamefont {Hongzheng}\ \bibnamefont {Zhao}}, \bibinfo {author} {\bibfnamefont {Marin}\ \bibnamefont {Bukov}}, \bibinfo {author} {\bibfnamefont {Markus}\ \bibnamefont {Heyl}}, \ and\ \bibinfo {author} {\bibfnamefont {Roderich}\ \bibnamefont {Moessner}},\ }\bibfield  {title} {\enquote {\bibinfo {title} {Making trotterization adaptive and energy-self-correcting for nisq devices and beyond},}\ }\href {https://link.aps.org/pdf/10.1103/PRXQuantum.4.030319} {\bibfield  {journal} {\bibinfo  {journal} {PRX Quantum}\ }\textbf {\bibinfo {volume} {4}},\ \bibinfo {pages} {030319} (\bibinfo {year} {2023})}\BibitemShut {NoStop}%
\bibitem [{\citenamefont {Mansuroglu}\ \emph {et~al.}(2023)\citenamefont {Mansuroglu}, \citenamefont {Fischer},\ and\ \citenamefont {Hartmann}}]{mansuroglu2023problem}%
  \BibitemOpen
  \bibfield  {author} {\bibinfo {author} {\bibfnamefont {Refik}\ \bibnamefont {Mansuroglu}}, \bibinfo {author} {\bibfnamefont {Felix}\ \bibnamefont {Fischer}}, \ and\ \bibinfo {author} {\bibfnamefont {Michael~J}\ \bibnamefont {Hartmann}},\ }\bibfield  {title} {\enquote {\bibinfo {title} {Problem-specific classical optimization of hamiltonian simulation},}\ }\href {https://link.aps.org/pdf/10.1103/PhysRevResearch.5.043035} {\bibfield  {journal} {\bibinfo  {journal} {Physical Review Research}\ }\textbf {\bibinfo {volume} {5}},\ \bibinfo {pages} {043035} (\bibinfo {year} {2023})}\BibitemShut {NoStop}%
\bibitem [{\citenamefont {Jern}\ \emph {et~al.}(2025)\citenamefont {Jern}, \citenamefont {Uotila}, \citenamefont {Yu},\ and\ \citenamefont {Zhao}}]{jern2025agentqfinetuninglargelanguage}%
  \BibitemOpen
  \bibfield  {author} {\bibinfo {author} {\bibfnamefont {Linus}\ \bibnamefont {Jern}}, \bibinfo {author} {\bibfnamefont {Valter}\ \bibnamefont {Uotila}}, \bibinfo {author} {\bibfnamefont {Cong}\ \bibnamefont {Yu}}, \ and\ \bibinfo {author} {\bibfnamefont {Bo}~\bibnamefont {Zhao}},\ }\href {https://arxiv.org/abs/2504.11109} {\enquote {\bibinfo {title} {Agent-q: Fine-tuning large language models for quantum circuit generation and optimization},}\ } (\bibinfo {year} {2025}),\ \Eprint {http://arxiv.org/abs/2504.11109} {arXiv:2504.11109 [quant-ph]} \BibitemShut {NoStop}%
\bibitem [{\citenamefont {Burum}\ \emph {et~al.}(1981{\natexlab{a}})\citenamefont {Burum}, \citenamefont {Linder},\ and\ \citenamefont {Ernst}}]{BURUM1981173}%
  \BibitemOpen
  \bibfield  {author} {\bibinfo {author} {\bibfnamefont {D.P}\ \bibnamefont {Burum}}, \bibinfo {author} {\bibfnamefont {M}~\bibnamefont {Linder}}, \ and\ \bibinfo {author} {\bibfnamefont {R.R}\ \bibnamefont {Ernst}},\ }\bibfield  {title} {\enquote {\bibinfo {title} {Low-power multipulse line narrowing in solid-state nmr},}\ }\href {\doibase https://doi.org/10.1016/0022-2364(81)90200-6} {\bibfield  {journal} {\bibinfo  {journal} {Journal of Magnetic Resonance (1969)}\ }\textbf {\bibinfo {volume} {44}},\ \bibinfo {pages} {173--188} (\bibinfo {year} {1981}{\natexlab{a}})}\BibitemShut {NoStop}%
\bibitem [{\citenamefont {Mart{\'\i}nez}\ \emph {et~al.}(2009)\citenamefont {Mart{\'\i}nez}, \citenamefont {Andrade}, \citenamefont {Birgin},\ and\ \citenamefont {Mart{\'\i}nez}}]{martinez2009packmol}%
  \BibitemOpen
  \bibfield  {author} {\bibinfo {author} {\bibfnamefont {Leandro}\ \bibnamefont {Mart{\'\i}nez}}, \bibinfo {author} {\bibfnamefont {Ricardo}\ \bibnamefont {Andrade}}, \bibinfo {author} {\bibfnamefont {Ernesto~G}\ \bibnamefont {Birgin}}, \ and\ \bibinfo {author} {\bibfnamefont {Jos{\'e}~Mario}\ \bibnamefont {Mart{\'\i}nez}},\ }\bibfield  {title} {\enquote {\bibinfo {title} {Packmol: A package for building initial configurations for molecular dynamics simulations},}\ }\href@noop {} {\bibfield  {journal} {\bibinfo  {journal} {Journal of computational chemistry}\ }\textbf {\bibinfo {volume} {30}},\ \bibinfo {pages} {2157--2164} (\bibinfo {year} {2009})}\BibitemShut {NoStop}%
\bibitem [{\citenamefont {Eastman}\ \emph {et~al.}(2023)\citenamefont {Eastman}, \citenamefont {Galvelis}, \citenamefont {Pel{\'a}ez}, \citenamefont {Abreu}, \citenamefont {Farr}, \citenamefont {Gallicchio}, \citenamefont {Gorenko}, \citenamefont {Henry}, \citenamefont {Hu}, \citenamefont {Huang} \emph {et~al.}}]{eastman2023openmm}%
  \BibitemOpen
  \bibfield  {author} {\bibinfo {author} {\bibfnamefont {Peter}\ \bibnamefont {Eastman}}, \bibinfo {author} {\bibfnamefont {Raimondas}\ \bibnamefont {Galvelis}}, \bibinfo {author} {\bibfnamefont {Ra{\'u}l~P}\ \bibnamefont {Pel{\'a}ez}}, \bibinfo {author} {\bibfnamefont {Charlles~RA}\ \bibnamefont {Abreu}}, \bibinfo {author} {\bibfnamefont {Stephen~E}\ \bibnamefont {Farr}}, \bibinfo {author} {\bibfnamefont {Emilio}\ \bibnamefont {Gallicchio}}, \bibinfo {author} {\bibfnamefont {Anton}\ \bibnamefont {Gorenko}}, \bibinfo {author} {\bibfnamefont {Michael~M}\ \bibnamefont {Henry}}, \bibinfo {author} {\bibfnamefont {Frank}\ \bibnamefont {Hu}}, \bibinfo {author} {\bibfnamefont {Jing}\ \bibnamefont {Huang}},  \emph {et~al.},\ }\bibfield  {title} {\enquote {\bibinfo {title} {Openmm 8: molecular dynamics simulation with machine learning potentials},}\ }\href@noop {} {\bibfield  {journal} {\bibinfo  {journal} {The Journal of Physical Chemistry B}\ }\textbf {\bibinfo {volume} {128}},\ \bibinfo {pages} {109--116} (\bibinfo
  {year} {2023})}\BibitemShut {NoStop}%
\bibitem [{\citenamefont {Wang}\ \emph {et~al.}(2004)\citenamefont {Wang}, \citenamefont {Wolf}, \citenamefont {Caldwell}, \citenamefont {Kollman},\ and\ \citenamefont {Case}}]{wang2004development}%
  \BibitemOpen
  \bibfield  {author} {\bibinfo {author} {\bibfnamefont {Junmei}\ \bibnamefont {Wang}}, \bibinfo {author} {\bibfnamefont {Romain~M}\ \bibnamefont {Wolf}}, \bibinfo {author} {\bibfnamefont {James~W}\ \bibnamefont {Caldwell}}, \bibinfo {author} {\bibfnamefont {Peter~A}\ \bibnamefont {Kollman}}, \ and\ \bibinfo {author} {\bibfnamefont {David~A}\ \bibnamefont {Case}},\ }\bibfield  {title} {\enquote {\bibinfo {title} {Development and testing of a general amber force field},}\ }\href@noop {} {\bibfield  {journal} {\bibinfo  {journal} {Journal of computational chemistry}\ }\textbf {\bibinfo {volume} {25}},\ \bibinfo {pages} {1157--1174} (\bibinfo {year} {2004})}\BibitemShut {NoStop}%
\bibitem [{\citenamefont {Seute}\ \emph {et~al.}(2025)\citenamefont {Seute}, \citenamefont {Hartmann}, \citenamefont {St{\"u}hmer},\ and\ \citenamefont {Gr{\"a}ter}}]{seute2025grappa}%
  \BibitemOpen
  \bibfield  {author} {\bibinfo {author} {\bibfnamefont {Leif}\ \bibnamefont {Seute}}, \bibinfo {author} {\bibfnamefont {Eric}\ \bibnamefont {Hartmann}}, \bibinfo {author} {\bibfnamefont {Jan}\ \bibnamefont {St{\"u}hmer}}, \ and\ \bibinfo {author} {\bibfnamefont {Frauke}\ \bibnamefont {Gr{\"a}ter}},\ }\bibfield  {title} {\enquote {\bibinfo {title} {Grappa--a machine learned molecular mechanics force field},}\ }\href@noop {} {\bibfield  {journal} {\bibinfo  {journal} {Chemical Science}\ }\textbf {\bibinfo {volume} {16}},\ \bibinfo {pages} {2907--2930} (\bibinfo {year} {2025})}\BibitemShut {NoStop}%
\bibitem [{\citenamefont {Zhang}\ \emph {et~al.}(2011)\citenamefont {Zhang}, \citenamefont {Su},\ and\ \citenamefont {Guo}}]{zhang2011atomistic}%
  \BibitemOpen
  \bibfield  {author} {\bibinfo {author} {\bibfnamefont {Jianguo}\ \bibnamefont {Zhang}}, \bibinfo {author} {\bibfnamefont {Jiaye}\ \bibnamefont {Su}}, \ and\ \bibinfo {author} {\bibfnamefont {Hongxia}\ \bibnamefont {Guo}},\ }\bibfield  {title} {\enquote {\bibinfo {title} {An atomistic simulation for 4-cyano-4'-pentylbiphenyl and its homologue with a reoptimized force field},}\ }\href@noop {} {\bibfield  {journal} {\bibinfo  {journal} {The Journal of Physical Chemistry B}\ }\textbf {\bibinfo {volume} {115}},\ \bibinfo {pages} {2214--2227} (\bibinfo {year} {2011})}\BibitemShut {NoStop}%
\bibitem [{\citenamefont {Horn}(1978)}]{horn1978refractive}%
  \BibitemOpen
  \bibfield  {author} {\bibinfo {author} {\bibfnamefont {Roger~G}\ \bibnamefont {Horn}},\ }\bibfield  {title} {\enquote {\bibinfo {title} {Refractive indices and order parameters of two liquid crystals},}\ }\href@noop {} {\bibfield  {journal} {\bibinfo  {journal} {Journal de physique}\ }\textbf {\bibinfo {volume} {39}},\ \bibinfo {pages} {105--109} (\bibinfo {year} {1978})}\BibitemShut {NoStop}%
\bibitem [{\citenamefont {Sherrell}\ and\ \citenamefont {Crellin}(1979)}]{sherrell1979susceptibilities}%
  \BibitemOpen
  \bibfield  {author} {\bibinfo {author} {\bibfnamefont {PL}~\bibnamefont {Sherrell}}\ and\ \bibinfo {author} {\bibfnamefont {DA}~\bibnamefont {Crellin}},\ }\bibfield  {title} {\enquote {\bibinfo {title} {Susceptibilities and order parameters of nematic liquid crystals},}\ }\href@noop {} {\bibfield  {journal} {\bibinfo  {journal} {Le Journal de Physique Colloques}\ }\textbf {\bibinfo {volume} {40}},\ \bibinfo {pages} {C3--211} (\bibinfo {year} {1979})}\BibitemShut {NoStop}%
\bibitem [{\citenamefont {Cacelli}\ \emph {et~al.}(2002)\citenamefont {Cacelli}, \citenamefont {Campanile}, \citenamefont {Prampolini},\ and\ \citenamefont {Tani}}]{cacelli2002stability}%
  \BibitemOpen
  \bibfield  {author} {\bibinfo {author} {\bibfnamefont {Ivo}\ \bibnamefont {Cacelli}}, \bibinfo {author} {\bibfnamefont {Silvio}\ \bibnamefont {Campanile}}, \bibinfo {author} {\bibfnamefont {Giacomo}\ \bibnamefont {Prampolini}}, \ and\ \bibinfo {author} {\bibfnamefont {Alessandro}\ \bibnamefont {Tani}},\ }\bibfield  {title} {\enquote {\bibinfo {title} {Stability of the nematic phase of 4-n-pentyl-4'-cyanobiphenyl studied by computer simulation using a hybrid model},}\ }\href@noop {} {\bibfield  {journal} {\bibinfo  {journal} {The Journal of chemical physics}\ }\textbf {\bibinfo {volume} {117}},\ \bibinfo {pages} {448--453} (\bibinfo {year} {2002})}\BibitemShut {NoStop}%
\bibitem [{\citenamefont {Mouret}\ and\ \citenamefont {Clune}(2015)}]{mouret2015illuminatingsearchspacesmapping}%
  \BibitemOpen
  \bibfield  {author} {\bibinfo {author} {\bibfnamefont {Jean-Baptiste}\ \bibnamefont {Mouret}}\ and\ \bibinfo {author} {\bibfnamefont {Jeff}\ \bibnamefont {Clune}},\ }\href {https://arxiv.org/abs/1504.04909} {\enquote {\bibinfo {title} {Illuminating search spaces by mapping elites},}\ } (\bibinfo {year} {2015}),\ \Eprint {http://arxiv.org/abs/1504.04909} {arXiv:1504.04909 [cs.AI]} \BibitemShut {NoStop}%
\bibitem [{\citenamefont {Bosse}\ \emph {et~al.}(2025)\citenamefont {Bosse}, \citenamefont {Childs}, \citenamefont {Derby}, \citenamefont {Gambetta}, \citenamefont {Montanaro},\ and\ \citenamefont {Santos}}]{Bosse2025-ha}%
  \BibitemOpen
  \bibfield  {author} {\bibinfo {author} {\bibfnamefont {Jan~Lukas}\ \bibnamefont {Bosse}}, \bibinfo {author} {\bibfnamefont {Andrew~M}\ \bibnamefont {Childs}}, \bibinfo {author} {\bibfnamefont {Charles}\ \bibnamefont {Derby}}, \bibinfo {author} {\bibfnamefont {Filippo~Maria}\ \bibnamefont {Gambetta}}, \bibinfo {author} {\bibfnamefont {Ashley}\ \bibnamefont {Montanaro}}, \ and\ \bibinfo {author} {\bibfnamefont {Raul~A}\ \bibnamefont {Santos}},\ }\bibfield  {title} {\enquote {\bibinfo {title} {Efficient and practical hamiltonian simulation from time-dependent product formulas},}\ }\href@noop {} {\bibfield  {journal} {\bibinfo  {journal} {Nat. Commun.}\ }\textbf {\bibinfo {volume} {16}},\ \bibinfo {pages} {2673} (\bibinfo {year} {2025})}\BibitemShut {NoStop}%
\bibitem [{\citenamefont {Tucci}(2005)}]{tucci2005introductioncartanskakdecomposition}%
  \BibitemOpen
  \bibfield  {author} {\bibinfo {author} {\bibfnamefont {Robert~R.}\ \bibnamefont {Tucci}},\ }\href {https://arxiv.org/abs/quant-ph/0507171} {\enquote {\bibinfo {title} {An introduction to cartan's kak decomposition for qc programmers},}\ } (\bibinfo {year} {2005}),\ \Eprint {http://arxiv.org/abs/quant-ph/0507171} {arXiv:quant-ph/0507171 [quant-ph]} \BibitemShut {NoStop}%
\bibitem [{\citenamefont {Acharya}\ \emph {et~al.}(2025)\citenamefont {Acharya}, \citenamefont {Aghababaie-Beni}, \citenamefont {Aleiner}, \citenamefont {Andersen} \emph {et~al.}}]{google2025QEC_bellow_threshold}%
  \BibitemOpen
  \bibfield  {author} {\bibinfo {author} {\bibfnamefont {Rajeev}\ \bibnamefont {Acharya}}, \bibinfo {author} {\bibfnamefont {Laleh}\ \bibnamefont {Aghababaie-Beni}}, \bibinfo {author} {\bibfnamefont {Igor}\ \bibnamefont {Aleiner}}, \bibinfo {author} {\bibfnamefont {Trond~I.}\ \bibnamefont {Andersen}},  \emph {et~al.},\ }\bibfield  {title} {\enquote {\bibinfo {title} {Quantum error correction below the surface code threshold},}\ }\href {\doibase 10.1038/s41586-024-08449-y} {\bibfield  {journal} {\bibinfo  {journal} {Nature}\ }\textbf {\bibinfo {volume} {638}},\ \bibinfo {pages} {920--926} (\bibinfo {year} {2025})}\BibitemShut {NoStop}%
\bibitem [{\citenamefont {Foxen}\ \emph {et~al.}(2020)\citenamefont {Foxen}, \citenamefont {Neill}, \citenamefont {Dunsworth}, \citenamefont {Roushan}, \citenamefont {Chiaro}, \citenamefont {Megrant}, \citenamefont {Kelly}, \citenamefont {Chen}, \citenamefont {Satzinger}, \citenamefont {Barends} \emph {et~al.}}]{foxen2020demonstrating}%
  \BibitemOpen
  \bibfield  {author} {\bibinfo {author} {\bibfnamefont {Brooks}\ \bibnamefont {Foxen}}, \bibinfo {author} {\bibfnamefont {Charles}\ \bibnamefont {Neill}}, \bibinfo {author} {\bibfnamefont {Andrew}\ \bibnamefont {Dunsworth}}, \bibinfo {author} {\bibfnamefont {Pedram}\ \bibnamefont {Roushan}}, \bibinfo {author} {\bibfnamefont {Ben}\ \bibnamefont {Chiaro}}, \bibinfo {author} {\bibfnamefont {Anthony}\ \bibnamefont {Megrant}}, \bibinfo {author} {\bibfnamefont {Julian}\ \bibnamefont {Kelly}}, \bibinfo {author} {\bibfnamefont {Zijun}\ \bibnamefont {Chen}}, \bibinfo {author} {\bibfnamefont {Kevin}\ \bibnamefont {Satzinger}}, \bibinfo {author} {\bibfnamefont {Rami}\ \bibnamefont {Barends}},  \emph {et~al.},\ }\bibfield  {title} {\enquote {\bibinfo {title} {Demonstrating a continuous set of two-qubit gates for near-term quantum algorithms},}\ }\href {https://journals.aps.org/prl/abstract/10.1103/PhysRevLett.125.120504} {\bibfield  {journal} {\bibinfo  {journal} {Physical Review Letters}\ }\textbf {\bibinfo {volume}
  {125}},\ \bibinfo {pages} {120504} (\bibinfo {year} {2020})}\BibitemShut {NoStop}%
\bibitem [{\citenamefont {Arute}\ \emph {et~al.}(2020)\citenamefont {Arute}, \citenamefont {Arya}, \citenamefont {Babbush}, \citenamefont {Bacon}, \citenamefont {Bardin} \emph {et~al.}}]{Arute2020SeparatedDynamics}%
  \BibitemOpen
  \bibfield  {author} {\bibinfo {author} {\bibfnamefont {Frank}\ \bibnamefont {Arute}}, \bibinfo {author} {\bibfnamefont {Kunal}\ \bibnamefont {Arya}}, \bibinfo {author} {\bibfnamefont {Ryan}\ \bibnamefont {Babbush}}, \bibinfo {author} {\bibfnamefont {Dave}\ \bibnamefont {Bacon}}, \bibinfo {author} {\bibfnamefont {Joseph~C.}\ \bibnamefont {Bardin}},  \emph {et~al.},\ }\bibfield  {title} {\enquote {\bibinfo {title} {Observation of separated dynamics of charge and spin in the {Fermi-Hubbard} model},}\ }\href {\doibase 10.1126/sciadv.aay5901} {\bibfield  {journal} {\bibinfo  {journal} {Science Advances}\ }\textbf {\bibinfo {volume} {6}},\ \bibinfo {pages} {eaay5901} (\bibinfo {year} {2020})},\ \Eprint {http://arxiv.org/abs/2010.07965} {arXiv:2010.07965 [quant-ph]} \BibitemShut {NoStop}%
\bibitem [{\citenamefont {Mi}\ \emph {et~al.}(2022{\natexlab{a}})\citenamefont {Mi}, \citenamefont {Ippoliti}, \citenamefont {Moessner}, \citenamefont {Sondhi} \emph {et~al.}}]{Mi2021TimeCrystal}%
  \BibitemOpen
  \bibfield  {author} {\bibinfo {author} {\bibfnamefont {Xiao}\ \bibnamefont {Mi}}, \bibinfo {author} {\bibfnamefont {Matteo}\ \bibnamefont {Ippoliti}}, \bibinfo {author} {\bibfnamefont {Roderich}\ \bibnamefont {Moessner}}, \bibinfo {author} {\bibfnamefont {S.~L.}\ \bibnamefont {Sondhi}},  \emph {et~al.},\ }\bibfield  {title} {\enquote {\bibinfo {title} {Observation of time-crystalline eigenstate order on a quantum processor},}\ }\href {\doibase 10.1038/s41586-021-04257-w} {\bibfield  {journal} {\bibinfo  {journal} {Nature}\ }\textbf {\bibinfo {volume} {601}},\ \bibinfo {pages} {53--57} (\bibinfo {year} {2022}{\natexlab{a}})},\ \Eprint {http://arxiv.org/abs/2107.13571} {arXiv:2107.13571 [quant-ph]} \BibitemShut {NoStop}%
\bibitem [{\citenamefont {Neill}\ \emph {et~al.}(2021)\citenamefont {Neill}, \citenamefont {McCourt}, \citenamefont {Mi}, \citenamefont {Jiang}, \citenamefont {Niu} \emph {et~al.}}]{Neill2021QuantumRing}%
  \BibitemOpen
  \bibfield  {author} {\bibinfo {author} {\bibfnamefont {Chris}\ \bibnamefont {Neill}}, \bibinfo {author} {\bibfnamefont {Tom}\ \bibnamefont {McCourt}}, \bibinfo {author} {\bibfnamefont {Xiao}\ \bibnamefont {Mi}}, \bibinfo {author} {\bibfnamefont {Zhang}\ \bibnamefont {Jiang}}, \bibinfo {author} {\bibfnamefont {Muqing~Y.}\ \bibnamefont {Niu}},  \emph {et~al.},\ }\bibfield  {title} {\enquote {\bibinfo {title} {Accurately computing the electronic properties of a quantum ring},}\ }\href {\doibase 10.1038/s41586-021-03576-2} {\bibfield  {journal} {\bibinfo  {journal} {Nature}\ }\textbf {\bibinfo {volume} {594}},\ \bibinfo {pages} {508--512} (\bibinfo {year} {2021})},\ \Eprint {http://arxiv.org/abs/2012.00921} {arXiv:2012.00921 [quant-ph]} \BibitemShut {NoStop}%
\bibitem [{\citenamefont {Mi}\ \emph {et~al.}(2022{\natexlab{b}})\citenamefont {Mi}, \citenamefont {Sonner}, \citenamefont {Niu}, \citenamefont {Lee}, \citenamefont {Foxen}, \citenamefont {Acharya}, \citenamefont {Aleiner}, \citenamefont {Andersen}, \citenamefont {Arute} \emph {et~al.}}]{Mi2022NoiseResilient}%
  \BibitemOpen
  \bibfield  {author} {\bibinfo {author} {\bibfnamefont {Xiao}\ \bibnamefont {Mi}}, \bibinfo {author} {\bibfnamefont {Marius}\ \bibnamefont {Sonner}}, \bibinfo {author} {\bibfnamefont {Muqing~Y.}\ \bibnamefont {Niu}}, \bibinfo {author} {\bibfnamefont {Kang~W.}\ \bibnamefont {Lee}}, \bibinfo {author} {\bibfnamefont {Benjamin}\ \bibnamefont {Foxen}}, \bibinfo {author} {\bibfnamefont {Rajeev}\ \bibnamefont {Acharya}}, \bibinfo {author} {\bibfnamefont {Igor}\ \bibnamefont {Aleiner}}, \bibinfo {author} {\bibfnamefont {Trond~I.}\ \bibnamefont {Andersen}}, \bibinfo {author} {\bibfnamefont {Frank}\ \bibnamefont {Arute}},  \emph {et~al.},\ }\bibfield  {title} {\enquote {\bibinfo {title} {Noise-resilient edge modes on a chain of superconducting qubits},}\ }\href {\doibase 10.1126/science.abq5769} {\bibfield  {journal} {\bibinfo  {journal} {Science}\ }\textbf {\bibinfo {volume} {378}},\ \bibinfo {pages} {785--790} (\bibinfo {year} {2022}{\natexlab{b}})},\ \Eprint {http://arxiv.org/abs/2204.11372} {arXiv:2204.11372
  [quant-ph]} \BibitemShut {NoStop}%
\bibitem [{\citenamefont {Morvan}\ \emph {et~al.}(2022)\citenamefont {Morvan}, \citenamefont {Andersen}, \citenamefont {Mi}, \citenamefont {Neill} \emph {et~al.}}]{Morvan2022BoundStates}%
  \BibitemOpen
  \bibfield  {author} {\bibinfo {author} {\bibfnamefont {Alexis}\ \bibnamefont {Morvan}}, \bibinfo {author} {\bibfnamefont {Trond~I.}\ \bibnamefont {Andersen}}, \bibinfo {author} {\bibfnamefont {Xiao}\ \bibnamefont {Mi}}, \bibinfo {author} {\bibfnamefont {Charles}\ \bibnamefont {Neill}},  \emph {et~al.},\ }\bibfield  {title} {\enquote {\bibinfo {title} {Formation of robust bound states of interacting microwave photons},}\ }\href {\doibase 10.1038/s41586-022-05348-y} {\bibfield  {journal} {\bibinfo  {journal} {Nature}\ }\textbf {\bibinfo {volume} {612}},\ \bibinfo {pages} {240--245} (\bibinfo {year} {2022})},\ \Eprint {http://arxiv.org/abs/2206.05254} {arXiv:2206.05254 [quant-ph]} \BibitemShut {NoStop}%
\bibitem [{\citenamefont {Temme}\ \emph {et~al.}(2016)\citenamefont {Temme}, \citenamefont {Bravyi},\ and\ \citenamefont {Gambetta}}]{Temme2016ZNEPEC}%
  \BibitemOpen
  \bibfield  {author} {\bibinfo {author} {\bibfnamefont {Kristan}\ \bibnamefont {Temme}}, \bibinfo {author} {\bibfnamefont {Sergey}\ \bibnamefont {Bravyi}}, \ and\ \bibinfo {author} {\bibfnamefont {Jay~M.}\ \bibnamefont {Gambetta}},\ }\bibfield  {title} {\enquote {\bibinfo {title} {Error mitigation for short-depth quantum circuits},}\ }\href {\doibase 10.1103/PhysRevLett.119.180509} {\bibfield  {journal} {\bibinfo  {journal} {Physical Review Letters}\ }\textbf {\bibinfo {volume} {119}} (\bibinfo {year} {2016}),\ 10.1103/PhysRevLett.119.180509}\BibitemShut {NoStop}%
\bibitem [{\citenamefont {Li}\ and\ \citenamefont {Benjamin}(2017{\natexlab{b}})}]{Li2017ZNE}%
  \BibitemOpen
  \bibfield  {author} {\bibinfo {author} {\bibfnamefont {Ying}\ \bibnamefont {Li}}\ and\ \bibinfo {author} {\bibfnamefont {Simon~C.}\ \bibnamefont {Benjamin}},\ }\bibfield  {title} {\enquote {\bibinfo {title} {Efficient variational quantum simulator incorporating active error minimization},}\ }\href {\doibase 10.1103/PHYSREVX.7.021050/FIGURES/5/MEDIUM} {\bibfield  {journal} {\bibinfo  {journal} {Physical Review X}\ }\textbf {\bibinfo {volume} {7}},\ \bibinfo {pages} {021050} (\bibinfo {year} {2017}{\natexlab{b}})}\BibitemShut {NoStop}%
\bibitem [{\citenamefont {Cai}\ \emph {et~al.}(2023)\citenamefont {Cai}, \citenamefont {Babbush}, \citenamefont {Benjamin}, \citenamefont {Endo}, \citenamefont {Huggins}, \citenamefont {Li}, \citenamefont {McClean},\ and\ \citenamefont {O'Brien}}]{Cai2022EMReview}%
  \BibitemOpen
  \bibfield  {author} {\bibinfo {author} {\bibfnamefont {Zhenyu}\ \bibnamefont {Cai}}, \bibinfo {author} {\bibfnamefont {Ryan}\ \bibnamefont {Babbush}}, \bibinfo {author} {\bibfnamefont {Simon~C.}\ \bibnamefont {Benjamin}}, \bibinfo {author} {\bibfnamefont {Suguru}\ \bibnamefont {Endo}}, \bibinfo {author} {\bibfnamefont {William~J.}\ \bibnamefont {Huggins}}, \bibinfo {author} {\bibfnamefont {Ying}\ \bibnamefont {Li}}, \bibinfo {author} {\bibfnamefont {Jarrod~R.}\ \bibnamefont {McClean}}, \ and\ \bibinfo {author} {\bibfnamefont {Thomas~E.}\ \bibnamefont {O'Brien}},\ }\bibfield  {title} {\enquote {\bibinfo {title} {Quantum error mitigation},}\ }\href {\doibase 10.1103/RevModPhys.95.045005} {\bibfield  {journal} {\bibinfo  {journal} {Reviews of Modern Physics}\ }\textbf {\bibinfo {volume} {95}},\ \bibinfo {eid} {045005} (\bibinfo {year} {2023})},\ \Eprint {http://arxiv.org/abs/2210.00921} {arXiv:2210.00921 [quant-ph]} \BibitemShut {NoStop}%
\bibitem [{\citenamefont {Minev}\ \emph {et~al.}(2025)\citenamefont {Minev}, \citenamefont {Najafi}, \citenamefont {Majumder}, \citenamefont {Wang}, \citenamefont {Stern}, \citenamefont {Kim}, \citenamefont {Jian},\ and\ \citenamefont {Zhu}}]{Minev2025ZNE}%
  \BibitemOpen
  \bibfield  {author} {\bibinfo {author} {\bibfnamefont {Zlatko~K.}\ \bibnamefont {Minev}}, \bibinfo {author} {\bibfnamefont {Khadijeh}\ \bibnamefont {Najafi}}, \bibinfo {author} {\bibfnamefont {Swarnadeep}\ \bibnamefont {Majumder}}, \bibinfo {author} {\bibfnamefont {Juven}\ \bibnamefont {Wang}}, \bibinfo {author} {\bibfnamefont {Ady}\ \bibnamefont {Stern}}, \bibinfo {author} {\bibfnamefont {Eun-Ah}\ \bibnamefont {Kim}}, \bibinfo {author} {\bibfnamefont {Chao-Ming}\ \bibnamefont {Jian}}, \ and\ \bibinfo {author} {\bibfnamefont {Guanyu}\ \bibnamefont {Zhu}},\ }\bibfield  {title} {\enquote {\bibinfo {title} {{Realizing string-net condensation: Fibonacci anyon braiding for universal gates and sampling chromatic polynomials}},}\ }\href {\doibase 10.1038/s41467-025-61493-8} {\bibfield  {journal} {\bibinfo  {journal} {Nature Communications}\ }\textbf {\bibinfo {volume} {16}},\ \bibinfo {pages} {6225} (\bibinfo {year} {2025})}\BibitemShut {NoStop}%
\bibitem [{\citenamefont {Bennett}\ \emph {et~al.}(1995)\citenamefont {Bennett}, \citenamefont {Brassard}, \citenamefont {Popescu}, \citenamefont {Schumacher}, \citenamefont {Smolin},\ and\ \citenamefont {Wootters}}]{Bennett1995Twirling}%
  \BibitemOpen
  \bibfield  {author} {\bibinfo {author} {\bibfnamefont {Charles~H.}\ \bibnamefont {Bennett}}, \bibinfo {author} {\bibfnamefont {Gilles}\ \bibnamefont {Brassard}}, \bibinfo {author} {\bibfnamefont {Sandu}\ \bibnamefont {Popescu}}, \bibinfo {author} {\bibfnamefont {Benjamin}\ \bibnamefont {Schumacher}}, \bibinfo {author} {\bibfnamefont {John~A.}\ \bibnamefont {Smolin}}, \ and\ \bibinfo {author} {\bibfnamefont {William~K.}\ \bibnamefont {Wootters}},\ }\bibfield  {title} {\enquote {\bibinfo {title} {Purification of noisy entanglement and faithful teleportation via noisy channels},}\ }\href {\doibase 10.1103/PhysRevLett.76.722} {\bibfield  {journal} {\bibinfo  {journal} {Physical Review Letters}\ }\textbf {\bibinfo {volume} {76}},\ \bibinfo {pages} {722--725} (\bibinfo {year} {1995})}\BibitemShut {NoStop}%
\bibitem [{\citenamefont {Knill}(2004)}]{Knill2004Twirling}%
  \BibitemOpen
  \bibfield  {author} {\bibinfo {author} {\bibfnamefont {E.}~\bibnamefont {Knill}},\ }\bibfield  {title} {\enquote {\bibinfo {title} {Fault-tolerant postselected quantum computation: Threshold analysis},}\ }\href {\doibase 10.48550/arXiv.quant-ph/0404104} {\bibfield  {journal} {\bibinfo  {journal} {arXiv e-prints}\ ,\ \bibinfo {eid} {quant-ph/0404104}} (\bibinfo {year} {2004})},\ \Eprint {http://arxiv.org/abs/quant-ph/0404104} {arXiv:quant-ph/0404104 [quant-ph]} \BibitemShut {NoStop}%
\bibitem [{\citenamefont {Viola}\ and\ \citenamefont {Lloyd}(1998)}]{Viola1998}%
  \BibitemOpen
  \bibfield  {author} {\bibinfo {author} {\bibfnamefont {Lorenza}\ \bibnamefont {Viola}}\ and\ \bibinfo {author} {\bibfnamefont {Seth}\ \bibnamefont {Lloyd}},\ }\bibfield  {title} {\enquote {\bibinfo {title} {{Dynamical suppression of decoherence in two-state quantum systems}},}\ }\href {\doibase 10.1103/PhysRevA.58.2733} {\bibfield  {journal} {\bibinfo  {journal} {Phys. Rev. A}\ }\textbf {\bibinfo {volume} {58}},\ \bibinfo {pages} {2733--2744} (\bibinfo {year} {1998})},\ \Eprint {http://arxiv.org/abs/quant-ph/9803057} {quant-ph/9803057} \BibitemShut {NoStop}%
\bibitem [{\citenamefont {Viola}\ \emph {et~al.}(1999)\citenamefont {Viola}, \citenamefont {Knill},\ and\ \citenamefont {Lloyd}}]{Viola1999}%
  \BibitemOpen
  \bibfield  {author} {\bibinfo {author} {\bibfnamefont {Lorenza}\ \bibnamefont {Viola}}, \bibinfo {author} {\bibfnamefont {Emanuel}\ \bibnamefont {Knill}}, \ and\ \bibinfo {author} {\bibfnamefont {Seth}\ \bibnamefont {Lloyd}},\ }\bibfield  {title} {\enquote {\bibinfo {title} {{Dynamical Decoupling of Open Quantum Systems}},}\ }\href {\doibase 10.1103/PhysRevLett.82.2417} {\bibfield  {journal} {\bibinfo  {journal} {Phys. Rev. Lett.}\ }\textbf {\bibinfo {volume} {82}},\ \bibinfo {pages} {2417--2421} (\bibinfo {year} {1999})},\ \Eprint {http://arxiv.org/abs/quant-ph/9809071} {quant-ph/9809071} \BibitemShut {NoStop}%
\bibitem [{\citenamefont {van~den Berg}\ \emph {et~al.}(2022)\citenamefont {van~den Berg}, \citenamefont {Minev},\ and\ \citenamefont {Temme}}]{BergMinev2022TREX}%
  \BibitemOpen
  \bibfield  {author} {\bibinfo {author} {\bibfnamefont {Ewout}\ \bibnamefont {van~den Berg}}, \bibinfo {author} {\bibfnamefont {Zlatko~K.}\ \bibnamefont {Minev}}, \ and\ \bibinfo {author} {\bibfnamefont {Kristan}\ \bibnamefont {Temme}},\ }\bibfield  {title} {\enquote {\bibinfo {title} {Model-free readout-error mitigation for quantum expectation values},}\ }\href {\doibase 10.1103/PhysRevA.105.032620} {\bibfield  {journal} {\bibinfo  {journal} {Phys. Rev. A}\ }\textbf {\bibinfo {volume} {105}},\ \bibinfo {pages} {032620} (\bibinfo {year} {2022})}\BibitemShut {NoStop}%
\bibitem [{\citenamefont {Rudolph}\ \emph {et~al.}(2025)\citenamefont {Rudolph}, \citenamefont {Jones}, \citenamefont {Teng}, \citenamefont {Angrisani},\ and\ \citenamefont {Holmes}}]{rudolph2025pauli}%
  \BibitemOpen
  \bibfield  {author} {\bibinfo {author} {\bibfnamefont {Manuel~S}\ \bibnamefont {Rudolph}}, \bibinfo {author} {\bibfnamefont {Tyson}\ \bibnamefont {Jones}}, \bibinfo {author} {\bibfnamefont {Yanting}\ \bibnamefont {Teng}}, \bibinfo {author} {\bibfnamefont {Armando}\ \bibnamefont {Angrisani}}, \ and\ \bibinfo {author} {\bibfnamefont {Zo{\"e}}\ \bibnamefont {Holmes}},\ }\bibfield  {title} {\enquote {\bibinfo {title} {Pauli propagation: A computational framework for simulating quantum systems},}\ }\href {https://arxiv.org/pdf/2505.21606} {\bibfield  {journal} {\bibinfo  {journal} {arXiv preprint arXiv:2505.21606}\ } (\bibinfo {year} {2025})}\BibitemShut {NoStop}%
\bibitem [{\citenamefont {Nielsen}\ and\ \citenamefont {Chuang}(2010)}]{nielsen2010quantum}%
  \BibitemOpen
  \bibfield  {author} {\bibinfo {author} {\bibfnamefont {Michael~A}\ \bibnamefont {Nielsen}}\ and\ \bibinfo {author} {\bibfnamefont {Isaac~L}\ \bibnamefont {Chuang}},\ }\href@noop {} {\emph {\bibinfo {title} {Quantum computation and quantum information}}}\ (\bibinfo  {publisher} {Cambridge university press},\ \bibinfo {year} {2010})\BibitemShut {NoStop}%
\bibitem [{\citenamefont {Saupe}(1964)}]{saupe1964kernresonanzen}%
  \BibitemOpen
  \bibfield  {author} {\bibinfo {author} {\bibfnamefont {Alfred}\ \bibnamefont {Saupe}},\ }\bibfield  {title} {\enquote {\bibinfo {title} {Kernresonanzen in kristallinen fl{\"u}ssigkeiten und in kristallinfl{\"u}ssigen l{\"o}sungen. teil i},}\ }\href@noop {} {\bibfield  {journal} {\bibinfo  {journal} {Zeitschrift f{\"u}r Naturforschung A}\ }\textbf {\bibinfo {volume} {19}},\ \bibinfo {pages} {161--171} (\bibinfo {year} {1964})}\BibitemShut {NoStop}%
\bibitem [{\citenamefont {Baum}\ \emph {et~al.}(1985{\natexlab{b}})\citenamefont {Baum}, \citenamefont {Munowitz}, \citenamefont {Garroway},\ and\ \citenamefont {Pines}}]{baum1985multiple}%
  \BibitemOpen
  \bibfield  {author} {\bibinfo {author} {\bibfnamefont {Jean}\ \bibnamefont {Baum}}, \bibinfo {author} {\bibfnamefont {Michael}\ \bibnamefont {Munowitz}}, \bibinfo {author} {\bibfnamefont {Allen~N}\ \bibnamefont {Garroway}}, \ and\ \bibinfo {author} {\bibfnamefont {Alex}\ \bibnamefont {Pines}},\ }\bibfield  {title} {\enquote {\bibinfo {title} {Multiple-quantum dynamics in solid state nmr},}\ }\href@noop {} {\bibfield  {journal} {\bibinfo  {journal} {The Journal of chemical physics}\ }\textbf {\bibinfo {volume} {83}},\ \bibinfo {pages} {2015--2025} (\bibinfo {year} {1985}{\natexlab{b}})}\BibitemShut {NoStop}%
\bibitem [{\citenamefont {Haeberlen}\ and\ \citenamefont {Waugh}(1968)}]{haeberlen1968coherent}%
  \BibitemOpen
  \bibfield  {author} {\bibinfo {author} {\bibfnamefont {Ulrich}\ \bibnamefont {Haeberlen}}\ and\ \bibinfo {author} {\bibfnamefont {John~S}\ \bibnamefont {Waugh}},\ }\bibfield  {title} {\enquote {\bibinfo {title} {Coherent averaging effects in magnetic resonance},}\ }\href@noop {} {\bibfield  {journal} {\bibinfo  {journal} {Physical Review}\ }\textbf {\bibinfo {volume} {175}},\ \bibinfo {pages} {453} (\bibinfo {year} {1968})}\BibitemShut {NoStop}%
\bibitem [{\citenamefont {Hohwy}\ and\ \citenamefont {Nielsen}(1998)}]{hohwy1998systematic}%
  \BibitemOpen
  \bibfield  {author} {\bibinfo {author} {\bibfnamefont {M}~\bibnamefont {Hohwy}}\ and\ \bibinfo {author} {\bibfnamefont {NC}~\bibnamefont {Nielsen}},\ }\bibfield  {title} {\enquote {\bibinfo {title} {Systematic design and evaluation of multiple-pulse experiments in nuclear magnetic resonance spectroscopy using a semi-continuous baker--campbell--hausdorff expansion},}\ }\href@noop {} {\bibfield  {journal} {\bibinfo  {journal} {The Journal of chemical physics}\ }\textbf {\bibinfo {volume} {109}},\ \bibinfo {pages} {3780--3791} (\bibinfo {year} {1998})}\BibitemShut {NoStop}%
\bibitem [{\citenamefont {Wei}\ \emph {et~al.}(2018)\citenamefont {Wei}, \citenamefont {Ramanathan},\ and\ \citenamefont {Cappellaro}}]{wei2018exploring}%
  \BibitemOpen
  \bibfield  {author} {\bibinfo {author} {\bibfnamefont {Ken~Xuan}\ \bibnamefont {Wei}}, \bibinfo {author} {\bibfnamefont {Chandrasekhar}\ \bibnamefont {Ramanathan}}, \ and\ \bibinfo {author} {\bibfnamefont {Paola}\ \bibnamefont {Cappellaro}},\ }\bibfield  {title} {\enquote {\bibinfo {title} {Exploring localization in nuclear spin chains},}\ }\href@noop {} {\bibfield  {journal} {\bibinfo  {journal} {Physical review letters}\ }\textbf {\bibinfo {volume} {120}},\ \bibinfo {pages} {070501} (\bibinfo {year} {2018})}\BibitemShut {NoStop}%
\bibitem [{\citenamefont {Guenneugues}\ \emph {et~al.}(1999)\citenamefont {Guenneugues}, \citenamefont {Berthault},\ and\ \citenamefont {Desvaux}}]{guenneugues1999method}%
  \BibitemOpen
  \bibfield  {author} {\bibinfo {author} {\bibfnamefont {Marc}\ \bibnamefont {Guenneugues}}, \bibinfo {author} {\bibfnamefont {Patrick}\ \bibnamefont {Berthault}}, \ and\ \bibinfo {author} {\bibfnamefont {Herv{\'e}}\ \bibnamefont {Desvaux}},\ }\bibfield  {title} {\enquote {\bibinfo {title} {A method for determiningb1field inhomogeneity. are the biases assumed in heteronuclear relaxation experiments usually underestimated?}}\ }\href@noop {} {\bibfield  {journal} {\bibinfo  {journal} {Journal of Magnetic Resonance}\ }\textbf {\bibinfo {volume} {136}},\ \bibinfo {pages} {118--126} (\bibinfo {year} {1999})}\BibitemShut {NoStop}%
\bibitem [{\citenamefont {Burum}\ \emph {et~al.}(1981{\natexlab{b}})\citenamefont {Burum}, \citenamefont {Linder},\ and\ \citenamefont {Ernst}}]{burum1981low}%
  \BibitemOpen
  \bibfield  {author} {\bibinfo {author} {\bibfnamefont {DP}~\bibnamefont {Burum}}, \bibinfo {author} {\bibfnamefont {M}~\bibnamefont {Linder}}, \ and\ \bibinfo {author} {\bibfnamefont {RR}~\bibnamefont {Ernst}},\ }\bibfield  {title} {\enquote {\bibinfo {title} {Low-power multipulse line narrowing in solid-state nmr},}\ }\href@noop {} {\bibfield  {journal} {\bibinfo  {journal} {Journal of Magnetic Resonance (1969)}\ }\textbf {\bibinfo {volume} {44}},\ \bibinfo {pages} {173--188} (\bibinfo {year} {1981}{\natexlab{b}})}\BibitemShut {NoStop}%
\bibitem [{\citenamefont {Hong}\ \emph {et~al.}(1996)\citenamefont {Hong}, \citenamefont {Pines},\ and\ \citenamefont {Caldarelli}}]{hong1996measurement}%
  \BibitemOpen
  \bibfield  {author} {\bibinfo {author} {\bibfnamefont {Mei}\ \bibnamefont {Hong}}, \bibinfo {author} {\bibfnamefont {Alexander}\ \bibnamefont {Pines}}, \ and\ \bibinfo {author} {\bibfnamefont {Stefano}\ \bibnamefont {Caldarelli}},\ }\bibfield  {title} {\enquote {\bibinfo {title} {Measurement and assignment of long-range c- h dipolar couplings in liquid crystals by two-dimensional nmr spectroscopy},}\ }\href@noop {} {\bibfield  {journal} {\bibinfo  {journal} {The Journal of Physical Chemistry}\ }\textbf {\bibinfo {volume} {100}},\ \bibinfo {pages} {14815--14822} (\bibinfo {year} {1996})}\BibitemShut {NoStop}%
\bibitem [{\citenamefont {Weitekamp}\ \emph {et~al.}(1982)\citenamefont {Weitekamp}, \citenamefont {Garbow},\ and\ \citenamefont {Pines}}]{weitekamp1982determination}%
  \BibitemOpen
  \bibfield  {author} {\bibinfo {author} {\bibfnamefont {DP}~\bibnamefont {Weitekamp}}, \bibinfo {author} {\bibfnamefont {JR}~\bibnamefont {Garbow}}, \ and\ \bibinfo {author} {\bibfnamefont {A}~\bibnamefont {Pines}},\ }\bibfield  {title} {\enquote {\bibinfo {title} {Determination of dipole coupling constants using heteronuclear multiple quantum nmr},}\ }\href@noop {} {\bibfield  {journal} {\bibinfo  {journal} {The Journal of Chemical Physics}\ }\textbf {\bibinfo {volume} {77}},\ \bibinfo {pages} {2870--2883} (\bibinfo {year} {1982})}\BibitemShut {NoStop}%
\bibitem [{\citenamefont {Schaefer}\ \emph {et~al.}(1983)\citenamefont {Schaefer}, \citenamefont {Peeling},\ and\ \citenamefont {Penner}}]{schaefer1983spin}%
  \BibitemOpen
  \bibfield  {author} {\bibinfo {author} {\bibfnamefont {Ted}\ \bibnamefont {Schaefer}}, \bibinfo {author} {\bibfnamefont {James}\ \bibnamefont {Peeling}}, \ and\ \bibinfo {author} {\bibfnamefont {Glenn~H}\ \bibnamefont {Penner}},\ }\bibfield  {title} {\enquote {\bibinfo {title} {The spin--spin coupling mechanism for 5 j (c-4, ch$\alpha$) in toluene derivatives and its conformational applications},}\ }\href@noop {} {\bibfield  {journal} {\bibinfo  {journal} {Canadian Journal of Chemistry}\ }\textbf {\bibinfo {volume} {61}},\ \bibinfo {pages} {2773--2776} (\bibinfo {year} {1983})}\BibitemShut {NoStop}%
\bibitem [{\citenamefont {Drobny}\ \emph {et~al.}(1979)\citenamefont {Drobny}, \citenamefont {Pines}, \citenamefont {Sinton}, \citenamefont {Weitekamp},\ and\ \citenamefont {Wemmer}}]{drobny1979faraday}%
  \BibitemOpen
  \bibfield  {author} {\bibinfo {author} {\bibfnamefont {G}~\bibnamefont {Drobny}}, \bibinfo {author} {\bibfnamefont {A}~\bibnamefont {Pines}}, \bibinfo {author} {\bibfnamefont {S}~\bibnamefont {Sinton}}, \bibinfo {author} {\bibfnamefont {D}~\bibnamefont {Weitekamp}}, \ and\ \bibinfo {author} {\bibfnamefont {D}~\bibnamefont {Wemmer}},\ }\bibfield  {title} {\enquote {\bibinfo {title} {Faraday div. c’hem},}\ }in\ \href@noop {} {\emph {\bibinfo {booktitle} {SW. Symp}}},\ Vol.~\bibinfo {volume} {13}\ (\bibinfo {year} {1979})\ p.~\bibinfo {pages} {49}\BibitemShut {NoStop}%
\bibitem [{\citenamefont {Kim}\ \emph {et~al.}(2016)\citenamefont {Kim}, \citenamefont {Thiessen}, \citenamefont {Bolton}, \citenamefont {Chen}, \citenamefont {Fu}, \citenamefont {Gindulyte}, \citenamefont {Han}, \citenamefont {He}, \citenamefont {He}, \citenamefont {Shoemaker} \emph {et~al.}}]{kim2016pubchem}%
  \BibitemOpen
  \bibfield  {author} {\bibinfo {author} {\bibfnamefont {Sunghwan}\ \bibnamefont {Kim}}, \bibinfo {author} {\bibfnamefont {Paul~A}\ \bibnamefont {Thiessen}}, \bibinfo {author} {\bibfnamefont {Evan~E}\ \bibnamefont {Bolton}}, \bibinfo {author} {\bibfnamefont {Jie}\ \bibnamefont {Chen}}, \bibinfo {author} {\bibfnamefont {Gang}\ \bibnamefont {Fu}}, \bibinfo {author} {\bibfnamefont {Asta}\ \bibnamefont {Gindulyte}}, \bibinfo {author} {\bibfnamefont {Lianyi}\ \bibnamefont {Han}}, \bibinfo {author} {\bibfnamefont {Jane}\ \bibnamefont {He}}, \bibinfo {author} {\bibfnamefont {Siqian}\ \bibnamefont {He}}, \bibinfo {author} {\bibfnamefont {Benjamin~A}\ \bibnamefont {Shoemaker}},  \emph {et~al.},\ }\bibfield  {title} {\enquote {\bibinfo {title} {Pubchem substance and compound databases},}\ }\href@noop {} {\bibfield  {journal} {\bibinfo  {journal} {Nucleic acids research}\ }\textbf {\bibinfo {volume} {44}},\ \bibinfo {pages} {D1202--D1213} (\bibinfo {year} {2016})}\BibitemShut {NoStop}%
\bibitem [{\citenamefont {Bengs}\ \emph {et~al.}(2025)\citenamefont {Bengs}, \citenamefont {Zhang},\ and\ \citenamefont {Ajoy}}]{bengs2025fundamental}%
  \BibitemOpen
  \bibfield  {author} {\bibinfo {author} {\bibfnamefont {Christian}\ \bibnamefont {Bengs}}, \bibinfo {author} {\bibfnamefont {Chongwei}\ \bibnamefont {Zhang}}, \ and\ \bibinfo {author} {\bibfnamefont {Ashok}\ \bibnamefont {Ajoy}},\ }\bibfield  {title} {\enquote {\bibinfo {title} {Fundamental bounds on many-body spin cluster intensities},}\ }\href@noop {} {\bibfield  {journal} {\bibinfo  {journal} {The Journal of Chemical Physics}\ }\textbf {\bibinfo {volume} {162}} (\bibinfo {year} {2025})}\BibitemShut {NoStop}%
\bibitem [{\citenamefont {Sinton}\ \emph {et~al.}(1984)\citenamefont {Sinton}, \citenamefont {Zax}, \citenamefont {Murdoch},\ and\ \citenamefont {Pines}}]{sinton1984multiple}%
  \BibitemOpen
  \bibfield  {author} {\bibinfo {author} {\bibfnamefont {SW}~\bibnamefont {Sinton}}, \bibinfo {author} {\bibfnamefont {DB}~\bibnamefont {Zax}}, \bibinfo {author} {\bibfnamefont {JB}~\bibnamefont {Murdoch}}, \ and\ \bibinfo {author} {\bibfnamefont {A}~\bibnamefont {Pines}},\ }\bibfield  {title} {\enquote {\bibinfo {title} {Multiple-quantum nmr study of molecular structure and ordering in a liquid crystal},}\ }\href@noop {} {\bibfield  {journal} {\bibinfo  {journal} {Molecular Physics}\ }\textbf {\bibinfo {volume} {53}},\ \bibinfo {pages} {333--362} (\bibinfo {year} {1984})}\BibitemShut {NoStop}%
\bibitem [{\citenamefont {Zhou}\ \emph {et~al.}(2020)\citenamefont {Zhou}, \citenamefont {Xu}, \citenamefont {Chen}, \citenamefont {Guo},\ and\ \citenamefont {Swingle}}]{zhou20operator}%
  \BibitemOpen
  \bibfield  {author} {\bibinfo {author} {\bibfnamefont {Tianci}\ \bibnamefont {Zhou}}, \bibinfo {author} {\bibfnamefont {Shenglong}\ \bibnamefont {Xu}}, \bibinfo {author} {\bibfnamefont {Xiao}\ \bibnamefont {Chen}}, \bibinfo {author} {\bibfnamefont {Andrew}\ \bibnamefont {Guo}}, \ and\ \bibinfo {author} {\bibfnamefont {Brian}\ \bibnamefont {Swingle}},\ }\bibfield  {title} {\enquote {\bibinfo {title} {The operator l\'{e}vy flight: light cones in chaotic long-range interacting systems},}\ }\href {https://arxiv.org/abs/1909.08646} {\bibfield  {journal} {\bibinfo  {journal} {Phys. Rev. Lett.}\ }\textbf {\bibinfo {volume} {124}},\ \bibinfo {pages} {180601} (\bibinfo {year} {2020})}\BibitemShut {NoStop}%
\bibitem [{\citenamefont {Pizzirusso}\ \emph {et~al.}(2014)\citenamefont {Pizzirusso}, \citenamefont {Di~Pietro}, \citenamefont {De~Luca}, \citenamefont {Celebre}, \citenamefont {Longeri}, \citenamefont {Muccioli},\ and\ \citenamefont {Zannoni}}]{pizzirusso2014order}%
  \BibitemOpen
  \bibfield  {author} {\bibinfo {author} {\bibfnamefont {Antonio}\ \bibnamefont {Pizzirusso}}, \bibinfo {author} {\bibfnamefont {Maria~Enrica}\ \bibnamefont {Di~Pietro}}, \bibinfo {author} {\bibfnamefont {Giuseppina}\ \bibnamefont {De~Luca}}, \bibinfo {author} {\bibfnamefont {Giorgio}\ \bibnamefont {Celebre}}, \bibinfo {author} {\bibfnamefont {Marcello}\ \bibnamefont {Longeri}}, \bibinfo {author} {\bibfnamefont {Luca}\ \bibnamefont {Muccioli}}, \ and\ \bibinfo {author} {\bibfnamefont {Claudio}\ \bibnamefont {Zannoni}},\ }\bibfield  {title} {\enquote {\bibinfo {title} {Order and conformation of biphenyl in cyanobiphenyl liquid crystals: a combined atomistic molecular dynamics and 1h nmr study},}\ }\href@noop {} {\bibfield  {journal} {\bibinfo  {journal} {ChemPhysChem}\ }\textbf {\bibinfo {volume} {15}},\ \bibinfo {pages} {1356--1367} (\bibinfo {year} {2014})}\BibitemShut {NoStop}%
\bibitem [{\citenamefont {Guo}\ and\ \citenamefont {Fung}(1991)}]{guo1991determination}%
  \BibitemOpen
  \bibfield  {author} {\bibinfo {author} {\bibfnamefont {Wen}\ \bibnamefont {Guo}}\ and\ \bibinfo {author} {\bibfnamefont {BM}~\bibnamefont {Fung}},\ }\bibfield  {title} {\enquote {\bibinfo {title} {Determination of the order parameters of liquid crystals from carbon-13 chemical shifts},}\ }\href@noop {} {\bibfield  {journal} {\bibinfo  {journal} {The Journal of chemical physics}\ }\textbf {\bibinfo {volume} {95}},\ \bibinfo {pages} {3917--3923} (\bibinfo {year} {1991})}\BibitemShut {NoStop}%
\bibitem [{\citenamefont {Sandstr{\"o}m}\ \emph {et~al.}(1996)\citenamefont {Sandstr{\"o}m}, \citenamefont {Komolkin},\ and\ \citenamefont {Maliniak}}]{sandstrom1996orientational}%
  \BibitemOpen
  \bibfield  {author} {\bibinfo {author} {\bibfnamefont {Dick}\ \bibnamefont {Sandstr{\"o}m}}, \bibinfo {author} {\bibfnamefont {Andrei~V}\ \bibnamefont {Komolkin}}, \ and\ \bibinfo {author} {\bibfnamefont {Arnold}\ \bibnamefont {Maliniak}},\ }\bibfield  {title} {\enquote {\bibinfo {title} {Orientational order in a liquid crystalline mixture studied by molecular dynamics simulation and nmr},}\ }\href@noop {} {\bibfield  {journal} {\bibinfo  {journal} {The Journal of chemical physics}\ }\textbf {\bibinfo {volume} {104}},\ \bibinfo {pages} {9620--9628} (\bibinfo {year} {1996})}\BibitemShut {NoStop}%
\bibitem [{\citenamefont {Nogueira}(2014--)}]{bayesoptpython}%
  \BibitemOpen
  \bibfield  {author} {\bibinfo {author} {\bibfnamefont {Fernando}\ \bibnamefont {Nogueira}},\ }\href {https://github.com/bayesian-optimization/BayesianOptimization} {\enquote {\bibinfo {title} {{Bayesian Optimization}: Open source constrained global optimization tool for {Python}},}\ } (\bibinfo {year} {2014--})\BibitemShut {NoStop}%
\bibitem [{\citenamefont {H\'emery}\ \emph {et~al.}(2019)\citenamefont {H\'emery}, \citenamefont {Pollmann},\ and\ \citenamefont {Luitz}}]{PhysRevB.100.104303}%
  \BibitemOpen
  \bibfield  {author} {\bibinfo {author} {\bibfnamefont {K\'evin}\ \bibnamefont {H\'emery}}, \bibinfo {author} {\bibfnamefont {Frank}\ \bibnamefont {Pollmann}}, \ and\ \bibinfo {author} {\bibfnamefont {David~J.}\ \bibnamefont {Luitz}},\ }\bibfield  {title} {\enquote {\bibinfo {title} {Matrix product states approaches to operator spreading in ergodic quantum systems},}\ }\href {\doibase 10.1103/PhysRevB.100.104303} {\bibfield  {journal} {\bibinfo  {journal} {Phys. Rev. B}\ }\textbf {\bibinfo {volume} {100}},\ \bibinfo {pages} {104303} (\bibinfo {year} {2019})}\BibitemShut {NoStop}%
\bibitem [{\citenamefont {Xu}\ and\ \citenamefont {Swingle}(2020)}]{xu2020accessing}%
  \BibitemOpen
  \bibfield  {author} {\bibinfo {author} {\bibfnamefont {Shenglong}\ \bibnamefont {Xu}}\ and\ \bibinfo {author} {\bibfnamefont {Brian}\ \bibnamefont {Swingle}},\ }\bibfield  {title} {\enquote {\bibinfo {title} {Accessing scrambling using matrix product operators},}\ }\href@noop {} {\bibfield  {journal} {\bibinfo  {journal} {Nature Physics}\ }\textbf {\bibinfo {volume} {16}},\ \bibinfo {pages} {199--204} (\bibinfo {year} {2020})}\BibitemShut {NoStop}%
\bibitem [{\citenamefont {Lopez-Piqueres}\ \emph {et~al.}(2021)\citenamefont {Lopez-Piqueres}, \citenamefont {Ware}, \citenamefont {Gopalakrishnan},\ and\ \citenamefont {Vasseur}}]{PhysRevB.104.104307}%
  \BibitemOpen
  \bibfield  {author} {\bibinfo {author} {\bibfnamefont {Javier}\ \bibnamefont {Lopez-Piqueres}}, \bibinfo {author} {\bibfnamefont {Brayden}\ \bibnamefont {Ware}}, \bibinfo {author} {\bibfnamefont {Sarang}\ \bibnamefont {Gopalakrishnan}}, \ and\ \bibinfo {author} {\bibfnamefont {Romain}\ \bibnamefont {Vasseur}},\ }\bibfield  {title} {\enquote {\bibinfo {title} {Operator front broadening in chaotic and integrable quantum chains},}\ }\href {\doibase 10.1103/PhysRevB.104.104307} {\bibfield  {journal} {\bibinfo  {journal} {Phys. Rev. B}\ }\textbf {\bibinfo {volume} {104}},\ \bibinfo {pages} {104307} (\bibinfo {year} {2021})}\BibitemShut {NoStop}%
\bibitem [{\citenamefont {Xu}\ and\ \citenamefont {Swingle}(2024)}]{PRXQuantum.5.010201}%
  \BibitemOpen
  \bibfield  {author} {\bibinfo {author} {\bibfnamefont {Shenglong}\ \bibnamefont {Xu}}\ and\ \bibinfo {author} {\bibfnamefont {Brian}\ \bibnamefont {Swingle}},\ }\bibfield  {title} {\enquote {\bibinfo {title} {Scrambling dynamics and out-of-time-ordered correlators in quantum many-body systems},}\ }\href {\doibase 10.1103/PRXQuantum.5.010201} {\bibfield  {journal} {\bibinfo  {journal} {PRX Quantum}\ }\textbf {\bibinfo {volume} {5}},\ \bibinfo {pages} {010201} (\bibinfo {year} {2024})}\BibitemShut {NoStop}%
\bibitem [{\citenamefont {Zaletel}\ \emph {et~al.}(2015)\citenamefont {Zaletel}, \citenamefont {Mong}, \citenamefont {Karrasch}, \citenamefont {Moore},\ and\ \citenamefont {Pollmann}}]{PhysRevB.91.165112}%
  \BibitemOpen
  \bibfield  {author} {\bibinfo {author} {\bibfnamefont {Michael~P.}\ \bibnamefont {Zaletel}}, \bibinfo {author} {\bibfnamefont {Roger S.~K.}\ \bibnamefont {Mong}}, \bibinfo {author} {\bibfnamefont {Christoph}\ \bibnamefont {Karrasch}}, \bibinfo {author} {\bibfnamefont {Joel~E.}\ \bibnamefont {Moore}}, \ and\ \bibinfo {author} {\bibfnamefont {Frank}\ \bibnamefont {Pollmann}},\ }\bibfield  {title} {\enquote {\bibinfo {title} {Time-evolving a matrix product state with long-ranged interactions},}\ }\href {\doibase 10.1103/PhysRevB.91.165112} {\bibfield  {journal} {\bibinfo  {journal} {Phys. Rev. B}\ }\textbf {\bibinfo {volume} {91}},\ \bibinfo {pages} {165112} (\bibinfo {year} {2015})}\BibitemShut {NoStop}%
\bibitem [{\citenamefont {Parker}\ \emph {et~al.}(2020)\citenamefont {Parker}, \citenamefont {Cao},\ and\ \citenamefont {Zaletel}}]{PhysRevB.102.035147}%
  \BibitemOpen
  \bibfield  {author} {\bibinfo {author} {\bibfnamefont {Daniel~E.}\ \bibnamefont {Parker}}, \bibinfo {author} {\bibfnamefont {Xiangyu}\ \bibnamefont {Cao}}, \ and\ \bibinfo {author} {\bibfnamefont {Michael~P.}\ \bibnamefont {Zaletel}},\ }\bibfield  {title} {\enquote {\bibinfo {title} {Local matrix product operators: Canonical form, compression, and control theory},}\ }\href {\doibase 10.1103/PhysRevB.102.035147} {\bibfield  {journal} {\bibinfo  {journal} {Phys. Rev. B}\ }\textbf {\bibinfo {volume} {102}},\ \bibinfo {pages} {035147} (\bibinfo {year} {2020})}\BibitemShut {NoStop}%
\bibitem [{\citenamefont {Arute}\ \emph {et~al.}(2019)\citenamefont {Arute}, \citenamefont {Arya}, \citenamefont {Babbush}, \citenamefont {Bacon}, \citenamefont {Bardin}, \citenamefont {Barends}, \citenamefont {Biswas}, \citenamefont {Boixo}, \citenamefont {Brandao} \emph {et~al.}}]{Arute2019Supremacy}%
  \BibitemOpen
  \bibfield  {author} {\bibinfo {author} {\bibfnamefont {Frank}\ \bibnamefont {Arute}}, \bibinfo {author} {\bibfnamefont {Kunal}\ \bibnamefont {Arya}}, \bibinfo {author} {\bibfnamefont {Ryan}\ \bibnamefont {Babbush}}, \bibinfo {author} {\bibfnamefont {Dave}\ \bibnamefont {Bacon}}, \bibinfo {author} {\bibfnamefont {Joseph~C.}\ \bibnamefont {Bardin}}, \bibinfo {author} {\bibfnamefont {Rami}\ \bibnamefont {Barends}}, \bibinfo {author} {\bibfnamefont {Rupak}\ \bibnamefont {Biswas}}, \bibinfo {author} {\bibfnamefont {Sergio}\ \bibnamefont {Boixo}}, \bibinfo {author} {\bibfnamefont {Fernando G. S.~L.}\ \bibnamefont {Brandao}},  \emph {et~al.},\ }\bibfield  {title} {\enquote {\bibinfo {title} {Quantum supremacy using a programmable superconducting processor},}\ }\href {\doibase 10.1038/s41586-019-1666-5} {\bibfield  {journal} {\bibinfo  {journal} {Nature}\ }\textbf {\bibinfo {volume} {574}},\ \bibinfo {pages} {505--510} (\bibinfo {year} {2019})}\BibitemShut {NoStop}%
\bibitem [{\citenamefont {Rosenberg}\ \emph {et~al.}(2024)\citenamefont {Rosenberg}, \citenamefont {Andersen}, \citenamefont {Samajdar}, \citenamefont {Petukhov}, \citenamefont {Hoke}, \citenamefont {Abanin}, \citenamefont {Bengtsson}, \citenamefont {Drozdov}, \citenamefont {Erickson}, \citenamefont {Klimov} \emph {et~al.}}]{rosenberg2024dynamics}%
  \BibitemOpen
  \bibfield  {author} {\bibinfo {author} {\bibfnamefont {Eliott}\ \bibnamefont {Rosenberg}}, \bibinfo {author} {\bibfnamefont {TI}~\bibnamefont {Andersen}}, \bibinfo {author} {\bibfnamefont {Rhine}\ \bibnamefont {Samajdar}}, \bibinfo {author} {\bibfnamefont {Andre}\ \bibnamefont {Petukhov}}, \bibinfo {author} {\bibfnamefont {JC}~\bibnamefont {Hoke}}, \bibinfo {author} {\bibfnamefont {Dmitry}\ \bibnamefont {Abanin}}, \bibinfo {author} {\bibfnamefont {Andreas}\ \bibnamefont {Bengtsson}}, \bibinfo {author} {\bibfnamefont {IK}~\bibnamefont {Drozdov}}, \bibinfo {author} {\bibfnamefont {Catherine}\ \bibnamefont {Erickson}}, \bibinfo {author} {\bibfnamefont {PV}~\bibnamefont {Klimov}},  \emph {et~al.},\ }\bibfield  {title} {\enquote {\bibinfo {title} {Dynamics of magnetization at infinite temperature in a heisenberg spin chain},}\ }\href@noop {} {\bibfield  {journal} {\bibinfo  {journal} {Science}\ }\textbf {\bibinfo {volume} {384}},\ \bibinfo {pages} {48--53} (\bibinfo {year} {2024})}\BibitemShut {NoStop}%
\bibitem [{\citenamefont {Isakov}\ \emph {et~al.}(2021)\citenamefont {Isakov}, \citenamefont {Kafri}, \citenamefont {Martin}, \citenamefont {Heidweiller}, \citenamefont {Mruczkiewicz}, \citenamefont {Harrigan}, \citenamefont {Rubin}, \citenamefont {Thomson}, \citenamefont {Broughton}, \citenamefont {Kissell}, \citenamefont {Peters}, \citenamefont {Gustafson}, \citenamefont {Li}, \citenamefont {Lamm}, \citenamefont {Perdue}, \citenamefont {Ho}, \citenamefont {Strain},\ and\ \citenamefont {Boixo}}]{isakov2021simulationsquantumcircuitsapproximate}%
  \BibitemOpen
  \bibfield  {author} {\bibinfo {author} {\bibfnamefont {Sergei~V.}\ \bibnamefont {Isakov}}, \bibinfo {author} {\bibfnamefont {Dvir}\ \bibnamefont {Kafri}}, \bibinfo {author} {\bibfnamefont {Orion}\ \bibnamefont {Martin}}, \bibinfo {author} {\bibfnamefont {Catherine~Vollgraff}\ \bibnamefont {Heidweiller}}, \bibinfo {author} {\bibfnamefont {Wojciech}\ \bibnamefont {Mruczkiewicz}}, \bibinfo {author} {\bibfnamefont {Matthew~P.}\ \bibnamefont {Harrigan}}, \bibinfo {author} {\bibfnamefont {Nicholas~C.}\ \bibnamefont {Rubin}}, \bibinfo {author} {\bibfnamefont {Ross}\ \bibnamefont {Thomson}}, \bibinfo {author} {\bibfnamefont {Michael}\ \bibnamefont {Broughton}}, \bibinfo {author} {\bibfnamefont {Kevin}\ \bibnamefont {Kissell}}, \bibinfo {author} {\bibfnamefont {Evan}\ \bibnamefont {Peters}}, \bibinfo {author} {\bibfnamefont {Erik}\ \bibnamefont {Gustafson}}, \bibinfo {author} {\bibfnamefont {Andy C.~Y.}\ \bibnamefont {Li}}, \bibinfo {author} {\bibfnamefont {Henry}\ \bibnamefont {Lamm}}, \bibinfo {author}
  {\bibfnamefont {Gabriel}\ \bibnamefont {Perdue}}, \bibinfo {author} {\bibfnamefont {Alan~K.}\ \bibnamefont {Ho}}, \bibinfo {author} {\bibfnamefont {Doug}\ \bibnamefont {Strain}}, \ and\ \bibinfo {author} {\bibfnamefont {Sergio}\ \bibnamefont {Boixo}},\ }\href {https://arxiv.org/abs/2111.02396} {\enquote {\bibinfo {title} {Simulations of quantum circuits with approximate noise using qsim and cirq},}\ } (\bibinfo {year} {2021}),\ \Eprint {http://arxiv.org/abs/2111.02396} {arXiv:2111.02396 [quant-ph]} \BibitemShut {NoStop}%
\end{thebibliography}
